%% file: article.tex
\documentclass[aps,a4paper,12pt,tightenlines,nofootinbib,longbibliography,notitlepage]{revtex4-1}

\usepackage{mathptmx}
\usepackage[utf8]{inputenc}
\usepackage[english]{babel}
\usepackage{latexsym}
\usepackage{amsmath}
\usepackage{amsfonts}
\usepackage{amssymb}
\usepackage{graphicx}

\usepackage{lipsum}
\usepackage{colortbl}
\usepackage{xcolor}
\usepackage{booktabs}
\usepackage{rotating}
\usepackage{relsize}
\usepackage{tabularx}
\usepackage{overpic}

\usepackage{hyperref}
\usepackage{framed}
\usepackage[section]{placeins}

\usepackage{setspace}
\usepackage[multiple]{footmisc}
\usepackage{footnote}

\usepackage{longtable}
\usepackage{multirow}

\usepackage{tikz}
\usetikzlibrary{backgrounds}
\usepackage{pgfplots}

\newcommand{\includegraphicslp}[4]{
  \begin{tikzpicture}
    \node[anchor=south west,inner sep=0] (image) at (0,0) {\includegraphics[#3]{#4}};
    \begin{scope}[x={(image.south east)},y={(image.north west)}]
      \node[anchor=south west,inner sep=0] at (0,#2) {#1};
    \end{scope}

  \end{tikzpicture}
}

\newcommand{\includegraphicsl}[3]{\includegraphicslp{#1}{.95}{#2}{#3}}

\newcommand{\avg}[1]{{\left<#1\right>}}

\newcommand{\dd}{\mathrm{d}}

\newcommand{\A}{\bm{A}}
\newcommand{\bb}{\bm{b}}
\newcommand{\e}{\bm{e}}

\newcommand{\ee}{\mathrm{e}}
\usepackage{mathtools}
\usepackage{bm}\def\multiset#1#2{\ensuremath{\left(\kern-.3em\left(\genfrac{}{}{0pt}{}{#1}{#2}\right)\kern-.3em\right)}}

\renewcommand{\index}[2]{}

\begin{document}

\title{Descriptive vs. inferential community detection in networks: pitfalls, myths, and half-truths}
\author{Tiago P. Peixoto}
\email{peixotot@ceu.edu}
\affiliation{Department of Network and Data Science, Central European University, Vienna, Austria}

\begin{abstract}

  Community detection is one of the most important methodological fields
  of network science, and one which has attracted a significant amount
  of attention over the past decades. This area deals with the automated
  division of a network into fundamental building blocks, with the
  objective of providing a summary of its large-scale structure.
  Despite its importance and widespread adoption, there is a noticeable
  gap between what is arguably the state-of-the-art and the methods that
  are actually used in practice in a variety of fields. Here we attempt
  to address this discrepancy by dividing existing methods according to
  whether they have a ``descriptive'' or an ``inferential'' goal. While
  descriptive methods find patterns in networks based on
  context-dependent notions of community structure, inferential methods
  articulate generative models, and attempt to fit them to data. In this
  way, they are able to provide insights into the mechanisms of network
  formation, and separate structure from randomness in a manner
  supported by statistical evidence. We review how employing descriptive
  methods with inferential aims is riddled with pitfalls and misleading
  answers, and thus should be in general avoided. We argue that
  inferential methods are more typically aligned with clearer scientific
  questions, yield more robust results, and should be in many cases
  preferred. We attempt to dispel some myths and half-truths often
  believed when community detection is employed in practice, in an
  effort to improve both the use of such methods as well as the
  interpretation of their results.
\end{abstract}

\maketitle
\newpage
\tableofcontents

\makeatletter
\let\toc@pre\relax
\let\toc@post\relax
\makeatother

\newpage

\input{main-text.tex}

\end{document}

%% file: main-text.tex
\section{Introduction}

Community detection is the task of dividing a network --- typically one
which is large --- into many smaller groups of nodes that have a similar
contribution to the overall network structure. With such a division, we
can better summarize the large-scale structure of a network by
describing how these groups are connected, instead of each individual
node. This simplified description can be used to digest an otherwise
intractable representation of a large system, providing insight into its
most important patterns, how they relate to its function, and the
underlying mechanisms responsible for its formation.

Because of its important role in network science, community detection
has attracted substantial attention from researchers, specially in the
last 20 years, culminating in an abundant literature (see
Refs.~\cite{fortunato_community_2010,fortunato_community_2016} for a
review). This field has developed significantly from its early days,
specially over the last 10 years, during which the focus has been
shifting towards methods that are based on statistical inference (see
e.g. Refs.~\cite{moore_computer_2017,abbe_community_2017,peixoto_bayesian_2019}).

Despite this shift in the state-of-the-art, there remains a significant
gap between the best practices and the adopted practices in the use of
community detection for the analysis of network data. It is still the
case that some of the earliest methods proposed remain in widespread
use, despite their many serious shortcomings that have been uncovered
over the years. Most of these problems have been addressed with more
recent methods, that also contributed to a much deeper theoretical
understanding of the problem of community
detection~\cite{decelle_asymptotic_2011,zdeborova_statistical_2016,moore_computer_2017,abbe_community_2017}.

Nevertheless, some misconceptions remain and are still promoted. Here we
address some of the more salient ones, in an effort to dispel
them. These misconceptions are not uniformly shared; and those that pay
close attention to the literature will likely find few surprises
here. However, it is possible that many researchers employing community
detection are simply unaware of the issues with the methods being used.
Perhaps even more commonly, there are those that are in fact aware of
them, but not of their actual solutions, or the fact that some supposed
countermeasures are ineffective.

Throughout the following we will avoid providing ``black box'' recipes to be
followed uncritically, and instead try as much as possible to frame the issues within
a theoretical framework, such that the criticisms and solutions can be
justified in a principled manner.

We will set the stage by making a fundamental distinction between
``descriptive'' and ``inferential'' community detection approaches. As
others have emphasized before~\cite{schaub_many_2017}, community
detection can be performed with many goals in mind, and this will
dictate which methods are most appropriate. We will provide a simple
``litmus test'' that can be used to determine which overall approach is
more adequate, based on whether our goal is to seek inferential
interpretations.  We will then move to a more focused critique of the
method that is arguably the most widely employed
--- modularity maximization. This method has an emblematic character,
since it contains all possible pitfalls of using descriptive methods for
inferential aims. We will then follow with a discussion of myths,
pitfalls, and half-truths that obstruct a more effective analysis of
community structure in networks.

(We will not give a throughout technical introduction to inferential
community detection methods, which can be obtained instead in
Ref.~\cite{peixoto_bayesian_2019}. For a practical guide on how to use
various inferential methods, readers are referred to the detailed
HOWTO\footnote{Available at
\url{https://graph-tool.skewed.de/static/doc/demos/inference/inference.html}.}
available as part of the \texttt{graph-tool} Python
library~\cite{peixoto_graph-tool_2014}.)

\section{Descriptive vs. inferential community detection}

At a very fundamental level, community detection methods can be divided
into two main categories: ``descriptive'' and ``inferential.''

\textbf{Descriptive methods} attempt to find communities according to
some context-dependent notion of a good division of the network into
groups.  These notions are based on the patterns that can be identified
in the network via an exhaustive algorithm, but without taking into
consideration the possible rules that were used to create the uncovered patterns. These
patterns are used only to \emph{describe} the network, not to explain
it. Usually, these approaches do not articulate precisely what
constitutes community structure to begin with, and focus instead only on
how to detect it.  For this kind of method, concepts of statistical
significance, parsimony and generalizability are usually not evoked.

\textbf{Inferential methods}, on the other hand, start with an explicit
definition of what constitutes community structure, via a generative
model for the network. This model describes how a \emph{latent}
(i.e. not observed) partition of the nodes would affect the placement of
the edges. The inference consists on reversing this procedure to
determine which node partitions are more likely to have been responsible
for the observed network. The result of this is a ``fit'' of a model to
data, that can be used as a tentative explanation of how the network came to
be. The concepts of statistical significance, parsimony and
generalizability arise naturally and can be quantitatively assessed in
this context.

Descriptive community detection methods are by far the most numerous,
and those that are in most widespread use. However, this contrasts with
the current state-of-the-art, which is composed in large part of
inferential approaches. Here we point out the major differences between
them and discuss how to decide which is more appropriate, and also why
one should in general favor the inferential varieties whenever the
objective is to derive generative interpretations from data.

\subsection{Describing vs. explaining}

We begin by observing that descriptive clustering approaches are the
methods of choice in certain contexts. For instance, such approaches
arise naturally when the objective is to divide a network into two or
more parts as a means to solve a variety of optimization
problems. Arguably, the most classic example of this is the design of
Very Large Scale Integrated (VLSI) circuits~\cite{baker_cmos_2010}. The
task is to combine from up to billions of transistors into a single
physical microprocessor chip. Transistors that connect to each other
must be placed together to take less space, consume less power, reduce
latency, and reduce the risk of cross-talk with other nearby
connections. To achieve this, the initial stage of a VLSI process
involves the partitioning of the circuit into many smaller modules with
few connections between them, in a manner that enables their efficient
spatial placement, i.e. by positioning the transistors in each module
close together and those in different modules farther apart.

Another notable example is parallel task scheduling, a problem that
appears in computer science and operations research. The objective is to
distribute processes (i.e. programs, or tasks in general) between
different processors, so they can run at the same time.  Since processes
depend on the partial results of other processes, this forms a
dependency network, which then needs to be divided such that the number
of dependencies across processors is minimized. The optimal division is
the one where all tasks are able to finish in the shortest time
possible.

Both examples above, and others, have motivated a large literature on
``graph partitioning'' dating back to the
70s~\cite{kernighan_graph_1969,kernighan_efficient_1970,bichot_graph_2013},
which covers a family of problems that play an important role in
computer science and algorithmic complexity theory.

Although reminiscent of graph partitioning, and sharing with it many
algorithmic similarities, community detection is used more broadly with
a different
goal~\cite{fortunato_community_2010,fortunato_community_2016}. Namely,
the objective is to perform \emph{data analysis}, where one wants to
extract scientific understanding from empirical observations. The
communities identified are usually directly used for representation
and/or interpretation of the data, rather than as a mere device to solve
a particular optimization problem. In this context, a merely descriptive
approach will fail at giving us a meaningful insight into the data, and
can be misleading, as we will discuss in the following.

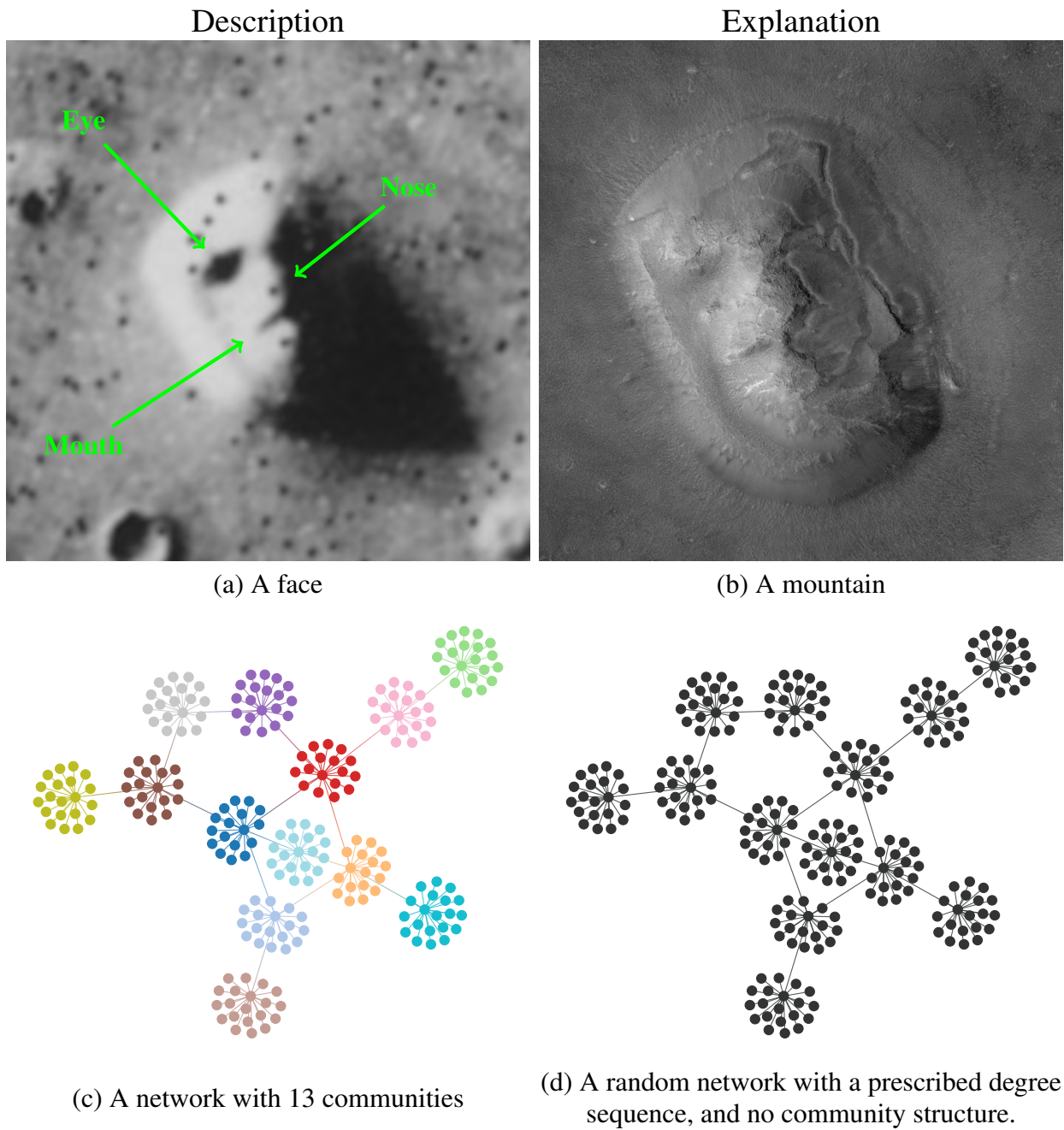
\begin{figure}[t!]
  \begin{tabular}{cc}
    {\larger Description} & {\larger Explanation} \\
    \begin{tikzpicture}
      \node[anchor=south west,inner sep=0] (image) at (0,0) {\includegraphics[width=.49\textwidth]{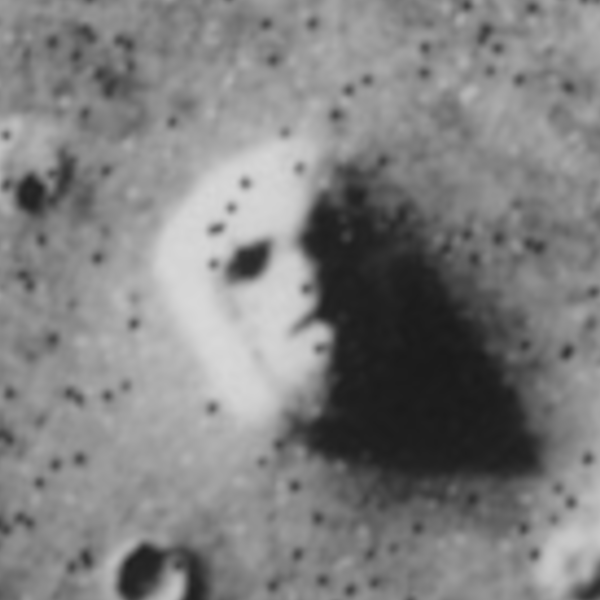}};
      \node[green] (eyes) at (3.2,4.7){};
      \node[green] (eyest) at (1.2,6.8){\textbf{Eye}};
      \draw[green,ultra thick,-to] (eyest) -> (eyes);

      \node[green] (nose) at (4.3,4.3){};
      \node[green] (noset) at (6.2,5.8){\textbf{Nose}};
      \draw[green,ultra thick,-to] (noset) -> (nose);

      \node[green] (mouth) at (3.8,3.5){};
      \node[green] (moutht) at (1.2,1.8){\textbf{Mouth}};
      \draw[green,ultra thick,-to] (moutht) -> (mouth);

    \end{tikzpicture}
      &
    \includegraphics[width=.49\textwidth]{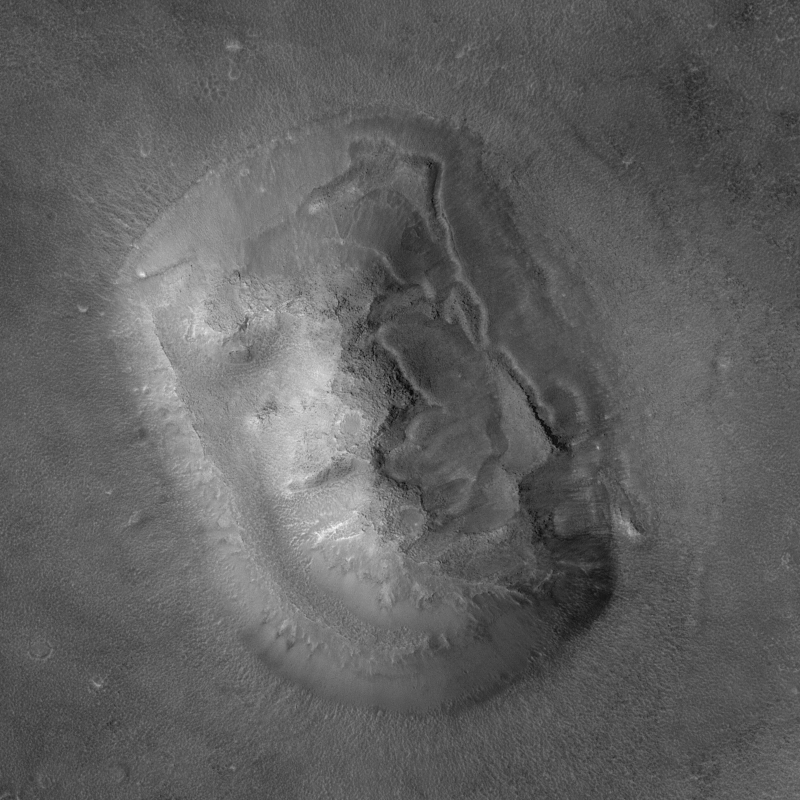}\\
    (a) A face & (b) A mountain\\

    \includegraphics[width=.49\textwidth]{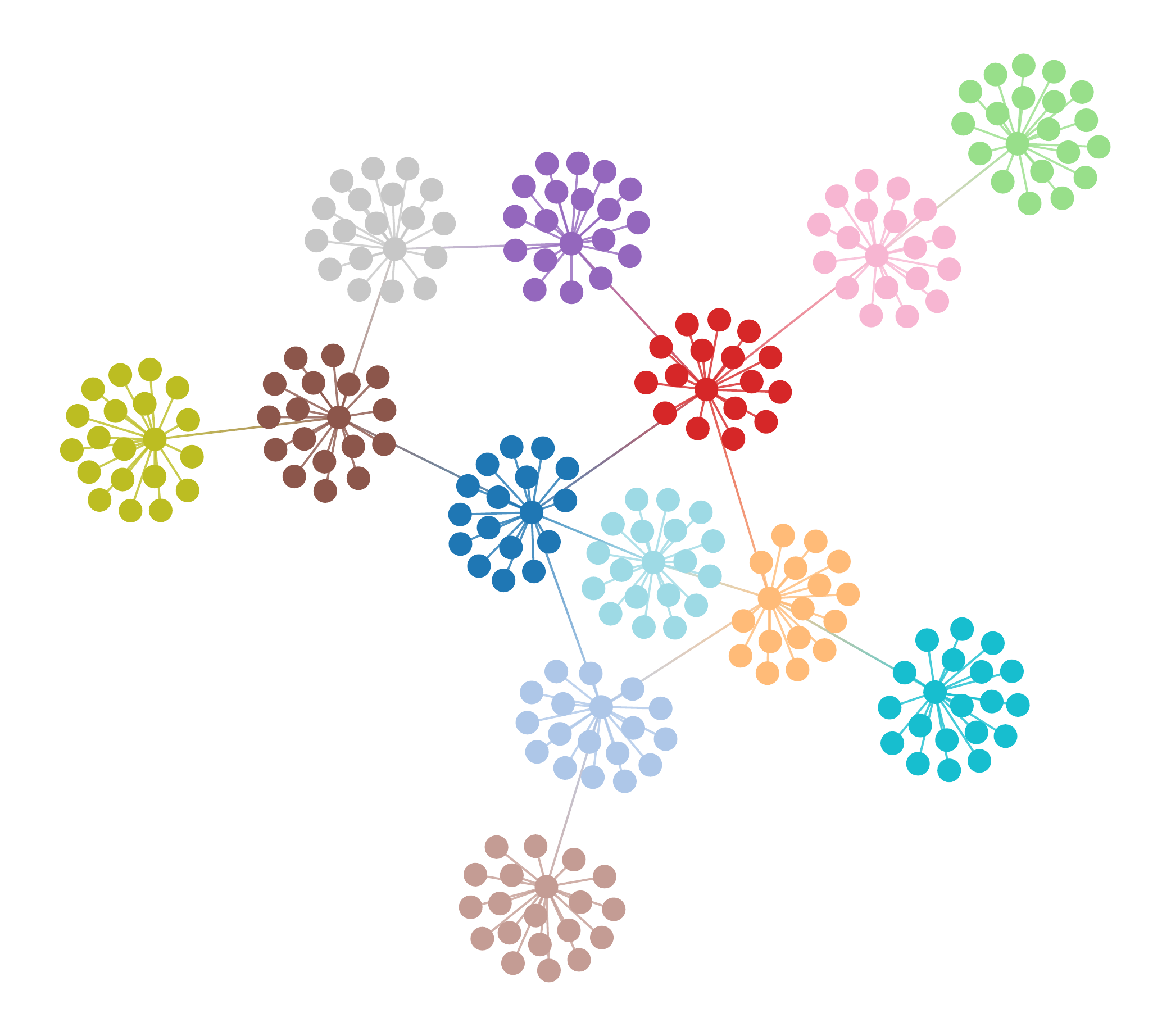} &
    \includegraphics[width=.49\textwidth]{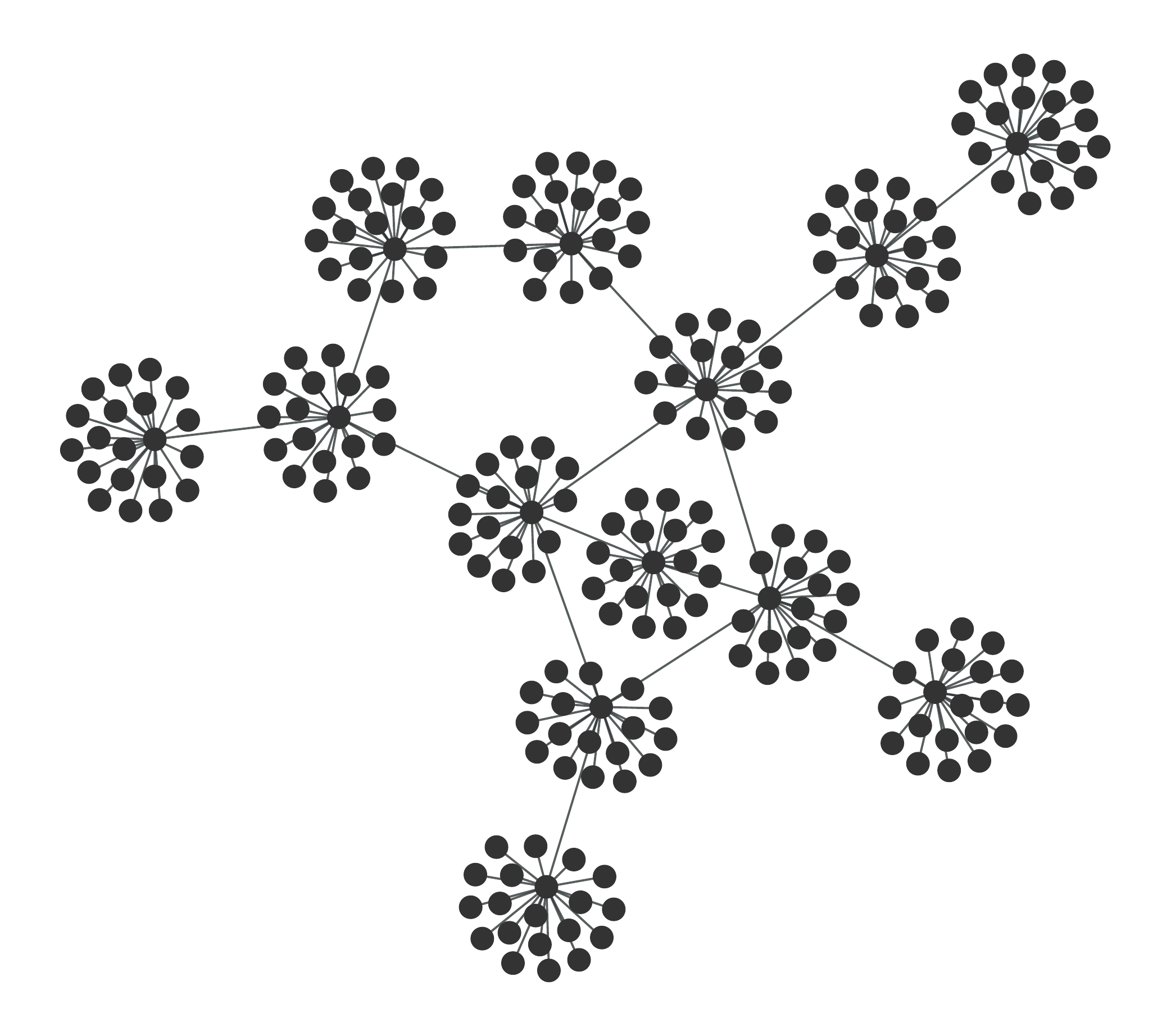} \\
    (c) A network with 13 communities &
    \begin{minipage}{.49\textwidth}
    (d)  A random network with a prescribed degree sequence, and no
      community structure.
    \end{minipage}
  \end{tabular}

  \caption{Difference between descriptive and inferential approaches to
  data analysis. As an analogy, in panels (a) and (b) we see two
  representations of the \emph{Cydonia Mensae} region on Mars. On panel
  (a) top left is a descriptive account of what we see in the picture,
  namely a face. On panel (b) is an inferential of representation of
  what lies behind it, namely a mountain. (We show a more recent image
  of the same region with a higher resolution to represent an
  inferential interpretation of the figure on the left.) More concretely
  for the problem of community detection, on panels (c) and (d) we see
  two representations of the same network. On panel (c) see a
  descriptive division into 13 assortative communities. On panel (d) we
  see an inferential representation as a degree-constrained random
  network, with no communities, since this is a more likely model of how
  this network was formed (see
  Fig.~\ref{fig:descriptive}). \label{fig:infvsdesc}}
\end{figure}

We illustrate the difference between descriptive and inferential
approaches in Fig.~\ref{fig:infvsdesc}. We first make an analogy with
the famous ``face'' seen on images of the \emph{Cydonia Mensae} region
of the planet Mars. A merely descriptive account of the image can be
made by identifying the facial features seen, which most people
immediately recognize. However, an inferential description of the same
image would seek instead to \emph{explain} what is being seen. The
process of explanation must invariably involve at its core an
application of the law of parsimony, or \textbf{Occam's razor}. This
principle predicates that when considering two hypotheses compatible
with an observation, the simplest one must prevail. Employing this logic
results in the conclusion that what we are seeing is in fact a regular
mountain, without denying that it looks like a face in that picture and
instead acknowledging that it does so accidentally. In other words, the
``facial'' description is not useful as an explanation, as it emerges
out of random features rather than exposing any underlying mechanism.

Going out of the analogy and back to the problem of community detection,
in Fig.~\ref{fig:infvsdesc} (b) and (c) we see a descriptive and an
inferential account of an example network. The descriptive one is a
division of the nodes into 13 assortative communities, which would be
identified with many descriptive community detection methods available
in the literature. Indeed, we can inspect visually that these groups
form assortative communities,\footnote{See Sec.~\ref{sec:obvious} for
possible pitfalls with relying on visual inspections.} and most people
would agree that these communities are really there, according to most
definitions in use: these are groups of nodes with many more internal
edges than external ones.  However, an inferential account of the same
network would reveal something else altogether. Specifically, it would
explain this network as the outcome of a process where the edges are
placed at random, without the existence of any communities. The
communities that we see in Fig.~\ref{fig:infvsdesc}(a) are just a
byproduct of this random process, and therefore carry no explanatory
power. In fact, this is exactly how the network in this example was
generated,
i.e. by choosing a specific degree sequence and connecting the edges
uniformly at random.

\begin{figure}[t]
  \begin{tabular}{cc}
    \multicolumn{2}{c}{(a) Generative process (random stub matching)}\\
    \multicolumn{2}{c}{\smaller $13$ nodes with degree $20$ and $230$ nodes with degree $1$}\\
    \multicolumn{2}{c}{\includegraphics[width=\textwidth]{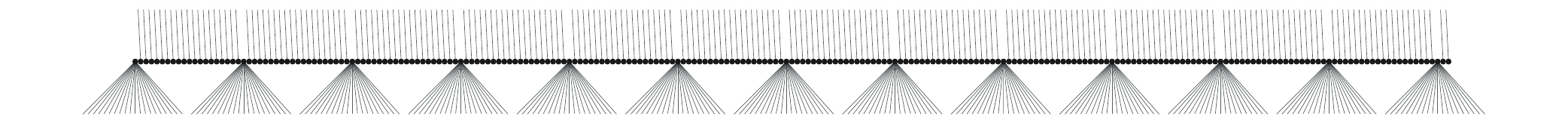}} \\
    \multicolumn{2}{c}{\smaller Stubs paired uniformly at random}\\
    \multicolumn{2}{c}{\includegraphics[width=\textwidth]{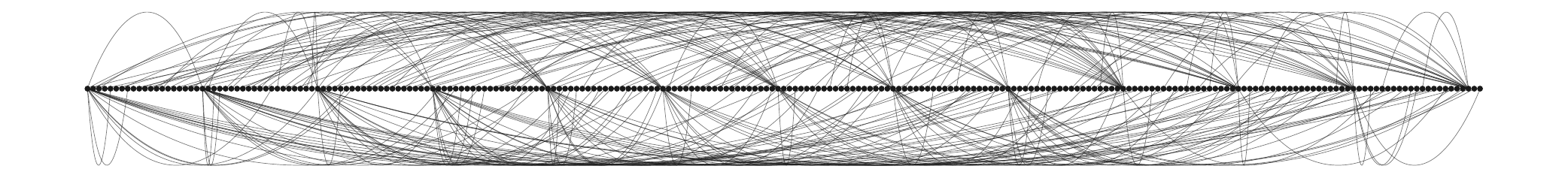}} \\
    (b) Observed network & (c) New sample \\
    \includegraphics[width=.49\textwidth]{figs/descriptive.pdf} &
    \includegraphics[width=.49\textwidth]{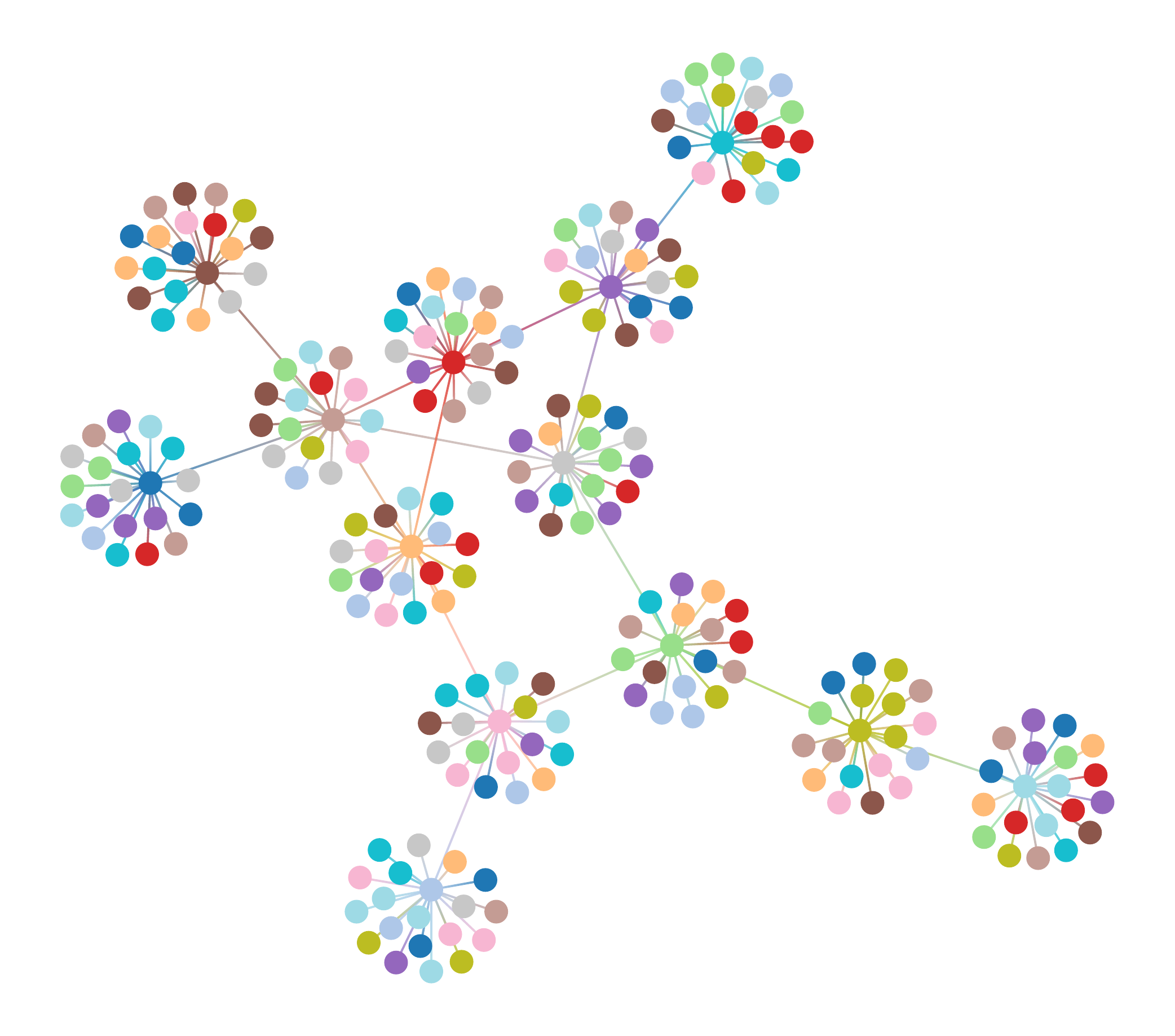}\\
  \end{tabular}

  \caption{Descriptive community detection finds a partition of the
  network according to an arbitrary criterion that bears in general no
  relation to the rules that were used to generate it.  In (a) is shown
  the generative model we consider, where first a degree sequence is
  given to the nodes (forming ``stubs'', or ``half-edges'') which then
  are paired uniformly at random, forming a graph. In (b) is shown a
  realization of this model. The node colors show the partition found
  with virtually any descriptive community detection method. In
  (c) is shown another network sampled from the same model, together
  with the same partition found in (b), which is completely uncorrelated
  with the new apparent communities seen, since they are the mere
  byproduct of the random placement of the edges. An inferential
  approach would find only a single community in both (b) and (c), since
  no partition of the nodes is relevant for the underlying generative
  model.\label{fig:descriptive}}
\end{figure}

In Fig.~\ref{fig:descriptive}(a) we illustrate in more detail how the
network in Fig.~\ref{fig:infvsdesc} was generated: The degrees of the
nodes are fixed, forming ``stubs'' or ``half-edges,'' which are then
paired uniformly at random forming the edges of the
network.\footnote{This uniform pairing will typically also result in the
occurrence of pairs of nodes of degree one connected together in their
own connected component. We consider an instance of the process where
this does not happen for visual clarity in the figure, but without
sacrificing its main message.} In Fig.~\ref{fig:descriptive}(b), like in
Fig.~\ref{fig:infvsdesc}, the node colors show the partition found with
descriptive community detection methods. However, this network division
carries no explanatory power beyond what is contained in the degree
sequence of the network, since it is generated otherwise uniformly at
random. This becomes evident in Fig.~\ref{fig:descriptive}(c), where we
show another network sampled from the same generative process,
i.e. another random pairing, but partitioned according to the same
division as in Fig.~\ref{fig:descriptive}(b). Since the nodes are paired
uniformly at random, constrained only by their degree, this will create
new apparent ``communities'' that are always uncorrelated with one
another.  Like the ``face'' on Mars, they can be seen and described, but
they cannot (plausibly) explain how the network came to be.

We emphasize that the communities found in Fig.~\ref{fig:descriptive}(b)
are indeed really there from a descriptive point of view, and they can
in fact be useful for a variety of tasks. For example, the \emph{cut}
given by the partition, i.e. the number of edges that go between
different groups, is only 13, which means that we need only to remove
this number of edges to break the network into (in this case) 13 smaller
components. Depending on context, this kind of information can be used
to prevent a widespread epidemic, hinder undesired communication, or, as
we have already discussed, distribute tasks among processors and design
a microchip. However, what these communities \emph{cannot} be used for
is to \emph{explain} the data. In particular, a conclusion that would be
completely incorrect is that the nodes that belong to the same group
would have a larger probability of being connected between
themselves. As shown in Fig.~\ref{fig:descriptive}(a), this is clearly not
the case, as the observed ``communities'' arise by pure chance, without
any preference between the nodes.

\subsection{To infer or to describe? A litmus test}\label{sec:infer}

Given the above differences, and the fact that both inferential and
descriptive approaches have their uses depending on context, we are left
with the question: Which approach is more appropriate for a given task
at hand?  In order to help answering this question, for any given
context, it is useful to consider the following ``litmus test'':\\

\tikzstyle{background rectangle}=[thin,draw=black]
\begin{savenotes}
\noindent\begin{tikzpicture}[show background rectangle]
\node[align=justify, text width=.92\linewidth, inner sep=1em]{
  Q: ``Would the usefulness of our conclusions change if we learn, after
  obtaining the communities, that the network being analyzed is
  maximally random?''\\

  If the answer is ``yes,'' then an inferential approach is needed.\\

  If the answer is ``no,''  then an inferential approach is not required.
};

\node[xshift=3ex, yshift=-0.7ex, overlay, fill=white, draw=white, above 
right] at (current bounding box.north west) {
  \textit{\textbf{Litmus test: to infer or to describe?}}
};
\end{tikzpicture} 
\end{savenotes}
If the answer to the above question is ``yes,'' then an inferential
approach is warranted, since the conclusions depend on an interpretation
of how the data were generated. Otherwise, a purely descriptive approach
may be appropriate since considerations about generative processes are
not relevant.

It is important to understand that the relevant question in this context
is not whether the network being analyzed is \emph{actually} maximally
random,\footnote{``Maximally random'' here means that, conditioned on
some global or local constraints, like the number of edges or the node
degrees, the placement of the edges is done in uniformly at random. In
other words, the network is sampled from a maximum-entropy model
constrained in a manner unrelated to community structure, such that
whatever communities we may ascribe to the nodes could have played no
role in the placement of the edges.} since this is rarely the case for
 empirical
networks. Instead, considering this hypothetical scenario serves as a
test to evaluate if our task requires us to separate between actual
latent community structures (i.e. those that are responsible for the
network formation), from those that arise completely out of random
fluctuations, and hence carry no explanatory power. Furthermore, most
empirical networks, even if not maximally random, like most interesting
data, are better explained by a mixture of structure and randomness, and
a method that cannot tell those apart cannot be used for inferential
purposes.

Returning to the VLSI and task scheduling examples we considered in the
previous section, it is clear that the answer to the litmus test above
would be ``no,'' since it hardly matters how the network was generated
and how we should interpret the partition found, as long as the
integrated circuit can be manufactured and function efficiently, or the
tasks finish in the minimal time. Interpretation and explanations are
simply not the primary goals in these cases.\footnote{Although this is
certainly true at a first instance, we can also argue that properly
understanding \emph{why} a certain partition was possible in the first
place would be useful for reproducibility and to aid the design of
future instances of the problem. For these purposes, an inferential
approach would be more appropriate.}

However, it is safe to say that in network data analyses very often the
answer to the question above question would be ``yes.'' Typically,
community detection methods are used to try to understand the overall
large-scale network structure, determine the prevalent mixing patterns,
make simplifications and generalizations, all in a manner that relies on
statements about what lies behind the data, e.g. whether nodes were more
or less likely to be connected to begin with. A majority of conclusions
reached would be severely undermined if one would discover that the
underlying network is in fact completely random. This means that these
analyses suffer the substantial risk of yielding misleading answers when
using purely descriptive methods, since they are likely to be
\emph{overfitting} the data ---
i.e. confusing randomness with underlying generative structure.\footnote{
We emphasize that the concept of overfitting is intrinsically tied with
an inferential goal, i.e.  one that involves interpretations about an
underlying distribution of probability relating to the network
structure. The partitioning of a graph with the objective of producing
an efficient chip design cannot overfit, because it does not elicit
an inferential interpretation. Therefore, whenever we mention that a
method overfits, we refer only to the situation where it is being
employed with an inferential goal, and that it incorporates a level of
detail that cannot be justified by the statistical evidence available in
the data.}

\subsection{Inferring, explaining, and compressing}\label{sec:inference}

\begin{figure}[t]
  \begin{tabular}[t]{cccc}
    \multicolumn{4}{c}{(a) Generative process}\\[.5em]
    \includegraphics[width=.25\textwidth]{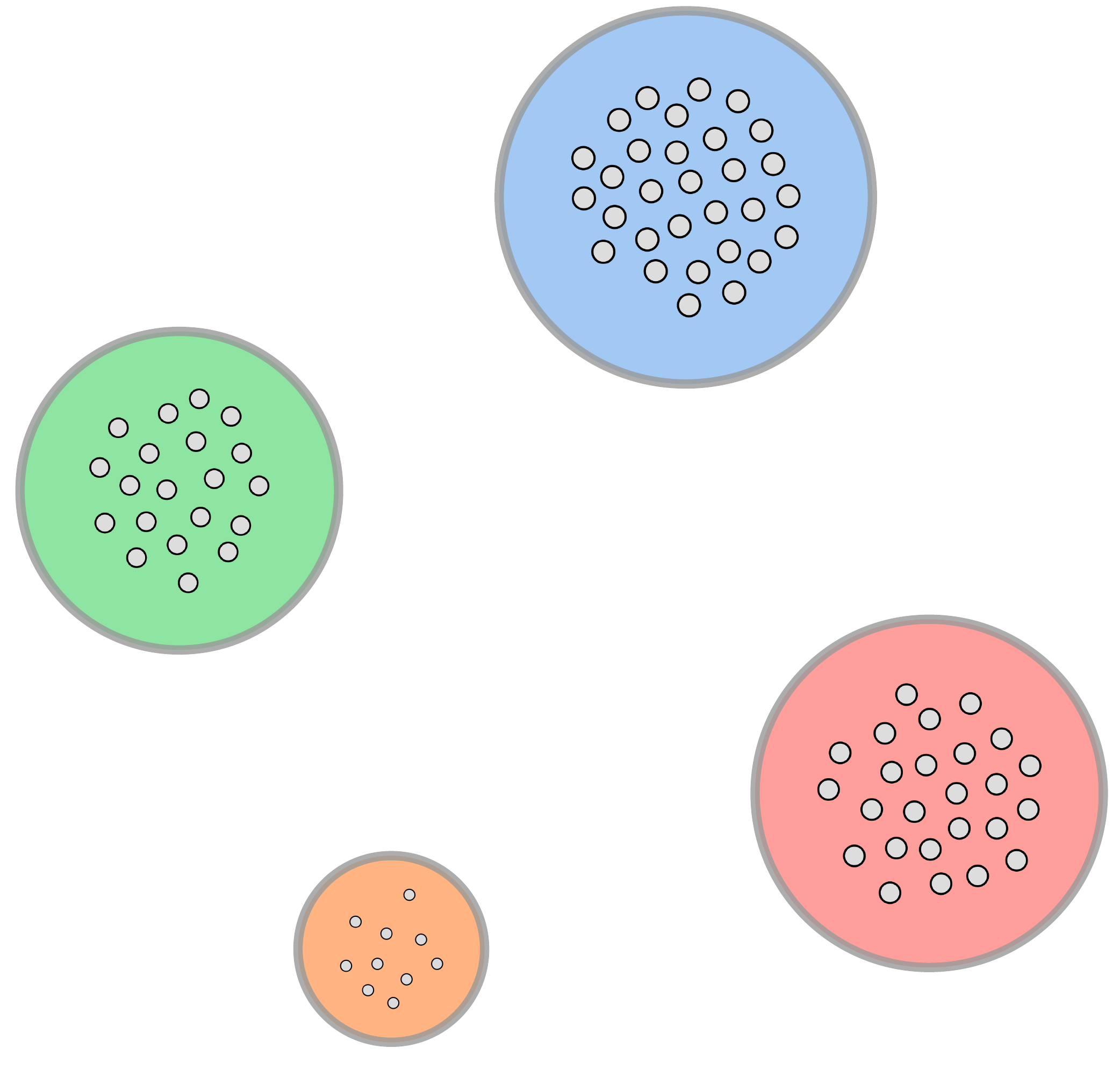}&
    \includegraphics[width=.25\textwidth]{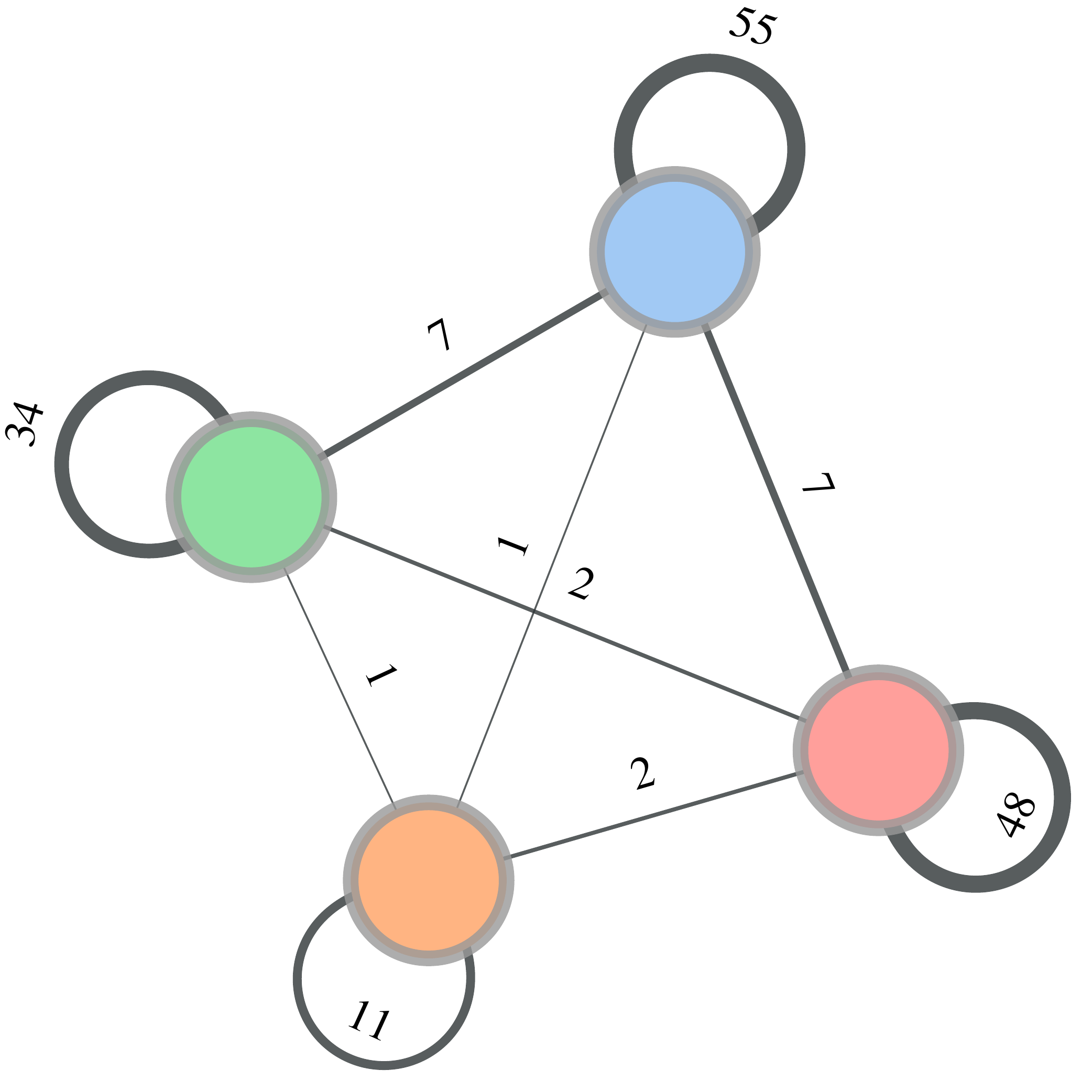}&
    \includegraphics[width=.25\textwidth]{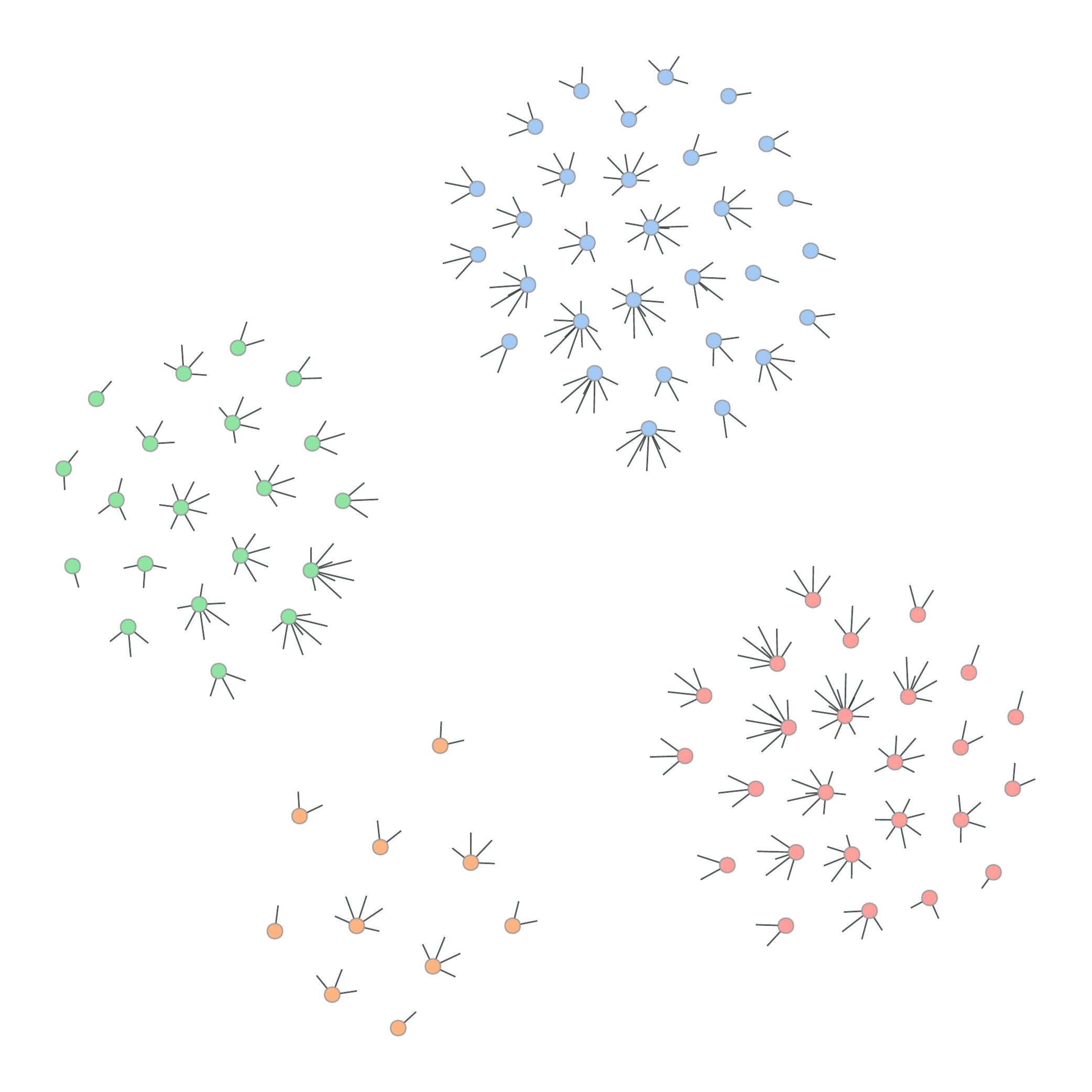}&
    \includegraphics[width=.25\textwidth]{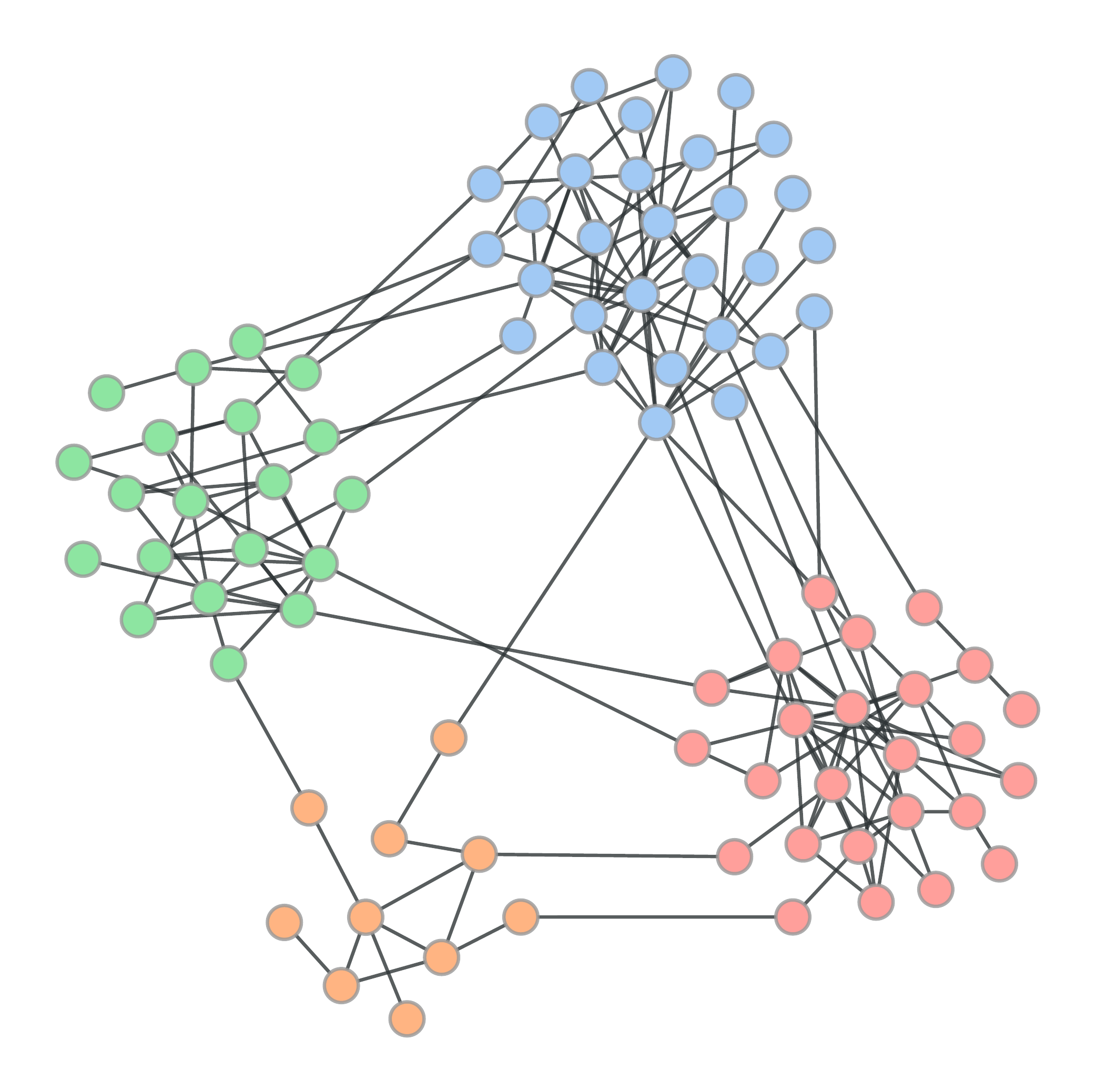}\\
    \smaller Node partition, $P(\bb)$ &
    \smaller Edges between groups, $P(\e|\bb)$ &
    \smaller Degrees, $P(\bm{k}|\e,\bb)$ &
    \smaller Network, $P(\A|\bm{k},\e,\bb)$
  \end{tabular}
  \vspace{1em}

  \begin{tabular}[c]{ccccc}
    \multicolumn{5}{c}{(b) Inference procedure}\\[.5em]
    \multirow[t]{2}{*}[-4.5em]{\includegraphics[width=.25\textwidth]{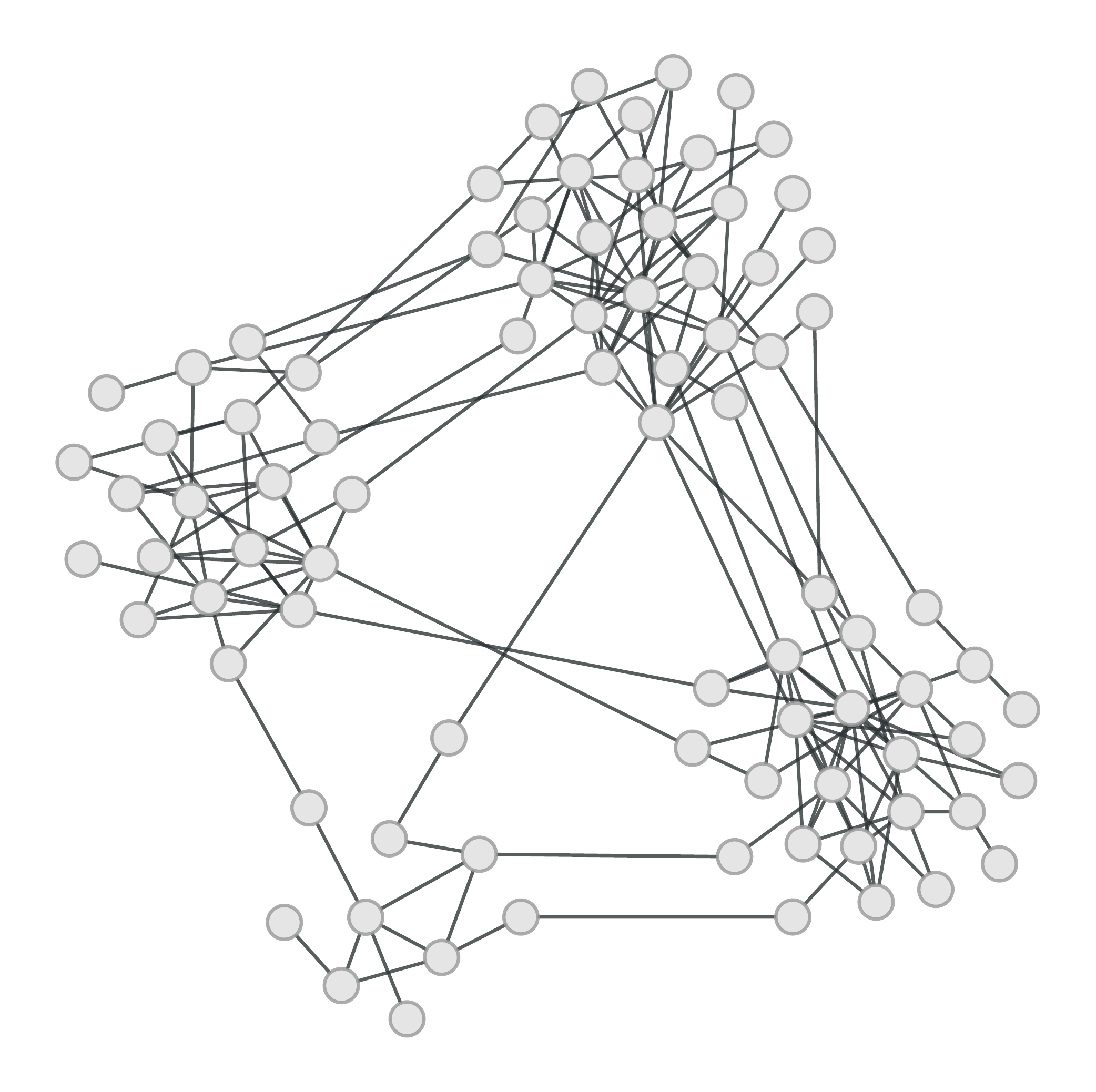}}&
    \includegraphics[width=.16\textwidth]{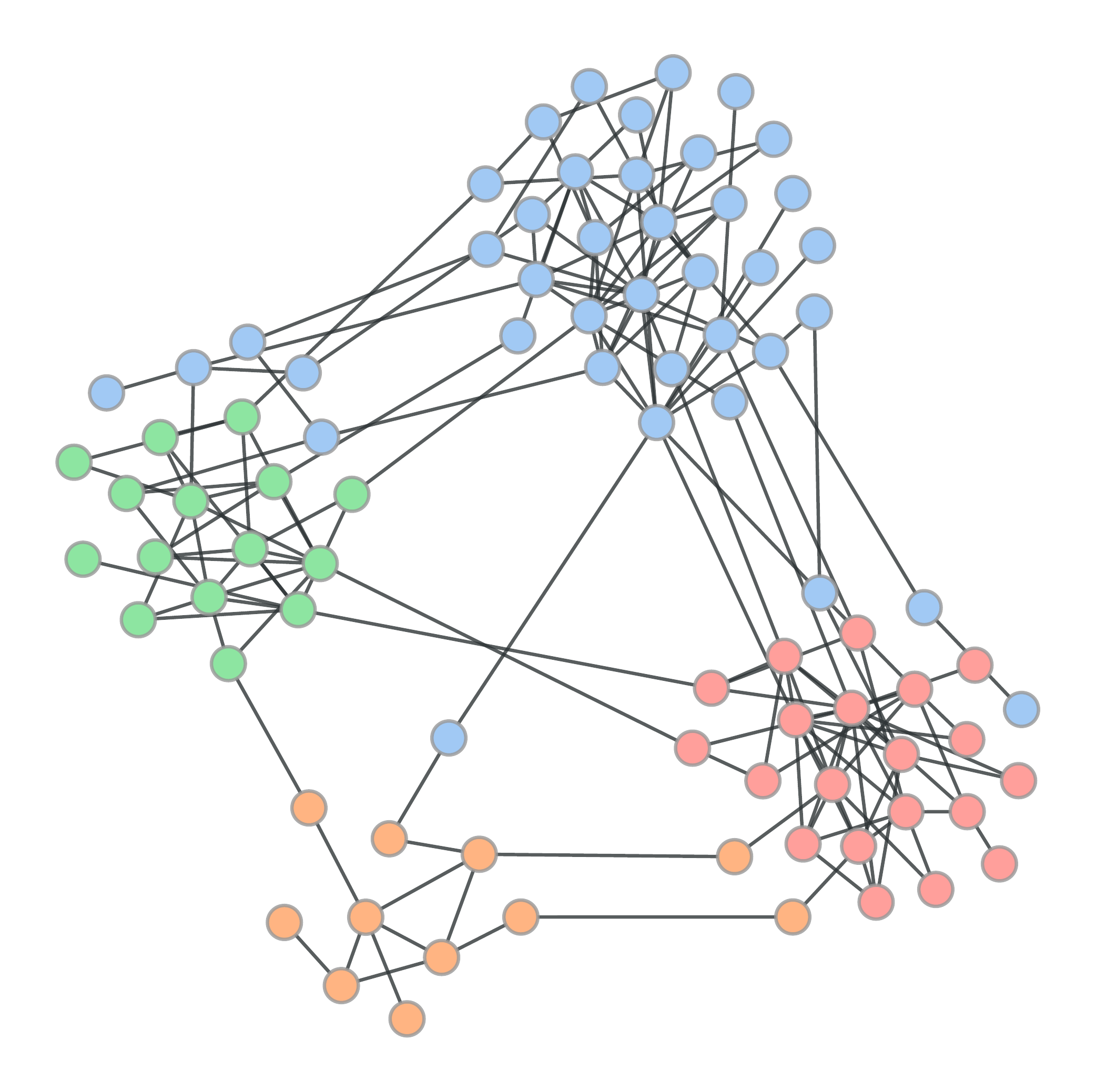}&
    \includegraphics[width=.16\textwidth]{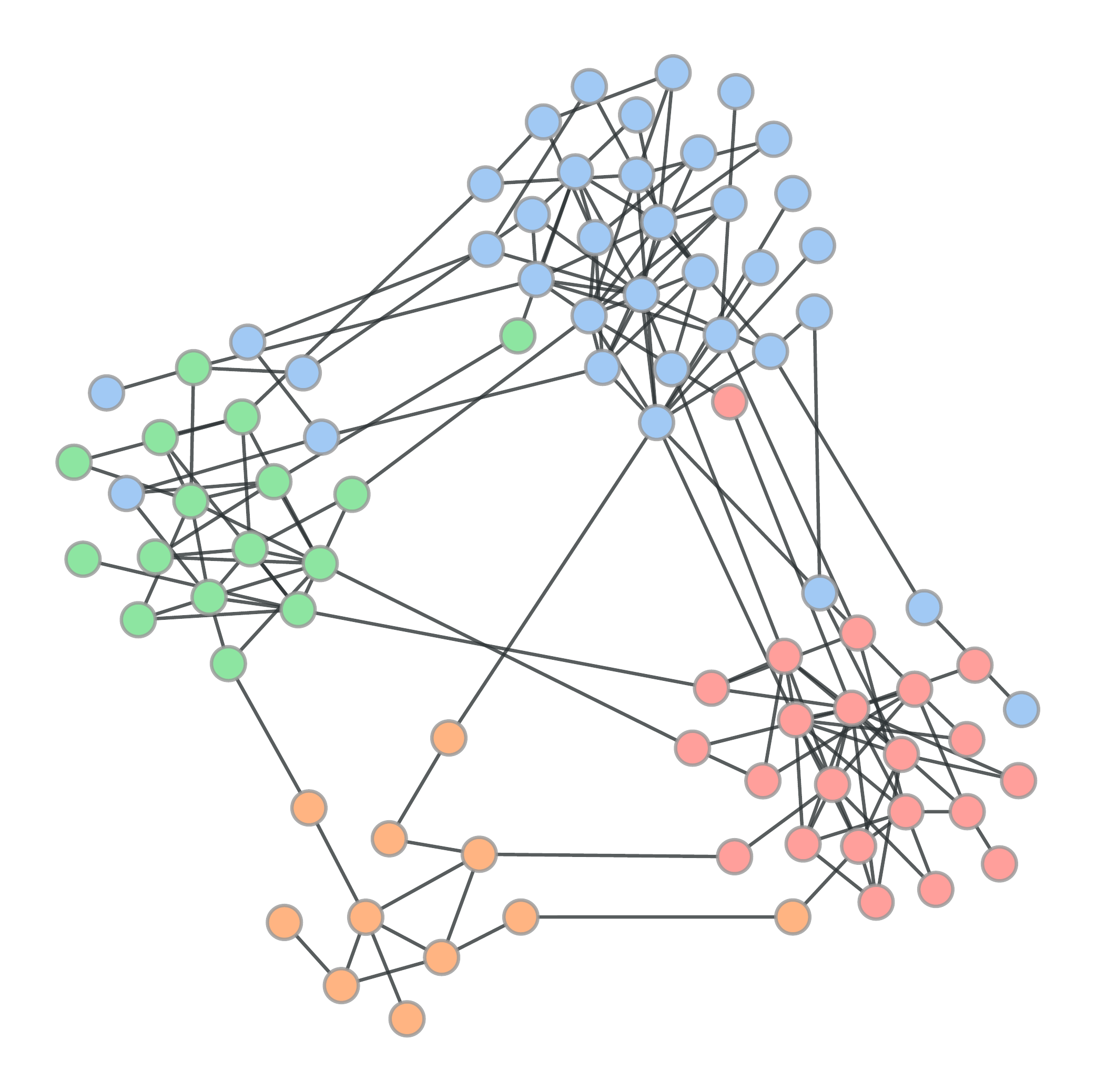}&
    \includegraphics[width=.16\textwidth]{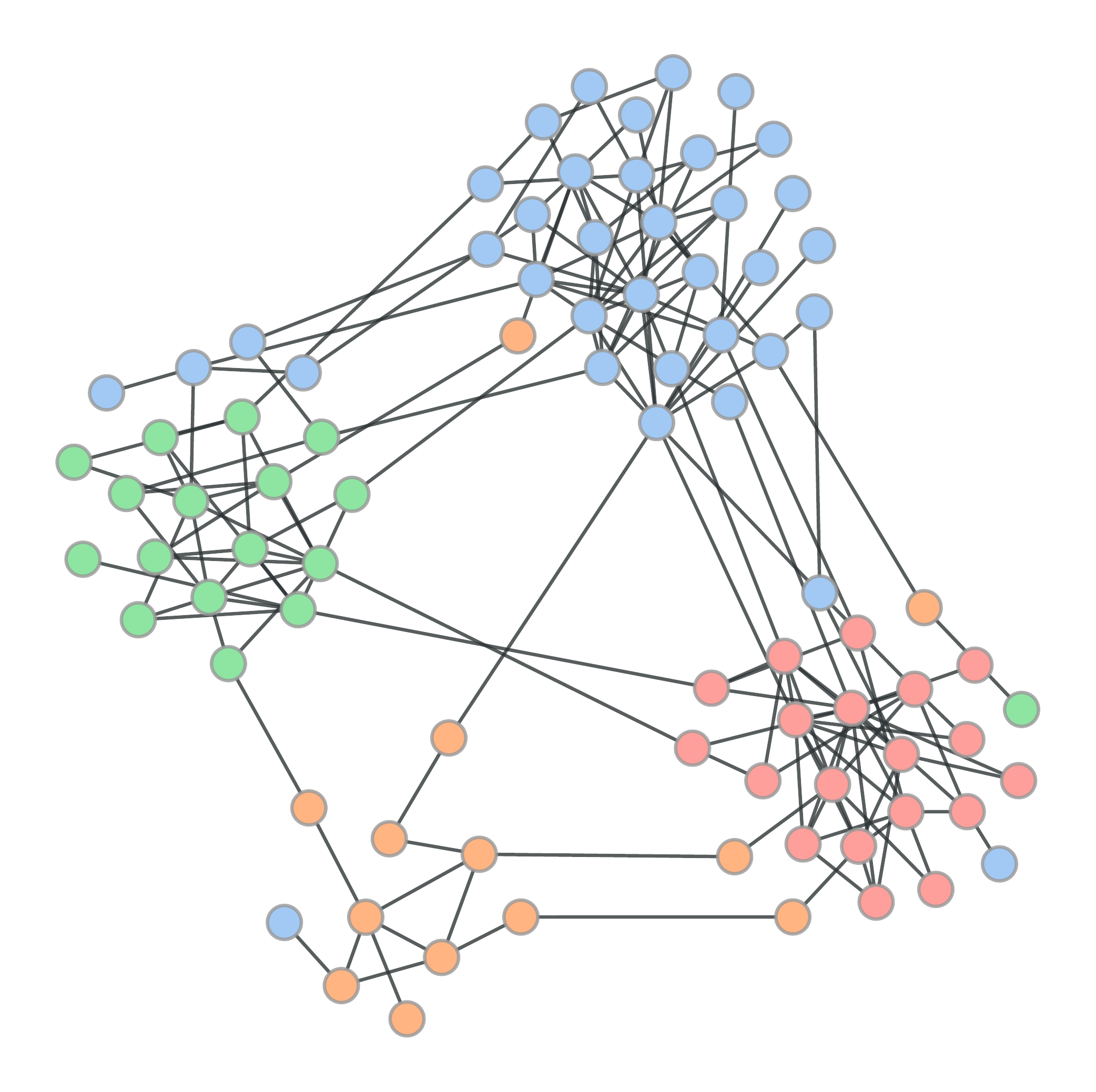}&
    \multirow[t]{2}{*}[-4.5em]{\includegraphics[width=.25\textwidth]{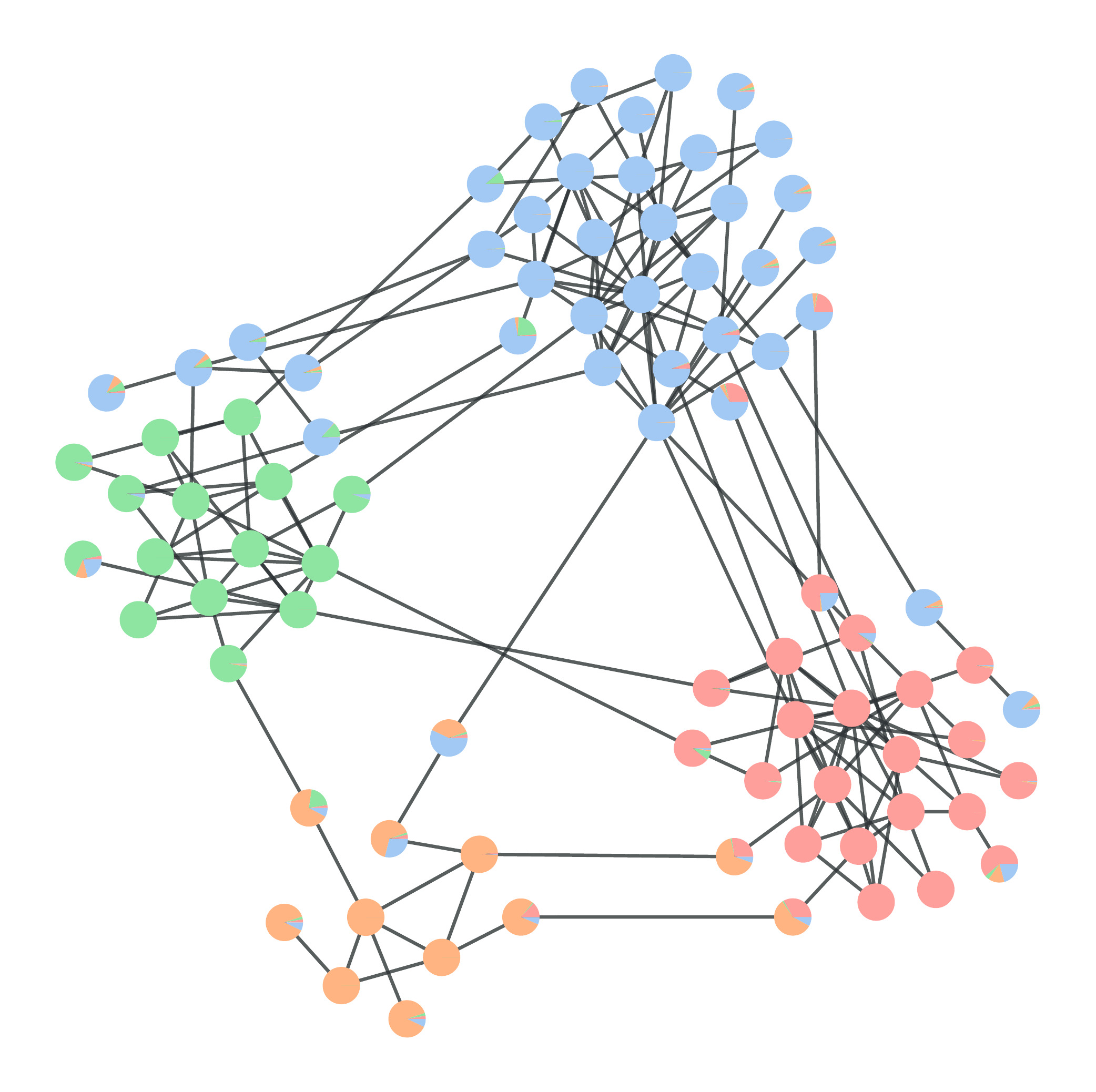}}\\
    &
    \includegraphics[width=.16\textwidth]{figs/example-post-0.pdf}&
    \includegraphics[width=.16\textwidth]{figs/example-post-1.pdf}&
    \includegraphics[width=.16\textwidth]{figs/example-post-2.pdf}&
    \\
    \smaller Observed network $\A$ &
    \multicolumn{3}{c}{\smaller Posterior distribution $P(\bb|\A)$} &
    \smaller Marginal probabilities
  \end{tabular}

  \caption{Inferential community detection considers a generative
  process (a), where the unobserved model parameters are sampled from
  prior distributions. In the case of the DC-SBM, these are the priors
  for the partition $P(\bb)$, the number of edges between groups
  $P(\e|\bb)$, and the node degrees, $P(\bm{k}|\e,\bb)$. Finally, the
  network itself is sampled from its model, $P(\A|\bm{k},\e,\bb)$. The
  inference procedure (b) consists on inverting the generative process
  given an observed network $\A$, corresponding to a posterior
  distribution $P(\bb|\A)$, which then can be summarized by a marginal
  probability that a node belongs to a given group (represented as pie
  charts on the nodes).\label{fig:dcsbm}}
\end{figure}

Inferential approaches to community detection (see
Ref.~\cite{peixoto_bayesian_2019} for a detailed introduction) are
designed to provide explanations for network data in a principled
manner. They are based on the formulation of generative models that
include the notion of community structure in the rules of how the edges
are placed. More formally, they are based on the definition of a
likelihood $P(\A|\bb)$ for the network $\A$ conditioned on a partition
$\bb$, which describes how the network could have been generated,
and the inference is obtained via the posterior distribution,
according to Bayes' rule, i.e.
\begin{equation}\label{eq:bayes}
  P(\bb|\A) = \frac{P(\A|\bb)P(\bb)}{P(\A)},
\end{equation}
where $P(\bb)$ is the prior probability for a partition $\bb$. The
inference procedure consists in sampling from or maximizing this
distribution, which yields the most likely division(s) of the network
into groups, according to the statistical evidence available in the data
(see Fig.~\ref{fig:dcsbm}).

Overwhelmingly, the models used to infer communities are variations of
the stochastic block model (SBM)~\cite{holland_stochastic_1983}, where
in addition to the node partition, it takes the probability of edges
being placed between the different groups as an additional set of
parameters. A particularly expressive variation is the degree-corrected
SBM (DC-SBM)~\cite{karrer_stochastic_2011}, with a marginal likelihood
given by~\cite{peixoto_nonparametric_2017}
\begin{equation}\label{eq:dcsbm-marginal}
  P(\A|\bb) = \sum_{\e, \bm k}P(\A|\bm{k},\e,\bb)P(\bm{k}|\e,\bb)P(\e|\bb),
\end{equation}
where $\e=\{e_{rs}\}$ is a matrix with elements $e_{rs}$ specifying how
many edges go between groups $r$ and $s$, and $\bm k=\{k_i\}$ are the
degrees of the nodes. Therefore, this model specifies that, conditioned
on a partition $\bb$, first the edge counts $\e$ are sampled from a
prior distribution $P(\e|\bb)$, followed by the degrees from the prior
$P(\bm{k}|\e,\bb)$, and finally the network is wired together according
to the probability $P(\A|\bm{k},\e,\bb)$, which respects the constraints
given by $\bm k$, $\e$, and $\bb$. See Fig.~\ref{fig:dcsbm}(a) for a
illustration of this process.

This model formulation includes maximally random networks as special
cases
--- indeed the model we considered in Fig.~\ref{fig:descriptive}
corresponds exactly to the DC-SBM with a single group. Together with the
Bayesian approach, the use of this model will inherently favor a more
parsimonious account of the data, whenever it does not warrant a more
complex description
--- amounting to a formal implementation of Occam's razor. This is best
seen by making a formal connection with information theory, and noticing
that we can write the numerator of Eq.~\ref{eq:bayes} as
\begin{equation}
  P(\A|\bb)P(\bb) = 2^{-\Sigma(\A,\bb)},
\end{equation}
where the quantity $\Sigma(\A,\bb)$ is known as the \emph{description
length}~\cite{rissanen_modeling_1978,grunwald_minimum_2007,rissanen_information_2010}
of the network. It is computed as\footnote{Note that the sum in
Eq.~\ref{eq:dcsbm-marginal} vanishes because only one term is non-zero
given a fixed network $\A$.}
\begin{equation}\label{eq:dl_dcsbm}
  \Sigma(\A,\bb) = \underset{\mathcal{D}(\A|\bm{k},\e,\bb)}{\underbrace{-\log_2P(\A|\bm{k},\e,\bb)}}\, 
  \underset{\mathcal{M}(\bm{k},\e,\bb)}{\underbrace{-\log_2P(\bm{k}|\e,\bb) - \log_2 P(\e|\bb) - \log_2P(\bb)}}.
\end{equation}
The second set of terms $\mathcal{M}(\bm{k},\e,\bb)$ in the above
equation quantifies the amount of information in bits necessary to
encode the parameters of the model.\footnote{If a value $x$ occurs with
probability $P(x)$, this means that in order to transmit it in a
communication channel we need to answer at least $-\log_2P(x)$ yes-or-no
questions to decode its value exactly. Therefore we need to answer one
yes-or-no question for a value with $P(x)=1/2$, zero questions for
$P(x)=1$, and $\log_2N$ questions for uniformly distributed values with
$P(x)=1/N$. This value is called ``information content,'' and
essentially measures the degree of ``surprise'' when encountering a
value sampled from a distribution. See
Ref.~\cite{mackay_information_2003} for a thorough but accessible
introduction to information theory and its relation to inference.} The
first term $\mathcal{D}(\A|\bm{k},\e,\bb)$ determines how many bits are
necessary to encode the network itself, once the model parameters are
known. This means that if Bob wants to communicate to Alice the
structure of a network $\A$, he first needs to transmit
$\mathcal{M}(\bm{k},\e,\bb)$ bits of information to describe the
parameters $\bb$, $\e$, and $\bm{k}$, and then finally transmit the
remaining $\mathcal{D}(\A|\bm{k},\e,\bb)$ bits to describe the network
itself.  Then, Alice will be able to understand the message by first
decoding the parameters $(\bm{k},\e,\bb)$ from the first part of the
message, and using that knowledge to obtain the network $\A$ from the
second part, without any errors.

What the above connection shows is that there is a formal equivalence
between \emph{inferring} the communities of a network and
\emph{compressing} it. This happens because finding the most likely
partition $\bb$ from the posterior $P(\bb|\A)$ is equivalent to
minimizing the description length $\Sigma(\A,\bb)$ used by Bob to
transmit a message to Alice containing the whole network.

Data compression amounts to a formal implementation of Occam's razor
because it penalizes models that are too complicated: if we want to
describe a network using many communities, then the model part of the
description length $\mathcal{M}(\bm{k},\e,\bb)$ will be large, and Bob
will need many bits to transmit the model parameters to Alice. However,
increasing the complexity of the model will also \emph{reduce} the first
term $\mathcal{D}(\A|\bm{k},\e,\bb)$, since there are fewer networks
that are compatible with the bigger set of constraints, and hence the
second part of Bob;s message will need to be shorter to convey the
network itself once the parameters are known. Compression (and hence
inference), therefore, is a balancing act between model complexity and
quality of fit, where an increase in the former is \emph{only} justified
when it results in \emph{an even larger} increase of the second, such
that the total description length is minimized.

The reason why the compression approach avoids overfitting the data is
due to a powerful fact from information theory, known as Shannon's
source coding theorem~\cite{shannon_mathematical_1948}, which states
that it is impossible to compress data sampled from a distribution
$P(x)$ using fewer bits per symbol than the entropy of the distribution,
$H=-\sum_xP(x)\log_2P(x)$ --- indeed, it's a remarkable fact from
Shannon's theory that a statement about a single sample (how many bits
we need to describe it) is intrinsically connected to the distribution
from which it came. Therefore, as the data become large, it also becomes
impossible to compress it more than it can be achieved by using a code
that is optimal according to its true distribution. In our context, this
means that it is impossible, for example, to compress a maximally random
network using a SBM with more than one group.\footnote{More accurately,
this becomes impossible only when the network becomes asymptotically
infinite; for finite networks the probability of compression is only
vanishingly small.} This means, for example, that when encountering an
example like in Fig.~\ref{fig:descriptive}, inferential methods will
detect a single community comprising all nodes in the network, since any
further division does not provide any increased compression, or
equivalently, no augmented explanatory power. From the inferential point
of view, a partition like Fig.~\ref{fig:descriptive}(b) \emph{overfits}
the data, since it incorporates irrelevant random features ---
a.k.a. ``noise''
--- into its description.

\begin{figure}[h]
  \begin{tabular}{cc}
    (a) Observed network & (b) New sample \\
    \includegraphics[width=.49\textwidth]{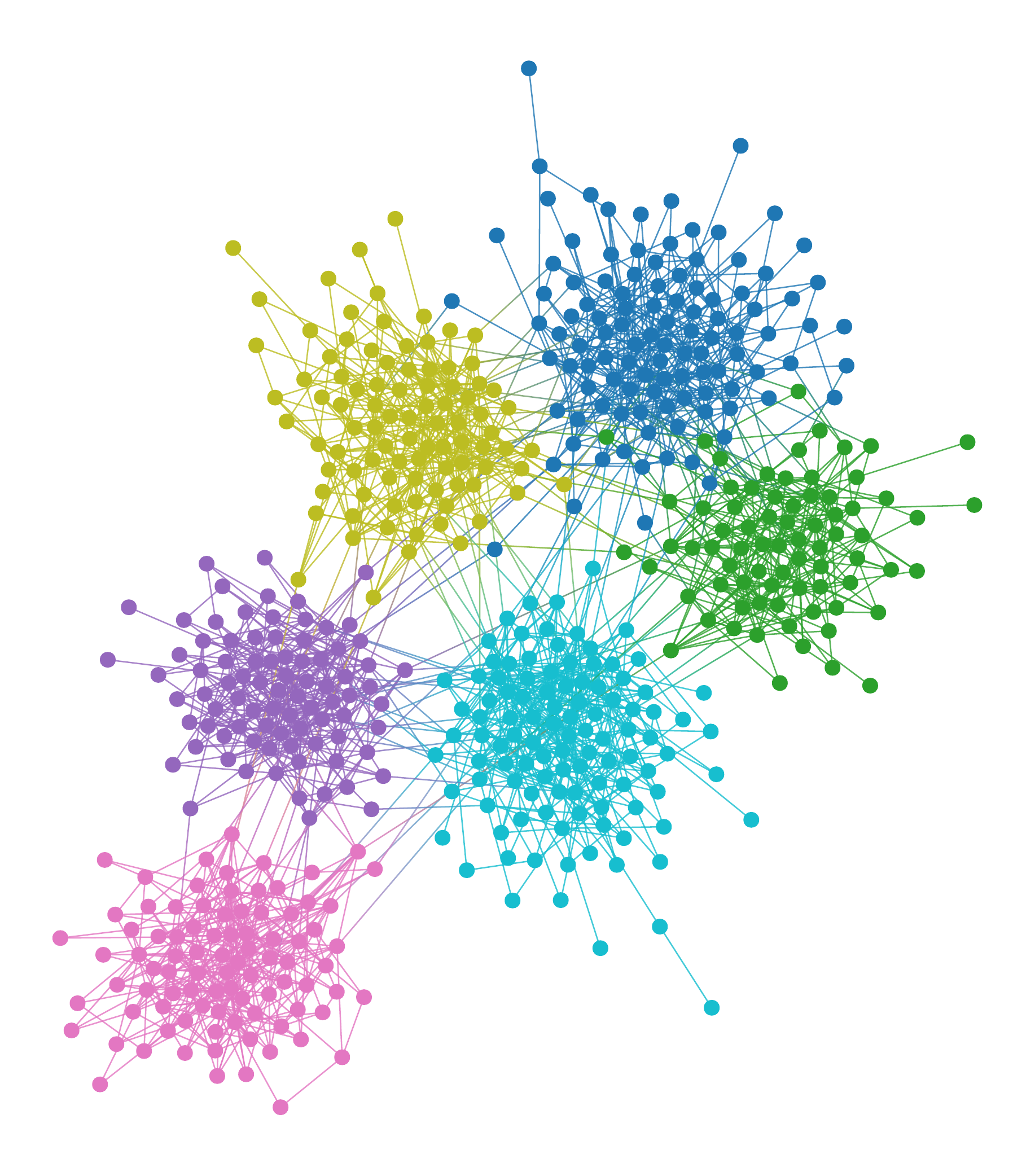} &
    \includegraphics[width=.49\textwidth]{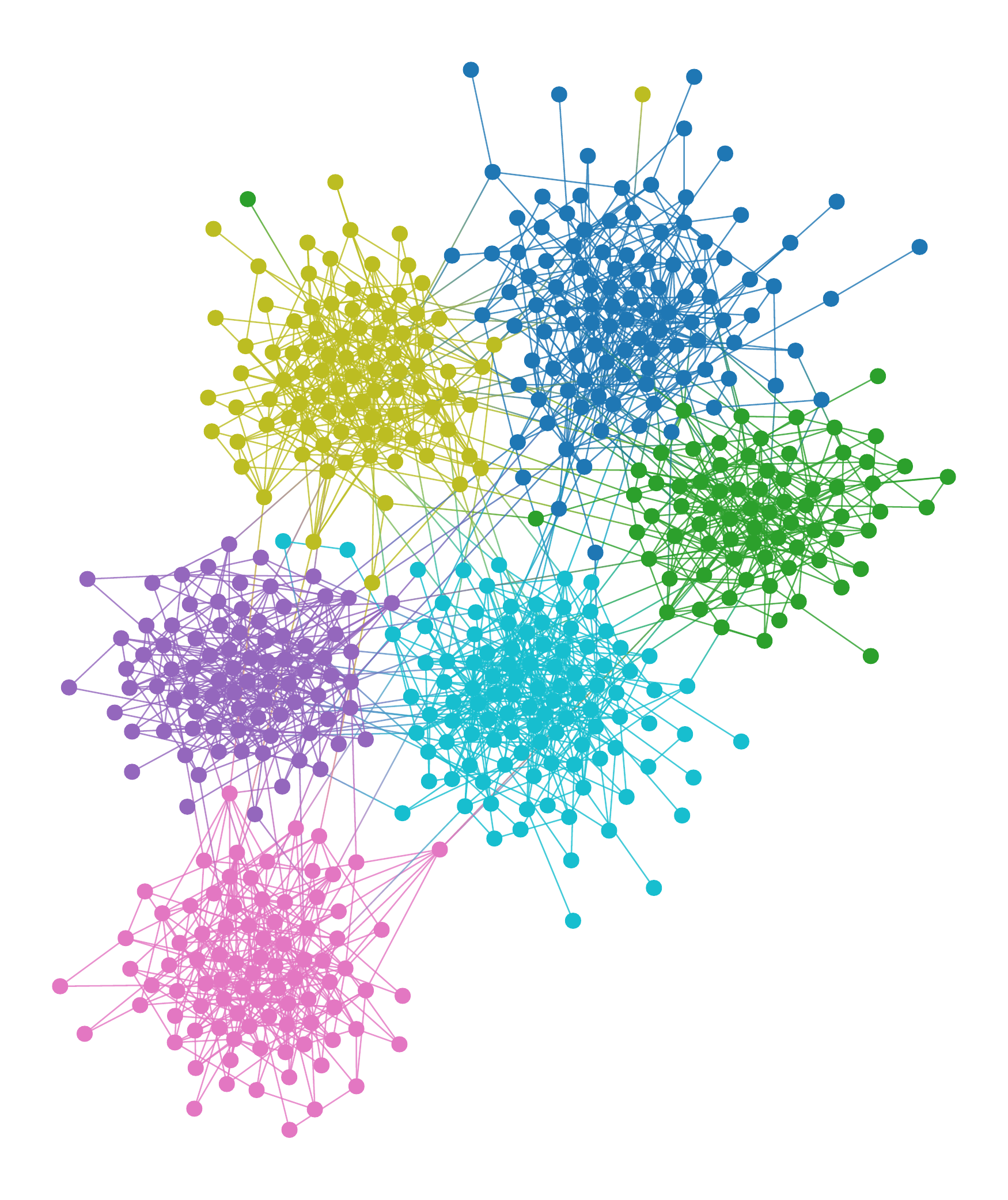}
  \end{tabular}

  \caption{Inferential community detection aims to find a partition of
  the network according to a fit of a generative model that can explain
  its structure. In (a) is shown a network sampled from a stochastic
  block model (SBM) with 6 groups, and where the group assignments were
  hidden from view. The node colors show the groups found via Bayesian
  inference of the SBM. In
  (b) is shown another network sampled from same SBM, together
  with the same partition found in (a), showing that it carries a
  substantial explanatory power --- very differently from the example in
  Fig.~\ref{fig:descriptive} (c).\label{fig:inferential}}
\end{figure}

In Fig.~\ref{fig:inferential}(a) is shown an example of the results
obtained with an inferential community detection algorithm, for a
network sampled from the SBM. As shown in Fig.~\ref{fig:inferential}(b),
the obtained partitions are still valid when carried over to an
independent sample of the model, because the algorithm is capable of
separating the general underlying pattern from the random
fluctuations. As a consequence of this separability, this kind of
algorithm does not find communities in maximally random networks, which
are composed only of ``noise.''

The concept of compression is more generally useful than just avoiding
overfitting within a class of models. In fact, the description length
gives us a model-agnostic objective criterion to compare different
hypotheses for the data generating process according to their
plausibility. Namely, since Shannon's theorem tells us that the best
compression can be achieved asymptotically only with the true model,
then if we are able to find a description length for a network using a
particular model, regardless of how it is parametrized, this also means
that we have automatically found an \emph{upper bound} on the optimal
compression achievable. By formulating different generative models and
computing their description length, we have not only an objective
criterion to compare them against each other, but we also have a way to
limit further what can be obtained with any other model. The result is
an overall scale on which different models can be compared, as we move
closer to the limit of what can be uncovered for a particular network at
hand.

\begin{figure}[h]
  \begin{tabular}{c}
    \begin{overpic}[width=.8\textwidth]{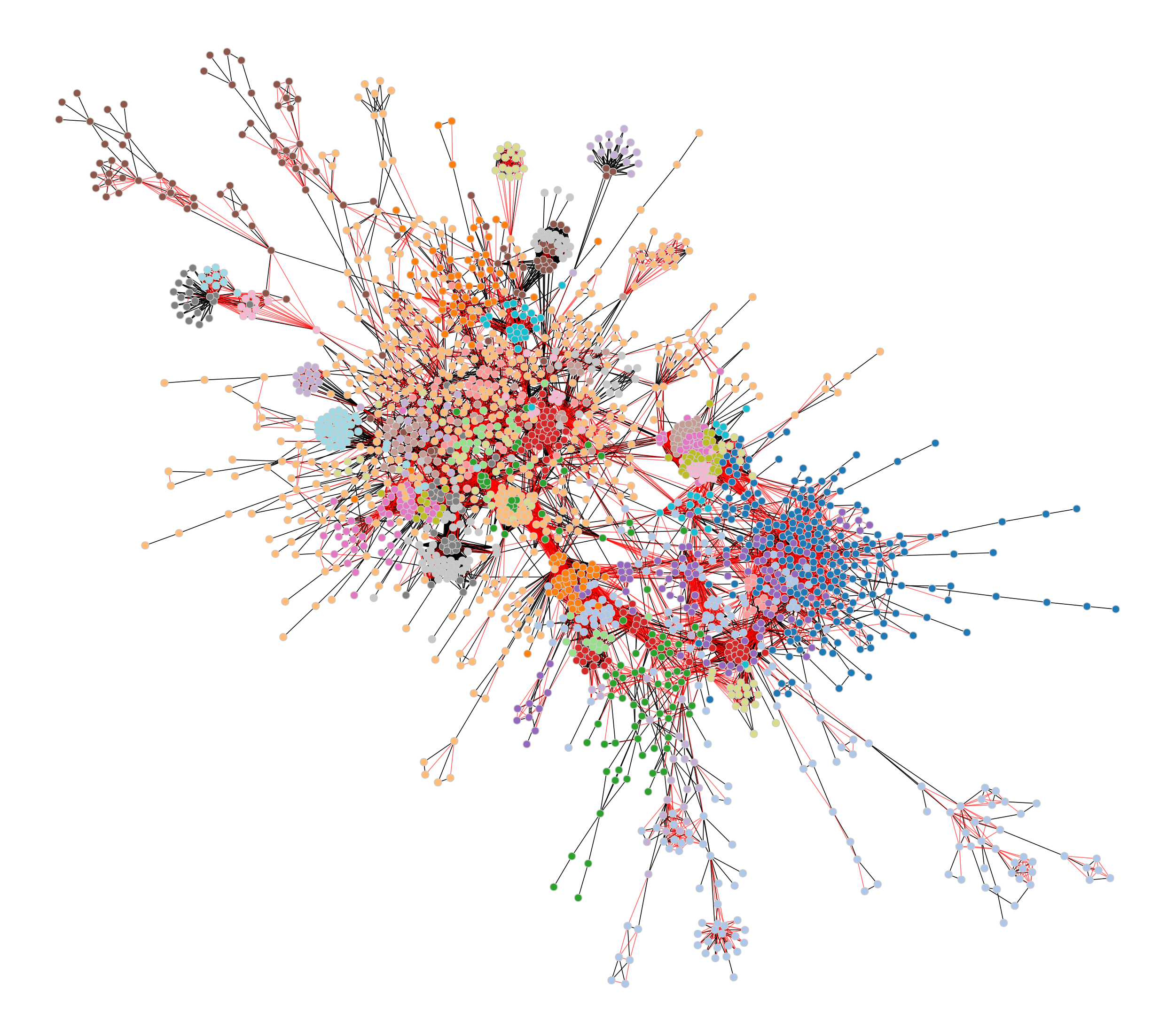}
      \put(-5,85){(a)}
    \end{overpic}
    \\
    \begin{overpic}[width=.8\textwidth]{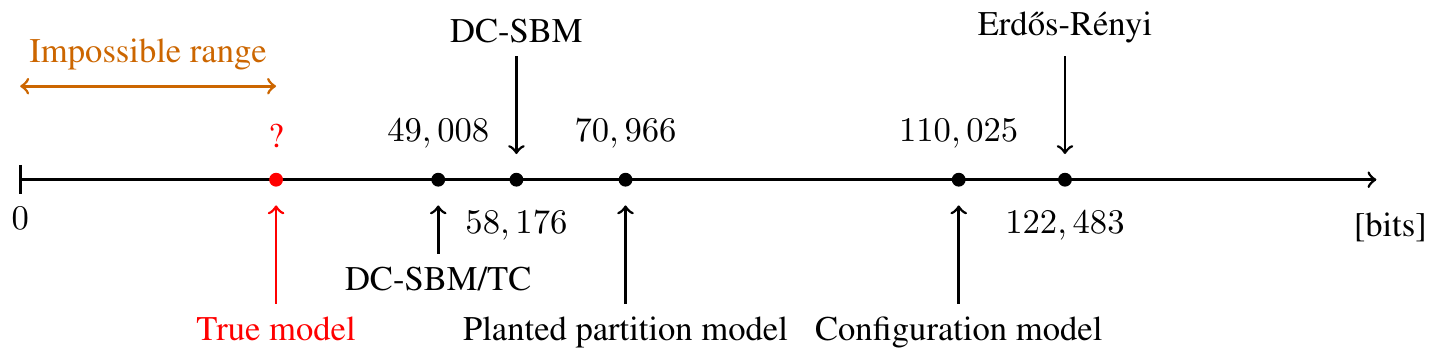}
      \put(-5,25){(b)}
    \end{overpic}
  \end{tabular}
 \caption{Compression points towards the true model. (a)
 Protein-protein interaction network for the organism \emph{Meleagris
 gallopavo}~\cite{zitnik_evolution_2019}. The node colors indicate the
 best partition found with the
 DC-SBM/TC~\cite{peixoto_disentangling_2022} (there are more groups than
 colors, so some colors are repeated), and the edge colors indicate
 whether they are attributed to triadic closure (red) or the DC-SBM
 (black). (b) Description length values according to
 different models. The unknown true model must yield a description
 length value smaller than the DC-SBM/TC, and no other model should be able
 to provide a superior compression that is statistically
 significant. \label{fig:compressed}}
\end{figure}

As an example, in Fig.~\ref{fig:compressed} we show the description
length values with some models obtained for a protein-protein
interaction network for the organism \emph{Meleagris gallopavo} (wild
turkey)~\cite{zitnik_evolution_2019}. In particular, we can see that
with the DC-SBM/TC (a version of the model with the addition of triadic
closure edges~\cite{peixoto_disentangling_2022}) we can achieve a
description length that is far smaller than what would be possible with
networks sampled from either the Erd\H{o}s-Rényi, configuration, or
planted partition (a SBM with strictly assortative
communities~\cite{zhang_statistical_2020}) models, meaning that the
inferred model is much closer to the true process that actually
generated this network than the alternatives.  Naturally, the actual
process that generated this network is different from the DC-SBM/TC, and
it likely involves, for example, mechanisms of node duplication which
are not incorporated into this rather simple
model~\cite{pastor-satorras_evolving_2003}. However, to the extent that
the true process leaves statistically significant traces in the network
structure,\footnote{Visually inspecting Fig.~\ref{fig:compressed}
reveals what seems to be local symmetries in the network structure,
presumably due to gene duplication. These patterns are not exploited by
the SBM description, and points indeed to a possible path for further
compression.} computing the description length according to it should
provide further compression when compared to the
alternatives.\footnote{In Sec.~\ref{sec:believe} we discuss further the
usefulness of models like the SBM despite the fact we know they are not
the true data generating process.} Therefore, we can try to extend or
reformulate our models to incorporate features that we hypothesize to be
more realistic, and then verify if this in fact the case, knowing that
whenever we find a more compressive model, it is moving closer to the
true model --- or at least to what remains detectable from it for the
finite data.

\FloatBarrier

%\FloatBarrier

The discussion above glosses over some important technical aspects. For
example, it is possible for two (or, in fact, many) models to have the
same or very similar description length values. In this case, Occam's
razor fails as a criterion to select between them, and we need to
consider them collectively as equally valid hypotheses. This means, for
example, that we would need to average over them when making specific
inferential statements~\cite{peixoto_revealing_2021} --- selecting
between them arbitrarily can be interpreted as a form of
overfitting. Furthermore, there is obviously no guarantee that the true
model can actually be found for any particular data. This is only
possible in the asymptotic limit of ``sufficient data,'' which will vary
depending on the actual model. Outside of this limit (which is the
typical case in empirical settings, in particular when dealing with
\emph{sparse} networks~\cite{yan_model_2014}), fundamental limits to
inference are unavoidable,\footnote{A very important result in the
context of community detection is the detectability limit of the SBM. As
discovered by Decelle et
al~\cite{decelle_phase_2011,decelle_asymptotic_2011}, if a sufficiently
large network is sampled from a SBM with a sufficiently weak but
nontrivial structure below a specific threshold, it becomes strictly
impossible to uncover the true model from this sample.} which means in
practice that we will always have limited accuracy and some amount of
error in our conclusions. However, when employing compression, these
potential errors tend towards overly simple explanations, rather than
overly complex ones. Whenever perfect accuracy is not possible, it is
difficult to argue in favor of a bias in the opposite direction.

We emphasize that it is not possible to ``cheat'' when doing
compression. For any particular model, the description length will have
the same form
\begin{equation}
  \Sigma(\A,\bm\theta) = \mathcal{D}(\A|\bm\theta) + \mathcal{M}(\bm\theta),
\end{equation}
where $\bm\theta$ is some arbitrary set of parameters. If we constrain
the model such that it becomes possible to describe the data with a
number of bits $\mathcal{D}(\A|\bm\theta)$ that is very small, this can
only be achieved, in general, by increasing the number of parameters
$\bm\theta$, such that the number of bits $\mathcal{M}(\bm\theta)$
required to describe them will also increase. Therefore, there is no
generic way to achieve compression that bypasses actually formulating a
meaningful hypothesis that matches statistically significant patterns
seen in the data. One may wonder, therefore, if there is an automatized
way of searching for hypotheses in a manner that guarantees optimal
compression. The most fundamental way to formulate this question is to
generalize the concept of minimum description length as follows: for any
binary string $\bm x$ (representing any measurable data), we define
$L(\bm x)$ as the length in bits of the shortest computer program that
yields $\bm x$ as an output. The quantity $L(\bm x)$ is known as
Kolmogorov complexity~\cite{cover_elements_1991,li_introduction_2008},
and if we would be able to compute it for a binary string representing
an observed network, we would be able to determine the ``true model'' value
in Fig.~\ref{fig:compressed}, and hence know how far we are from the
optimum.\footnote{As mentioned before, this would not necessarily mean
that we would be able to find the actual true model in a practical
setting with perfect accuracy, since for a finite $\bm x$ there could be
many programs of the same minimal length (or close) that generate it.}

Unfortunately, an important result in information theory is that $L(\bm
x)$ is not computable~\cite{li_introduction_2008}. This means that it is
strictly impossible to write a computer program that computes $L(\bm x)$
for any string $\bm x$.\footnote{There are two famous ways to prove
this. One is by contradiction: if we assume that we have a program that
computes $L(\bm x)$, then we could use it as subroutine to write another
program that outputs $\bm x$ with a length smaller than $L(\bm x)$. The
other involves undecidabilty: if we enumerate all possible computer
programs in order of increasing length and check if their outputs match
$\bm x$, we will eventually find programs that loop
indefinitely. Deciding whether a program finishes in finite time is
known as the ``halting problem,'' which has been proved to be impossible
to solve. In general, it cannot be determined if a program reaches an
infinite loop in a manner that avoids actually running the program and
waiting for it to finish. Therefore, this rather intuitive algorithm to
determine $L(\bm x)$ will not necessarily finish for any given string
$\bm x$. For more details see
e.g. Refs~\cite{cover_elements_1991,li_introduction_2008}} This does not
invalidate using the description length as a criterion to select among
alternative models, but it dashes any hope of fully automatizing the
discovery of optimal hypotheses.

\subsection{Role of inferential approaches in community detection}

Inferential approaches based on the SBM have an old history, and were
introduced for the study of social networks in the early
80's~\cite{holland_stochastic_1983}. But despite such an old age, and
having appeared repeatedly in the literature over the
years~\cite{snijders_estimation_1997,nowicki_estimation_2001,tallberg_bayesian_2004,hastings_community_2006,
rosvall_information-theoretic_2007, airoldi_mixed_2008,
clauset_hierarchical_2008,hofman_bayesian_2008, morup_learning_2009}
(also under different names in other contexts
e.g.~\cite{boguna_class_2003,bollobas_phase_2007}), they entered the
mainstream community detection literature rather late, arguably after
the influential paper by Karrer and Newman that introduced the
DC-SBM~\cite{karrer_stochastic_2011} in 2011, at a point where
descriptive approaches were already dominating. However, despite the
dominance of descriptive methods, the existence of inferential
\emph{criteria} was already long noticeable.  In fact, in a well-known
attempt to systematically compare the quality of a variety of
descriptive community detection methods, the authors of
Ref.~\cite{lancichinetti_benchmark_2008} proposed the now so-called LFR
benchmark, offered as a more realistic alternative to the simpler
Newman-Girvan benchmark~\cite{girvan_community_2002} introduced
earlier. Both are in fact generative models, essentially particular
cases of the DC-SBM, containing a ``ground truth'' community label
assignment, against which the results of various algorithms are supposed
to be compared. Clearly, this is an inferential evaluation criterion,
although, historically, virtually all of the methods compared against
that benchmark are descriptive in
nature~\cite{lancichinetti_community_2009} (these studies were conducted
mostly before inferential approaches had gained more traction). The use
of such a criterion already betrays that the answer to the litmus test
considered previously would be ``yes,'' and therefore descriptive
approaches are fundamentally unsuitable for the task. In contrast,
methods based on statistical inference are not only more principled, but
in fact provably optimal in the inferential scenario: an estimation
based on the posterior distribution obtained from the true generative
model is called ``Bayes optimal,'' since there is no procedure that can,
on average, produce results with higher accuracy. The use of this
inferential formalism has led to the development of asymptotically
optimal algorithms and the identification of sharp transitions in the
detectability of planted community
structure~\cite{decelle_asymptotic_2011,decelle_inference_2011}.

The conflation one often finds between descriptive and inferential goals
in the literature of community detection likely stems from the fact that
while it is easy to define benchmarks in the inferential setting, it is
substantially more difficult to do so in a descriptive setting. Given
any descriptive method (modularity
maximization~\cite{newman_modularity_2006},
Infomap~\cite{rosvall_maps_2008}, Markov
stability~\cite{lambiotte_random_2014}, etc.) it is usually problematic
to determine for which network these methods are optimal (or even if one
exists), and what would be a canonical output that would be
unambiguously correct. In fact, the difficulty with establishing these
fundamental references already serve as evidence that the task itself is
ill-defined. On the other hand, taking an inferential route forces one
to \emph{start with the right answer}, via a well-specified generative
model that articulates what \emph{the communities actually mean} with
respect to the network structure. Based on this precise definition, one
then \emph{derives} the optimal detection method by employing Bayes'
rule.

It is also useful to observe that inferential analyses of aspects of the
network other than directly its structure might still be only
descriptive of the structure itself. A good example of this is the
modelling of dynamics that take place on a network, such as a random
walk. This is precisely the case of the Infomap
method~\cite{rosvall_maps_2008}, which models a simulated
random walk on a network in an inferential manner, using for that a
division of the network into groups. While this approach can be
considered inferential with respect to an artificial dynamics, it is
still only descriptive when it comes to the actual network structure
(and will suffer the same problems, such a finding communities in
maximally random networks). Communities found in this way could be
useful for particular tasks, such as to identify groups of nodes that
would be similarly affected by a diffusion process. This could be used,
for example, to prevent or facilitate the diffusion by removing or
adding edges between the identified groups. In this setting, the answer
to the litmus test above would also be ``no,'' since what is important
is how the network ``is'' (i.e. how a random walk behaves on it), not
how it came to be, or if its features are there by chance alone. Once
more, the important issue to remember is that the groups identified in
this manner cannot be interpreted as having any explanatory power about
the network structure itself, and cannot be used reliably to extract
inferential conclusions about it. We are firmly in a descriptive, not
inferential setting with respect to the network structure.

Another important difference between inferential and descriptive
approaches is worth mentioning. Descriptive approaches are often tied to
very particular contexts, and cannot be directly compared to one
another. This has caused great consternation in the literature, since
there is a vast number of such methods, and little robust methodology on
how to compare them. Indeed, why should we expect that the modules found
by optimizing task scheduling should be comparable to those that
optimize the description of a dynamics? In contrast, inferential
approaches all share the same underlying context: they attempt to
explain the network structure; they vary only in how this is done. They
are, therefore, amenable to principled \emph{model selection}
procedures~\cite{gelman_bayesian_2013,bishop_pattern_2011,mackay_information_2003},
designed to evaluate which is the most appropriate fit for any
particular network, even if the models used operate with very different
parametrizations, as we discussed already in
Sec.~\ref{sec:inference}. In this situation, the multiplicity of
different models available becomes a boon rather than a hindrance, since
they all contribute to a bigger toolbox we have at our disposal when
trying to understand empirical observations.

Finally, inferential approaches offer additional advantages that make them more
suitable as part of a scientific pipeline. In particular, they can be naturally
extended to accommodate measurement uncertainties~\cite{newman_network_2018-1,
  martin_structural_2016,peixoto_reconstructing_2018} --- an unavoidable
property of empirical data, which descriptive methods almost universally fail to
consider. This information can be used not only to propagate the uncertainties
to the community assignments~\cite{peixoto_revealing_2021} but also to
reconstruct the missing or noisy measurements of the network
itself~\cite{clauset_hierarchical_2008, guimera_missing_2009}. Furthermore,
inferential approaches can be coupled with even more indirect observations such
as time-series on the nodes~\cite{hoffmann_community_2020}, instead of a direct
measurement of the edges of the network, such that the network itself is
reconstructed, not only the community structure~\cite{peixoto_network_2019}. All
these extensions are possible because inferential approaches give us more than
just a division of the network into groups; they give us a model estimate of the
network, containing insights about its formation mechanism.

\subsection{Behind every description there is an implicit generative model}\label{sec:implicit}

Descriptive methods of community detection --- such as graph
partitioning for VLSI~\cite{kernighan_graph_1969} or
Infomap~\cite{rosvall_maps_2008} --- are not designed to produce
inferential statements about the network structure. They do not need to
explicitly articulate a generative model, and the quality of their
results should be judged solely against their manifestly noninferential
goals, e.g. whether a chip design can be efficiently manufactured in the
case of graph partitioning.

Nevertheless, descriptive methods are often employed with inferential
aims in practice. This happens, for example, when modularity
maximization is used to discover homophilic patterns in a social
network, or when Infomap is used to uncover latent communities generated
by the LFR benchmark. In these situations, it is useful to consider to
what extent can we expect any of these methods reveal meaningful
inferential results, despite their intended use.

From a purely mathematical perspective, there is actually no formal
distinction between descriptive and inferential methods, because every
descriptive method can be mapped to an inferential one, according to
some implicit model. Therefore, whenever we are attempting to interpret
the results of a descriptive community detection method in an
inferential way --- i.e. make a statement about how the network came to
be --- we cannot in fact avoid making \emph{implicit} assumptions about
the data generating process that lies behind it. (At first this
statement seems to undermine the distinction we have been making between
descriptive and inferential methods, but in fact this is not the case,
as we will see below.)

It is not difficult to demonstrate that it is possible to formulate any
conceivable community detection method as a particular inferential
method. Let us consider an arbitrary quality function
\begin{equation}
  W(\A, \bb) \in \mathbb{R}
\end{equation}
which is used to perform community detection via the optimization
\begin{equation}\label{eq:opt}
  \bb^* = \underset{\bb}{\operatorname{argmax}}\; W(\A, \bb).
\end{equation}
We can then interpret the quality function $W(\A, \bb)$ as the
``Hamiltonian'' of a posterior distribution
\begin{equation}
  P(\bb|\A) = \frac{\ee^{\beta W(\A,\bb)}}{Z(\A)},
\end{equation}
with normalization $Z(\A)=\sum_{\bb}\ee^{\beta W(\A,\bb)}$. By
making $\beta\to\infty$ we recover the optimization of Eq.~\ref{eq:opt},
or we may simply try to find the most likely partition according to the
posterior, in which case $\beta>0$ remains an arbitrary
parameter. Therefore, employing Bayes' rule in the opposite direction,
we obtain the following effective generative model:
\begin{align}
  P(\A|\bb)
  &= \frac{P(\bb|\A)P(\A)}{P(\bb)},\\
  &= \frac{\ee^{\beta W(\A,\bb)}}{Z(\A)}\frac{P(\A)}{P(\bb)},
\end{align}
where $P(\A) = \sum_{\bb}P(\A|\bb)P(\bb)$ is the marginal distribution
over networks, and $P(\bb)$ is the prior distribution for the
partition. Due to the normalization of $P(\A|\bb)$ we have the following
constraint that needs to be fulfilled:
\begin{equation}\label{eq:wconstraint}
  \sum_{\A}\frac{\ee^{\beta W(\A,\bb)}}{Z(\A)}P(\A) = P(\bb).
\end{equation}
Therefore, not all choices of $P(\A)$ and $P(\bb)$ are compatible with
the posterior distribution and the exact possibilities will depend on
the actual shape of $W(\A,\bb)$. However, one choice that is always
possible is a maximum-entropy one,
\begin{equation}\label{eq:eq_marg}
  P(\A) = \frac{Z(\A)}{\Xi},\qquad P(\bb) = \frac{\Omega(\bb)}{\Xi},
\end{equation}
with $\Omega(\bb)=\sum_{\A}\ee^{\beta W(\A,\bb)}$ and
$\Xi=\sum_{\A,\bb}\ee^{\beta W(\A,\bb)}$. Taking this choice leads to
the effective generative model
\begin{equation}\label{eq:eq_l}
  P(\A|\bb) = \frac{\ee^{\beta W(\A,\bb)}}{\Omega(\bb)}.
\end{equation}
Therefore, inferentially interpreting a community detection algorithm
with a quality function $W(\A,\bb)$ is equivalent to assuming the
generative model $P(\A|\bb)$ and prior $P(\bb)$ of Eqs.~\ref{eq:eq_l}
and \ref{eq:eq_marg} above. Furthermore, this also means that any arbitrary
community detection algorithm implies a description length given (in
nats) by\footnote{The description length of Eq.~\ref{eq:dl_W} is only
valid if there are no further parameters in the quality function
$W(\A,\bb)$ other than $\bb$ that are being optimized.}
\begin{equation}\label{eq:dl_W}
  \Sigma(\A,\bb) = -\beta W(\A,\bb) + \ln\sum_{\A',\bb'}\ee^{\beta W(\A',\bb')}.
\end{equation}
What the above results show is that there is no such thing as a
``model-free'' community detection method, since they are all equivalent
to the inference of \emph{some} generative model. The only difference to
a direct inferential method is that in that case the modelling
assumptions are made explicitly, inviting rather than preventing
scrutiny. Most often, the effective model and prior that are equivalent
to an \emph{ad hoc} community detection method will be difficult to
interpret, justify, or even compute (in general, Eq.~\ref{eq:dl_W}
cannot be written in closed form).

Furthermore there is no guarantee that the obtained description length
of Eq.~\ref{eq:dl_W} will yield a competitive or even meaningful
compression. In particular, there is no guarantee that this effective
inference will not overfit the data. Although we mentioned in the
previous section that inference and compression are equivalent, the
compression achieved when considering a particular generative model is
constrained by the assumptions encoded in its likelihood and prior. If
these are poorly chosen, no actual compression might be achieved, for
example when comparing to the one obtained with a maximally random
model. This is precisely what happens with descriptive community
detection methods: they overfit because their implicit modelling
assumptions do not accommodate the possibility that a network may be
maximally random, or contain a balanced mixture of structure and
randomness.

Since we can always interpret any community detection method as
inferential, is it still meaningful to categorize some methods as
descriptive? Arguably yes, because directly inferential approaches make
their generative models and priors explicit, while for a descriptive
method we need to extract them from reverse engineering. Explicit modelling
allows us to make judicious choices about the model and prior that
reflect the kinds of structures we want to detect, relevant scales or
lack thereof, and many other aspects that improve their performance in
practice, and our understanding of the results. With implicit
assumptions we are ``flying blind,'' relying substantially on
serendipity and trial-and-error --- not always with great success.

It is not uncommon to find criticisms of inferential methods due to a
perceived implausibility of the generative models used --- such as the
conditional independence of the placement of the edges present in the
SBM~\cite{schaub_many_2017} --- although these assumptions are also
present, but only \emph{implicitly}, in other methods, like modularity
maximization (see Sec.~\ref{sec:equivalence}). We discuss this issue
further in Sec.~\ref{sec:believe}.

The above inferential interpretation is not specific to community
detection, but is in fact valid for any learning problem. The set of
explicit or implicit assumptions that must come with any learning
algorithm is called an ``inductive bias.'' An algorithm is expected to
function optimally only if its inductive bias agrees with the actual
instances of the problems encountered. It is important to emphasize that
no algorithm can be free of an inductive bias, we can only chose
\emph{which} intrinsic assumptions we make about how likely we are to
encounter a particular kind of data, not \emph{whether} we are making an
assumption. Therefore, it is particularly problematic when a method does
not articulate explicitly what these assumptions are, since even if they
are hidden from view, they exist nonetheless, and still need to be
scrutinized and justified. This means we should be particularly
skeptical of the impossible claim that a learning method is
``model-free,'' since this denomination is more likely to signal an
inability or unwillingness to expose the underlying modelling
assumptions, which could potentially be revealed as unappealing and
fragile when eventually forced to come under scrutiny.

\subsection{Caveats and challenges with inferential methods}

Inferential community detection is a challenging task, and is not
without its caveats. One aspect they share with descriptive approaches
is algorithmic complexity (see Sec.~\ref{sec:performance}), and the fact
that they in general try to solve NP-hard problems. This means that
there is no known algorithm that is guaranteed to produce exact results
in a reasonable amount of time, except for very small networks. That
does not mean that every instance of the problem is hard to answer, in
fact it can be shown that in key cases robust answers can be
obtained~\cite{decelle_inference_2011}, but in general all existing
methods are approximative, with the usual trade-off between accuracy and
speed. The quest for general approaches that behave well while being
efficient is still ongoing and is unlikely to exhausted soon.

Furthermore, employing statistical inference is not a ``silver bullet''
that automatically solves every problem. If our models are
``misspecified,'' i.e. represent very poorly the structure present in
the data, then our inferences using them will be very limited and
potentially misleading (see Sec.~\ref{sec:believe}) --- the most we can
expect from our methodology in this case is to obtain good diagnostics
of when this is happening~\cite{peixoto_revealing_2021}. There is also a
typical trade-off between realism and simplicity, such that models that
more closely match reality are more difficult to express in simple terms
with tractable models. Usually, the more complex a model is, the more
difficult becomes its inference. The technical task of using algorithms
such as Markov chain Monte Carlo (MCMC) to produce reliable inferences
for a complex model is nontrivial and requires substantial expertise,
and is likely to be a long-living field of research.

In general it can be said that, although statistical inference does not
provide automatic answers, it gives us an invaluable platform where the
questions can be formulated more clearly, and allows us to navigate the
space of answers using more robust methods and theory.

\section{Modularity maximization considered harmful}\label{sec:modularity}

The most widespread method for community detection is modularity
maximization~\cite{newman_modularity_2006}, which happens also to be one
the most problematic. This method is based on the modularity function,
\begin{equation}\label{eq:Q}
  Q(\A,\bb) = \frac{1}{2E}\sum_{ij}\left(A_{ij} - \frac{k_ik_j}{2E}\right)\delta_{b_i,b_j},
\end{equation}
where $A_{ij}\in\{0,1\}$ is an entry of the adjacency matrix,
$k_i=\sum_jA_{ij}$ is the degree of node $i$, $b_i$ is the group
membership of node $i$, and $E$ is the total number of edges. The method
consists in finding the partition $\hat\bb$ that maximizes $Q(\A,\bb)$,
\begin{equation}\label{eq:qmax}
  \hat\bb = \underset{\bb}{\operatorname{argmax}}\; Q(\A,\bb).
\end{equation}
The motivation behind the modularity function is that it compares the
existence of an edge $(i,j)$ to the probability of it existing according
to a null model, $P_{ij} = k_ik_j/2E$, namely that of the configuration
model~\cite{fosdick_configuring_2018} (or more precisely, the Chung-Lu
model~\cite{chung_connected_2002}). The motivation for this method is
that we should consider a partition of the network meaningful if the
occurrence of edges between nodes of the same group exceeds what we
would expect with a random null model without communities.

Despite its widespread adoption, this approach suffers from a variety of
serious conceptual and practical flaws, which have been documented
extensively~\cite{guimera_modularity_2004,fortunato_resolution_2007,
good_performance_2010,fortunato_community_2010,fortunato_community_2016}. The
most problematic one is that it \emph{purports} to use an inferential
criterion --- a deviation from a null generative model --- but is in
fact merely descriptive. As has been recognized very early, this method
categorically fails in its own stated goal, since it always finds
high-scoring partitions in networks sampled from its own null
model~\cite{guimera_modularity_2004}. Indeed, the generative model we
used in Fig.~\ref{fig:descriptive}(a) is exactly the null model considered
in the modularity function, which if maximized yields the partition seen
in Fig.~\ref{fig:descriptive}(a). As we already discussed, this result
bears no relevance to the underlying generative process, and overfits
the data.

The reason for this failure is that the method does not take into
account the deviation from the null model in a statistically consistent
manner. The modularity function is just a re-scaled version of the
assortativity coefficient~\cite{newman_mixing_2003}, a correlation
measure of the community assignments seen at the endpoints of edges in
the network. We should expect such a correlation value to be close to
zero for a partition that is determined \emph{before} the edges of the
network are placed according to the null model, or equivalently, for a
partition chosen at random. However, it is quite a different matter to
find a partition that \emph{optimizes} the value of $Q(\A,\bb)$, after
the network is observed. The deviation from a null model computed in
Eq.~\ref{eq:Q} completely ignores the optimization step of
Eq.~\ref{eq:qmax}, although it is a crucial part of the algorithm. As a
result, the method of modularity maximization tends to massively
overfit, and find spurious communities even in networks sampled from its
null model. If we search for patterns of correlations in a random
graph, most of the time we will find them. This is a pitfall known
as ``data dredging'' or ``$p$-hacking,'' where one searches exhaustively
for different patterns in the same data and reports only those that are
deemed significant, according to a criterion that does not take into
account the fact that we are doing this search in the first place.

\begin{figure}[b]
  \resizebox{\textwidth}{!}{
  \begin{tabular}{ccc}
    & \smaller\hspace{2em} Random partition & \smaller\hspace{2em} Maximum modularity\\
    \includegraphicslp{(a)}{.98}{width=.42\textwidth}{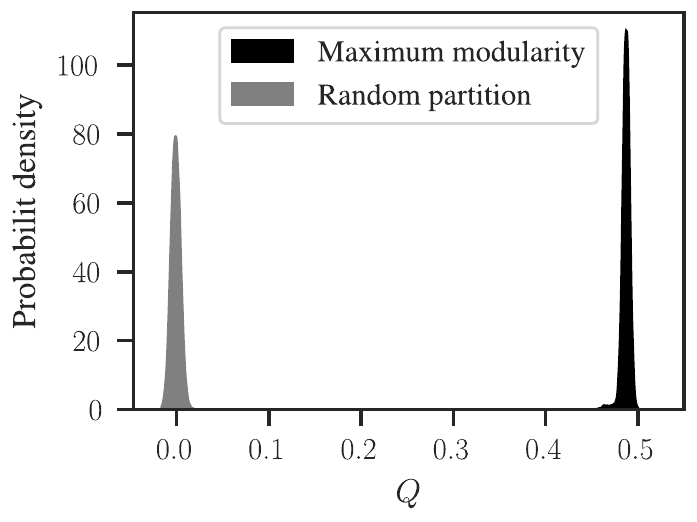}&
    \includegraphicslp{(b)}{.98}{width=.29\textwidth,trim=0 -.65cm 0 0}{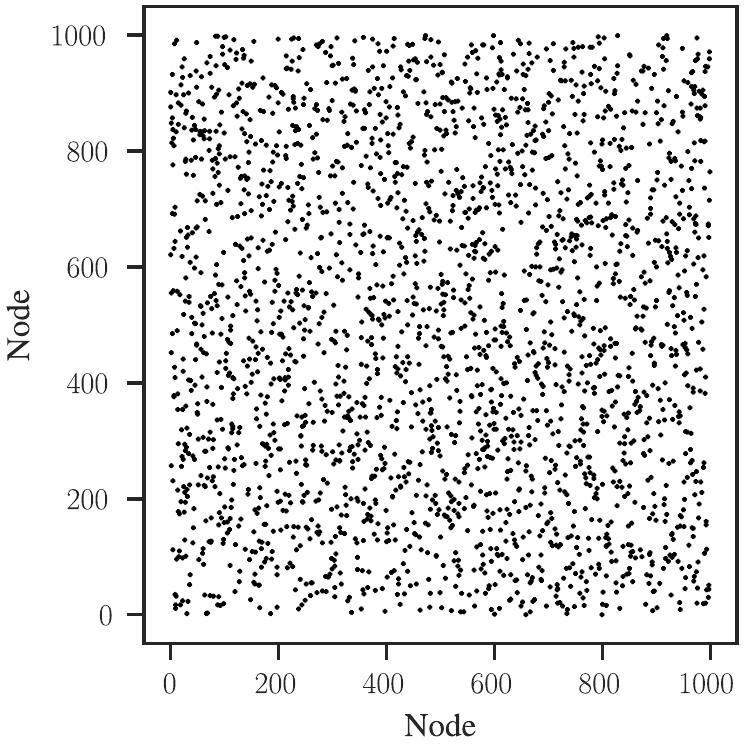}&
    \includegraphicslp{(c)}{.98}{width=.29\textwidth,trim=0 -.65cm 0 0}{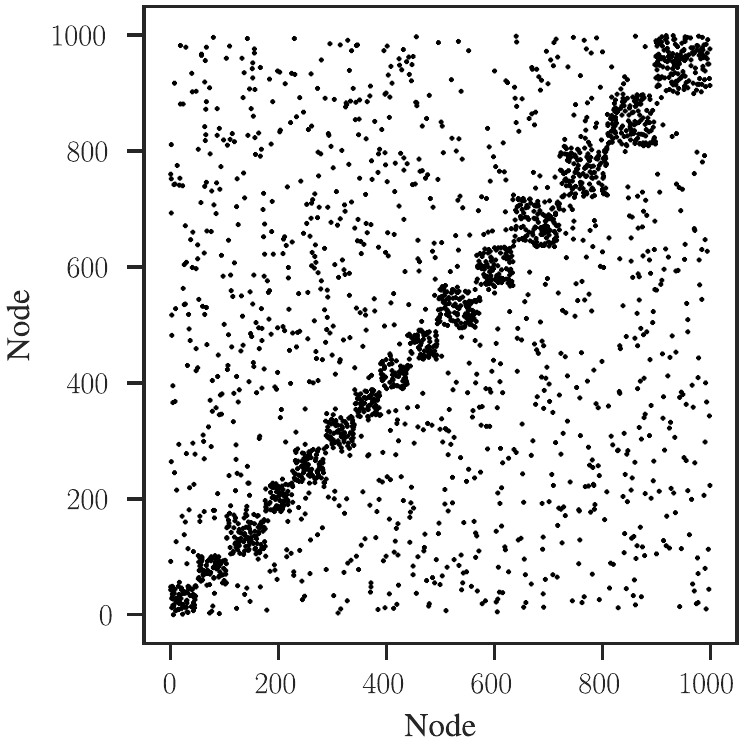}
  \end{tabular}} \caption{Modularity maximization systematically
  overfits, and finds spurious structures even its own null model. In
  this example we consider a random network model with $N=10^3$ nodes,
  with every node having degree $5$. (a) Distribution of modularity
  values for a partition into 15 groups chosen at random, and for the
  optimized value of modularity, for $5000$ networks sampled from the
  same model. (b) Adjacency matrix of a sample from the model, with the
  nodes ordered according to a random partition. (c) Same as (b), but
  with the nodes ordered according to the partition that maximizes
  modularity.\label{fig:randomQ}}
\end{figure}

We demonstrate this problem in Fig.~\ref{fig:randomQ}, where we show the
distribution of modularity values obtained with a uniform configuration
model with $k_i=5$ for every node $i$, considering both a random
partition and the one that maximizes $Q(\A,\bb)$. While for a random
partition we find what we would expect, i.e. a value of $Q(\A,\bb)$
close to zero, for the optimized partition the value is substantially
larger. Inspecting the optimized partition in Fig.~\ref{fig:randomQ}(c),
we see that it corresponds indeed to 15 seemingly clear assortative
communities --- which by construction bear no relevance to how the
network was generated. They have been dredged out of randomness by the
optimization procedure.

\begin{figure}[h!]
  \begin{tabular}{cc}
    Modularity maximization & SBM inference\\
    \includegraphicsl{(a)}{width=.5\textwidth}{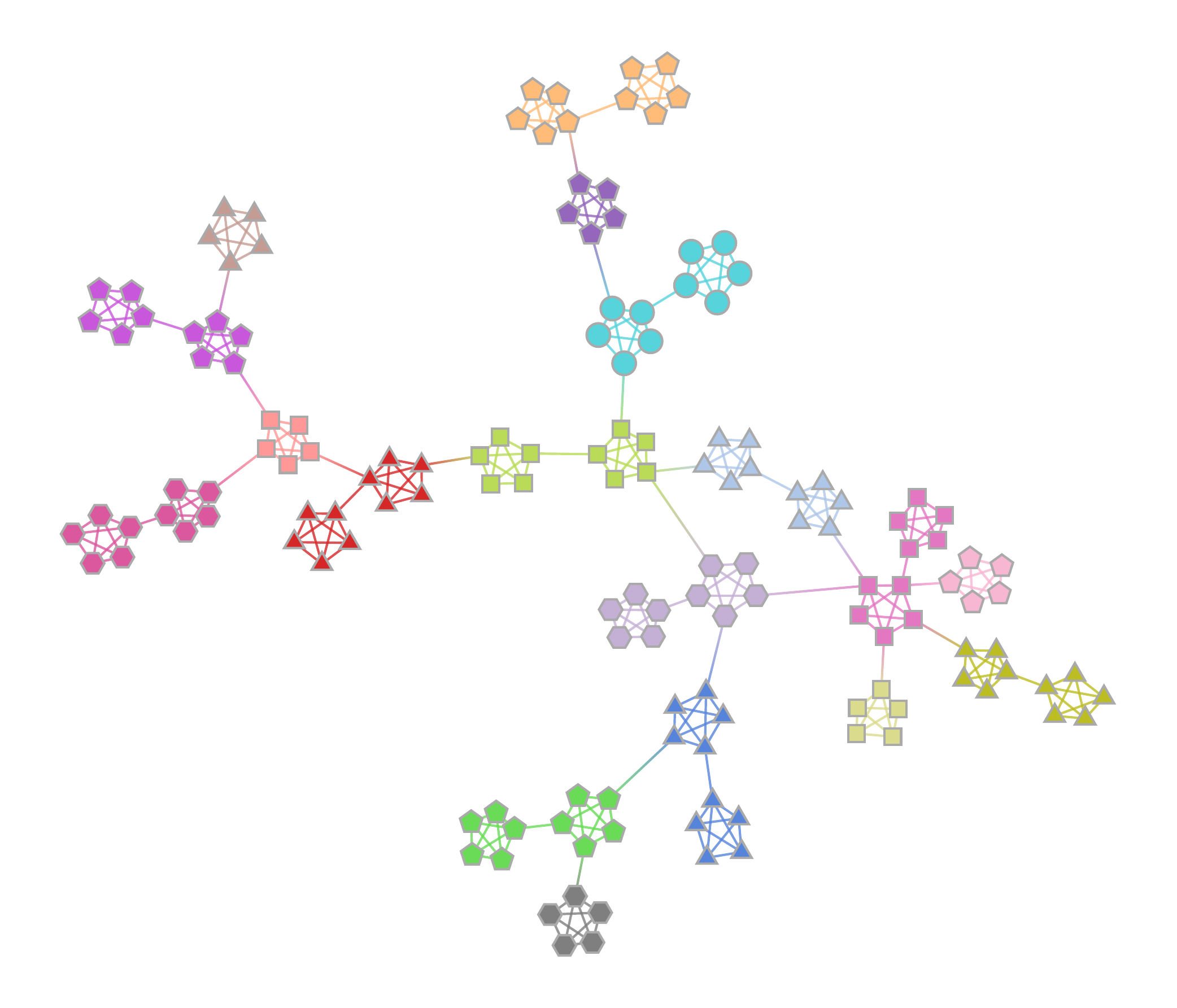}&
    \includegraphicsl{(b)}{width=.5\textwidth}{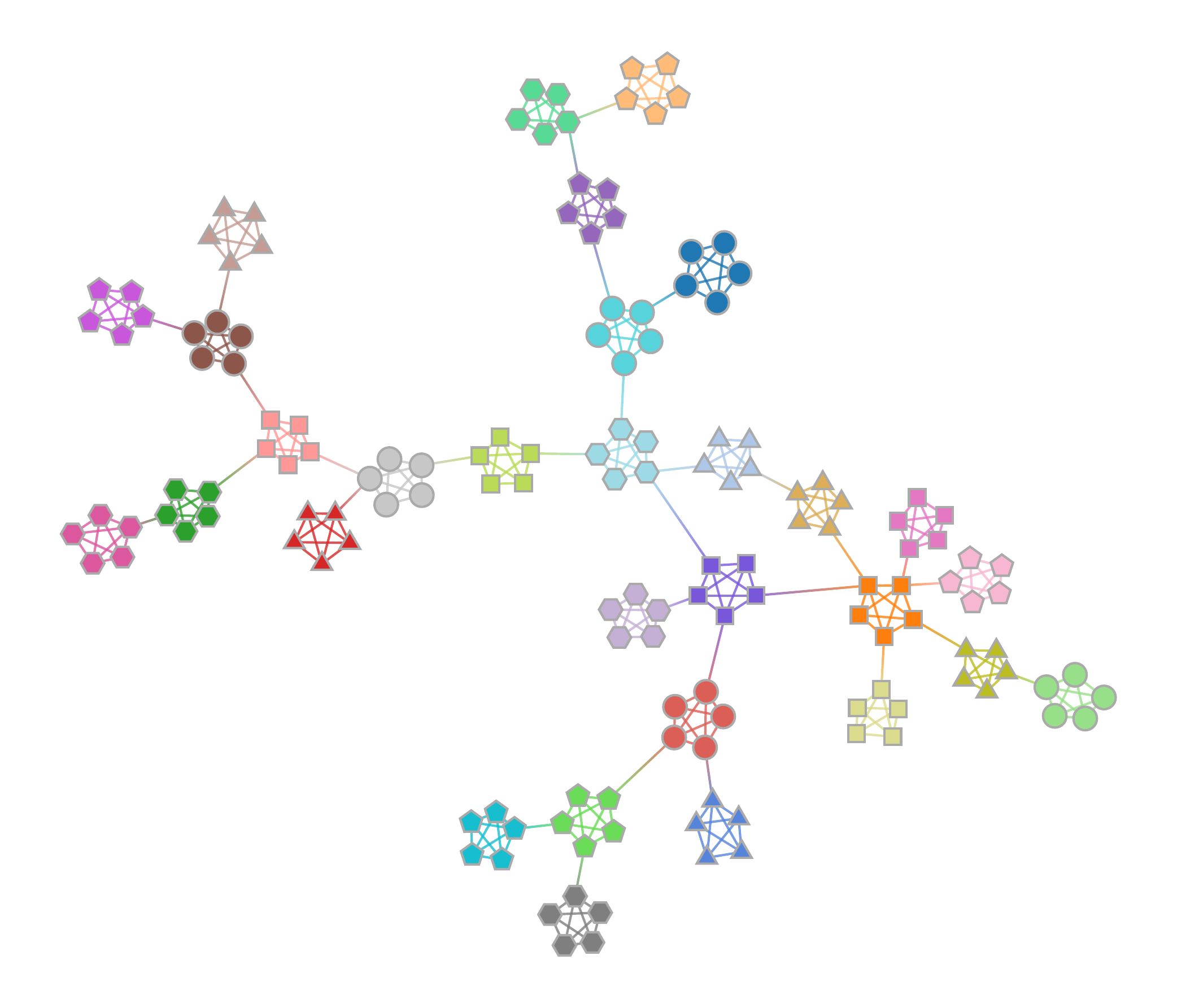}\\
    \includegraphicsl{(c)}{width=.5\textwidth}{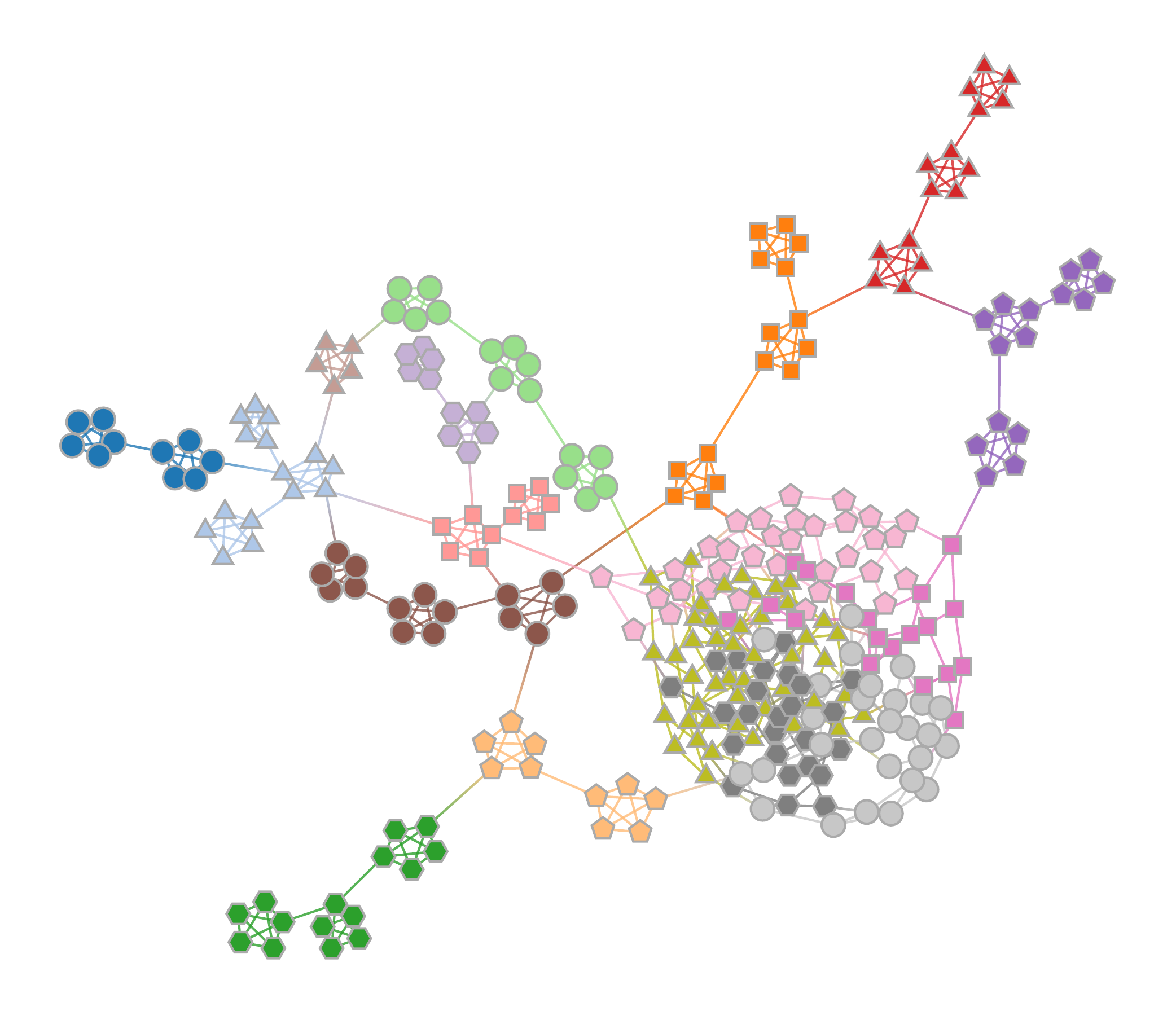}&
    \includegraphicsl{(d)}{width=.5\textwidth}{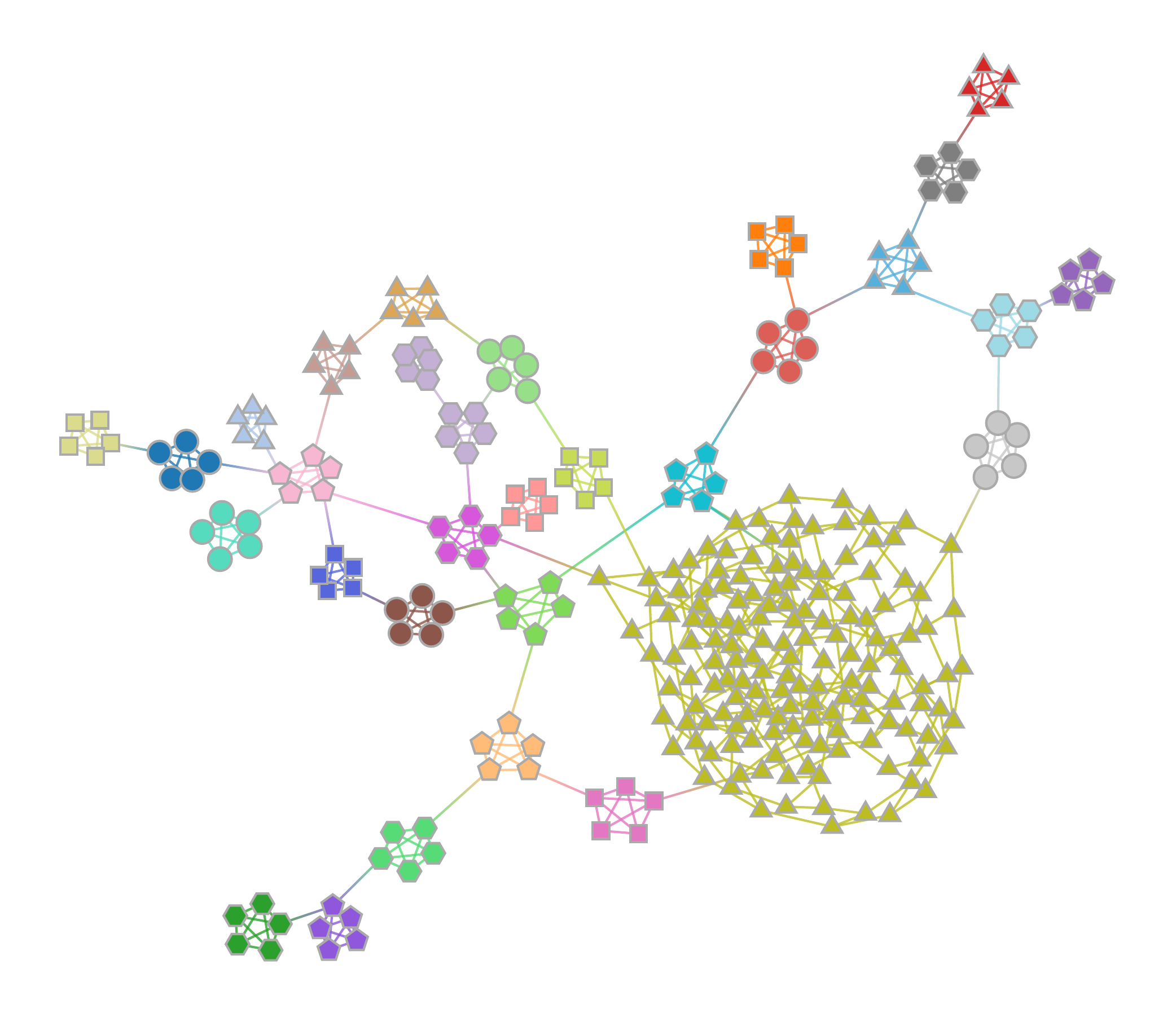}
  \end{tabular} \caption{The resolution limit of modularity maximization
  prevents small communities from being identified, even if there is
  sufficient statistical evidence to support them. Panel (a) shows a
  network with $B=30$ communities sampled from an assortative SBM
  parametrization. The colors indicate the $18$ communities found with
  modularity maximization, where several pairs of true communities are
  merged together. Panel (b) shows the inference result of an
  assortative SBM~\cite{zhang_statistical_2020}, recovering the true
  communities with perfect accuracy. Panels (c) and (d) show the
  results for a similar model where a larger community has been
  introduced. In (c) we see the results of modularity maximization,
  which not only merges the smaller communities together, but also
  splits the larger community into several spurious ones --- thus both
  underfitting and overfitting different parts of the network at the
  same time. In (d) we see the result obtained by inferring the SBM,
  which once again finds the correct answer.\label{fig:resolution}}
\end{figure}

Somewhat paradoxically, another problem with modularity maximization is
that in addition to systematically overfitting, it also systematically
\emph{underfits}. This occurs via the so-called \emph{resolution limit}:
in a connected network\footnote{Modularity maximization, like many
descriptive community detection methods, will always place connected
components in different communities. This is another clear distinction
with inferential approaches, since maximally random models --- without
latent community structure --- can generate disconnected networks if
they are sufficiently sparse. From an inferential point of view, it is
therefore incorrect to assume that every connected component must belong
to a different community.} the method cannot find more than $\sqrt{2E}$
communities~\cite{fortunato_resolution_2007}, even if they seem
intuitive or can be found by other methods. An example of this is shown
in Fig.~\ref{fig:resolution}, where for a network generated with the SBM
containing 30 communities, modularity maximization finds only 18, while
an inferential approach has no problems finding the true
structure. There are attempts to counteract the resolution limit by
introducing a ``resolution parameter'' to the modularity function, but
as we discuss in Sec.~\ref{sec:resolution} they are in general
ineffective.

These two problems --- overfitting and underfitting --- can occur in
tandem, such that portions of the network dominated by randomness are
spuriously revealed to contain communities, whereas other portions with
clear modular structure can have those obstructed. The result is a very
unreliable method to capture the structure of heterogeneous networks. We
demonstrate this in Fig.~\ref{fig:resolution}(c) and~(d)

In addition to these major problems, modularity maximization also often
possesses a degenerate landscape of solutions, with very different
partitions having similar values of
$Q(\A,\bb)$~\cite{good_performance_2010}. In these situations the
partition with maximum value of modularity can be a poor representative
of the entire set of high-scoring solutions and depend on idiosyncratic
details of the data rather than general patterns --- which can be
interpreted as a different kind of overfitting.\footnote{This kind of
degeneracy in the solution landscape can also occur in an inferential
setting~\cite{riolo_consistency_2020,peixoto_revealing_2021}. However, there it
can be interpreted as the existence of competing hypotheses for the same data,
whose relative plausibility can be quantitatively assessed via their
posterior probability. In case the multiplicity of alternative
hypotheses is too large, this would be indicative of poor fit, or a
misspecification of the model, i.e. a general inadequacy of the model
structure to capture the structure in the data for any possible choice
of parameters.}

\begin{figure}
  \begin{tabular}{ccc}
    \includegraphicslp{(a)}{.99}{width=.33\textwidth}{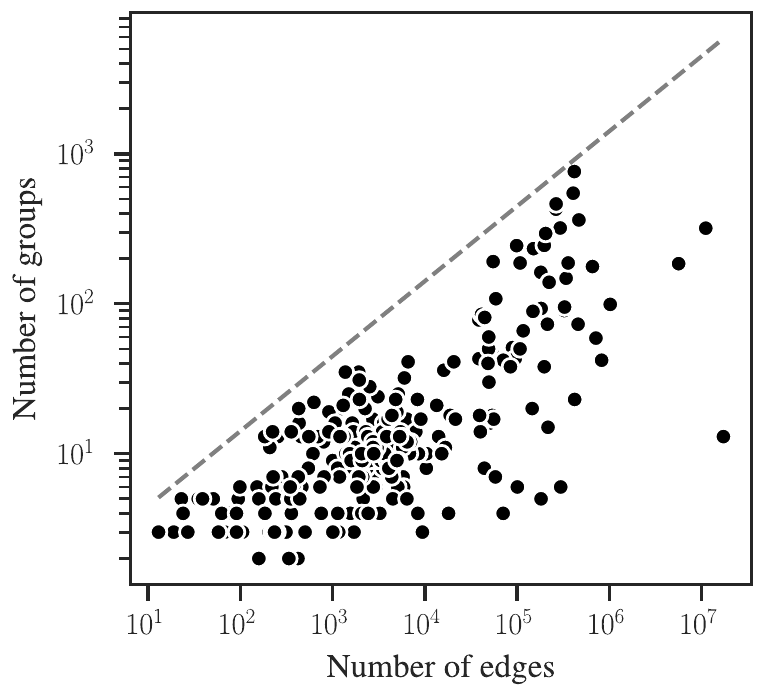}&
    \includegraphicslp{(b)}{.99}{width=.33\textwidth}{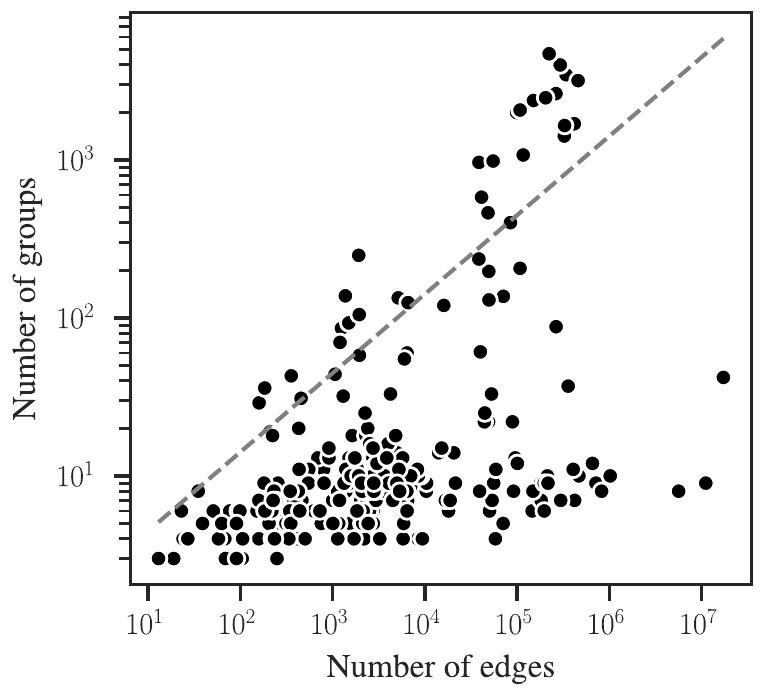}&
    \includegraphicslp{(c)}{.99}{width=.33\textwidth}{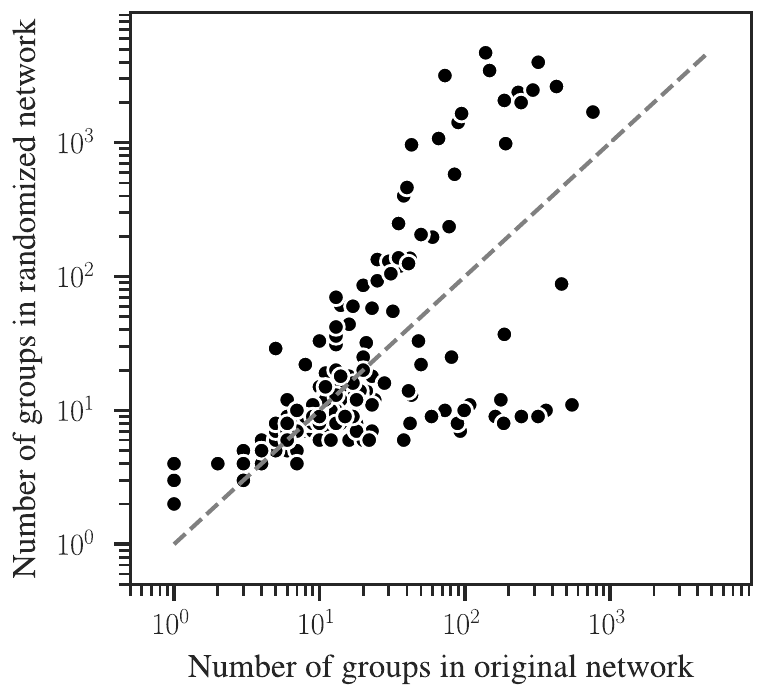}\\
    \smaller Original networks &
    \smaller Randomized networks &
  \end{tabular} \caption{Modularity maximization incorporates a
  substantial amount of noise into its results. (a) Number of groups found using
  modularity maximization for 263 empirical networks as a function of
  the number of edges. The dashed line corresponds to the $\sqrt{2E}$
  upper bound due to the resolution limit.
  (b) The same as in (a) but with randomized versions of each
  network. (c) Correspondence between the number of groups of the
  original and randomized network. The dashed line shows the diagonal. \label{fig:Qrand}}
\end{figure}

The combined effects of underfitting and overfitting can make the
results obtained with the method unreliable and difficult to
interpret. As a demonstration of the systematic nature of the problem,
in Fig.~\ref{fig:Qrand}(a) we show the number of communities obtained
using modularity maximization for 263 empirical networks of various
sizes and belonging to different domains~\cite{zhang_preparation},
obtained from the Netzschleuder
catalogue~\cite{peixoto_netzschleuder_2020}. Since the networks
considered are all connected, the values are always below $\sqrt{2E}$,
due to the resolution limit; but otherwise they are well distributed
over the allowed range. However, in Fig.~\ref{fig:Qrand}(b) we show the
same analysis, but for a version of each network that is fully
randomized, while preserving the degree sequence. In this case, the
number of groups remains distributed in the same range (sometimes even
exceeding the resolution limit, because the randomized versions can end
up disconnected). As Fig.~\ref{fig:Qrand}(c) shows, the number of groups
found for the randomized networks is strongly correlated with the
original ones, despite the fact that the former have no latent community
structure. This is a strong indication of the substantial amount of
noise that is incorporated into the partitions found with the method.

The systematic overfitting of modularity maximization --- as well as
other descriptive methods such as Infomap --- has been also
demonstrated recently in Ref.~\cite{ghasemian_evaluating_2019}, from the
point of view of edge prediction, on a separate empirical dataset of 572
networks from various domains.

Although many of the problems with modularity maximization were long
known, for some time there were no principled solutions to them, but
this is no longer the case. In the table below we summarize some of the
main problems with modularity and how they are solved with inferential
approaches.

%\newpage
\begin{longtable}{p{.45\textwidth}@{\hskip 1em}p{.45\textwidth}}
  \textbf{Problem} & \textbf{Principled solution via inference}\\ \hline Modularity maximization
  overfits, and finds modules in maximally random
  networks.~\cite{guimera_modularity_2004} & Bayesian inference of the
  SBM is designed from the ground up to avoid this problem in a
  principled way and systematically succeeds~\cite{peixoto_bayesian_2019}.\\[1em]

  Modularity maximization has a resolution limit, and finds at most
  $\sqrt{2E}$ groups in connected
  networks~\cite{fortunato_resolution_2007}. & Inferential approaches
  with hierarchical
  priors~\cite{peixoto_hierarchical_2014,peixoto_nonparametric_2017} or
  strictly assortative structures~\cite{zhang_statistical_2020} do not
  have any appreciable resolution limit, and can find a maximum number
  of groups that scales as $O(N/\log N)$. Importantly, this is achieved
  without sacrificing the robustness against overfitting.\\[1em]

  Modularity maximization has a characteristic scale, and tends to find
  communities of similar size; in particular with the same sum of
  degrees (see Sec.~\ref{sec:resolution}). & Hierarchical priors can be
  specifically chosen to be \emph{a priori} agnostic about
  characteristic sizes, densities of groups and degree
  sequences~\cite{peixoto_nonparametric_2017}, such that these are not
  imposed, but instead obtained from inference, in an unbiased way.\\[1em]

  Modularity maximization can only find strictly assortative
  communities. & Inferential approaches can be based on any generative
  model. The general SBM will find any kind of mixing pattern in an
  unbiased way, and has no problems identifying modular structure in
  bipartite networks, core-periphery networks, and any mixture of these
  or other patterns. There are also specialized versions for
  bipartite~\cite{larremore_efficiently_2014},
  core-periphery~\cite{zhang_identification_2015}, and assortative
  patterns~\cite{zhang_statistical_2020}, if these are being searched
  exclusively.  \\[1em]

  The solution landscape of modularity maximization is often degenerate,
  with many different solutions with close to the same modularity
  value~\cite{good_performance_2010}, and with no clear way of how to
  select between them. & Inferential methods are characterized by a
  posterior distribution of partitions. The consensus or dissensus
  between the different solutions~\cite{peixoto_revealing_2021} can be
  used to determine how many cohesive hypotheses can be extracted from
  inference, and to what extent is the model being used a poor or a good
  fit for the network.

  \\\hline
\end{longtable}

Because of the above problems, the use of modularity maximization should
be discouraged, since it is demonstrably not fit for purpose as an
inferential method. As a consequence, the use of modularity maximization
in any recent network analysis that relies on inferential conclusions
can be arguably considered a ``red flag'' that strongly indicates
methodological inappropriateness. In the absence of secondary evidence
supporting the alleged community structures found, or extreme care to
counteract the several limitations of the method (see
Secs.~\ref{sec:consensus}, \ref{sec:significance}
and~\ref{sec:resolution} for how typical attempts usually fail), the
safest assumption is that the results obtained with that method tend to
contain a substantial amount of noise, rendering any inferential
conclusion derived from them highly suspicious.

As a final note, we focus on modularity here not only for its widespread
adoption but also because of its emblematic character. At a fundamental
level, all of its shortcoming are shared with any descriptive method in
the literature --- to varied but always non-negligible degrees.

\section{Myths, pitfalls, and half-truths}

In this section we focus on assumed or asserted statements about how to
circumvent pitfalls in community detection, which are in fact better
characterized as myths or half-truths, since they are either misleading,
or obstruct a more careful assessment of the true underlying nature of
the problem.

\subsection{``Modularity maximization and SBM inference are equivalent methods.''}\label{sec:equivalence}

As we have discussed in Sec.~\ref{sec:implicit}, it is possible to
interpret \emph{any} community detection algorithm as the inference of
\emph{some} generative model. Because of this, the mere fact that an
equivalence with an inferential approach exists cannot be used to
justify the inferential use of a descriptive method, or to use it as a
criterion to distinguish between approaches that are statistically
principled or not. To this aim, we need to ask instead whether the
modelling assumptions that are \emph{implicit} in the descriptive
approach can be meaningfully justified, and whether they can be used to
consistently infer structures from networks.

Some recent works have detailed some specific equivalences of modularity
maximization with statistical
inference~\cite{zhang_scalable_2014,newman_equivalence_2016}. As we will
discuss below, these equivalences are far more limited than commonly
interpreted. They serve mostly to understand in more detail the reasons
why modularity maximization fails as a reliable method, but do not
prevent it from failing --- they expose more clearly its sins, but offer
no redemption.

We start with a very interesting connection revealed by Zhang and
Moore~\cite{zhang_scalable_2014} between the effective posterior
distribution we obtain when using the modularity function as a
Hamiltonian,
\begin{equation}\label{eq:Qgibbs}
  P(\bb|\A) = \frac{\ee^{\beta E Q(\A,\bb)}}{Z(\A)},
\end{equation}
and the posterior distribution of the strictly assortative DC-SBM, which
we refer here as the degree-corrected planted partition model (DC-PP),
\begin{equation}\label{eq:dcpp_post}
  P(\bb|\A,\omega_{\text{in}},\omega_{\text{out}},\bm{\theta}) =
  \frac{P(\A|\omega_{\text{in}},\omega_{\text{out}},\bm{\theta},\bb)P(\bb)}
  {P(\A|\omega_{\text{in}},\omega_{\text{out}},\bm{\theta})},
\end{equation}
which has a likelihood given by
\begin{equation}\label{eq:dcpp}
  P(\A|\omega_{\text{in}},\omega_{\text{out}},\bm{\theta},\bb)
  = \prod_{i<j}\frac{\ee^{-\omega_{b_i,b_j}\theta_i\theta_j}\left(\omega_{b_i,b_j}\theta_i\theta_j\right)^{A_{ij}}}{A_{ij}!},
\end{equation}
where
\begin{equation}
  \omega_{rs} = \omega_{\text{in}}\delta_{rs} + \omega_{\text{out}}(1-\delta_{rs}).
\end{equation}
This model assumes that there are constant rates $\omega_{\text{in}}$
and $\omega_{\text{out}}$ controlling the number of edges that connect
to nodes of the same and different communities, respectively. In
addition, each node has its own propensity $\theta_i$, which determines
the relative probability it has of receiving an edge, such that nodes
inside the same community are allowed to have very different
degrees. This is a far more restrictive version of the full DC-SBM we
considered before, since it not only assumes assortativity as the only
mixing pattern, but also that all communities share the same rate
$\omega_{\text{in}}$, which imposes a rather unrealistic similarity
between the different groups.

Before continuing, it is important to emphasize that the posterior of
Eq.~\ref{eq:dcpp_post} corresponds to the situation where the number of
communities and all parameters of the model, except the partition
itself, are known \emph{a priori}. This does not correspond to any
typical empirical setting where community detection is employed, since
we do not often have such detailed information about the community
structure, and in fact no good reason to even use this particular
parametrization to begin with. The equivalences that we are about to
consider apply only in very idealized scenarios, and are not expected to
hold in practice.

\begin{figure}
  \begin{tabular}{cc}
    \includegraphicsl{(a)}{width=.5\textwidth}{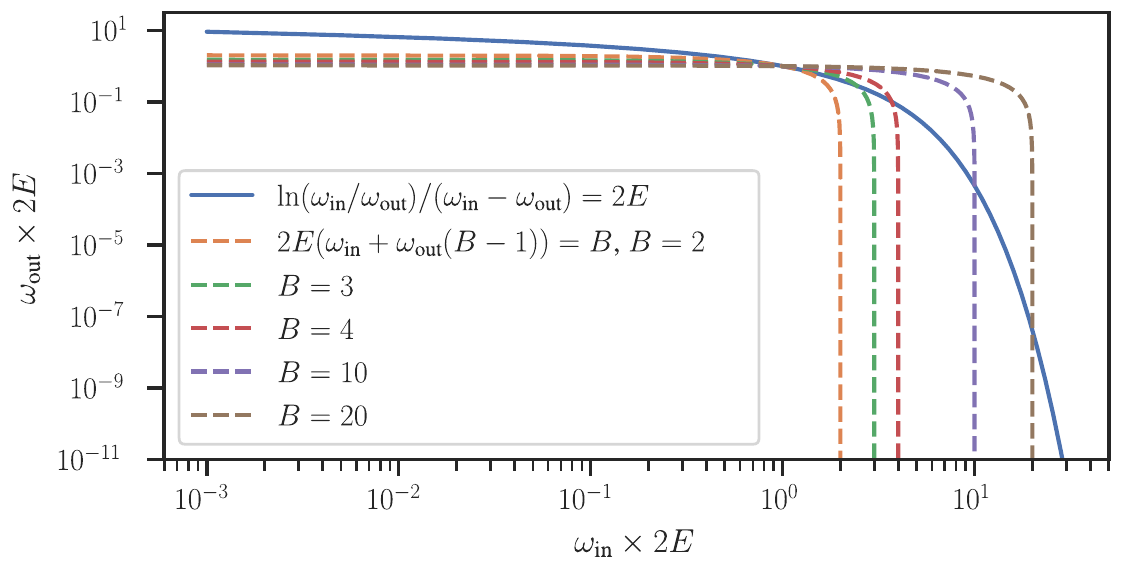}&
    \includegraphicsl{(b)}{width=.5\textwidth}{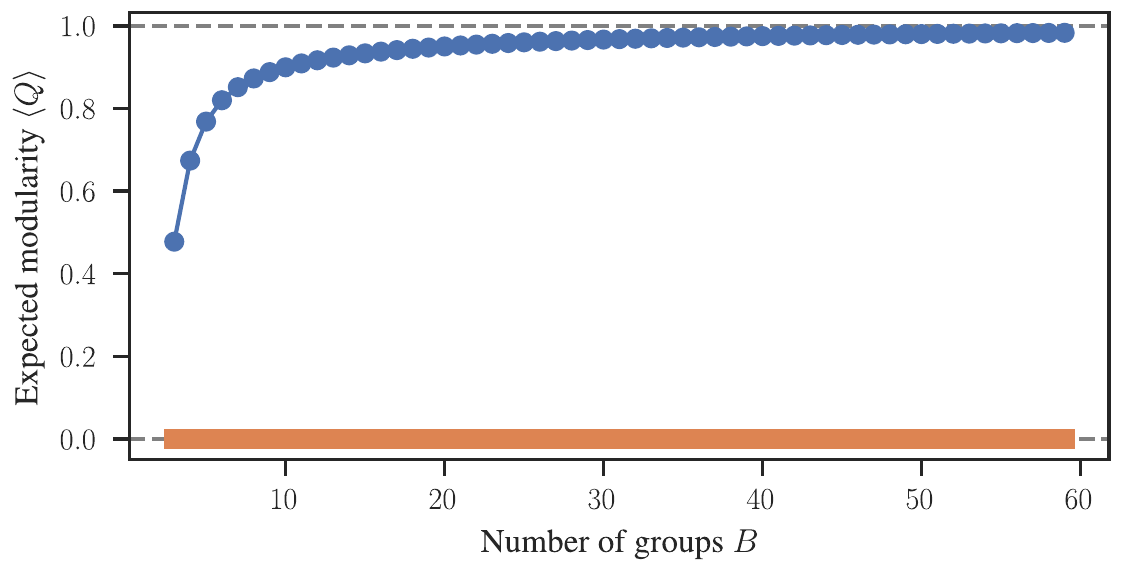}

  \end{tabular} \caption{Using modularity maximization is equivalent to
  performing a maximum likelihood estimate of the DC-PP model with very
  specific parameter choices, that depend on the number of edges $E$ in
  the network and the number of communities $B$. In (a) we show the
  valid choices of $\omega_{\text{in}}$ and $\omega_{\text{out}}$
  obtained when the solid and dashed lines cross, corresponding
  respectively to Eqs.~\ref{eq:PPQ} and~\ref{eq:PPE}, where we can see
  that for $B=2$ no solution is possible where the expected modularity
  is positive. In (b) we show the two possible values for the expected
  modularity that are consistent with the implicit model assumptions, as
  a function of the number of groups.\label{fig:Qconstraint}}
\end{figure}

Taking the logarithm of both sides of Eq.~\ref{eq:dcpp}, and ignoring
constant terms with respect to the model parameters we have
\begin{multline}\label{eq:pp_L}
  \ln P(\A|\omega_{\text{in}},\omega_{\text{out}},\bm{\theta},\bb)
  = \left(\ln \frac{\omega_{\text{in}}}{\omega_{\text{out}}}\right)
  \left[\sum_{i<j}\left(A_{ij}-\frac{\omega_{\text{in}}-\omega_{\text{out}}}
    {\ln (\omega_{\text{in}}/\omega_{\text{out}})}\theta_i\theta_j\right)\delta_{b_i,b_j}\right] +\\
  \sum_{i<j}\left[A_{ij}\ln(\theta_i\theta_j\omega_{\text{out}})-\theta_i\theta_j\omega_{\text{out}}\right].
\end{multline}
Therefore, ignoring additive terms that do not depend on $\bb$ (since
they become irrelevant after normalization in Eq.~\ref{eq:Qgibbs}) and
making the arbitrary choices (we will inspect these in detail soon),
\begin{equation}
  \beta = \ln \frac{\omega_{\text{in}}}{\omega_{\text{out}}},\qquad \frac
    {\ln (\omega_{\text{in}}/\omega_{\text{out}})}{\omega_{\text{in}}-\omega_{\text{out}}} = 2E, \qquad \theta_i=k_i,
\end{equation}
we obtain the equivalence,
\begin{equation}
  \ln P(\A|\omega_{\text{in}},\omega_{\text{out}},\bm{\theta},\bb) =  \beta E Q(\A,\bb),
\end{equation}
allowing us to equate Eqs.~\ref{eq:Qgibbs} and~\ref{eq:dcpp_post} (there
is a methodological problem with the choice $\theta_i=k_i$, as we will
see later, but we will ignore this for the time being).  Therefore, for
particular choices of the model parameters, one recovers modularity
optimization from the maximum likelihood estimation of the DC-PP model
with respect to $\bb$. Indeed, this allows us to understand more clearly
what \emph{implicit} assumptions go behind using modularity for
inferential aims.  For example, besides making very specific prior
assumptions about the model parameters $\omega_{\text{in}}$,
$\omega_{\text{out}}$ and $\bm{\theta}$, this posterior also assumes
that all partitions are equally likely \emph{a priori},
\begin{equation}
  P(\bb) \propto 1.
\end{equation}
We can in fact write this uniform prior more precisely as
\begin{equation}
  P(\bb) = \left[\sum_{B=1}^{N}\genfrac\{\}{0pt}{}{N}{B}B!\right]^{-1},
\end{equation}
where $\genfrac\{\}{0pt}{}{N}{B}B!$ is the number of labelled partitions
of a set $N$ into $B$ groups. This number reaches a maximum at $B\approx
.72\times N$, and decays fast from there, meaning that such a uniform
prior is in fact very concentrated on a very large number of groups ---
partially explaining the tendency of the modularity posterior to
overfit. Let us examine now the prior assumption
\begin{equation}\label{eq:PPQ}
 \frac{\ln (\omega_{\text{in}}/\omega_{\text{out}})}{\omega_{\text{in}}-\omega_{\text{out}}} = 2E.
\end{equation}
For any value of $E$ the above condition admits many solutions. However,
not all of them are consistent with the \emph{expected} number of edges
in the network according to the DC-PP model. Assuming, for simplicity,
that all $B$ groups have the same size $N/B$, and that all nodes have
the same degree $2E/N$, then the expected number of edges according to
the assumed DC-PP model is given by
\begin{equation}
  2\avg{E} =  (2E)^2\left(\frac{\omega_{\text{in}}}{B} + \frac{\omega_{\text{out}}(B-1)}{B}\right).
\end{equation}
Equating the expected with the observed value, $\avg{E}=E$, leads to
\begin{equation}\label{eq:PPE}
  \omega_{\text{in}} + \omega_{\text{out}}(B-1) = \frac{B}{2E}.
\end{equation}
Combining Eqs.~\ref{eq:PPQ} and~\ref{eq:PPE} gives us at most only two
values of $\omega_{\text{in}}$ and $\omega_{\text{out}}$ that are
compatible with the expected density of the network and the modularity
interpretation of the likelihood, as seen in
Fig.~\ref{fig:Qconstraint}(a), and therefore only two possible values
for the expected modularity, computed as
\begin{equation}
  \avg{Q} = \frac{1}{B}\left(2E\omega_{\text{in}} - 1\right).
\end{equation}
One possible solution is always
$\omega_{\text{in}}=\omega_{\text{out}}=1/2E$, which leads to
$\avg{Q}=0$. The other solution is only possible for $Q>2$, and yields a
specific expected value of modularity which approaches $\avg{Q}\to 1$ as
$B$ increases (see Fig.~\ref{fig:Qconstraint}(b)). This yields an
implausibly narrow range for the consistency of modularity maximization
with the inference of the DC-PP model. The bias towards larger values of
$Q(\A,\bb)$ as the number of groups increases is not an inherent
property of the DC-PP model, as it accommodates any expected value of
modularity by properly choosing its parameters. Instead, this is an
arbitrary implicit assumption baked in $Q(\A,\bb)$, which further
explains why maximizing it will tend to find many groups even on random
networks.

In a later work~\cite{newman_equivalence_2016}, Newman relaxed the above
connection with modularity by using instead its generalized
version~\cite{reichardt_statistical_2006,arenas_analysis_2008},
\begin{equation}\label{eq:Qgamma}
  Q(\A,\bb,\gamma) = \frac{1}{2E}\sum_{ij}\left(A_{ij} - \gamma\frac{k_ik_j}{2E}\right)\delta_{b_i,b_j},
\end{equation}
where $\gamma$ is the so-called ``resolution'' parameter. With this
additional parameter, we have more freedom about the implicit
assumptions of the DC-PP model. Newman in fact showed that if we make
the choices,
\begin{equation}\label{eq:choices}
  \beta = \ln \frac{\omega_{\text{in}}}{\omega_{\text{out}}},\qquad
  \gamma=\frac{\omega_{\text{in}}-\omega_{\text{out}}}
    {\ln (\omega_{\text{in}}/\omega_{\text{out}})}, \qquad \theta_i =\frac{k_i}{\sqrt{2E}},
\end{equation}
then we recover the Gibbs distribution with the generalized modularity
from the DC-PP likelihood of Eq.~\ref{eq:dcpp}. Due to the independent
parameter $\gamma$, now the assumed values of $\omega_{\text{in}}$ and
$\omega_{\text{out}}$ are no longer constrained by $E$ alone, and can
take any value. Therefore, if we knew the correct value of the model
parameters, we could use them to choose the appropriate value of
$\gamma$ and hence maximize $Q(\A, \bb, \gamma)$, yielding the same
answer as maximizing $\ln
P(\A|\omega_{\text{in}},\omega_{\text{out}},\bm{\theta},\bb)$ with the
same parameters.

There are, however, serious problems remaining that prevent this
equivalence from being true in general, or in fact even typically. For
the equivalence to hold, we need the number of groups $B$ and all
parameters to be known a priori and to be equal to
Eq.~\ref{eq:choices}. However, the choice $\theta_i = k_i/\sqrt{2E}$
involves information about the observed network, namely the actual
degrees seen --- and therefore is not just a prior assumption, but one
done \emph{a posteriori}, and hence needs to be justified via a
consistent estimation that respects the likelihood principle. When we
perform a maximum likelihood estimate of the parameters
$\omega_{\text{in}}$, $\omega_{\text{out}}$, and $\bm\theta$, we obtain
the following system of nonlinear
equations~\cite{zhang_statistical_2020},
\begin{align}
  \omega_{\text{in}}^* &= \frac{2e_{\text{in}}}{\sum_r\hat\theta_r^2}\label{eq:mle_lin}\\
  \omega_{\text{out}}^* &= \frac{e_{\text{out}}}{\sum_{r<s}\hat\theta_r\hat\theta_s}\label{eq:mle_lout}\\
  \theta_i^* &= k_i \left[\frac{2e_{\text{in}}\hat\theta_{b_i}}{\sum_r\hat\theta_r^2}+\frac{e_{\text{out}}\sum_{r\neq b_i}\hat\theta_r}{\sum_{r<s}\hat\theta_r\hat\theta_s}\right]^{-1},\label{eq:mle_theta}
\end{align}
where $e_{\text{in}}=\sum_{i<j}A_{ij}\delta_{b_i,b_j}$,
$e_{\text{out}}=\sum_{i<j}A_{ij}(1-\delta_{b_i,b_j})$, and
$\hat\theta_r=\sum_i\theta_i^*\delta_{b_i,r}$. The above system admits
$\theta^*_i = k_i/\sqrt{2E}$ as a solution only if the following
condition is met for every group $r$:
\begin{equation}\label{eq:symmetry}
  \sum_ik_i\delta_{b_i,r} = \frac{2E}{B}.
\end{equation}
In other words, the sum of degrees inside each group must be the same
for every group. Note also that the expected degrees according to the
DC-PP model will be inconsistent with Eq.~\ref{eq:choices} if the above
condition is not met, i.e.
\begin{equation}
  \avg{k_i} = \theta_i\left[\omega_{\text{in}}\sum_j\theta_j\delta_{b_i,b_j} + \omega_{\text{out}}\sum_j\theta_j(1-\delta_{b_i,b_j})\right].
\end{equation}
Substituting $\theta_i=k_i/\sqrt{2E}$ in the above equation will yield
in general $\avg{k_i}\ne k_i$, as long as Eq.~\ref{eq:symmetry} is not
fulfilled, \emph{regardless of how we choose $\omega_{\text{in}}$ and
$\omega_{\text{out}}$}.

\begin{figure}
  \begin{tabular}{cc}
    \includegraphicsl{(a)}{width=.5\textwidth}{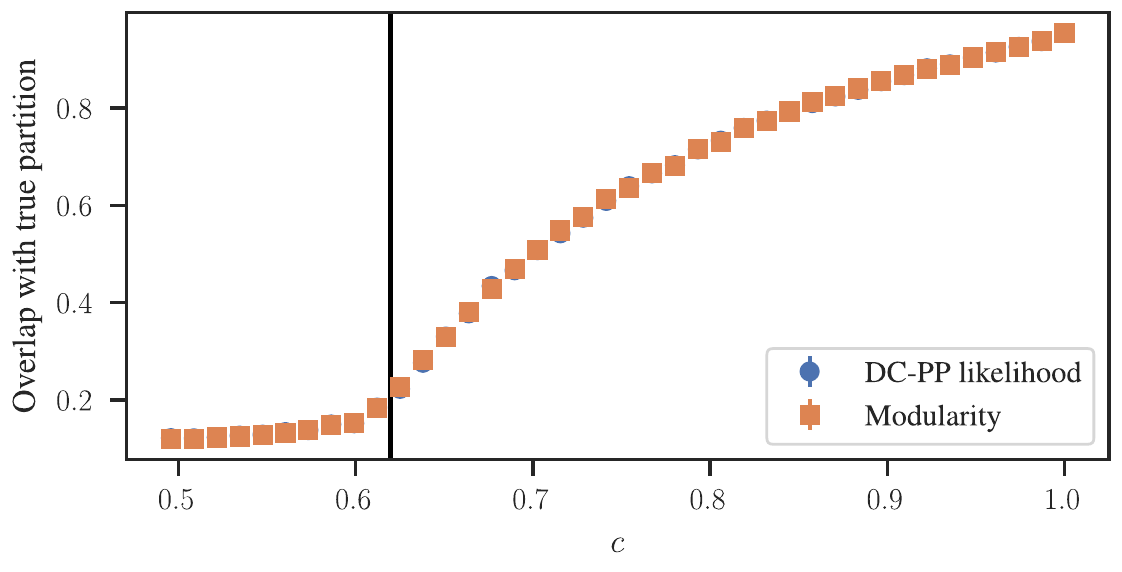}&
    \includegraphicsl{(b)}{width=.5\textwidth}{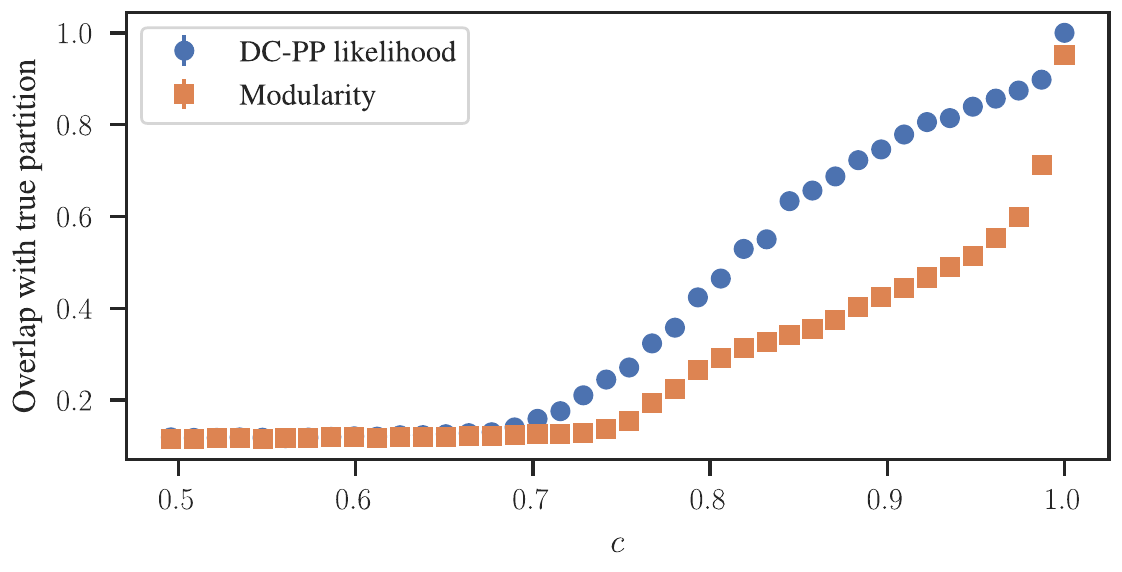}
  \end{tabular}
  \caption{Generalized modularity and the DC-PP model are only
    equivalent if the symmetry of Eq.~\ref{eq:symmetry} is
    preserved. Here we consider an instance of the DC-PP model with
    $\omega_{\text{in}}=2Ec/N$, $\omega_{\text{out}}=2E(1-c)/\sum_{r\ne
    s}\sqrt{n_rn_s}$, and $\theta_i=1/\sqrt{n_{b_i}}$, where $n_r$ is
    the number of nodes in group $r$. The parameter $c\in[0,1]$ controls
    the degree of assortativity. For non-uniform group sizes, the
    symmetry of Eq.~\ref{eq:symmetry} is not preserved with this choice
    of parameters. We use the parametrization $n_r =
    N\alpha^{r-1}(1-\alpha)/(1-\alpha^B)$, where $\alpha > 0$ controls
    the group size heterogeneity. When employing generalized modularity,
    we choose the closest possible parameter choice with
    $\omega_{\text{in}}=2Ec/({\sum_re_r^2/2E})$ and
    $\omega_{\text{out}}=2E(1-c)/(2E - {\sum_re_r^2/2E})$, where
    $e_r=\sum_ik_i\delta_{b_i,r}$. In (a) we show the inference results
    for the uniform case with $\alpha\to 1$, where both approaches are
    identical, performing equally well all the way down to the
    detectability threshold~\cite{decelle_asymptotic_2011} (vertical
    line). In (b) we show the result with $\alpha=2$, which leads to
    unequal group sizes, causing the behavior between both approaches to
    diverge. In all cases we consider averages over 5 networks with
    $N=10^4$ nodes, average degree $2E/N=3$, and $B=10$ groups.
    \label{fig:equivalence}}
\end{figure}

Framing it differently, for any choice of $\omega_{\text{in}}$,
$\omega_{\text{out}}$ and $\bm\theta$ such that the sums
$\sum_i\theta_i\delta_{b_i,r}$ are not identical for every group $r$,
the DC-SBM likelihood $\ln
P(\A|\omega_{\text{in}},\omega_{\text{out}},\bm{\theta},\bb)$ is not
captured by $Q(\A,\bb,\gamma)$ for any value of $\gamma$, and therefore
maximizing both functions will not yield the same results. That is, the
equivalence is only valid for special cases of the model \emph{and}
data. We show in Fig.~\ref{fig:equivalence} an example of an instance of
the DC-PP model where the generalized modularity yields results which
are inconsistent with using likelihood of the DC-PP model directly.

Because of the above caveats, we have to treat the claimed equivalence
with a grain of salt. In general there are only three scenarios we may
consider when analysing a network:
\begin{enumerate}
\item We know that the network has been sampled from the DC-PP model, as
well as
  the correct number of groups $B$ and the values of the
  parameters $\omega_{\text{in}}$,
  $\omega_{\text{out}}$, and $\bm\theta$, and the following symmetry
  exists:
  \begin{equation}\label{eq:symmetry_case}
    \sum_i\theta_i\delta_{b_i,r} = C,
  \end{equation}
  where $C$ is a constant.
\item Like the first case, but where the symmetry of
  Eq.~\ref{eq:symmetry_case} does not exist.
\item Every other situation.
\end{enumerate}
Cases 1 and 2 are highly idealized and are not expected to be
encountered in practice, which almost always falls in case
3. Nevertheless, the equivalence between the DC-PP model and generalized
modularity is only valid in case 1. In case 2, as we already discussed,
the use of generalized modularity will be equivalent to \emph{some}
generative model --- as all methods are --- but which cannot be expressed
within the DC-PP parametrization.

Because of the above problems, the relevance of this partial equivalence
between these approaches in practical scenarios is arguably dubious. It
serves only to demonstrate how the implicit assumptions behind
modularity maximization are hard to justify.

We emphasize also the obvious fact that even if the equivalence with the
DC-PP model were to hold more broadly, this would not make the
pathological behavior of modularity described in
Sec.~\ref{sec:modularity} disappear. Instead, it would only show that
this particular inferential method would \emph{also} be pathological. In
fact, it is well understood that maximum likelihood is not in general an
appropriate inferential approach for models with an arbitrarily large
number of degrees of freedom, since it lacks the regularization
properties of Bayesian methods~\cite{peixoto_bayesian_2019}, such as the
one we described in Sec.~\ref{sec:infer}, where instead of considering
point estimates of the parameters, we integrate over all possibilities,
weighted according to their prior probability. In this way, it is
possible to \emph{infer} the number of communities, instead of assuming
it \emph{a priori}, together with all other model parameters. In fact,
when such a Bayesian approach is employed for the DC-PP model, one
obtains the following marginal likelihood~\cite{zhang_statistical_2020},
\begin{align}
  P(\A|\bb) &= \int P(\A|\omega_{\text{in}},\omega_{\text{out}},\bm\theta,\bb)P(\omega_{\text{in}})P(\omega_{\text{out}})P(\bm\theta|\bb)\,\dd\omega_{\text{in}}\dd\omega_{\text{out}}\dd\bm\theta\\
  &=   \frac{e_{\text{in}}!e_{\text{out}}!}
  {\left(\frac{B}{2}\right)^{e_{\text{in}}}{B\choose 2}^{e_{\text{out}}}(E+1)^{1-\delta_{B,1}}}
  \prod_r\frac{(n_r-1)!}{(e_r+n_r-1)!}\times\frac{\prod_ik_i!}{\prod_{i<j}A_{ij}!\prod_i A_{ii}!!},\label{eq:pp_marginal}
\end{align}
where $e_{\text{in}}=\sum_{i<j}(a)_{ij}\delta_{b_i,b_j}$ and
$e_{\text{out}}=E-e_{\text{in}}$.  As demonstrated in
Ref.~\cite{zhang_statistical_2020}, this approach allows us to detect
purely assortative community structures in a nonparametric way, in a
manner that prevents both overfitting and underfitting --- i.e. the
resolution limit \emph{vanishes} since we inherently consider every
possible value of the parameters $\omega_{\text{in}}$ and
$\omega_{\text{out}}$ --- thus lifting two major limitations of
modularity. Note also that Eq.~\ref{eq:pp_marginal} (or its logarithm)
does not bear any direct resemblance to the modularity function, and
therefore it does not seem possible to reproduce its behavior via a
simple modification of the latter.\footnote{There is also no need to
``fix'' modularity. We can simply use Eq.~\ref{eq:pp_marginal} in its
place for most algorithms, which incurs almost no additional
computational overhead.}

We also mention briefly a result obtained by Bickel and
Chen~\cite{bickel_nonparametric_2009}, which states that modularity
maximization can consistently identify the community assignments of
networks generated by the SBM in the dense limit. This limit corresponds
to networks where the average number of neighbors is comparable to the
total number of nodes. In this situation, the community detection
problem becomes substantially easier, and many algorithms, including
e.g. unregularized spectral clustering, can do just as well as
modularity maximization. This result tells us more about how easy it is
to find communities in dense networks than about the quality of the
algorithms compared. The dense scenario does not represent well the
difficulty of finding communities in real networks, which are
overwhelmingly sparse, with an average degree much smaller than the
total number of nodes.  In the sparse case, likelihood-based inferential
approaches are optimal and outperform
modularity~\cite{bickel_nonparametric_2009,
  decelle_asymptotic_2011}. Comparable equivalences have also been
encountered with spectral methods~\cite{newman_spectral_2013}, but they
also rely on particular realizations of the community detection problem,
and do not hold in general.

In short, if the objective is to infer the DC-PP model, there is no
reason to do it via the maximization of $Q(\A,\bb,\gamma)$, nor is it in
general equivalent to any consistent inference approach such as maximum
likelihood or Bayesian posterior inference. Even in the unlikely case
where the true number of communities is known, the implicit assumptions
of modularity correspond to the DC-PP model not only with uniform
probabilities between communities but also uniform sums of degrees for
every community. If these properties are not present in the network, the
method offers no inherent diagnostic, and will find spurious structures
that tend to match it, regardless of their statistical
significance. Combined with the overall lack of regularization, these
features render the method substantially prone to distortion and
overfitting. Ultimately, the use of any form of modularity maximization
fails the litmus test we considered earlier, and should be considered a
purely descriptive community detection method. Whenever the objective is
to understand network structure, it needs to be replaced with a flexible
and robust inferential procedure.

\subsection{``Consensus clustering can eliminate overfitting.''}\label{sec:consensus}

\begin{figure}
  \begin{tabular}{c}
    \includegraphics[width=.5\textwidth]{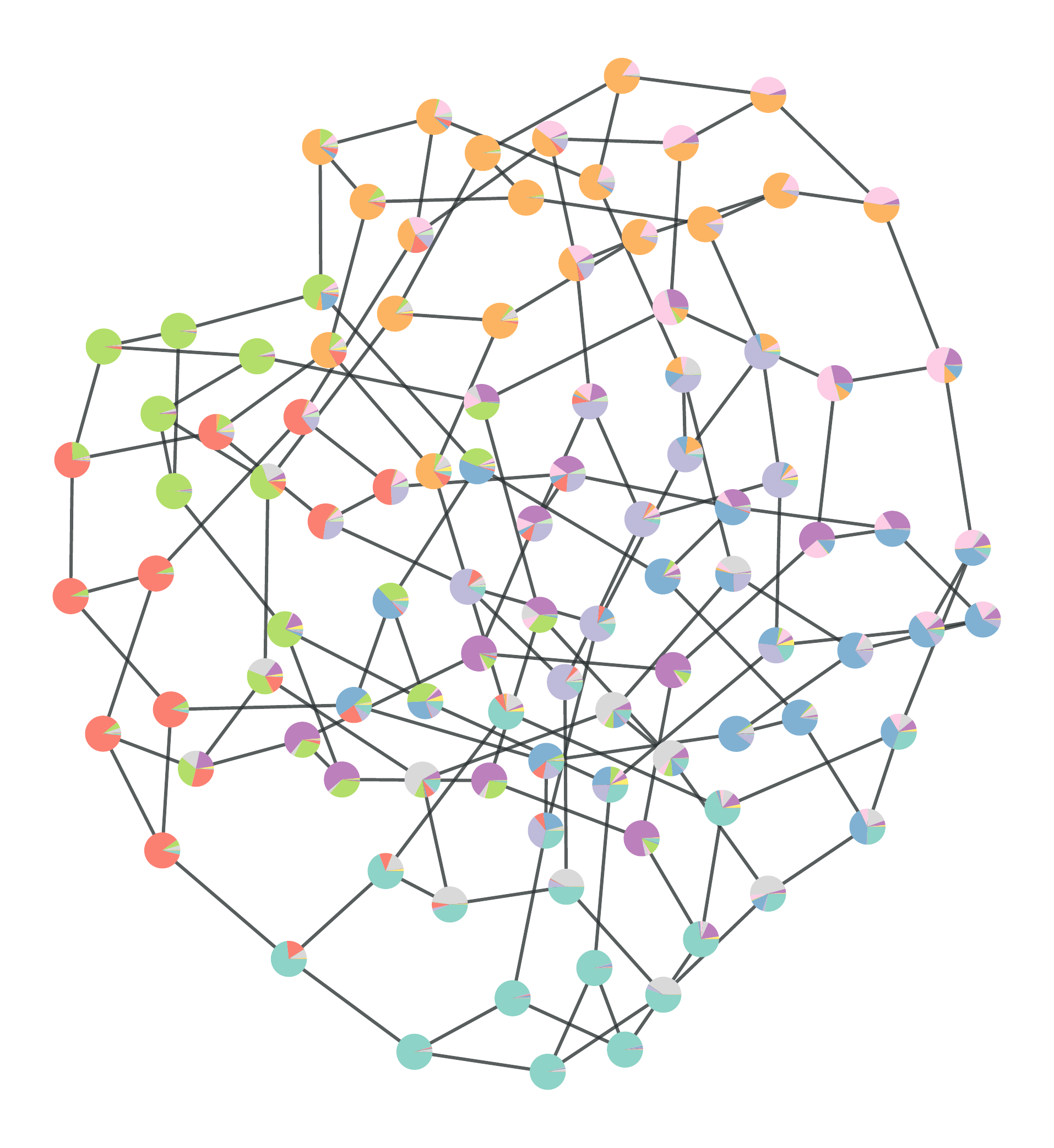}
  \end{tabular}
  \caption{Consensus clustering of a maximally random network,
    sampled from the Erd\H{os}-Rényi model, that combines $10^5$
    solutions of the maximum modularity method. On each node there is a
    pie chart describing the frequencies with which it was observed in a
    given community, obtained using the approach described in
    Ref.~\cite{peixoto_revealing_2021}. Despite the lack of latent
    communities, there is a substantial agreement between the different
    answers.
    \label{fig:consensus}}
\end{figure}

As mentioned in Sec.~\ref{sec:modularity}, methods like modularity
maximization tend to have a degenerate solution landscape. One strategy
proposed to tackle this problem is to obtain a \emph{consensus
clustering}, i.e. leverage the entire landscape of solutions to produce
a single partition that points in a cohesive direction, representative
of the whole
ensemble~\cite{massen_thermodynamics_2006,lancichinetti_consensus_2012,riolo_consistency_2020}. If
no cohesive direction exists, one could then conclude that no actual
community structure exists, and therefore solve the overfitting problem
of finding communities in maximally random networks. In reality, however, a
descriptive community detection method can in fact display a cohesive
set of solutions on a maximally random network. We demonstrate this in
Fig.~\ref{fig:consensus} which shows the consensus between $10^5$
different maximum modularity solutions for a small random network, using
the method of Ref.~\cite{peixoto_revealing_2021} to obtain the
consensus. Although we can notice a significant variability between the
different partitions, there is also substantial agreement. In
particular, there is no clear indication from the consensus that the
underlying network is maximally random. The reason for this that the
randomness of the network is \emph{quenched}, and does indeed point to a
specific community structure with the highest modularity. The ideas of
solution heterogeneity and overfitting are, in general, orthogonal
concepts.
\FloatBarrier

With care, it is possible to probe the solution landscape in a manner
that reveals a signal of the randomness of the underlying network. For
this purpose, some authors have proposed that instead of finding the
maximum modularity partition, one instead samples them from the Gibbs
distribution~\cite{massen_thermodynamics_2006,reichardt_when_2006,hu_phase_2012,zhang_scalable_2014},
\begin{equation}
  P(\bb) = \frac{\ee^{\beta Q(\A,\bb)}}{Z(\A)},
\end{equation}
with normalization $Z(\A)=\sum_{\bb}\ee^{\beta Q(\A,\bb)}$, effectively
considering $Q(\A,\bb)$ as the Hamiltonian of a spin system with an
inverse temperature parameter $\beta$. For a sufficiently large random
network, there is a particular value $\beta=\beta^*$, below which
samples from the distribution become uncorrelated, forming a lack of
consensus~\cite{zhang_scalable_2014}. There is a problem, however: there
is no guarantee that if a lack of consensus exists for $\beta<\beta^*$,
then the network must be random; only the reverse is true. In general,
while statements can be made about the behavior of the modularity
landscape for maximally random and sufficiently large networks, or even for
networks sampled from a SBM, very little can be said about its behavior
on real, finite networks. Since real networks are likely to contain a
heterogeneous mixture of randomness and structure (e.g. as illustrated in
Fig.~\ref{fig:resolution}(c)) this kind of approach becomes ultimately
unreliable. One fundamental problem here is that these approaches
attempt to reach an inferential conclusion (``is the network sampled
from a random model?'') without fully going through Bayes' formula of
Eq.~\ref{eq:bayes}, and reasoning about model assumptions, prior
information and compressibility. We currently lack a principled
methodology to reach such a conclusion while avoiding these crucial
steps.

Another aspect of the relationship between consensus clustering and
overfitting is worth mentioning. In an inferential setting, if we wish
to obtain an estimator for the true partition $\hat\bb$, this will in
general depend on how we evaluate its accuracy. In other words, we must
define an error function $\epsilon(\bb',\bb)$ such that
\begin{equation}
  \bb = \underset{\bb'}{\operatorname{argmin}}\; \epsilon(\bb',\bb).
\end{equation}
Based on this, our best possible estimate is the one which minimizes the
average error over the entire posterior distribution,
\begin{equation}
  \hat\bb = \underset{\bb'}{\operatorname{argmin}}\; \sum_{\bb} \epsilon(\bb',\bb) P(\bb|\A).
\end{equation}
Note that in general this estimator will be different from the most
likely partition, i.e.
\begin{equation}
  \hat\bb \neq \underset{\bb}{\operatorname{argmax}}\; P(\bb|\A).
\end{equation}
The optimal estimator $\hat\bb$ will indeed correspond to a consensus
over all possible partitions, weighted according to their
plausibility. In situations where the posterior distribution is
concentrated on a single partition, both estimators will
coincide. Otherwise, the most likely partition might in fact be less
accurate and incorporate more noise than the consensus estimator, which
might be seen as a form of overfitting. This kind of overfitting is of a
different nature than the one we have considered so far, since it
amounts to a residual loss of accuracy, where an (often small) fraction
of the nodes end up incorrectly classified, instead of spurious groups
being identified. However, there are many caveats to this kind of
analysis. First, it will be sensitive to the error function chosen,
which needs to be carefully justified. Second, there might be no
cohesive consensus, in situations where the posterior distribution is
composed of several distinct ``modes,'' each corresponding to a
different hypothesis for the network. In such a situation the consensus
between them might be unrepresentative of the ensemble of
solutions. There are principled approaches to deal with this problem, as
described in
Refs.~\cite{peixoto_revealing_2021,kirkley_representative_2022}.

\subsection{``Overfitting can be tackled by doing a statistical significance test of the quality function.''}\label{sec:significance}

Sometimes practitioners are aware that non-inferential methods like
modularity maximization can find communities in random networks. In an
attempt to extract an inferential conclusion from their results, they
compare the value of the quality function with a randomized version of
the network --- and if a significant discrepancy is found, they conclude
that the community structure is statistically
meaningful~\cite{reichardt_when_2006}. Unfortunately, this approach is
as fundamentally flawed as it is straightforward to implement.

The reason why the test fails is because in reality it answers a
question that is different from the one intended. When we compare the
value of the quality function obtained from a network and its randomized
counterpart, we can use this information to answer \emph{only} the
following question: ``Can we reject the hypothesis that the observed
network was sampled from a random null model?'' No other information can
be obtained from this test, including whether the \emph{network
partition} we obtained is significant. All we can determine is if the
optimized value of the quality function is significant or not. The
distinction between the significance of the quality function value and
the network partition itself is subtle but crucial.  \FloatBarrier

\begin{figure}
  \begin{tabular}{ccc}
    \includegraphicsl{(a)}{width=.33\textwidth}{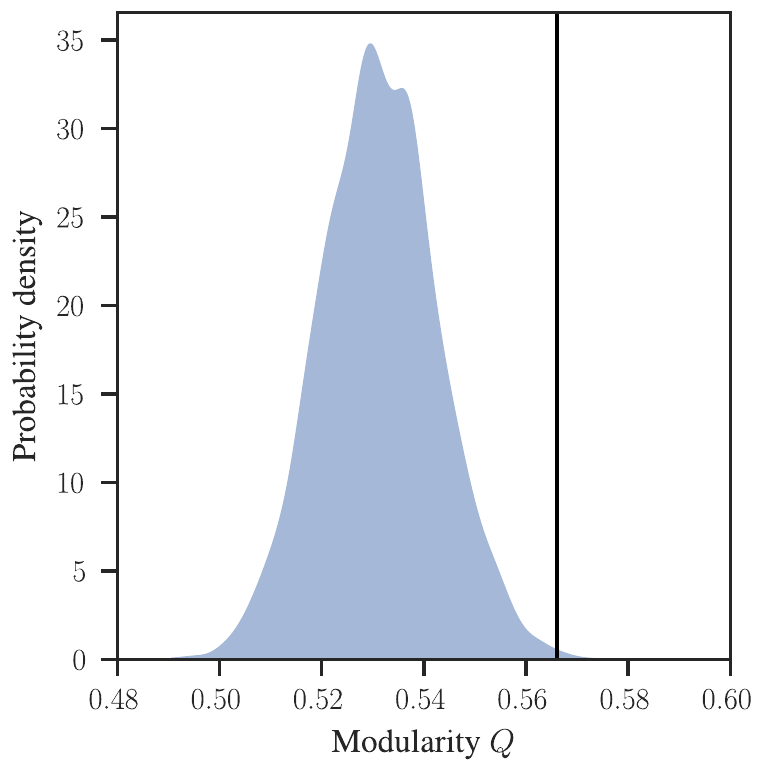}&
    \includegraphicslp{(b)}{.87}{width=.33\textwidth}{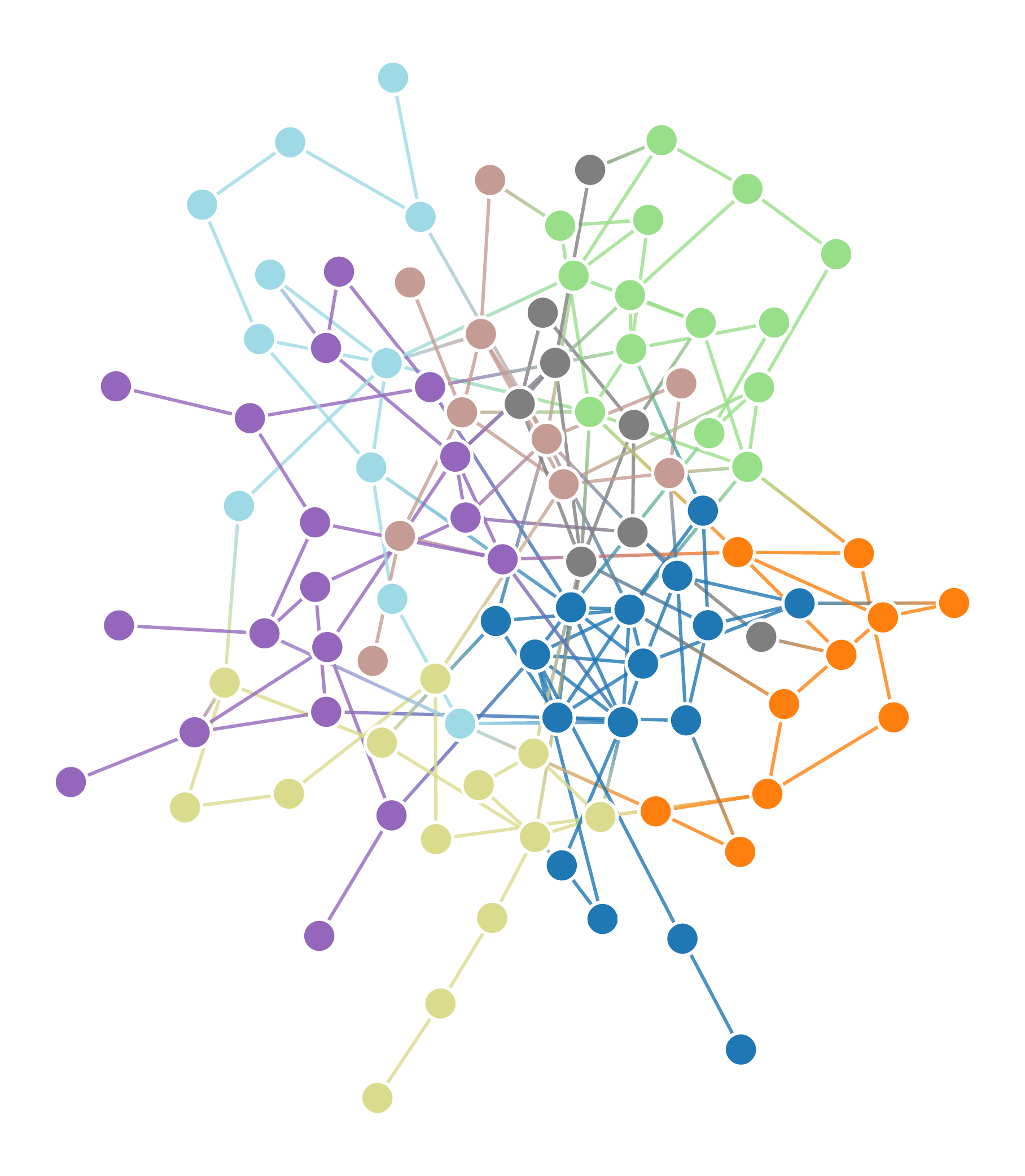}&
    \includegraphicslp{(c)}{.87}{width=.33\textwidth}{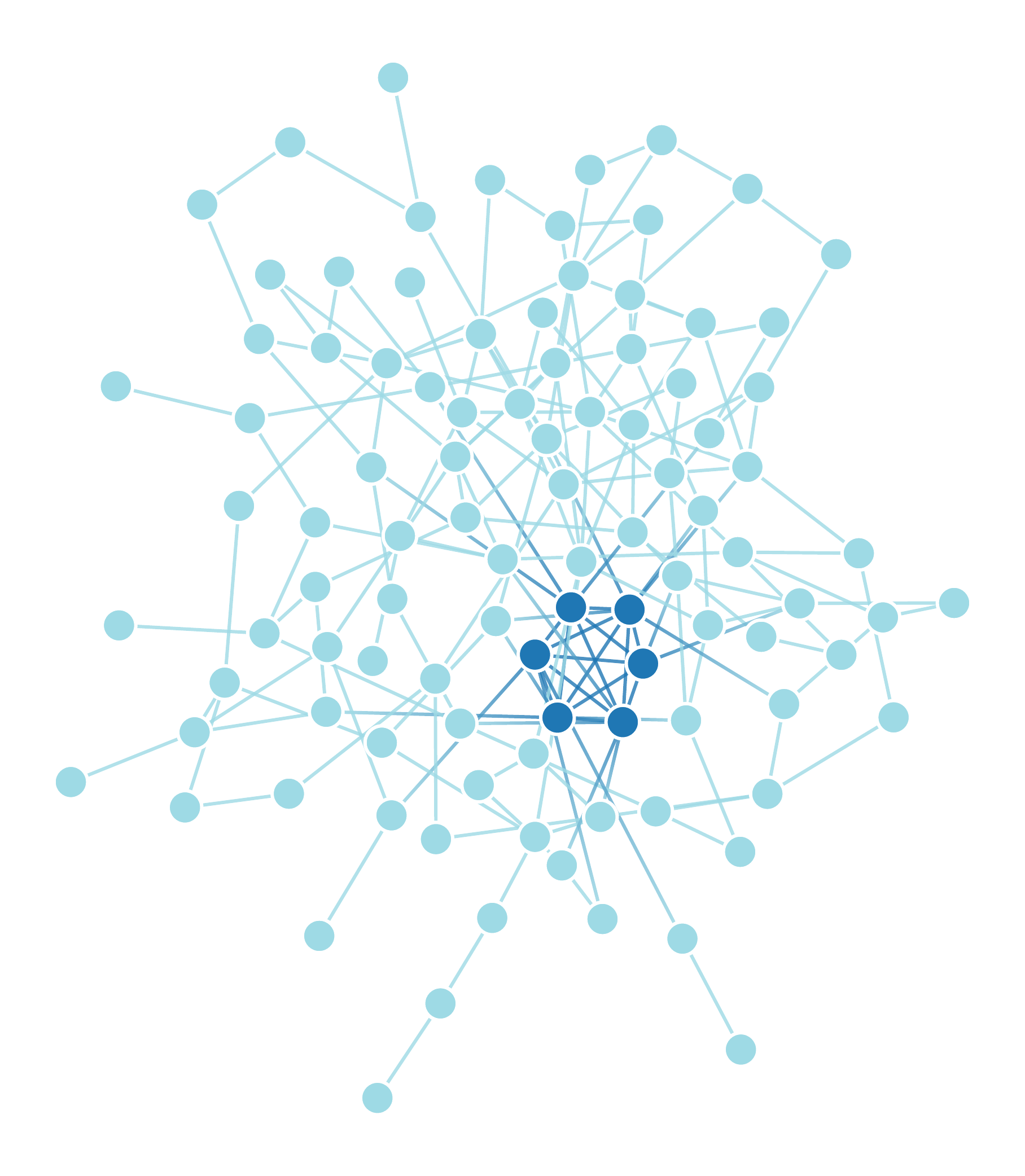}
  \end{tabular} \caption{The statistical significance of the maximum
  modularity value is not informative of the significance of the
  community structure. In (a) we show the distribution of optimized
  values of modularity for networks sampled from the Erd\H{o}s-Rényi
  (ER) model with the same number of nodes and edges as the network
  shown in (b) and (c). The vertical line shows the value obtained for
  the partition shown in (b), indicating that the network is very
  unlikely to have been sampled from the ER model ($P=0.002$). However,
  what sets this network apart from typical samples is the existence of
  a small clique of six nodes that would not occur in the ER model. The
  remaining communities found in (b) are entirely meaningless. In (c) we
  show the result of inferring the SBM on this network, which perfectly
  identifies the planted clique without overfitting the rest of the
  network.
    \label{fig:modularity_null}}
\end{figure}

We illustrate the above difference with an example in
Fig.~\ref{fig:modularity_null}(b). This network is created by starting
with a maximally random Erd\H{o}s-Rényi (ER) network, and adding to it a
few more edges so that it has an embedded clique of six nodes. The
occurrence of such a clique from an ER model is very unlikely, so if we
perform a statistical test on this network that is powerful enough, we
should be able to rule out that it came from the ER model with good
confidence. Indeed, if we use the value of maximum modularity for this
test, and compare with the values obtained for the ER model with the
name number of nodes and edges (see Fig.~\ref{fig:modularity_null}(a)),
we are able to reach the correct conclusion that the null model should
be rejected, since the optimized value of modularity is significantly
higher for the observed network. Should we conclude therefore that the
communities found in the network are significant? If we inspect
Fig.~\ref{fig:modularity_null}(b), we see that the maximum value of
modularity indeed corresponds to a more-or-less decent detection of the
planted clique. However, it also finds another seven completely spurious
communities in the random part of the network. What is happening is
clear --- the planted clique is enough to increase the value of $Q$ such
that it becomes a suitable test to reject the null model,\footnote{Note
that it is possible to construct alternative examples, where instead of
planting a clique, we introduce the placement of triangles, or other
features that are known to increase the value of modularity, but that do
not correspond to an actual community
structure~\cite{foster_clustering_2011}.} but the test is not able to
determine that the communities themselves are statistically
meaningful. In short, the statement ``the value of $Q$ is significant''
is not synonymous with ``the network partition is significant.''
Conflating the two will lead to the wrong conclusion about the
significance of the communities uncovered.

In Fig.~\ref{fig:modularity_null}(c) we show the result of a more
appropriate inferential approach, based on the SBM as described in
Sec~\ref{sec:inference}, that attempts to answer a much more relevant
question: ``which partition of the network into groups is more likely?''
The result is able to cleanly separate the planted clique from the rest
of the network, which is grouped into a single community.

This example also shows how the task of rejecting a null model is very
oblique to Bayesian inference of generative models. The former attempts
to determine what the network \emph{is not}, while the latter what
\emph{it is}.  The first task tends to be easy --- we usually do not
need very sophisticated approaches to determine that our data did not
come from a null model, specially if our data is complex. On the other
hand, the second task is far more revealing, constructive, and arguably
more useful in general.

\subsection{``Setting the resolution parameter of modularity maximization can remove the resolution limit.''}\label{sec:resolution}

\begin{figure}
  \begin{tabular}{ccc}
    \multirow{4}{*}[10em]{\includegraphicslp{(a)}{.95}{width=.33\textwidth}{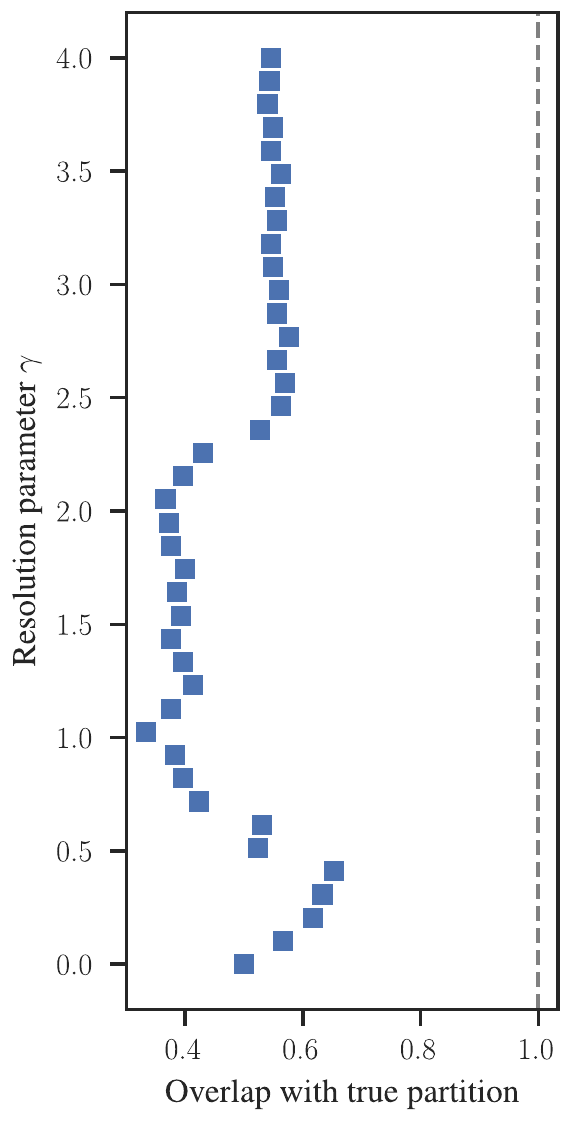}}&
    \includegraphicslp{(b)}{.87}{width=.33\textwidth}{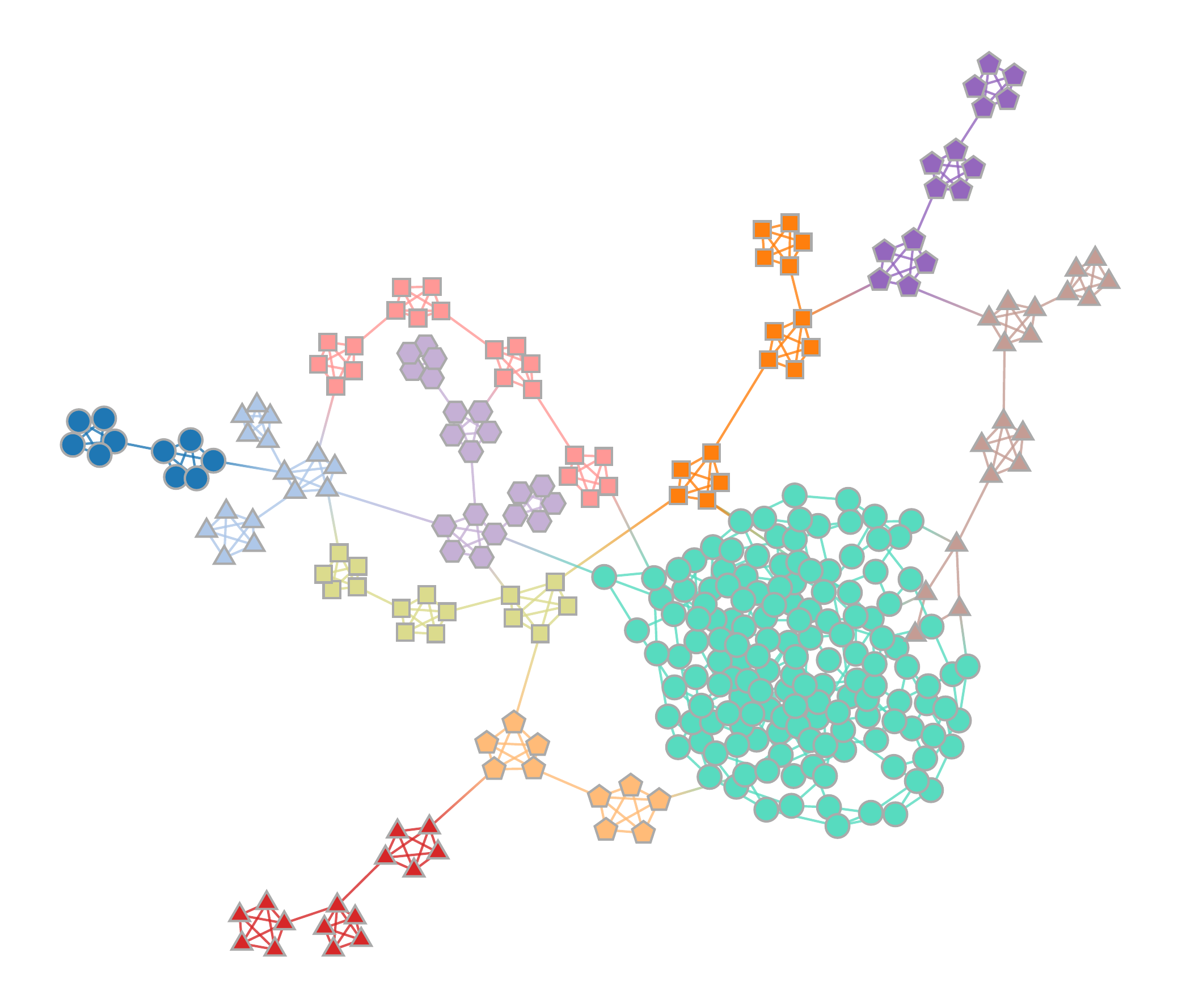}&
    \includegraphicslp{(c)}{.87}{width=.33\textwidth}{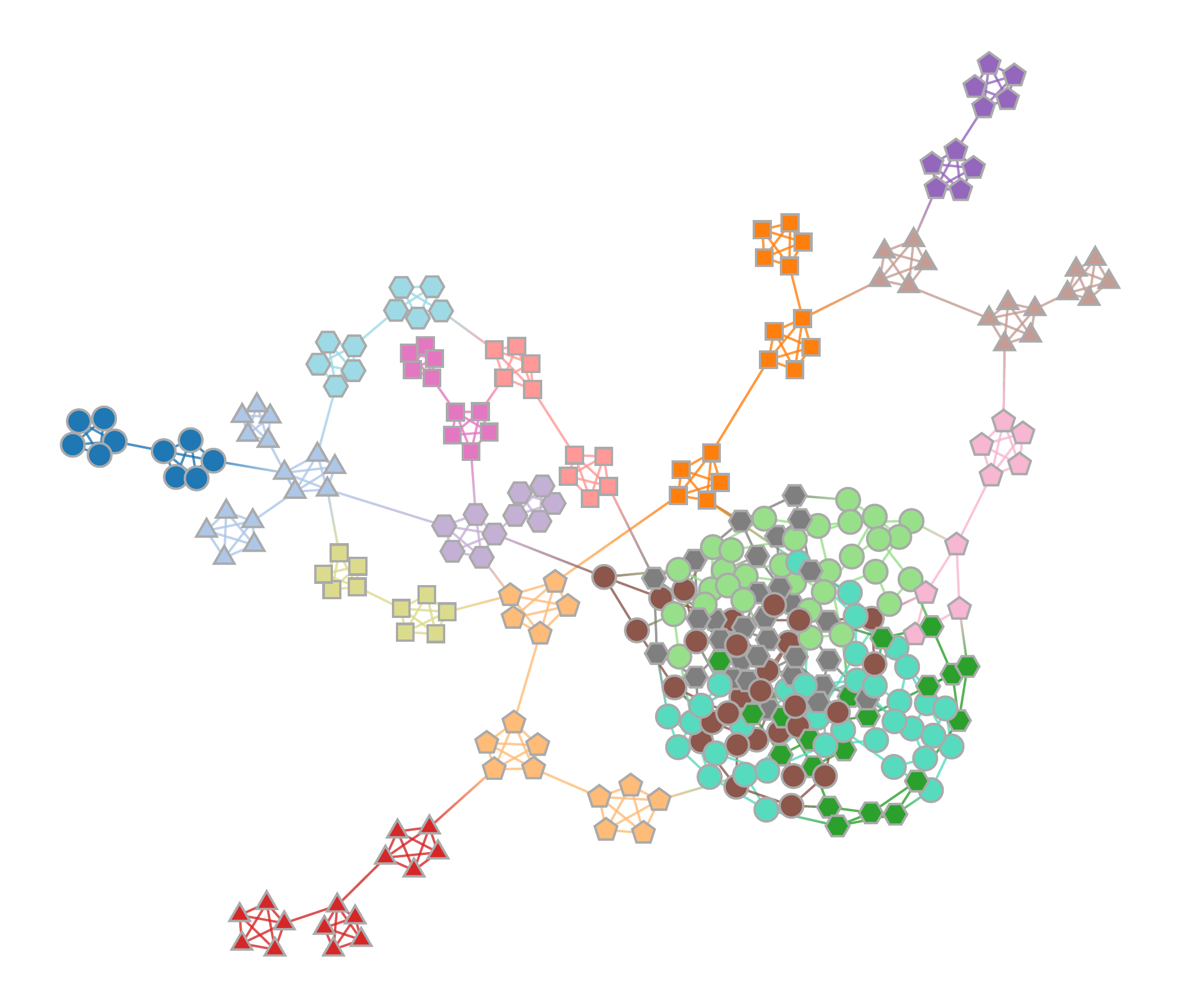}\\
    &\smaller$\gamma=0.41$ & \smaller$\gamma=1.02$\\
    &\includegraphicslp{(d)}{.87}{width=.33\textwidth}{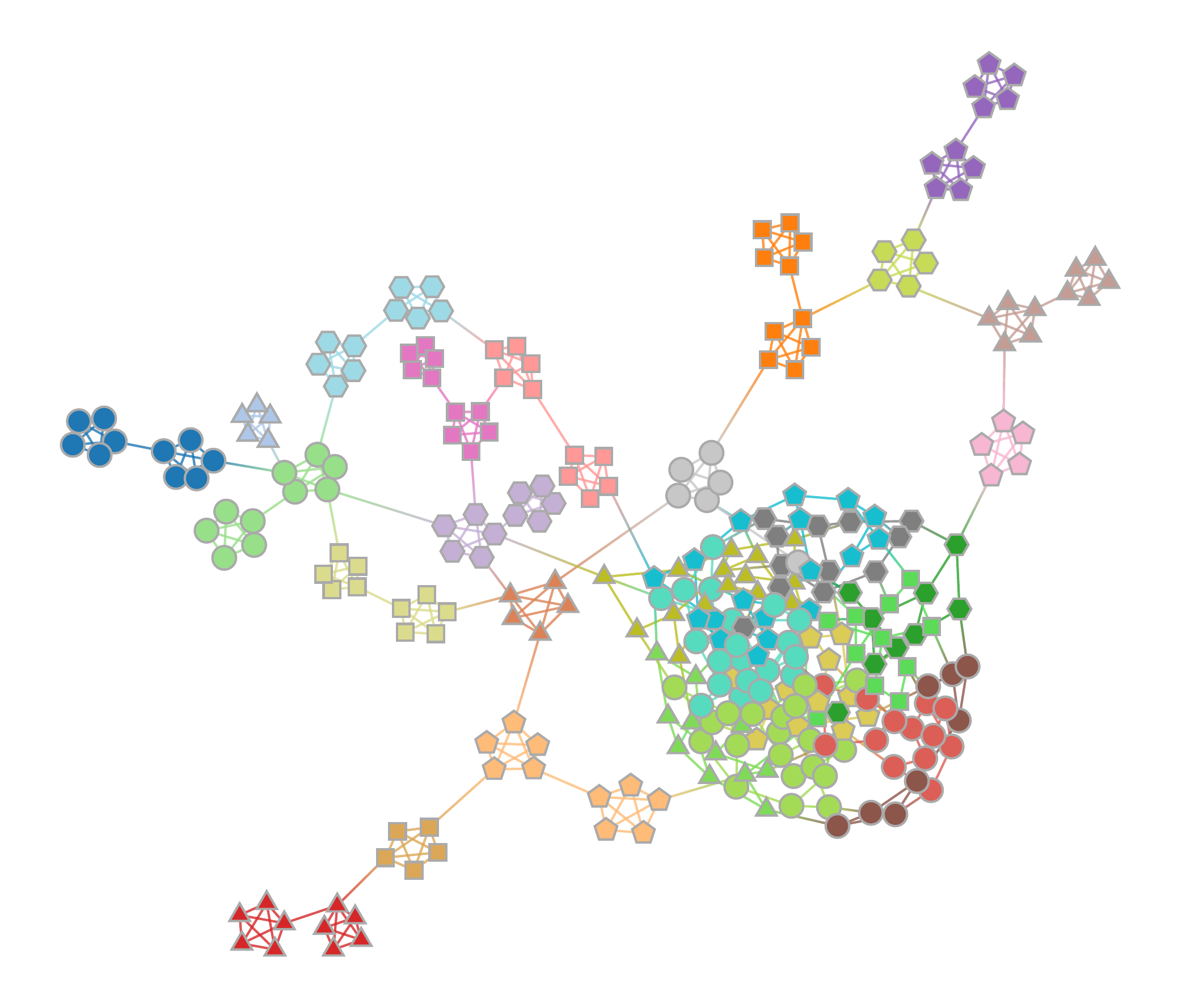}&
    \includegraphicslp{(e)}{.87}{width=.33\textwidth}{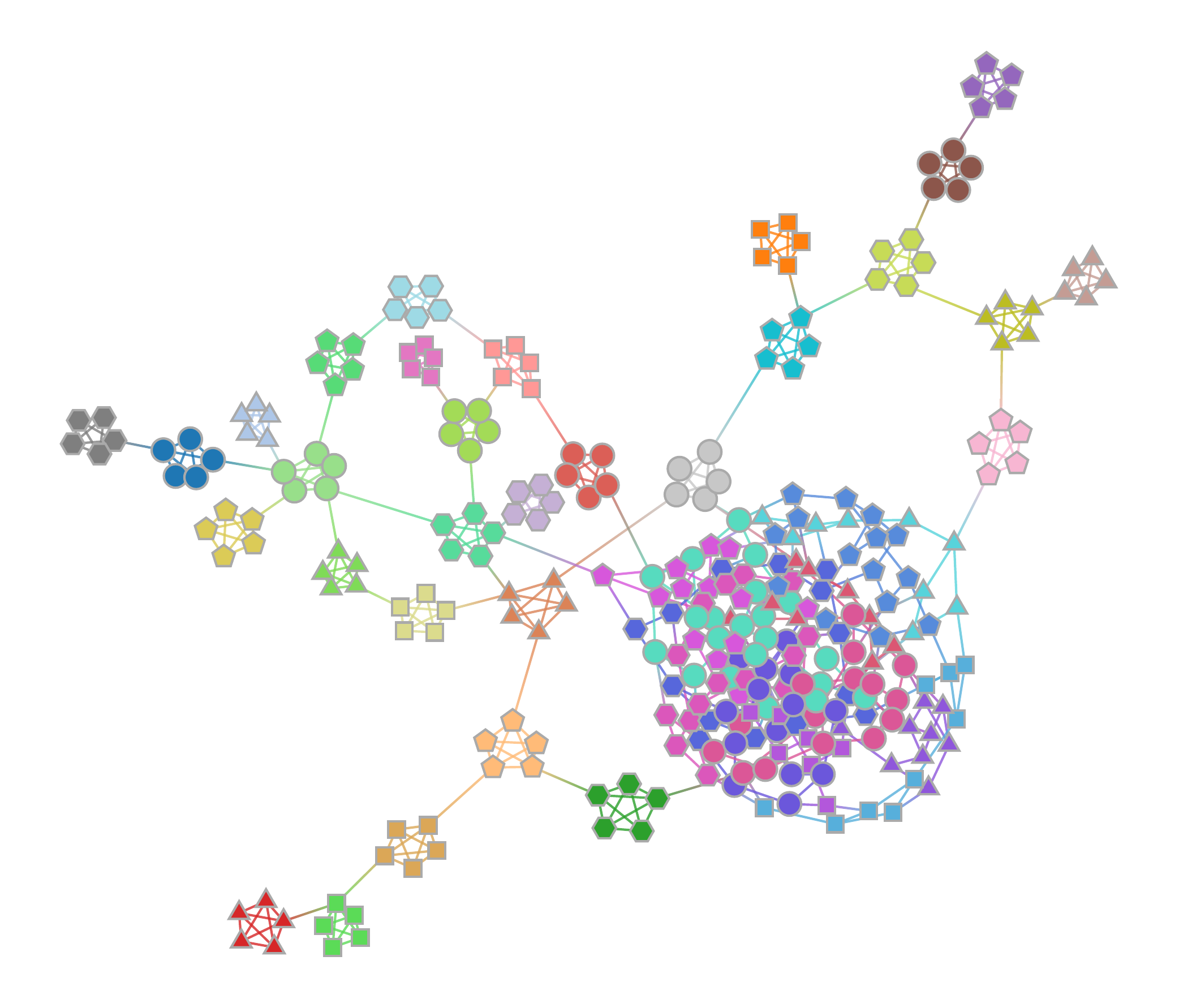}\\
    &\smaller$\gamma=2.05$ & \smaller$\gamma=2.56$
  \end{tabular}

  \caption{Modularity maximization imposes characteristic community
  sizes in a manner that hides heterogeneity. Panel (a) shows the
  overlap between the true and obtained partition for the network
  described in Fig.~\ref{fig:resolution}, as a function of the
  resolution parameter $\gamma$. Panels (b) to (e) show the partitions
  found for different values of $\gamma$, where we see that as smaller
  groups are uncovered, bigger ones are spuriously split. The result is
  that no value of $\gamma$ allows the true communities to be
  uncovered.\label{fig:resolution_param}}
\end{figure}

The resolution limit of the generalized modularity of
Eq.~\ref{eq:Qgamma} is such that, in a connected network, no more than
$\sqrt{\gamma 2E}$ communities can be found, with $\gamma$ being the
resolution
parameter~\cite{reichardt_statistical_2006,arenas_analysis_2008}. Therefore,
by changing the value of $\gamma$, we can induce the discovery of
modules of arbitrary size, at least in principle. However, there are
several underlying problems with tuning the value of $\gamma$ for the
purpose of counter-acting the resolution limit. The first is that it
requires a specific prior knowledge about what would be the relevant
scale for a particular network --- which is typically unavailable ---
turning an otherwise nonparametric approach into one which is
parametric.\footnote{We emphasize that the maximum likelihood approach
proposed in Ref.~\cite{newman_equivalence_2016} to determine $\gamma$,
even ignoring the caveats discussed in Sec.~\ref{sec:equivalence} that
render it invalid unless very specific conditions are met, is only
applicable for situations when the number of groups is known, directly
undermining its use to counteract the resolution limit.} The second
problem is even more serious: In many cases no single value of $\gamma$
is appropriate. This happens because, as we have seen in
Seq.~\ref{sec:equivalence}, generalized modularity comes with the
built-in assumption that the sum of degrees of every group should be the
same. The preservation of this homogeneity means that when the network
is composed of communities of different sizes, either the smaller ones
will be merged together or the bigger ones will be split into smaller
ones, regardless of the statistical
evidence~\cite{lancichinetti_limits_2011}. We show a simple example of
this in Fig.~\ref{fig:resolution_param}, where no value of $\gamma$ can
be used to recover the correct partition.

However, the most important problem with the analysis of the resolution
limit in the context of modularity maximization is that it is often
discussed in a manner that is largely decoupled from the issue of
statistical significance. Since we can interpret a limit on the maximum
number of groups as type of systematic underfitting, we can only
meaningfully discuss the removal of this limitation if we also do not
introduce a tendency to \emph{overfit}, i.e. find more groups than
justifiable by statistical evidence. This is precisely the problem with
``mutliresolution'' approaches~\cite{granell_hierarchical_2012}, or
analyses of quality functions other than
modularity~\cite{kawamoto_estimating_2015}, that claim a reduced or a
lack of resolution limit, but without providing a robustness against
overfitting. This one-sided evaluation is fundamentally incomplete, as
we may end up trading one serious limitation for another.

Methods based on the Bayesian inference of the SBM can tackle the issue
of over- and underfitting, as well as preferred sizes of communities at
the source. As was shown in Ref.~\cite{peixoto_parsimonious_2013}, a
uninformative assumption about the mixing patterns between groups leads
naturally to a resolution limit similar to the one existing for
modularity, where no more than $O(\sqrt{N})$ groups can be inferred for
sparse networks. However, since in an inferential context our
assumptions are made explicitly, we can analyse them more easily and
come up with more appropriate choices. In
Ref.~\cite{peixoto_hierarchical_2014} it was shown how replacing the
noninformative assumption by a Bayesian hierarchical model can
essentially remove the resolution limit, with a maximum number of groups
scaling as $O(N/\log N)$. That model is still unbiased with respect to
the expected mixing patterns, and incorporates only the assumption that
the patterns themselves are generated by another SBM, with its own
patterns generated by yet another SBM, and so on recursively. Another
model that has also been shown to be free of the resolution limit is the
assortative SBM of Ref.~\cite{zhang_statistical_2020}. Importantly, in
both these cases the removal of the resolution limit is achieved without
sacrificing the capacity of the method to avoid overfitting --- e.g. none of
these approaches will find spurious groups in random networks.

The issue with preferred group sizes can also be tackled in a principled
way in an inferential setting. As demonstrated in
Ref.~\cite{peixoto_nonparametric_2017}, we can also design Bayesian
prior hierarchies where the group size distribution is chosen in a
non-informative manner, before the partition itself is determined. This
results in an inference method that is by design agnostic with respect
to the distribution of group sizes, and will not prefer any of them in
particular. Such a method can then be safely employed on networks with
heterogeneous group sizes in an unbiased manner. In
Fig.~\ref{fig:resolution}(d) we show how such an approach can easily
infer groups of different sizes for the same example of
Fig.~\ref{fig:resolution_param}, in a completely nonparametric manner.

\subsection{``Modularity maximization can be fixed by replacing the null model.''}\label{sec:null}

Several variations of the method of modularity maximization have been
proposed, where instead of the configuration model, another null model
is used, in a manner that makes the method applicable in various
scenarios, e.g. with bipartite networks~\cite{barber_modularity_2007},
correlation matrices~\cite{macmahon_community_2015}, signed edge
weights~\cite{traag_community_2009}, networks embedded in euclidean
spaces~\cite{expert_uncovering_2011}, to name a few. While the choice of
null model has an important effect on what kind of structures are
uncovered, its choice does not address any of the statistical
shortcomings of modularity that we consider here. In general, just like
it happens for the configuration model, the approach will find spurious
communities in networks sampled from its null model, regardless of how
it is chosen. As as discussed in Sec.~\ref{sec:modularity}, this happens
because the measured deviation does not account for the optimization
procedure employed. Any method based on optimizing the modularity score
will amount to a data dredging procedure, independently of the null
model chosen, and are thus unsuitable for inferential aims.

\subsection{``Descriptive approaches are good enough when the community structure is obvious.''}\label{sec:obvious}

A common argument goes that, sometimes, the community structure of a
network is so ``obvious'' that it will survive whatever abuse we direct
at it, and it will be uncovered by a majority of community detection
methods that we employ. Therefore, if we are confident that our network
contains a clear signal of its community structure, specially if several
algorithms substantially agree with each other, or they agree with
metadata, then it does not matter very much which algorithm we use.

There are several problems with this argument. First, if an ``obvious''
structure exists, it does not necessarily mean that it is really
meaningful, or statistically significant. If ten algorithms overfit, and
one does not, the majority vote is incorrect, and we should prefer the
minority opinion. This is precisely the case we considered in
Fig.~\ref{fig:descriptive}, where virtually any descriptive method would
uncover the same 13 communities --- thus overfitting the network ---
while an inferential approach would not. And if a method agrees with
metadata, while another finds further structure not in agreement, what
is to say that this structure is not really there? (Metadata are not
``ground truth,'' they are only more
data~\cite{hric_network_2016,newman_structure_2016,peel_ground_2017},
and hence can have its own complex, incomplete, noisy, or even
irrelevant relationship with the network.)

Secondly, and even more importantly, how do we even define what is an
``obvious'' community structure? In general, networks are not low
dimensional objects, and we lack methods to inspect their structure
directly, a fact which largely motivates community detection in the
first place. Positing that we can just immediately determine community
structure largely undermines this fact. Often, structure which is deemed
``obvious'' at first glance, ceases to be so upon closer inspection. For
example, one can find claims in the literature that different connected
components must ``obviously'' correspond to different
communities. However, maximally random graphs can end up disconnected if
they are sufficiently sparse, which means that from an inferential point
of view different components can belong to the same community.

Another problem is that analyses of community detection results rely
frequently on visual inspections of graphical network layouts, where one
tries to evaluate if the community labels agree with the position of the
nodes. However, the positioning of the nodes and edges is not inherent
to the network itself, and needs to be obtained with some graph drawing
algorithm. A typical example are the so-called ``spring-block'' or
``force-directed'' layouts, where one considers attractive forces
between nodes connected by an edge (like a spring) and an overall
repulsive force between all nodes~\cite{hu_efficient_2005}. The final
layout is then obtained by minimizing the energy of the system,
resulting in edges that have similar length and as few crossings between
edges as possible (e.g. in Fig.~\ref{fig:infvsdesc} we used the
algorithm of Ref.~\cite{hu_efficient_2005}). This kind of drawing in
itself can be seen as a type of indirect descriptive community detection
method, since nodes belonging to the same assortative community will
tend to be placed close to each
other~\cite{noack_modularity_2009}. Based on this observation, when we
say that we ``see'' the communities in a drawing like in
Fig.~\ref{fig:infvsdesc}, we are in reality only seeing what the layout
algorithm is telling us. Therefore, we should always be careful when
comparing the results we get with a community detection algorithm to the
structure we see in these layouts, because there is no reason to assume
that the layout algorithm itself is doing a better job than the
clustering algorithm we are evaluating.\footnote{Indeed, if we inspect
Fig.~\ref{fig:consensus}, which shows the consensus clustering of a
maximally random network, we notice that nodes that are classified in the
same community end up close together in the drawing, i.e. the layout
algorithm also agrees with the modularity consensus. Therefore, it
should not be used as a ``confirmation'' of the structure any more than
the result of any other community detection algorithm, since it is also
overfitting from an inferential perspective.} In fact, this is often not
the case, since the actual community structures in many networks do not
necessarily have a sufficiently low-dimensional representation that is
required for this kind of visualization to be effective.

\subsection{``The no-free-lunch theorem means that every community detection method is equally good.''}

For a wide class of optimization and learning problems there exist
so-called ``no-free-lunch'' (NFL) theorems, which broadly state that
when averaged over all possible problem instances, all algorithms show
equivalent
performance~\cite{wolpert_no_1995,wolpert_lack_1996,wolpert_no_1997}. Peel
\emph{et al}~\cite{peel_ground_2017} have proved that this is also valid
for the problem of community detection, meaning that no single method
can perform systematically better than any other, when averaged over all
community detection problems. This has been occasionally interpreted as
a reason to reject the claim that we should systematically prefer
certain classes of algorithms over others. This is, however, a
misinterpretation of the theorem, as we will now discuss.

The NFL theorem for community detection is easy to state. Let us
consider a generic deterministic community detection algorithm indexed
by $f$, defined by the function $\hat\bb_f(\A)$, which ascribes a single
partition to a network $\A$. Peel \emph{et al}~\cite{peel_ground_2017}
consider an instance of the community detection problem to be an
arbitrary pair $(\A,\bb)$ composed of a network $\A$ and the correct
partition $\bb$ that one wants to find from $\A$. We can evaluate the
accuracy of the algorithm $f$ via an error (or ``loss'') function
\begin{equation}
  \epsilon (\bb, \hat\bb_f(\A)),
\end{equation}
which should take the smallest possible value if $\hat\bb_f(\A) =
\bb$. If the error function does not have an inherent preference for any
partition (it's ``homogeneous''), then the NFL theorem
states~\cite{wolpert_lack_1996,peel_ground_2017}
\begin{equation}\label{eq:nfl}
  \sum_{(\A, \bb)}\epsilon (\bb, \hat\bb_f(\A)) = \Lambda(\epsilon),
\end{equation}
where $\Lambda(\epsilon)$ is a value that depends only on the error
function chosen, but not on the community detection algorithm $f$. In
other words, when averaged over all problem instances, all algorithms
have the same accuracy. This implies, therefore, that in order for one
class of algorithms to perform systematically better than another, we
need to restrict the universe of problems to a particular subset. This
is a seemingly straightforward result, but which is unfortunately very
susceptible to misinterpretation and overstatement.

A common criticism of this kind of NFL theorem is that it is a poor
representation of the typical problems we may encounter in real domains
of application, which are unlikely to be uniformly distributed across
the entire problem space. Therefore, as soon as we constrain ourselves
to a subset of problems that are relevant to a particular domain, then
this will favor some algorithms over others --- but then no algorithm
will be superior for all domains. But since we are typically only
interested in some domains, the NFL theorem is then arguably
``theoretically sound, but practically
irrelevant''~\cite{schaffer_conservation_1994}. Although indeed correct,
in the case of community detection this logic is arguably an
understatement. This is because as soon as we restrict our domain to
community detection problems that reveal something \emph{informative}
about the network structure, then we are out of reach of the NFL
theorem, and some algorithms will do better than others, without evoking
any particular domain of application. We demonstrate this in the
following.

The framework of the NFL theorem of Ref.~\cite{peel_ground_2017}
operates on a liberal notion of what constitutes a community detection
problem and its solution, which means for an arbitrary pair $(\A,\bb)$
choosing the right $f$ such that $\hat\bb_f(\A)=\bb$. Under this
framework, algorithms are just arbitrary mappings from network to
partition, and there is no necessity to articulate more specifically how
they relate to the structure of the network --- community detection just
becomes an arbitrary game of ``guess the hidden node labels.'' This
contrasts with how actual community detection algorithms are proposed,
which attempt to match the node partitions to patterns in the network,
e.g. assortativity, general connection preferences between groups,
etc. Although the large variety of algorithms proposed for this task
already reveal a lack of consensus on how to precisely define it, few
would consider it meaningful to leave the class of community detection
problems so wide open as to accept any matching between an arbitrary
network and an arbitrary partition as a valid instance.

Even though we can accommodate any (deterministic) algorithm deemed
valid according to any criterion under the NFL framework, most
algorithms in this broader class do something else altogether. In fact,
the absolute vast majority of them correspond to a maximally random
matching between network and partition, which amounts to little more
than just randomly guessing a partition for any given network, i.e. they
return widely different partitions for inputs that are very similar, and
overall point to no correlation between input and output.~\footnote{An
interesting exercise is to count how many such algorithms exist. A given
community detection algorithm $f$ needs to map each of all
$\Omega(N)=2^{N\choose 2}$ networks of $N$ nodes to one of
$\Xi(N)=\sum_{B=1}^{N}\genfrac\{\}{0pt}{}{N}{B}B!$ labeled partitions of
its nodes. Therefore, if we restrict ourselves to a single value of $N$,
the total number of input-output tables is $\Xi(N)^{\Omega(N)}$. If we
sample one such table uniformly at random, it will be asymptotically
impossible to compress it using fewer than $\Omega(N)\log_2\Xi(N)$ bits
--- a number that grows super-exponentially with $N$. As an
illustration, a random community detection algorithm that works only
with $N=100$ nodes would already need $10^{1479}$ terabytes of
storage. Therefore, simply considering algorithms that humans can write
and use (together with their expected inputs and outputs) already pulls
us very far away from the general scenario considered by the NFL
theorem. } It is not difficult to accept that these random algorithms
perform equally ``well'' for any particular problem, or even all
problems, but the NFL theorem says that they have equivalent performance
even to algorithms that we may deem more meaningful. How do we make a
formal distinction between algorithms that are just randomly guessing
from those that are doing something coherent and trying to discover
actual network patterns? As it turns out, there is an answer to this
question that does not depend on particular domains of application: we
require the solutions found to be \emph{structured} and
\emph{compressive of the network}.

In order to interpret the statement of the NFL theorem in this vein, it
is useful to re-write Eq.~\ref{eq:nfl} using an equivalent probabilistic
language,
\begin{equation}\label{eq:nflp}
  \sum_{\A, \bb}P(\A,\bb)\epsilon (\bb, \hat\bb_f(\A)) = \Lambda'(\epsilon),
\end{equation}
where $\Lambda'(\epsilon)\propto \Lambda(\epsilon)$, and $P(\A,\bb)
\propto 1$ is the uniform probability of encountering a problem
instance. When writing the theorem statement in this way, we notice
immediately that instead of being agnostic about problem instances, it
implies a \emph{very specific} network generative model, which assumes a
complete independence between network and partition. Namely, if we
restrict ourselves to networks of $N$ nodes, we have then:\footnote{We
could easily introduce arbitrary constraints such as total number of
edges or degree distribution, which would change the form of
Eqs.~\ref{eq:uniform} and~\ref{eq:uniform_a}, but none of the ensuing
analysis.}
\begin{align}
  P(\A,\bb)&=P(\A)P(\bb),\label{eq:uniform}\\
  P(\A) &= 2^{-{N\choose 2}},\label{eq:uniform_a}\\
  P(\bb) &= \left[\sum_{B=1}^{N}\genfrac\{\}{0pt}{}{N}{B}B!\right]^{-1}.\label{eq:uniform_b}
\end{align}
Therefore, the NFL theorem states simply that if we sample networks and
partitions from a maximally random generative model, then all algorithms
will have the same average accuracy at inferring the partition from the
network. This is hardly a spectacular result --- indeed the
Bayes-optimal algorithm in this case, i.e. the one derived from the
posterior distribution of the true generative model and which guarantees
the best accuracy on average, consists of simply guessing partitions
uniformly at random, ignoring the network structure altogether.

\begin{figure}
  \begin{tabular}{cc}
    \includegraphicsl{(a)}{height=.5\textwidth}{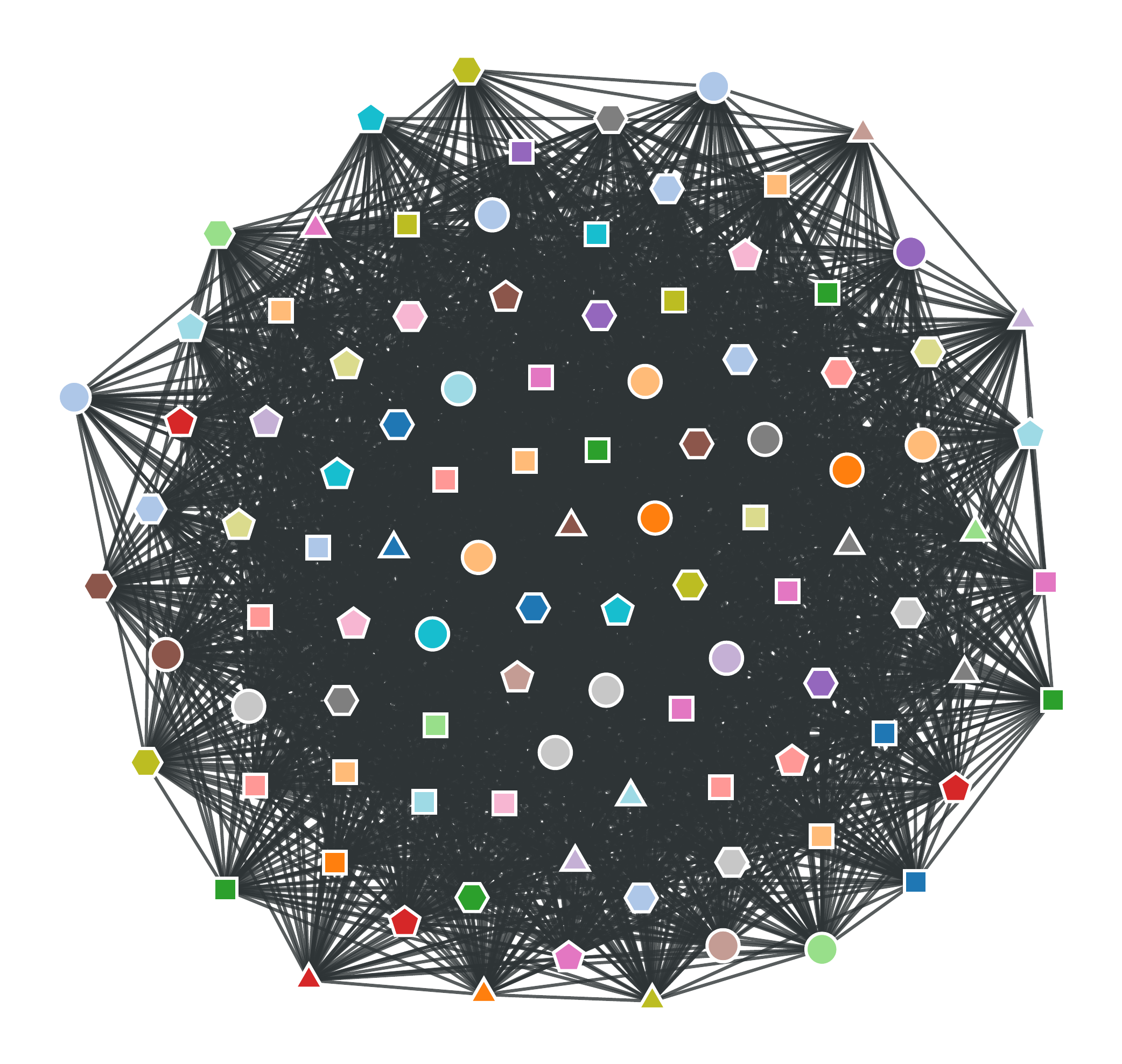}&
    \includegraphicsl{(b)}{height=.5\textwidth}{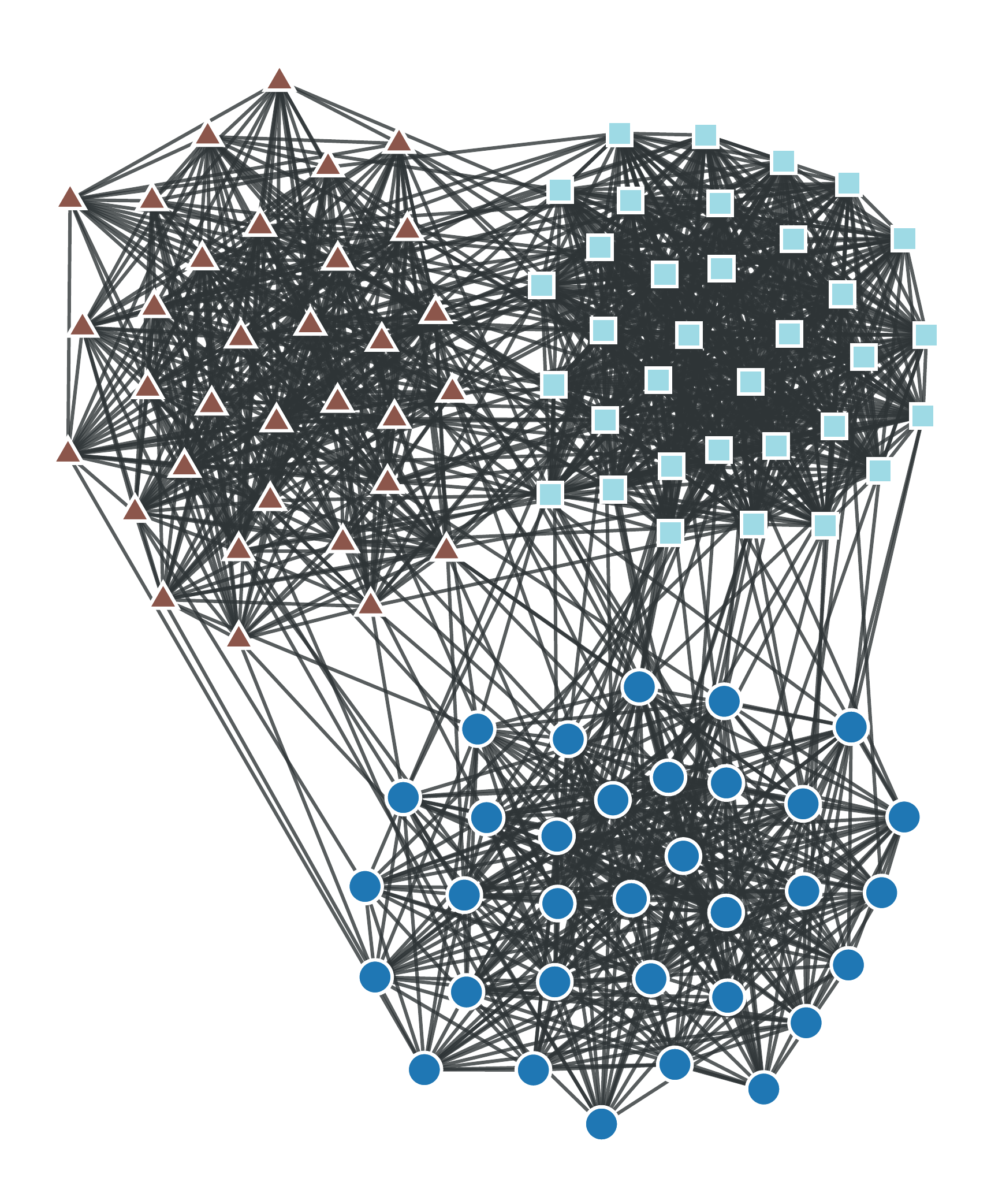}\\
    $\Sigma_{\text{min}}(\A|\bb) = 4950$ bits & \\
    $\Sigma_{\text{SBM}}(\A|\bb) = 6612$ bits & $\Sigma_{\text{SBM}}(\A|\bb) = 2280$ bits
  \end{tabular}

  \caption{The NFL theorem involves predominantly instances of the
    community detection problem that are strictly incompressible,
    i.e. the true partitions cannot be used to explain the network. In
    (a) we show a typical sample of the uniform problem space given by
    Eq.~\ref{eq:uniform}, for $N=100$ nodes, which yields a dense
    maximally random network, randomly divided into $B=72$ groups. It is
    asymptotically impossible to use this partition to compress this
    network into fewer than $\Sigma_{\text{min}}(\A|\bb) = {N \choose 2}
    = 4950$ bits, and therefore the partition is not learnable from the
    network alone with any inferential algorithm. We show also the
    description length of the SBM conditioned on the true partition,
    $\Sigma_{\text{SBM}}(\A|\bb)$, as a reference. In (b) we show an
    example of a community detection problem that is solvable, at least
    in principle, since $\Sigma_{\text{SBM}}(\A|\bb) <
    \Sigma_{\text{min}}(\A|\bb)$.  In this case, the partition can be
    used to inform the network structure, and potentially
    vice-versa. This class of problem instance has a negligible
    contribution to the sum in the NFL theorem in Eq.~\ref{eq:nfl},
    since it occurs only with an extremely small probability when
    sampled from the uniform model of Eq.~\ref{eq:uniform}. It is
    therefore more reasonable to state that the network in example (b)
    has an \emph{actual} community structure, while the one in
    (a) does not.
    \label{fig:nfl}}
\end{figure}

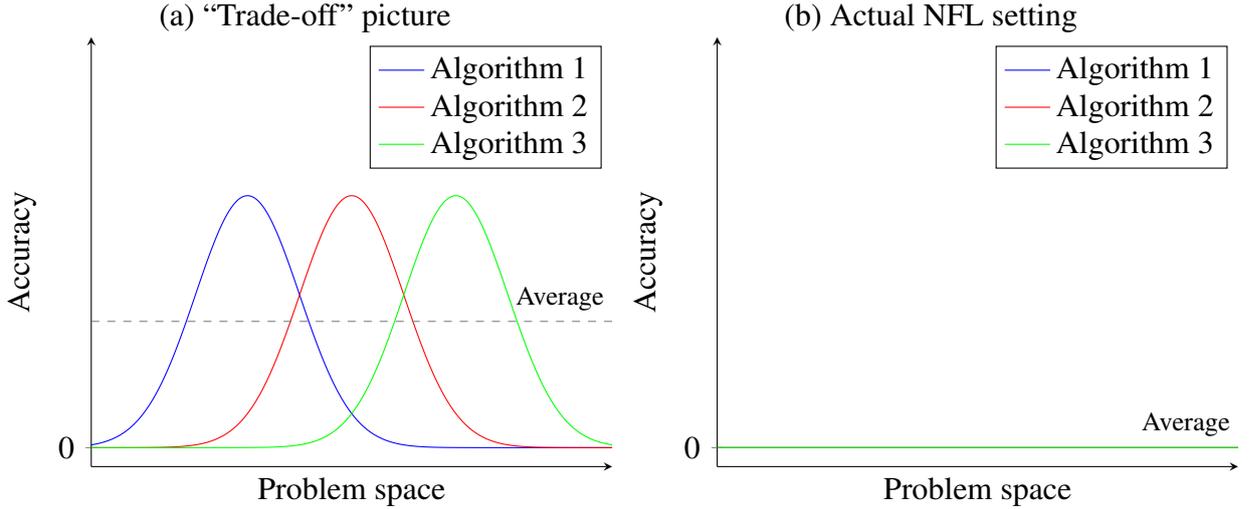
\begin{figure}
  \begin{tabular}{cc}
    (a) ``Trade-off'' picture & (b) Actual NFL setting\\
    \begin{tikzpicture}
      \begin{axis}[
          axis lines=left,
          %ticks=none,
          ylabel near ticks,
          xlabel near ticks,
          xlabel={Problem space},
          ylabel={Accuracy},
          legend entries={Algorithm 1, Algorithm 2, Algorithm 3},
          ymin=-0.03, ymax=.65,
          ytick={0},
          xmajorticks=false
        ]

        \addplot [blue,domain=-3:7,samples=401]
                 {exp(-x^2 / 2) / sqrt(2*pi)};
        \addplot [red,domain=-3:7,samples=401]
                 {exp(-(x-2)^2 / 2) / sqrt(2*pi)};
        \addplot [green,domain=-3:7,samples=401]
                 {exp(-(x-4)^2 / 2) / sqrt(2*pi)};
        \addplot [gray, dashed, domain=-3:7,samples=11]
                 {.2};
        \addplot [black, no marks] coordinates{(6,.2)} node[above] {\smaller Average};
      \end{axis}
    \end{tikzpicture} &
    \begin{tikzpicture}
      \begin{axis}[
          axis lines=left,
          %ticks=none,
          ylabel near ticks,
          xlabel near ticks,
          xlabel={Problem space},
          ylabel={Accuracy},
          legend entries={Algorithm 1, Algorithm 2, Algorithm 3},
          ymin=-0.03, ymax=.65,
          ytick={0},
          xmajorticks=false
        ]
        \addplot [blue,domain=-3:7,samples=10]
                 {0};
        \addplot [red,domain=-3:7,samples=10]
                 {0};
        \addplot [green,domain=-3:7,samples=10]
                 {0};
        \addplot [black, no marks] coordinates{(6,.0)} node[above] {\smaller Average};
      \end{axis}
    \end{tikzpicture}

  \end{tabular}

  \caption{A common interpretation of the NFL theorem for community
    detection is that it reveals a necessary trade-off between
    algorithms: since they all have the same average performance, if one
    algorithm does better than another in one set of instances, it must
    do worse on a equal number of different instances, as depicted in
    panel (a). However, in the actual setting considered by the NFL
    theorem there is no meaningful trade-off: asymptotically, all
    algorithms perform maximally poorly for the vast majority of
    instances, as depicted in panel (b), since in these cases the
    network structure is uninformative of the partition. If we constrain
    ourselves to informative problem instances (which compose only an
    infinitesimal fraction of all instances), the NFL theorem is no
    longer applicable.
    \label{fig:trade-off}}
\end{figure}

The probabilistic interpretation reveals that the NFL theorem involves a
very specific assumption about what kind of community detection problem
we are expecting. It is important to remember that it is not possible to
make ``no assumption'' about a problem; we are always forced to make
\emph{some} assumption, which even if implicit is not exempted from
justification, and the uniform assumption of Eqs.~\ref{eq:uniform}
to~\ref{eq:uniform_b} is no exception. In Fig.~\ref{fig:nfl}(a) we show
a typical sample from this ensemble of community detection problems. In
a very concrete sense, we can state that such problem instances are
\emph{unstructured} and contain \emph{no learnable community structure},
or in fact no learnable network structure \emph{at all}. We say that a
community structure is (in principle) learnable if the knowledge of the
partition $\bb$ can be used to compress the network $\A$, i.e. there
exists an encoding $\mathcal{H}$ (i.e. a generative model) such that
\begin{align}
  \Sigma(\A|\bb,\mathcal{H}) &< -\log_2P(\A),\\
  &< {N\choose 2},
\end{align}
where $\Sigma(\A|\bb,\mathcal{H}) = -\log_2P(\A|\bb,\mathcal{H})$ is the
description length of $\A$ according to model $\mathcal{H}$, conditioned
on the partition being known. However, it is a direct consequence of
Shannon's source coding
theorem~\cite{shannon_mathematical_1948,cover_elements_1991}, that for
the vast majority of networks sampled from the model of
Eq.~\ref{eq:uniform} the inequality above cannot be fulfilled as
$N\to\infty$, i.e. the networks are incompressible.\footnote{For finite
networks a positive compression might be achievable with small
probability, but due to chance alone, and not in a manner that makes its
structure learnable.}  This means that the true partition $\bb$ carries
no information about the network structure, and vice versa, i.e. the
partition is not learnable from the network. In view of this, the common
interpretation of the NFL theorem as ``all algorithms perform equally
well'' is in fact quite misleading, and should be more accurately
phrased as ``all algorithms perform equally \emph{poorly},'' since no
inferential algorithm can uncover the true community structure in most
cases, at least no better than by chance alone.  In other words, the
universe of community detection problems considered in the NFL theorem
is composed overwhelmingly of instances for which compression and
explanation are not possible.\footnote{One could argue that such a
uniform model is justified by the principle of maximum entropy, which
states that in the absence of prior knowledge about which problem
instances are more likely, we should assume they are all equally likely
\emph{a priori}. This argument fails precisely because we \emph{do} have
sufficient prior knowledge that empirical networks are not maximally
random --- specially those possessing community structure, according to
any meaningful definition of the term. Furthermore, it is easy to verify
for each particular problem instance that the uniform assumption does
not hold; either by compressing an observed network using any generative
model (which should be asymptotically impossible under the uniform
assumption~\cite{shannon_mathematical_1948}), or performing a
statistical test designed to be able to reject the uniform null
model. It is exceedingly difficult to find an empirical network for
which the uniform model cannot be rejected with near-absolute
confidence.\label{foot:uniform}} This uniformity between instances also
reveals that there is no meaningful trade-off between algorithms for
most instances, since all algorithms will yield the same negligible
asymptotic performance, with an accuracy tending towards zero as the
number of nodes increases. In this setting, there is not only no free
lunch, but in fact there is no lunch at all (see
Fig.~\ref{fig:trade-off}).

If we were to restrict the space of possible community detection
algorithms to those that provide actual explanations, then by definition
this would imply a positive correlation between network and
partition,\footnote{Note that Eq.~\ref{eq:solvable} is a necessary but
not sufficient condition for the community detection problem to be
solvable. An example of this are networks generated by the SBM, which
are solvable only if the strength of the community structure exceeds a
detectability threshold~\cite{decelle_asymptotic_2011}, even if
Eq.~\ref{eq:solvable} is fulfilled.}
 i.e.
\begin{align}
  P(\A,\bb) &= P(\A|\bb)P(\bb)\\
            &\neq P(\A)P(\bb).\label{eq:solvable}
\end{align}
Not only this implies a specific generative model but, as a consequence,
also an \emph{optimal} community detection algorithm, that operates
based on the posterior distribution
\begin{equation}
  P(\bb|\A) = \frac{P(\A|\bb)P(\bb)}{P(\A)}.
\end{equation}
Therefore, \emph{learnable} community detection problems are invariably
tied to an \emph{optimal} class of algorithms, undermining to a
substantial degree the relevance of the NFL theorem in practice. In
other words, whenever there is an actual community structure in the
network being considered --- i.e. due to a systematic correlation
between $\A$ and $\bb$, such that $P(\A,\bb)\ne P(\A)P(\bb)$ --- there
will be algorithms that can exploit this correlation better than others
(see Fig.~\ref{fig:nfl}(b) for an example of a learnable community
detection problem). Importantly, the set of learnable problems form only
an infinitesimal fraction of all problem instances, with a measure that
tends to zero as the number of nodes increases, and hence remain firmly
out of scope of the NFL theorem. This observation has been made before,
and is equally valid, in the wider context of NFL theorems beyond
community detection~\cite{streeter_two_2003,mcgregor_no_2006,everitt_universal_2013,lattimore_no_2013,schurz_humes_2019}.

Note that since there are many ways to choose a nonuniform model
according to Eq.~\ref{eq:solvable}, the optimal algorithms will still
depend on the particular assumptions made via the choice of $P(\A,\bb)$
and how it relates to the true distribution. However, this does not
imply that all algorithms have equal performance on compressible problem
instances. If we sample a problem from the universe $\mathcal{H}_1$,
with $P(\A,\bb|\mathcal{H}_1)$, but use instead two algorithms optimal
in $\mathcal{H}_2$ and $\mathcal{H}_3$, respectively, their relative
performances will depend on how close each of these universes is to
$\mathcal{H}_1$, and hence will not be in general the same. In fact, if
our space of universes is finite, we can compose them into a single
unified universe~\cite{jaynes_probability_2003} according to
\begin{equation}
  P(\A,\bb) = \sum_{i=1}^{M}P(\A,\bb|\mathcal{H}_i)P(\mathcal{H}_i),
\end{equation}
which will incur a compression penalty of at most $\log_2M$ bits added
to the description length of the optimal algorithm. This gives us a
path, based on hierarchical Bayesian models and minimum description
length, to achieve optimal or near-optimal performance on instances of
the community detection problem that are actually solvable, simply by
progressively expanding our set of hypotheses.

The idea that we can use compression as an inference criterion has been
formalized by Solomonoff's theory of inductive
inference~\cite{solomonoff_formal_1964}, which forms a rigorous
induction framework based on the principle of Occam's
razor. Importantly, the expected errors of predictions achieved under
this framework are provably upper-bounded by the Kolmogorov complexity
of the data generating process~\cite{hutter_universal_2007}, making the
induction framework consistent. As we mentioned already in
Sec.~\ref{sec:inference}, the Kolmogorov complexity is a generalization
of the description length we have been using, and it is defined by the
length of the shortest binary program that generates the data. The only
major limitation of Solomonoff's framework is its uncomputability,
i.e. the impossibility of determining Kolmogorov's complexity with any
algorithm~\cite{li_introduction_2008}. However, this impossibility does
not invalidate the framework, it only means that induction cannot be
fully automatized: we have a consistent criterion to compare hypotheses,
but no deterministic mechanism to produce directly the best
hypothesis. There are open philosophical questions regarding the
universality of this inductive
framework~\cite{hutter_open_2009,montanez_why_2017}, but whatever
fundamental limitations it may have do not follow directly from NFL
theorems such as the one from Ref.~\cite{peel_ground_2017}. In fact, as
mentioned in footnote~\ref{foot:uniform}, it is a rather simple task to
use compression to reject the uniform hypothesis forming the basis of
the NFL theorem for almost any network data.

Since compressive community detection problems are out of the scope of
the NFL theorem, it is not meaningful to use it to justify avoiding
comparisons between algorithms, on the grounds that all choices must be
equally ``good'' in a fundamental sense. In fact, we do not need much
sophistication to reject this line of argument, since the NFL theorem
applies also when we are considering trivially inane algorithms,
e.g. one that always returns the same partition for every network. The
only domain where such an algorithm is as good as any other is when we
have no community \emph{structure} to begin with, which is precisely
what the NFL theorem relies on.

Nevertheless, there are some lessons we can draw from the NFL
theorem. It makes it clear that the performances of algorithms are tied
directly to the inductive bias adopted, which should always be made
explicit. The superficial interpretation of the NFL theorem as an
inherent equity between all algorithms stems from the assumption that
considering all problem instances uniformly is equivalent to being free
of an inductive bias, but that is not possible. The uniform assumption
is itself an inductive bias, and one that it is hard to justify in
virtually any context, since it involves almost exclusively unsolvable
problems (from the point of view of compressibility). In contrast,
considering only \emph{compressible} problem instances is also an
inductive bias, but one that relies only on Occam's razor as a guiding
principle. The advantage of the latter is that it is independent of
domain of application, i.e. we are requiring only that an inferred
partition can help explaining the network in some manner, without having
to specify exactly how \emph{a priori}.

In view of the above observations, it becomes easier to understand
results such as of Ghasemian \emph{et
al}~\cite{ghasemian_evaluating_2019} who found that compressive
inferential community detection methods tend to systematically
outperform descriptive methods in empirical settings, when these are
employed for the task of edge prediction. Even though edge prediction
and community detection are not the same task, and using the former to
evaluate the latter can lead in some cases to
overfitting~\cite{valles-catala_consistencies_2018}, typically the most
compressive models will also lead to the best generalization. Therefore,
the superior performance of the inferential methods is understandable,
even though Ghasemian \emph{et al} also found a minority of instances
where some descriptive methods can outperform inferential ones. To the
extent that these minority results cannot be attributed to overfitting,
or technical issues such as insufficient MCMC equilibration, it could
simply mean that the structure of these networks fall sufficiently
outside of what is assumed by the inferential methods, but without it
being a necessary trade-off that comes as a consequence of the NFL
theorem --- after all, under the uniform assumption, edge prediction is
also strictly impossible, just like community detection.  In other
words, these results do not rule out the existence of an algorithm that
works better in all cases considered, at least if their number is not
too large.\footnote{It is important to distinguish the actual statement
of the NFL theorem
--- ``all algorithms perform equally well when averaged over all problem
instances'' --- from the alternative statement: ``No single algorithm
exhibits strictly better performance than all others over all
instances.'' Although the latter is a corollary of the former, it can
also be true when the former is false. In other words, a particular
algorithm can be better on average over relevant problem instances, but
still underperform for some of them. In fact, it would only be possible
for an algorithm to strictly dominate all others if it can always
achieve perfect accuracy for every instance. Otherwise, there will be at
least one algorithm (e.g. one that always returns the same partition)
that can achieve perfect accuracy for a single network where the optimal
algorithm does not (``even a broken clock is right twice a
day''). Therefore, sub-optimal algorithms can eventually outperform
optimal ones by chance when a sufficiently large number of instances is
encountered, even when the NFL theorem is not applicable (and therefore
this fact is not necessarily a direct consequence of it).}
 In fact, this is precisely what is
achieved in Ref.~\cite{ghasemian_stacking_2020} via model stacking,
i.e. a combination of several predictors into a meta-predictor that
achieves systematically superior performance. This points indeed to the
possibility of using universal methods to discover the latent
\emph{compressive} modular structure of networks, without any tension
with the NFL theorem.

\subsection{``Statistical inference requires us to believe the generative model being used.''}\label{sec:believe}

We have been advocating for the use of statistical inference for
community detection in networks, whenever our objective is of an
inferential nature.

One possible objection to the use of statistical inference is when the
generative models on which they are based are considered unrealistic for
a particular kind of network. Although this type of consideration is
ultimately important, it is not necessarily an obstacle. First we need
to remember that realism is a matter of degree, not kind, since no model
can be fully realistic, and therefore we should never be fully committed to
``believe'' any particular model. Because of this, an inferential
approach can be used to target a particular kind of structure, and the
corresponding model is formulated with this in mind, but without the
need to describe other properties of the data. The SBM is a good example
of this, since it is often used with the objective of finding
communities, rather than any kind of network structure. A model like the
SBM is a good way to offset the regularities that relate to the
community structure with the irregularities present in real networks,
without requiring us to believe that in fact it generated the network.

Furthermore, certain kinds of models are flexible enough so that they
can approximate other models. For example, a good analogy with fitting
the SBM to network data is to fit a histogram to numerical data, with
the node partitioning being analogous to the data binning. Although a
piecewise constant model is almost never the true underlying
distribution, it provides a reasonable approximation in a tractable,
nonparametric manner. Because of its capacity to approximate a wide
class of distributions, we certainly do not need to believe that a
histogram is the true data generating process to extract meaningful
inferences from it. In fact, the same can be said of the SBM in its
capacity to approximate a wide class of network
models~\cite{olhede_network_2014,young_universality_2018}.

The above means that we can extract useful, statistically meaningful
information from data even if the models we use are misspecified. For
example, if a network is generated by a latent space
model~\cite{hoff_latent_2002}, and we fit a SBM to it, the communities
that are obtained in this manner are not quite meaningless: they will
correspond to discrete spatial regions. Hence, the inference would yield
a caricature of the underlying latent space, amounting to a
discretization of the true model --- indeed, much like a histogram. This
is very different from, say, finding communities in an Erd\H{o}s-Rényi
graph, which bear no relation to the true underlying model, and would be
just overfitting the data. In contrast, the SBM fit to a spatial network
would be approximately capturing the true model structure, in a manner
that could be used to compress the data and make predictions (although
not optimally).

Furthermore, the associated description length of a network model is a
good criterion to tell whether the patterns we have found are actually
simplifying our network description, without requiring the underlying
model to be perfect. This happens in the same way as using a software
like \texttt{gzip} makes our files smaller, without requiring us to
believe that they are in fact generated by the Markov chain
underlying the Lempel-Ziv algorithm~\cite{ziv_universal_1977}.

Of course, realism becomes important as soon as we demand more from the
point of view of interpretation and prediction. Are the observed
community structures due to homophily or triadic
clusure~\cite{peixoto_disentangling_2022}? Or are they due to spatial
embedding~\cite{hoff_latent_2002}? What models are capable of
reproducing other network descriptors, together with the community
structure? Which models can better reconstruct incomplete
networks~\cite{guimera_missing_2009,peixoto_reconstructing_2018}? When
answering these questions, we are forced to consider more detailed
generative processes, and compare them. However, we are never required
to \emph{believe} them --- models are always tentative, and should
always be replaced by superior alternatives when these are
found. Indeed, criteria such as MDL serve precisely to implement such a
comparison between models, following the principle of Occam's
razor. Therefore, the lack of realism of any particular model cannot be
used to dismiss statistical inference as an underlying methodology. On
the contrary, the Bayesian workflow~\cite{gelman_bayesian_2020} enables
a continuous improvement of our modelling apparatus, via iterative model
building, model checking, and validation, all within a principled and
consistent framework.

It should be emphasized that, fundamentally, there is no
alternative. Rejecting an inferential approach based on the SBM on the
grounds that it is an unrealistic model (e.g. because of the conditional
independence of the edges being placed, or some other unpalatable
assumption), but instead preferring some other non-inferential community
detection method is incoherent: As we discussed in
Sec.~\ref{sec:implicit}, every descriptive method can be mapped to an
inferential analogue, with implicit assumptions that are hidden from
view. Unless one can establish that the implicit assumptions are in fact
more realistic, then the comparison cannot be justified. Unrealistic
assumptions should be replaced by more realistic ones, not by burying
one's head in the sand.

\subsection{``Inferential approaches are prohibitively expensive.''}\label{sec:performance}

One of the reasons why descriptive methods such as modularity
maximization are widely used is because of very efficient heuristics
that enable their application for very large networks. The most famous
of which is the Louvain algorithm~\cite{blondel_fast_2008}, touted for
its speed and good ability to find high-scoring partitions. A more
recent variation of this method is the Leiden
algorithm~\cite{traag_louvain_2019}, which is a refinement of the
Louvain approach, designed to achieve even more high-scoring partitions,
without sacrificing speed. None of these methods were developed with the
purpose of assessing the statistical evidence of the partitions found,
and since they are most often employed as modularity maximization
techniques, they suffer from all the shortcomings that come with it.

It is often perceived that principled inferential approaches based on
the SBM, designed to overcome all of the shortcomings of descriptive
methods including modularity maximization, are comparatively much
slower, often prohibitively so. However, we show here that this
perception is quite inaccurate, since modern inferential approaches can
be quite competitive. From the point of view of algorithmic complexity,
agglomerative~\cite{peixoto_efficient_2014} or merge-split
MCMC~\cite{peixoto_merge-split_2020} have at most a log-linear
complexity $O(E\log^2 N)$, where $N$ and $E$ are the number of nodes and
edges, respectively, when employed to find the most likely
partition. This means they belong to the same complexity class as the
Louvain and Leiden algorithms, despite the fact the SBM-based algorithms
are in fact more general, and do not attempt to find strictly
assortative structures --- and hence cannot make any optimizations that
are only applicable in this case, as done by Louvain and Leiden. In
practice, all these algorithms return results in comparable times.

\FloatBarrier

\begin{figure}
  \begin{tabular}{cc}
    \includegraphicsl{(a)}{width=.5\textwidth}{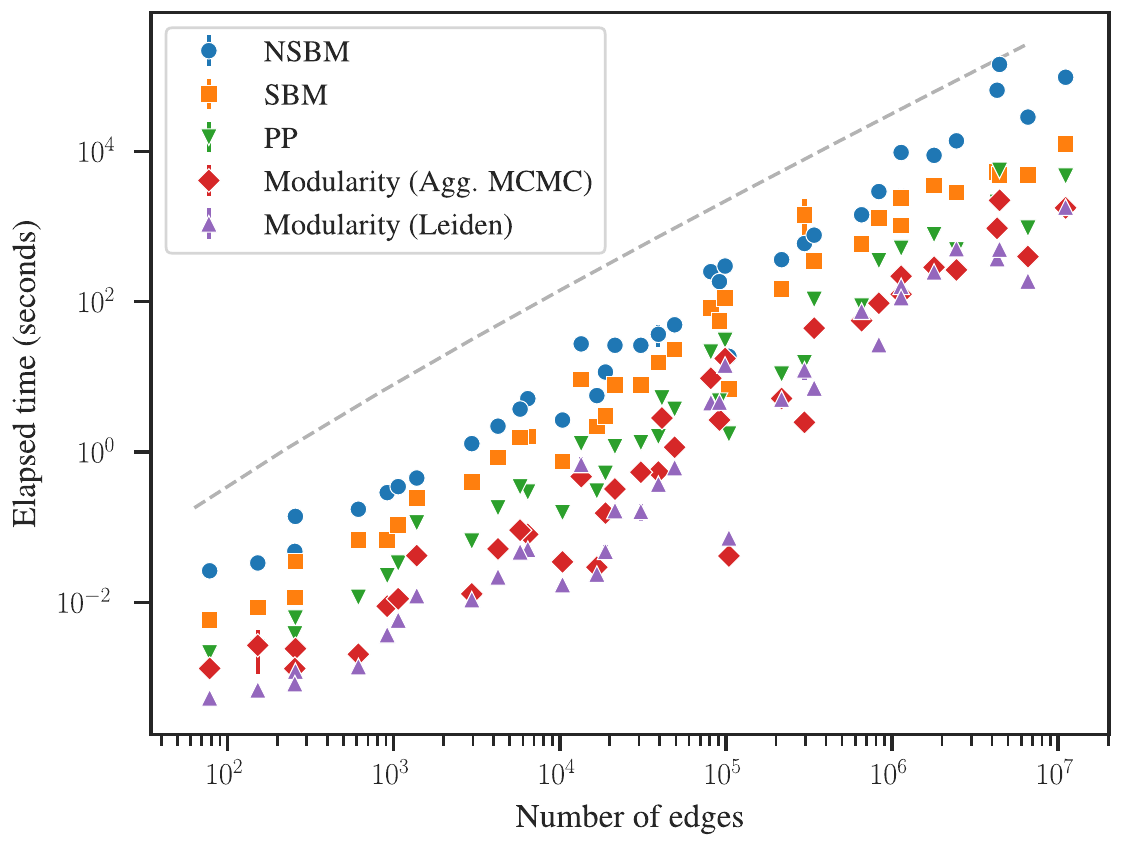}&
    \includegraphicsl{(b)}{width=.5\textwidth}{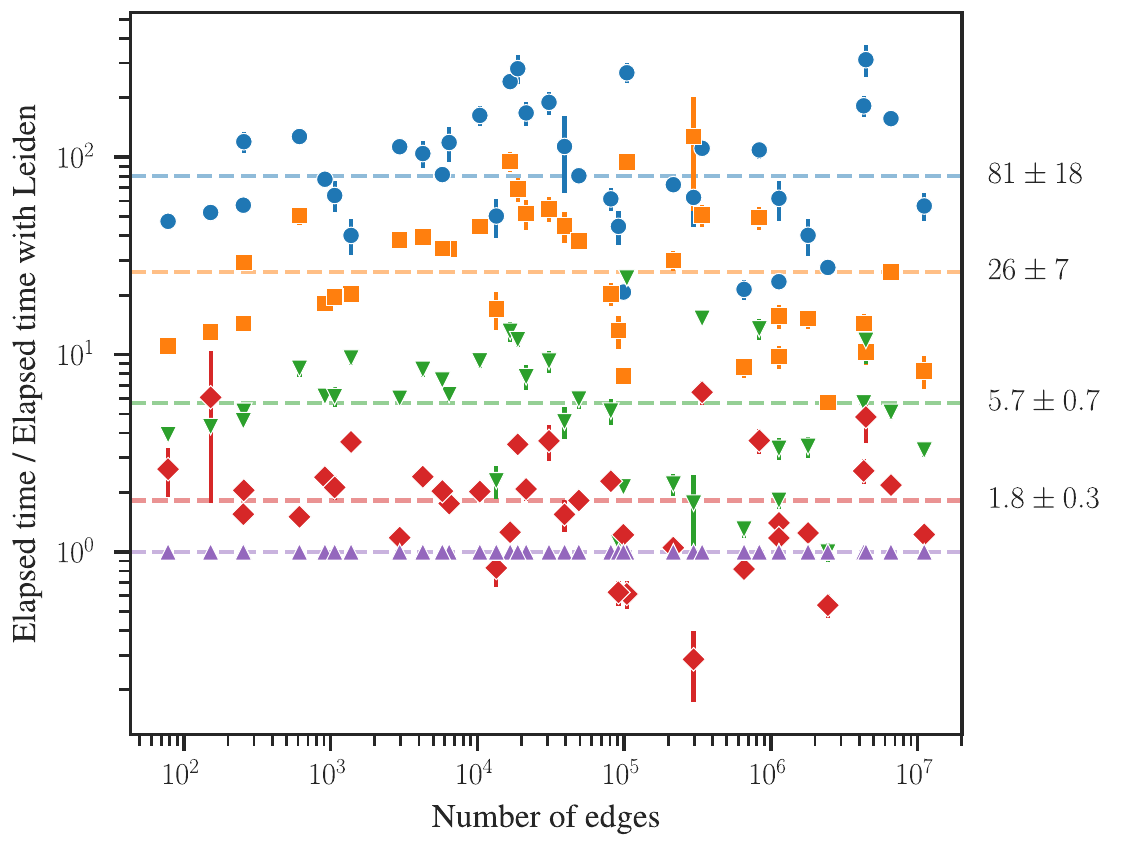}
  \end{tabular}

  \caption{Inferential algorithms show competitive performance with
  descriptive ones. In panel (a) is shown the run-time of the Leiden
  algorithm~\cite{traag_louvain_2019} and the agglomerative
  MCMC~\cite{peixoto_efficient_2014} for modularity, and three SBM
  parametrizations: planted partition (PP), degree-corrected SBM, and
  nested degree-corrected SBM, for 38 empirical
  networks~\cite{peixoto_netzschleuder_2020}. All experiments were done
  on a laptop with an i9-9980HK Intel CPU, and averaged over at least 10
  realizations. The dashed line shows an $O(E\log^2 E)$ scaling. In (b)
  are shown the same run times, but relative to the Leiden
  algorithm. The horizontal dashed lines show the median
  values. \label{fig:performance}}
\end{figure}

In Fig.~\ref{fig:performance} we show a performance comparison between
various algorithms on 38 empirical networks of various domains and
number of edges spanning six orders of magnitude, obtained from the
Netzschleuder repository~\cite{peixoto_netzschleuder_2020}. We used the
Leiden implementation provided by its authors,\footnote{Retreived from
\url{https://github.com/vtraag/leidenalg}.} and compared with various
SBM parametrizations implemented in the \texttt{graph-tool}
library~\cite{peixoto_graph-tool_2014}. In particular we consider the
agglomerative MCMC of Ref.~\cite{peixoto_efficient_2014} employed for
modularity maximization, the Bayesian planted partition (PP)
model~\cite{zhang_statistical_2020}, the degree-corrected SBM with
uniform priors~\cite{peixoto_nonparametric_2017} and the nested
SBM~\cite{peixoto_hierarchical_2014,peixoto_nonparametric_2017}. As seen
in Fig.~\ref{fig:performance}(a), all algorithms display the same
scaling with the number of edges, and differ only by an approximately
constant factor. This difference is speed is due to the more complex
likelihoods used by the SBM and additional data structures that are
needed for its computation. When the agglomerative
MCMC~\cite{peixoto_efficient_2014} is used with the simpler modularity
function, it comes very close to the Leiden algorithm, despite not
taking advantage of any custom optimization for that particular quality
function. When used with the strictly assortative PP model, the
algorithm slows down by a larger factor when compared to Leiden --- most
of which can be attributed to the increased complexity of the quality
function. For the general SBM and nested SBM the algorithm slows down
further, since now it is searching for arbitrary mixing patterns (not
only assortative ones) and entire modular hierarchies. Indeed the
performance difference between the most complex SBM and Leiden can be
substantial, but at this point it also becomes an apples-and-oranges
comparison, since the inferential method not only is not restricted to
assortative communities, but it also uncovers an entire hierarchy of
partitions in a nonparametric manner, while being unhindered by the
resolution limit and with protection against overfitting. Overall, if a
practitioner is considering modularity maximization, they should prefer
instead at least the Bayesian PP model, which solves the same kind of
problem but it is not marred by all the shortcomings of modularity,
including the resolution limit and systematic overfitting, while still
being comparatively fast. The more advanced SBM formulations allow the
researcher to probe a wider array of mixing patterns, without abdicating
from statistical robustness, at the expense of increased computation
time. As this analysis shows, all algorithms are accessible for fairly
large networks of up to $10^7$ edges on a laptop, but in fact can scale
to $10^9$ or more on HPC systems.

Based on the above, it becomes difficult to justify the use modularity
maximization based solely on performance concerns, even on very large
networks, since there are superior inferential approaches available with
comparable speed, and which achieve more meaningful results in
general.\footnote{In this comparison we consider only the task of
finding point estimates, i.e. best scoring partitions. This is done to
maintain an apples-to-apples comparison, since this all that can be
obtained with the Leiden and other modularity maximization
algorithms. To take full advantage of the Bayesian framework we would
need to characterize the full posterior distribution instead, and sample
partitions from it, instead of maximizing it, which incurs a larger
computational cost and requires a more detailed
analysis~\cite{peixoto_revealing_2021}. We emphasize, however, that the
point estimates obtained with the SBM posterior already contain a
substantial amount of regularization, and will not overfit the number of
communities, for example.}

\subsection{``Belief propagation outperforms MCMC.''}

The method of belief propagation (BP)~\cite{decelle_asymptotic_2011} is
an alternative algorithm to MCMC for inferring the partitions from the
posterior distribution of the SBM in the semi-parametric case where the
model parameters controlling the probability of connections between
groups and the expected sizes of the groups are known \emph{a
priori}. It relies on the assumption that the network analyzed was truly
sampled from the SBM, that the number of groups is much smaller than the
number of nodes, $B\ll N$, and the network is sufficiently large, $N\gg
1$. Even though none of these assumptions are likely to hold in
practice, BP is an extremely useful and powerful algorithm since it
returns an estimate of the marginal posterior probability that is not
stochastic, unlike MCMC. Furthermore, it is amenable to analytical
investigations, which was used to uncover the detectability threshold of
the SBM~\cite{decelle_inference_2011,decelle_asymptotic_2011}, and
important connections with spectral
clustering~\cite{krzakala_spectral_2013}. It is often claimed, however,
that it is also faster than MCMC when employed for the same task. This
is, however, not quite true in general, as we now discuss. The
complexity of BP is $O(\tau NB^2)$, where $\tau$ is the convergence
time, which is typically small compared to the other quantities [for the
DC-SBM the complexity becomes $O(\tau \ell NB^2)$, where $\ell$ is the
number of distinct degrees in the network~\cite{yan_model_2014}]. A MCMC
sweep of the SBM, i.e. the number of operations required to give a
chance of each node to be moved once from its current node membership,
can be implemented in time $O(N)$, independent of the number of groups
$B$~\cite{peixoto_efficient_2014,peixoto_merge-split_2020}, when using
the parametrization of
Refs.~\cite{peixoto_parsimonious_2013,peixoto_nonparametric_2017}. This
means that the performance difference between both approaches can be
substantial when the number of groups is large. In fact, if
$B=O(\sqrt{N})$ which a reasonable reference for empirical networks, BP
becomes $O(N^2)$ while MCMC remains $O(N)$. Agglomerative MCMC
initialization schemes, which can significantly improve the mixing time,
have themselves a complexity $O(N\log^2
N)$~\cite{peixoto_efficient_2014}, still significantly faster than BP
for large $B$.

\begin{figure}
  \includegraphics[width=.6\textwidth]{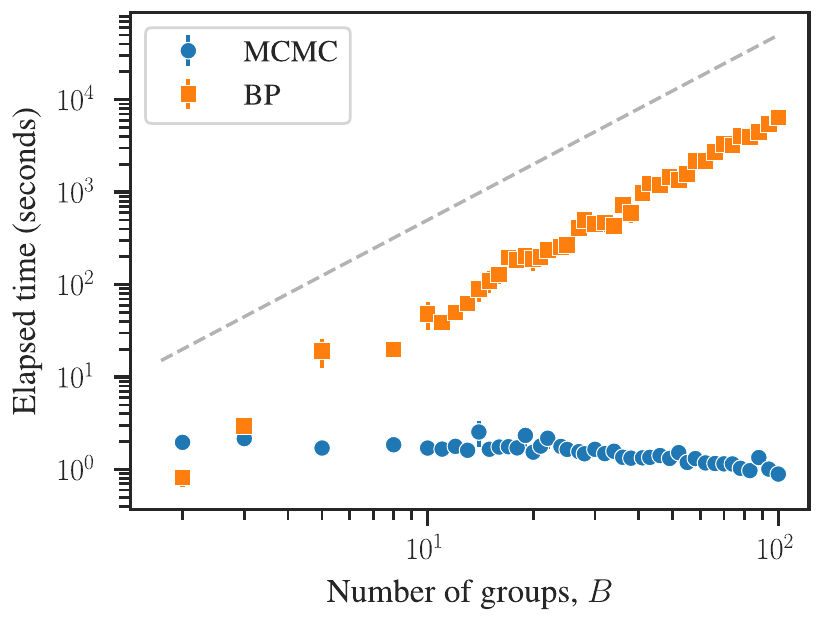} \caption{Comparison
  of run times between MCMC and BP on laptop with an i9-9980HK Intel
  CPU, for a network of flights between airports, with $N=3188$ nodes
  and $E=18833$. We used the agglomerative algorithm of
  Ref.~\cite{peixoto_efficient_2014}, and initialized BP with the model
  parameters found with MCMC. The dashes line shows a $B^2$
  slope.\label{fig:BP}}
\end{figure}

In Fig.~\ref{fig:BP} we show a run-time comparison between BP and MCMC
for an empirical network of flights between airports.\footnote{Obtained
from \url{https://openflights.org/data.html}.} As the number of groups
increases, the run-time of BP grows quadratically, as expected, while
for MCMC it remains constant. There are several caveats in this
comparison, which is somewhat apples-to-oranges: BP outputs a full
marginal distribution for every node, containing even probabilities that
are very low, while for MCMC we obtain anything from a point estimate to
full marginal or joint probabilities, at the expense of longer running
times, which is not revealed by the comparison in Fig.~\ref{fig:BP},
which corresponds only to a point estimate. On the other hand, BP
requires a value of the model parameters besides the partition itself,
which can in principle be obtained together with the marginals via
expectation-maximization (EM)~\cite{decelle_asymptotic_2011}, although a
meaningful convergence for complex problems cannot be guaranteed with
this algorithm~\cite{kawamoto_algorithmic_2018}. Overall, we can state
that some answers can be achieved in log-linear time with MCMC
independently from the number of groups (and requiring no particular
assumptions on the data), while with BP we can never escape the
quadratic dependence on $B$.

We emphasize that BP is only applicable in the semiparametric case,
where the number of groups and model parameters are known. The
nonparametric case considered in Sec.~\ref{sec:inference}, which is
arguably more relevant in practice, cannot be tackled using BP, leaving
MCMC as the only game in town, at least with the current
state-of-the-art.

\subsection{``Spectral clustering outperforms likelihood-based methods.''}

Spectral clustering methods divide a network into groups based on the
leading eigenvectors of a linear operator associated with the network
structure~\cite{spielman_spectral_2007,von_luxburg_tutorial_2007}. There
are important connections between spectral methods and statistical
inference, in particular there are certain linear operators that can be
shown to provide a consistent estimation of the
SBM~\cite{rohe_spectral_2011,krzakala_spectral_2013}. However, when
compared to likelihood-based methods, spectral methods are only
approximations, as they amount to a simplification of the
problem. Nevertheless, one of the touted advantages of this class of
methods is that they tend to be significantly faster than likelihood
based methods using MCMC. But like in the case of BP considered in the
previous section, the run-time of spectral methods is intimately related
to the number of groups one wishes to infer, unlike MCMC. Independently
of the operator being used, the clustering into $B$ groups requires the
computation of the first $B$ leading eigenvectors. The most efficient
algorithms for this purpose are based on the implicitly restarted
Arnoldi method~\cite{lehoucq_deflation_1996}, which has a worse-case
time complexity $O(NB^2)$ for sparse matrices. Therefore, for
sufficiently large number of groups they can cease to be faster than
MCMC, which has a run-time complexity independent of the number of
groups~\cite{peixoto_efficient_2014,peixoto_merge-split_2020}.

\FloatBarrier
\begin{figure}
  \includegraphics[width=.6\textwidth]{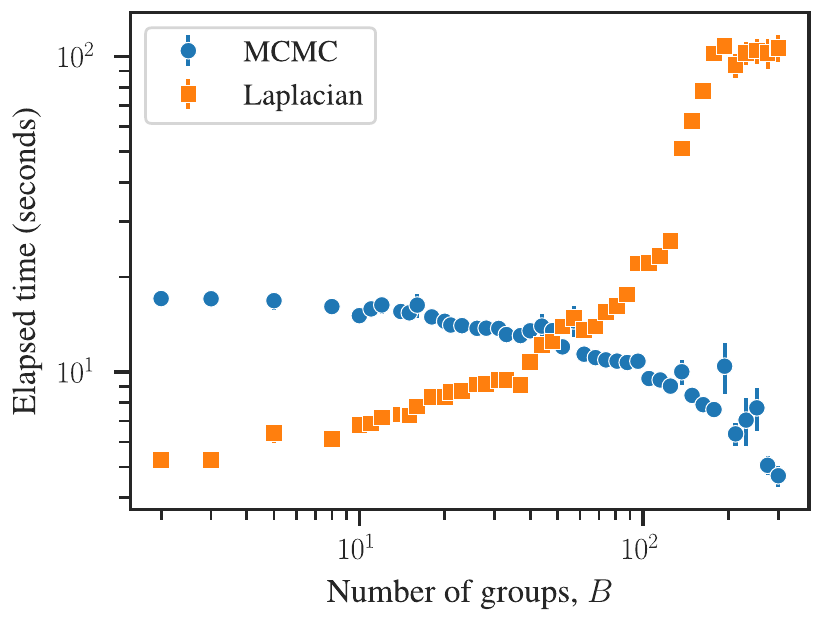}
  \caption{Comparison
  of run times between MCMC and spectral clustering using the Laplacian
  matrix, on a laptop with an i9-9980HK Intel CPU, for the Anybeat
  social network~\cite{fire_link_2013}, with $N=12645$ vertices and
  $E=49132$ edges. We used the agglomerative algorithm of
  Ref.~\cite{peixoto_efficient_2014} and the ARPACK eigenvector
  solver~\cite{lehoucq_ARPACK_1998}.\label{fig:spectral}}
\end{figure}

In Fig.~\ref{fig:spectral} we show a comparison of spectral clustering
and MCMC inference for the Anybeat social
network~\cite{fire_link_2013}. Indeed, for small number of groups
spectral clustering can be significantly faster, but eventually becomes
slower as the number of groups increases. The complexity of the spectral
algorithm does not scale exactly like the worse case $O(NB^2)$ in
practice, and the actual times will depend on the details of the
particular operator. The MCMC algorithm becomes slightly faster, on the
other hand, since the agglomerative initialization heuristic used
terminates sooner when more groups are
imposed~\cite{peixoto_efficient_2014}. As usual, there are caveats with
this comparison. First, the eigenvectors by themselves do not provide a
clustering of the network. Usually, these are given as input to a
general-purpose clustering algorithm, typically $k$-means, which itself
also has a complexity $O(NB^2)$, not included in the comparison of
Fig.~\ref{fig:spectral}. Furthermore, spectral clustering usually
requires the number of groups itself to be known in advance --- although
heuristics exist for spectral algorithms, but which usually require a
significant part of the entire spectrum to be
determined~\cite{krzakala_spectral_2013}. Likelihood-based methods, if
implemented as a nonparametric Bayesian posterior like done in
Sec.~\ref{sec:inference}, do not require this prior information. On the
other hand, spectral methods can be parallelized rather easily, unlike
MCMC, and hence can take advantage of multicore processors.

\subsection{``Bayesian posterior, MDL, BIC and AIC are different but equally valid model selection criteria.''}

One outstanding problem with using inferential community detection is
that the likelihood of a model like the SBM does not, by itself, offer a
principled way to determine the appropriate number of groups. This is
because if we maximize the likelihood directly, it will favor a number
of groups that is equal to the number of nodes, i.e. an extreme
overfitting. This is similar to what happens when we fit a polynomial to
a set of one-dimensional data points by varying its degree: for a degree
equal to the number of points we can fit any set of points perfectly,
but we are guaranteed to be overfitting the data. In other words, if we
do not account for model complexity explicitly, we cannot separate
randomness from structure.

In the literature we often see mentions of Bayesian posterior inference,
minimum description length
(MDL)~\cite{rissanen_modeling_1978,grunwald_minimum_2007}, as well as
likelihood penalty schemes such as Bayesian Information Criterion
(BIC)~\cite{schwarz_estimating_1978} and Akaike's Information Criterion
(AIC)\cite{akaike_new_1974}, as being equally valid alternatives that
can be used to solve this problem. It is sometimes said that the choice
between them is philosophical and often simply reflects the culture that
a researcher stems from. As we show here, this is demonstrably
incorrect, since Bayes, MDL, and BIC are in fact the same criterion,
where BIC is simply an (arguably crude) approximation of the first two,
which are in fact identical. AIC is indeed a different criterion, but,
like BIC, it involves approximations that are known to be invalid for
community detection.

The exact equivalence between MDL and Bayesian inference is easy to
demonstrate~\cite{peixoto_hierarchical_2014,peixoto_nonparametric_2017},
as we have already done already in Sec.~\ref{sec:inference}. Namely, the
posterior distribution of the community detection problem is given by
\begin{align}
  P(\bb|\A) &= \frac{P(\A|\bb)P(\bb)}{P(\A)},\label{eq:bayes_dl}\\
            &= \frac{2^{-\Sigma(\A,\bb)}}{P(\A)},
\end{align}
where the numerator of Eq.~\ref{eq:bayes_dl} is related to the description
length $\Sigma(\A,\bb)$ via
\begin{equation}\label{eq:dl}
  \Sigma(\A,\bb) = -\log_2P(\A|\bb)-\log_2P(\bb).
\end{equation}
Therefore, maximizing Eq.~\ref{eq:bayes_dl} is identical to minimizing
Eq.~\ref{eq:dl}. Although this is already sufficient to demonstrate
their equivalence, we can go in even more detail and show that the
marginal integrated likelihood,
\begin{equation}\label{eq:canonical_marginal}
  P(\A|\bb) = \int P(\A|\bm\omega,\bm\kappa,\bb)P(\bm\omega,\bm\kappa|\bb)\,\dd\bm\omega\,\dd\bm\kappa,
\end{equation}
where $\bm\omega$ and $\bm\kappa$ are the parameters of the canonical
DC-SBM~\cite{karrer_stochastic_2011}, is identical to the marginal
likelihood of the microcanonical SBM we have used in
Eq.~\ref{eq:dcsbm-marginal}. This is proved in
Ref.~\cite{peixoto_nonparametric_2017}. Therefore, the MDL criterion is
simply an information-theoretical interpretation of the Bayesian
approach, and the two methods coincide in their
implementation.\footnote{In general, it is possible to construct
particular MDL formulations of ``universal codes'' that do not have a
clear Bayesian interpretation~\cite{grunwald_minimum_2007}. However,
these formulations are typically intractable and seldom find an
application. All MDL uses encountered in practice for the community
detection problem are equivalent to Bayesian methods.}

The BIC criterion is based on the exact same framework, but it amounts
to an approximation of the integrated marginal likelihood of a generic
model $\mathcal M$, $ P(\bm D|\bm\theta,\mathcal M)$, where $\bm D$ is a
data vector of size $n$ and $\bm\theta$ is a parameter vector of size
$k$, given by
\begin{align}
  P(\bm D|\mathcal M) &= \int P(\bm D|\bm\theta,\mathcal M)P(\bm\theta)\,\dd\bm\theta,\\\label{eq:taylor}
  &\approx \left(\frac{2\pi}{n}\right)^{k/2}\left|I(\hat\theta)\right| \hat L\times P(\hat{\bm\theta}),\\
  &\approx \exp(-\text{BIC}/2),
\end{align}
where $[I(\theta)]_{ij}=\int(\partial\ln
P(D|\bm\theta)/\partial\theta_i)(\partial\ln
P(D|\bm\theta)/\partial\theta_j) P(D|\bm\theta)\,\dd\bm\theta$ is the
Fisher information matrix, and the values of the likelihood and parameters
are obtained at the maximum,
\begin{equation}
  \hat L = \underset{\theta}{\max}\;P(\bm D|\bm\theta,\mathcal M),\qquad \hat\theta =
  \underset{\theta}{\operatorname{argmax}}\;P(\bm D|\bm\theta,\mathcal M),
\end{equation}
and finally the BIC score is obtained from Eq.~\ref{eq:taylor} by assuming
$n\gg k$,
\begin{equation}
  \text{BIC} = k \ln n - 2 \ln \hat L.
\end{equation}
The BIC method consists of employing the equation above as criterion to
decide which model to select, applicable even if they have different
number of parameters $k$, with the first term functioning as penalty for
larger models. Eq.~\ref{eq:taylor} corresponds to an approximation of the
likelihood obtained via Laplace's method, which involves a second-order
Taylor expansion of the log-likelihood. Therefore, it requires the
likelihood function to be well approximated by a multivariate Gaussian
distribution with respect to the parameters at the vicinity of its
maximum. However, as demonstrated by Yan \emph{et
al}.~\cite{yan_model_2014}, this assumption is invalid for SBMs, however
large the networks are, as long as they are \emph{sparse},
i.e. with an average degree much smaller than the number of nodes. This
is because for sparse SBMs we have both the number of parameters
$k=O(N)$ [or even larger, since for $B$ groups we a matrix $\bm\omega$
  of size $O(B^2)$, and in principle we could have $B=O(N)$] and
effective data size $n=O(N)$ where $N$ is the number of nodes, therefore
the ``sufficient data'' limit required for the approximation to hold is
never realized for any $N$.  Furthermore, the BIC penalty completely
neglects the contribution of the prior $P(\bm\theta)$ in the
regularization, which cannot be ignored outside of this limit. Since the
vast majority of empirical networks of interest are sparse, this renders
this method unreliable, and in fact it will tend to overfit in most
cases when employed with the SBM. We emphasize that the approximation of
Eq.~\ref{eq:taylor} is unnecessary, since we can compute the marginal
likelihood of Eq.~\ref{eq:canonical_marginal} exactly for most versions
of the
SBM~\cite{guimera_missing_2009,peixoto_hierarchical_2014,come_model_2015,newman_estimating_2016,peixoto_nonparametric_2017}.
When we compare the BIC penalty with the exact values of the integrated
likelihoods we see that they in general produce significantly different
regularizations, even asymptotically, and also even if we add \emph{ad
hoc} parameters,
e.g. $\lambda k\ln n - 2\ln\hat L$.  This is because simply counting the
number of parameters is too crude an estimation of the model complexity,
since it is composed of different classes of parameters occupying
different volumes which need (and can) be more carefully
computed. Therefore the use of BIC for model selection in community
detection should be in general avoided.

Akaike's Information Criterion (AIC)~\cite{akaike_new_1974}, on the
other hand, actually starts out from a different framework. The idea is
assume that the data are sampled from a true generative model $P(\bm
D|\mathcal M_{\text{true}})$, and a candidate model $\mathcal M$ with
its parameter estimates $\hat{\bm\theta}(\bm D)$ is evaluated according
to its Kullback-Leibler (KL) divergence with respect to the true model,
\begin{equation}
  \int P(\bm D'|\mathcal M_{\text{true}}) \ln \frac{P(\bm D'|\hat{\bm\theta}(\bm D), \mathcal M)}{P(\bm D'|\mathcal M_{\text{true}})}\;\dd\bm D'.
\end{equation}
Of course, whenever it is relevant to employ model selection criteria we
do not have access to the true model, which means we cannot compute the
above quantity. We can, however, estimate the following upper bound,
corresponding to the average over all data $\bm D$,
\begin{equation}\label{eq:kld}
  \int P(\bm D|\mathcal M_{\text{true}})P(\bm D'|\mathcal M_{\text{true}}) \ln \frac{P(\bm D'|\hat{\bm\theta}(\bm D), \mathcal M)}{P(\bm D'|\mathcal M_{\text{true}})}\;\dd\bm D'\,\dd\bm D.
\end{equation}
In this case, for sufficiently large data $\bm D$, the above quantity
can be estimated making use of a series of Laplace
approximations~\cite{burnham_model_2002}, resulting in
\begin{equation}
  \ln P(\bm D|\hat{\bm\theta}(\bm D)) - \operatorname{tr}\left[J(\bm\theta_0)I(\bm\theta_0)^{-1}\right],
\end{equation}
where $\bm\theta_0$ is the point around which we compute the quadratic
approximation in Laplace's method, and $J_{ij}(\bm\theta_0) = \int P(\bm D'|\mathcal
M_{\text{true}})\mathcal{I}_{ij}(\bm D,\bm\theta_0)\,\dd\bm D$, $I_{ij}(\bm\theta_0) = \int P(\bm D'|\bm\theta_0,\mathcal M)\mathcal{I}_{ij}(\bm D,\bm\theta_0)\,\dd\bm D$, with
\begin{equation}
  \mathcal{I}_{ij}(\bm D,\hat{\bm\theta}) = \left.\frac{\partial}{\partial\theta_i}\ln P(\bm D'|\bm\theta,\mathcal M)\right|_{\theta_i = \hat\theta_i}\times\left.\frac{\partial}{\partial\theta_j}\ln P(\bm D'|\bm\theta,\mathcal M)\right|_{\theta_j = \hat\theta_j}.
\end{equation}
The AIC criterion is finally obtained
by heuristically assuming
$\operatorname{tr}\left[J(\bm\theta_0)I(\bm\theta_0)^{-1}\right] \approx k$,
yielding
\begin{equation}
  \text{AIC} = 2k - 2\ln P(\bm D|\hat{\bm\theta}(\bm D)),
\end{equation}
where the overall sign and multiplicative factor is a matter of
convention. It is also possible to recover AIC from BIC by making a
choice of prior $P(\mathcal M)\propto\exp(k\ln n/2 -
k)$~\cite{burnham_model_2002}, which makes it clear that it favors more
complex models over BIC. Independently of how one judges the suitability
of the fundamental criterion of Eq.~\ref{eq:kld}, just like BIC, AIC
involves several approximations that are known to be invalid for sparse
networks. Together with its heuristic nature and crude counting of
parameters, it is safe to conclude that the use of AIC is ill-advised
for community detection, specially considering the more principled and
exact alternatives of Bayes/MDL.

\section{Conclusion}

We have framed the problem of community detection under two different
paradigms, namely that of ``inference'' and ``description.'' We argued
that statistical inference is unavoidable when the objective is to draw
inferential interpretations from the communities found, and we provided
a simple ``litmus test'' to help deciding when this is indeed the
case. Under this framing, we showed that descriptive methods always come
with hidden inferential assumptions, and reviewed the dangers of
employing descriptive methods with inferential aims, focusing on
modularity maximization as a representative (and hence not unique) case.

We covered a series of pitfalls encountered in community detection,
as well as myths and half-truths commonly believed, and attempted to
clarify them under the same lenses, focusing on simple examples and
conceptual arguments.

Although it is true that community detection in general involves diverse
aims, and hence it is difficult to argue for an one-size-fits-all
approach, here we have taken a more opinionated stance, since it is also
not true that all approaches are used in a manner consistent with their
intended aims. We have clearly favored inferential methods, since they
are more theoretically grounded, are better aligned with well-defined
scientific questions (whenever those involve inferential queries), are
more widely applicable, and can be used to develop more robust
algorithms.

Inferential methodology for community detection has reached a level of
maturity, both in our understanding of them and in the efficiency of
available implementations, that should make it the preferred choice when
analysing network data, whenever the ultimate goal has an inferential
nature.

\bibliography{bib,extra}

%% file: article.bbl
%merlin.mbs apsrev4-1.bst 2010-07-25 4.21a (PWD, AO, DPC) hacked
%Control: key (0)
%Control: author (0) dotless jnrlst
%Control: editor formatted (1) identically to author
%Control: production of article title (0) allowed
%Control: page (1) range
%Control: year (0) verbatim
%Control: production of eprint (0) enabled
\begin{thebibliography}{132}%
\makeatletter
\providecommand \@ifxundefined [1]{%
 \@ifx{#1\undefined}
}%
\providecommand \@ifnum [1]{%
 \ifnum #1\expandafter \@firstoftwo
 \else \expandafter \@secondoftwo
 \fi
}%
\providecommand \@ifx [1]{%
 \ifx #1\expandafter \@firstoftwo
 \else \expandafter \@secondoftwo
 \fi
}%
\providecommand \natexlab [1]{#1}%
\providecommand \enquote  [1]{``#1''}%
\providecommand \bibnamefont  [1]{#1}%
\providecommand \bibfnamefont [1]{#1}%
\providecommand \citenamefont [1]{#1}%
\providecommand \href@noop [0]{\@secondoftwo}%
\providecommand \href [0]{\begingroup \@sanitize@url \@href}%
\providecommand \@href[1]{\@@startlink{#1}\@@href}%
\providecommand \@@href[1]{\endgroup#1\@@endlink}%
\providecommand \@sanitize@url [0]{\catcode `\\12\catcode `\$12\catcode
  `\&12\catcode `\#12\catcode `\^12\catcode `\_12\catcode `\%12\relax}%
\providecommand \@@startlink[1]{}%
\providecommand \@@endlink[0]{}%
\providecommand \url  [0]{\begingroup\@sanitize@url \@url }%
\providecommand \@url [1]{\endgroup\@href {#1}{\urlprefix }}%
\providecommand \urlprefix  [0]{URL }%
\providecommand \Eprint [0]{\href }%
\providecommand \doibase [0]{http://dx.doi.org/}%
\providecommand \selectlanguage [0]{\@gobble}%
\providecommand \bibinfo  [0]{\@secondoftwo}%
\providecommand \bibfield  [0]{\@secondoftwo}%
\providecommand \translation [1]{[#1]}%
\providecommand \BibitemOpen [0]{}%
\providecommand \bibitemStop [0]{}%
\providecommand \bibitemNoStop [0]{.\EOS\space}%
\providecommand \EOS [0]{\spacefactor3000\relax}%
\providecommand \BibitemShut  [1]{\csname bibitem#1\endcsname}%
\let\auto@bib@innerbib\@empty
%</preamble>
\bibitem [{\citenamefont {Fortunato}(2010)}]{fortunato_community_2010}%
  \BibitemOpen
  \bibfield  {author} {\bibinfo {author} {\bibfnamefont {Santo}\ \bibnamefont
  {Fortunato}},\ }\bibfield  {title} {\enquote {\bibinfo {title} {Community
  detection in graphs},}\ }\href {\doibase 16/j.physrep.2009.11.002} {\bibfield
   {journal} {\bibinfo  {journal} {Physics Reports}\ }\textbf {\bibinfo
  {volume} {486}},\ \bibinfo {pages} {75--174} (\bibinfo {year}
  {2010})}\BibitemShut {NoStop}%
\bibitem [{\citenamefont {Fortunato}\ and\ \citenamefont
  {Hric}(2016)}]{fortunato_community_2016}%
  \BibitemOpen
  \bibfield  {author} {\bibinfo {author} {\bibfnamefont {Santo}\ \bibnamefont
  {Fortunato}}\ and\ \bibinfo {author} {\bibfnamefont {Darko}\ \bibnamefont
  {Hric}},\ }\bibfield  {title} {\enquote {\bibinfo {title} {Community
  detection in networks: {A} user guide},}\ }\href {\doibase
  10.1016/j.physrep.2016.09.002} {\bibfield  {journal} {\bibinfo  {journal}
  {Physics Reports}\ } (\bibinfo {year} {2016}),\
  10.1016/j.physrep.2016.09.002}\BibitemShut {NoStop}%
\bibitem [{\citenamefont {Moore}(2017)}]{moore_computer_2017}%
  \BibitemOpen
  \bibfield  {author} {\bibinfo {author} {\bibfnamefont {Cristopher}\
  \bibnamefont {Moore}},\ }\bibfield  {title} {\enquote {\bibinfo {title} {The
  {Computer} {Science} and {Physics} of {Community} {Detection}: {Landscapes},
  {Phase} {Transitions}, and {Hardness}},}\ }\href
  {http://arxiv.org/abs/1702.00467} {\bibfield  {journal} {\bibinfo  {journal}
  {arXiv:1702.00467 [cond-mat, physics:physics]}\ } (\bibinfo {year}
  {2017})}\BibitemShut {NoStop}%
\bibitem [{\citenamefont {Abbe}(2017)}]{abbe_community_2017}%
  \BibitemOpen
  \bibfield  {author} {\bibinfo {author} {\bibfnamefont {Emmanuel}\
  \bibnamefont {Abbe}},\ }\bibfield  {title} {\enquote {\bibinfo {title}
  {Community detection and stochastic block models: recent developments},}\
  }\href {http://arxiv.org/abs/1703.10146} {\bibfield  {journal} {\bibinfo
  {journal} {arXiv:1703.10146 [cs, math, stat]}\ } (\bibinfo {year}
  {2017})}\BibitemShut {NoStop}%
\bibitem [{\citenamefont
  {Peixoto}(2019{\natexlab{a}})}]{peixoto_bayesian_2019}%
  \BibitemOpen
  \bibfield  {author} {\bibinfo {author} {\bibfnamefont {Tiago~P.}\
  \bibnamefont {Peixoto}},\ }\bibfield  {title} {{\selectlanguage
  {english}\enquote {\bibinfo {title} {Bayesian {Stochastic}
  {Blockmodeling}},}\ }}in\ \href {\doibase 10.1002/9781119483298.ch11}
  {{\selectlanguage {english}\emph {\bibinfo {booktitle} {Advances in {Network}
  {Clustering} and {Blockmodeling}}}}}\ (\bibinfo  {publisher} {John Wiley \&
  Sons, Ltd},\ \bibinfo {year} {2019})\ pp.\ \bibinfo {pages}
  {289--332}\BibitemShut {NoStop}%
\bibitem [{\citenamefont {Decelle}\ \emph
  {et~al.}(2011{\natexlab{a}})\citenamefont {Decelle}, \citenamefont
  {Krzakala}, \citenamefont {Moore},\ and\ \citenamefont
  {Zdeborová}}]{decelle_asymptotic_2011}%
  \BibitemOpen
  \bibfield  {author} {\bibinfo {author} {\bibfnamefont {Aurelien}\
  \bibnamefont {Decelle}}, \bibinfo {author} {\bibfnamefont {Florent}\
  \bibnamefont {Krzakala}}, \bibinfo {author} {\bibfnamefont {Cristopher}\
  \bibnamefont {Moore}}, \ and\ \bibinfo {author} {\bibfnamefont {Lenka}\
  \bibnamefont {Zdeborová}},\ }\bibfield  {title} {\enquote {\bibinfo {title}
  {Asymptotic analysis of the stochastic block model for modular networks and
  its algorithmic applications},}\ }\href {\doibase 10.1103/PhysRevE.84.066106}
  {\bibfield  {journal} {\bibinfo  {journal} {Physical Review E}\ }\textbf
  {\bibinfo {volume} {84}},\ \bibinfo {pages} {066106} (\bibinfo {year}
  {2011}{\natexlab{a}})}\BibitemShut {NoStop}%
\bibitem [{\citenamefont {Zdeborová}\ and\ \citenamefont
  {Krzakala}(2016)}]{zdeborova_statistical_2016}%
  \BibitemOpen
  \bibfield  {author} {\bibinfo {author} {\bibfnamefont {Lenka}\ \bibnamefont
  {Zdeborová}}\ and\ \bibinfo {author} {\bibfnamefont {Florent}\ \bibnamefont
  {Krzakala}},\ }\bibfield  {title} {\enquote {\bibinfo {title} {Statistical
  physics of inference: thresholds and algorithms},}\ }\href {\doibase
  10.1080/00018732.2016.1211393} {\bibfield  {journal} {\bibinfo  {journal}
  {Advances in Physics}\ }\textbf {\bibinfo {volume} {65}},\ \bibinfo {pages}
  {453--552} (\bibinfo {year} {2016})}\BibitemShut {NoStop}%
\bibitem [{\citenamefont {Schaub}\ \emph {et~al.}(2017)\citenamefont {Schaub},
  \citenamefont {Delvenne}, \citenamefont {Rosvall},\ and\ \citenamefont
  {Lambiotte}}]{schaub_many_2017}%
  \BibitemOpen
  \bibfield  {author} {\bibinfo {author} {\bibfnamefont {Michael~T.}\
  \bibnamefont {Schaub}}, \bibinfo {author} {\bibfnamefont {Jean-Charles}\
  \bibnamefont {Delvenne}}, \bibinfo {author} {\bibfnamefont {Martin}\
  \bibnamefont {Rosvall}}, \ and\ \bibinfo {author} {\bibfnamefont {Renaud}\
  \bibnamefont {Lambiotte}},\ }\bibfield  {title} {{\selectlanguage
  {english}\enquote {\bibinfo {title} {The many facets of community detection
  in complex networks},}\ }}\href {\doibase 10.1007/s41109-017-0023-6}
  {\bibfield  {journal} {\bibinfo  {journal} {Applied Network Science}\
  }\textbf {\bibinfo {volume} {2}},\ \bibinfo {pages} {1--13} (\bibinfo {year}
  {2017})}\BibitemShut {NoStop}%
\bibitem [{\citenamefont
  {Peixoto}(2014{\natexlab{a}})}]{peixoto_graph-tool_2014}%
  \BibitemOpen
  \bibfield  {author} {\bibinfo {author} {\bibfnamefont {Tiago~P.}\
  \bibnamefont {Peixoto}},\ }\bibfield  {title} {\enquote {\bibinfo {title}
  {The \texttt{graph-tool} python library},}\ }\href {\doibase
  10.6084/m9.figshare.1164194} {\bibfield  {journal} {\bibinfo  {journal}
  {figshare}\ } (\bibinfo {year} {2014}{\natexlab{a}}),\
  10.6084/m9.figshare.1164194},\ \bibinfo {note} {available at
  \url{https://graph-tool.skewed.de}.}\BibitemShut {Stop}%
\bibitem [{\citenamefont {Baker}(2010)}]{baker_cmos_2010}%
  \BibitemOpen
  \bibfield  {author} {\bibinfo {author} {\bibfnamefont {R.~Jacob}\
  \bibnamefont {Baker}},\ }\href@noop {} {{\selectlanguage {english}\emph
  {\bibinfo {title} {{CMOS}: {Circuit} {Design}, {Layout}, and
  {Simulation}}}}},\ \bibinfo {edition} {3rd}\ ed.\ (\bibinfo  {publisher}
  {Wiley-IEEE Press},\ \bibinfo {address} {Piscataway, NJ : Hoboken, NJ},\
  \bibinfo {year} {2010})\BibitemShut {NoStop}%
\bibitem [{\citenamefont {Kernighan}(1969)}]{kernighan_graph_1969}%
  \BibitemOpen
  \bibfield  {author} {\bibinfo {author} {\bibfnamefont {Brian~Wilson}\
  \bibnamefont {Kernighan}},\ }\href@noop {} {\emph {\bibinfo {title} {Some
  graph partitioning problems related to program segmentation}}}\ (\bibinfo
  {publisher} {Princeton University},\ \bibinfo {year} {1969})\BibitemShut
  {NoStop}%
\bibitem [{\citenamefont {Kernighan}\ and\ \citenamefont
  {Lin}(1970)}]{kernighan_efficient_1970}%
  \BibitemOpen
  \bibfield  {author} {\bibinfo {author} {\bibfnamefont {B.W.}\ \bibnamefont
  {Kernighan}}\ and\ \bibinfo {author} {\bibfnamefont {S.}~\bibnamefont
  {Lin}},\ }\bibfield  {title} {\enquote {\bibinfo {title} {An efficient
  heuristic procedure for partitioning graphs},}\ }\href@noop {} {\bibfield
  {journal} {\bibinfo  {journal} {Bell System Technical Journal}\ }\textbf
  {\bibinfo {volume} {49}},\ \bibinfo {pages} {291--307} (\bibinfo {year}
  {1970})}\BibitemShut {NoStop}%
\bibitem [{\citenamefont {Bichot}\ and\ \citenamefont
  {Siarry}(2013)}]{bichot_graph_2013}%
  \BibitemOpen
  \bibfield  {author} {\bibinfo {author} {\bibfnamefont {Charles-Edmond}\
  \bibnamefont {Bichot}}\ and\ \bibinfo {author} {\bibfnamefont {Patrick}\
  \bibnamefont {Siarry}},\ }\href@noop {} {\emph {\bibinfo {title} {Graph
  partitioning}}}\ (\bibinfo  {publisher} {John Wiley \& Sons},\ \bibinfo
  {year} {2013})\BibitemShut {NoStop}%
\bibitem [{\citenamefont {Holland}\ \emph {et~al.}(1983)\citenamefont
  {Holland}, \citenamefont {Laskey},\ and\ \citenamefont
  {Leinhardt}}]{holland_stochastic_1983}%
  \BibitemOpen
  \bibfield  {author} {\bibinfo {author} {\bibfnamefont {Paul~W.}\ \bibnamefont
  {Holland}}, \bibinfo {author} {\bibfnamefont {Kathryn~Blackmond}\
  \bibnamefont {Laskey}}, \ and\ \bibinfo {author} {\bibfnamefont {Samuel}\
  \bibnamefont {Leinhardt}},\ }\bibfield  {title} {\enquote {\bibinfo {title}
  {Stochastic blockmodels: {First} steps},}\ }\href {\doibase
  16/0378-8733(83)90021-7} {\bibfield  {journal} {\bibinfo  {journal} {Social
  Networks}\ }\textbf {\bibinfo {volume} {5}},\ \bibinfo {pages} {109--137}
  (\bibinfo {year} {1983})}\BibitemShut {NoStop}%
\bibitem [{\citenamefont {Karrer}\ and\ \citenamefont
  {Newman}(2011)}]{karrer_stochastic_2011}%
  \BibitemOpen
  \bibfield  {author} {\bibinfo {author} {\bibfnamefont {Brian}\ \bibnamefont
  {Karrer}}\ and\ \bibinfo {author} {\bibfnamefont {M.~E.~J.}\ \bibnamefont
  {Newman}},\ }\bibfield  {title} {\enquote {\bibinfo {title} {Stochastic
  blockmodels and community structure in networks},}\ }\href {\doibase
  10.1103/PhysRevE.83.016107} {\bibfield  {journal} {\bibinfo  {journal}
  {Physical Review E}\ }\textbf {\bibinfo {volume} {83}},\ \bibinfo {pages}
  {016107} (\bibinfo {year} {2011})}\BibitemShut {NoStop}%
\bibitem [{\citenamefont {Peixoto}(2017)}]{peixoto_nonparametric_2017}%
  \BibitemOpen
  \bibfield  {author} {\bibinfo {author} {\bibfnamefont {Tiago~P.}\
  \bibnamefont {Peixoto}},\ }\bibfield  {title} {\enquote {\bibinfo {title}
  {Nonparametric {Bayesian} inference of the microcanonical stochastic block
  model},}\ }\href {\doibase 10.1103/PhysRevE.95.012317} {\bibfield  {journal}
  {\bibinfo  {journal} {Physical Review E}\ }\textbf {\bibinfo {volume} {95}},\
  \bibinfo {pages} {012317} (\bibinfo {year} {2017})}\BibitemShut {NoStop}%
\bibitem [{\citenamefont {Rissanen}(1978)}]{rissanen_modeling_1978}%
  \BibitemOpen
  \bibfield  {author} {\bibinfo {author} {\bibfnamefont {J.}~\bibnamefont
  {Rissanen}},\ }\bibfield  {title} {\enquote {\bibinfo {title} {Modeling by
  shortest data description},}\ }\href {\doibase 10.1016/0005-1098(78)90005-5}
  {\bibfield  {journal} {\bibinfo  {journal} {Automatica}\ }\textbf {\bibinfo
  {volume} {14}},\ \bibinfo {pages} {465--471} (\bibinfo {year}
  {1978})}\BibitemShut {NoStop}%
\bibitem [{\citenamefont {Grünwald}(2007)}]{grunwald_minimum_2007}%
  \BibitemOpen
  \bibfield  {author} {\bibinfo {author} {\bibfnamefont {Peter~D.}\
  \bibnamefont {Grünwald}},\ }\href@noop {} {\emph {\bibinfo {title} {The
  {Minimum} {Description} {Length} {Principle}}}}\ (\bibinfo  {publisher} {The
  MIT Press},\ \bibinfo {year} {2007})\BibitemShut {NoStop}%
\bibitem [{\citenamefont {Rissanen}(2010)}]{rissanen_information_2010}%
  \BibitemOpen
  \bibfield  {author} {\bibinfo {author} {\bibfnamefont {Jorma}\ \bibnamefont
  {Rissanen}},\ }\href@noop {} {\emph {\bibinfo {title} {Information and
  {Complexity} in {Statistical} {Modeling}}}},\ \bibinfo {edition} {1st}\ ed.\
  (\bibinfo  {publisher} {Springer},\ \bibinfo {year} {2010})\BibitemShut
  {NoStop}%
\bibitem [{\citenamefont {MacKay}(2003)}]{mackay_information_2003}%
  \BibitemOpen
  \bibfield  {author} {\bibinfo {author} {\bibfnamefont {David J.~C.}\
  \bibnamefont {MacKay}},\ }\href@noop {} {\emph {\bibinfo {title} {Information
  {Theory}, {Inference} and {Learning} {Algorithms}}}},\ \bibinfo {edition}
  {first edition}\ ed.\ (\bibinfo  {publisher} {Cambridge University Press},\
  \bibinfo {year} {2003})\BibitemShut {NoStop}%
\bibitem [{\citenamefont {Shannon}(1948)}]{shannon_mathematical_1948}%
  \BibitemOpen
  \bibfield  {author} {\bibinfo {author} {\bibfnamefont {C.~E}\ \bibnamefont
  {Shannon}},\ }\bibfield  {title} {\enquote {\bibinfo {title} {A mathematical
  theory of communication},}\ }\href@noop {} {\bibfield  {journal} {\bibinfo
  {journal} {Bell Syst Tech. J}\ }\textbf {\bibinfo {volume} {27}},\ \bibinfo
  {pages} {623} (\bibinfo {year} {1948})}\BibitemShut {NoStop}%
\bibitem [{\citenamefont {Zitnik}\ \emph {et~al.}(2019)\citenamefont {Zitnik},
  \citenamefont {Sosič}, \citenamefont {Feldman},\ and\ \citenamefont
  {Leskovec}}]{zitnik_evolution_2019}%
  \BibitemOpen
  \bibfield  {author} {\bibinfo {author} {\bibfnamefont {Marinka}\ \bibnamefont
  {Zitnik}}, \bibinfo {author} {\bibfnamefont {Rok}\ \bibnamefont {Sosič}},
  \bibinfo {author} {\bibfnamefont {Marcus~W.}\ \bibnamefont {Feldman}}, \ and\
  \bibinfo {author} {\bibfnamefont {Jure}\ \bibnamefont {Leskovec}},\
  }\bibfield  {title} {{\selectlanguage {english}\enquote {\bibinfo {title}
  {Evolution of resilience in protein interactomes across the tree of life},}\
  }}\href {\doibase 10.1073/pnas.1818013116} {\bibfield  {journal} {\bibinfo
  {journal} {Proceedings of the National Academy of Sciences}\ }\textbf
  {\bibinfo {volume} {116}},\ \bibinfo {pages} {4426--4433} (\bibinfo {year}
  {2019})},\ \bibinfo {note} {publisher: National Academy of Sciences Section:
  PNAS Plus}\BibitemShut {NoStop}%
\bibitem [{\citenamefont {Peixoto}(2022)}]{peixoto_disentangling_2022}%
  \BibitemOpen
  \bibfield  {author} {\bibinfo {author} {\bibfnamefont {Tiago~P.}\
  \bibnamefont {Peixoto}},\ }\bibfield  {title} {\enquote {\bibinfo {title}
  {Disentangling {Homophily}, {Community} {Structure}, and {Triadic} {Closure}
  in {Networks}},}\ }\href {\doibase 10.1103/PhysRevX.12.011004} {\bibfield
  {journal} {\bibinfo  {journal} {Physical Review X}\ }\textbf {\bibinfo
  {volume} {12}},\ \bibinfo {pages} {011004} (\bibinfo {year}
  {2022})}\BibitemShut {NoStop}%
\bibitem [{\citenamefont {Zhang}\ and\ \citenamefont
  {Peixoto}(2020)}]{zhang_statistical_2020}%
  \BibitemOpen
  \bibfield  {author} {\bibinfo {author} {\bibfnamefont {Lizhi}\ \bibnamefont
  {Zhang}}\ and\ \bibinfo {author} {\bibfnamefont {Tiago~P.}\ \bibnamefont
  {Peixoto}},\ }\bibfield  {title} {\enquote {\bibinfo {title} {Statistical
  inference of assortative community structures},}\ }\href {\doibase
  10.1103/PhysRevResearch.2.043271} {\bibfield  {journal} {\bibinfo  {journal}
  {Physical Review Research}\ }\textbf {\bibinfo {volume} {2}},\ \bibinfo
  {pages} {043271} (\bibinfo {year} {2020})}\BibitemShut {NoStop}%
\bibitem [{\citenamefont {Pastor-Satorras}\ \emph {et~al.}(2003)\citenamefont
  {Pastor-Satorras}, \citenamefont {Smith},\ and\ \citenamefont
  {Solé}}]{pastor-satorras_evolving_2003}%
  \BibitemOpen
  \bibfield  {author} {\bibinfo {author} {\bibfnamefont {Romualdo}\
  \bibnamefont {Pastor-Satorras}}, \bibinfo {author} {\bibfnamefont {Eric}\
  \bibnamefont {Smith}}, \ and\ \bibinfo {author} {\bibfnamefont {Ricard~V.}\
  \bibnamefont {Solé}},\ }\bibfield  {title} {\enquote {\bibinfo {title}
  {Evolving protein interaction networks through gene duplication},}\ }\href
  {\doibase 10.1016/S0022-5193(03)00028-6} {\bibfield  {journal} {\bibinfo
  {journal} {Journal of Theoretical Biology}\ }\textbf {\bibinfo {volume}
  {222}},\ \bibinfo {pages} {199--210} (\bibinfo {year} {2003})}\BibitemShut
  {NoStop}%
\bibitem [{\citenamefont {Peixoto}(2021)}]{peixoto_revealing_2021}%
  \BibitemOpen
  \bibfield  {author} {\bibinfo {author} {\bibfnamefont {Tiago~P.}\
  \bibnamefont {Peixoto}},\ }\bibfield  {title} {\enquote {\bibinfo {title}
  {Revealing {Consensus} and {Dissensus} between {Network} {Partitions}},}\
  }\href {\doibase 10.1103/PhysRevX.11.021003} {\bibfield  {journal} {\bibinfo
  {journal} {Physical Review X}\ }\textbf {\bibinfo {volume} {11}},\ \bibinfo
  {pages} {021003} (\bibinfo {year} {2021})}\BibitemShut {NoStop}%
\bibitem [{\citenamefont {Yan}\ \emph {et~al.}(2014)\citenamefont {Yan},
  \citenamefont {Shalizi}, \citenamefont {Jensen}, \citenamefont {Krzakala},
  \citenamefont {Moore}, \citenamefont {Zdeborová}, \citenamefont {Zhang},\
  and\ \citenamefont {Zhu}}]{yan_model_2014}%
  \BibitemOpen
  \bibfield  {author} {\bibinfo {author} {\bibfnamefont {Xiaoran}\ \bibnamefont
  {Yan}}, \bibinfo {author} {\bibfnamefont {Cosma}\ \bibnamefont {Shalizi}},
  \bibinfo {author} {\bibfnamefont {Jacob~E.}\ \bibnamefont {Jensen}}, \bibinfo
  {author} {\bibfnamefont {Florent}\ \bibnamefont {Krzakala}}, \bibinfo
  {author} {\bibfnamefont {Cristopher}\ \bibnamefont {Moore}}, \bibinfo
  {author} {\bibfnamefont {Lenka}\ \bibnamefont {Zdeborová}}, \bibinfo
  {author} {\bibfnamefont {Pan}\ \bibnamefont {Zhang}}, \ and\ \bibinfo
  {author} {\bibfnamefont {Yaojia}\ \bibnamefont {Zhu}},\ }\bibfield  {title}
  {{\selectlanguage {english}\enquote {\bibinfo {title} {Model selection for
  degree-corrected block models},}\ }}\href {\doibase
  10.1088/1742-5468/2014/05/P05007} {\bibfield  {journal} {\bibinfo  {journal}
  {Journal of Statistical Mechanics: Theory and Experiment}\ }\textbf {\bibinfo
  {volume} {2014}},\ \bibinfo {pages} {P05007} (\bibinfo {year}
  {2014})}\BibitemShut {NoStop}%
\bibitem [{\citenamefont {Decelle}\ \emph
  {et~al.}(2011{\natexlab{b}})\citenamefont {Decelle}, \citenamefont
  {Krzakala}, \citenamefont {Moore},\ and\ \citenamefont
  {Zdeborová}}]{decelle_phase_2011}%
  \BibitemOpen
  \bibfield  {author} {\bibinfo {author} {\bibfnamefont {Aurelien}\
  \bibnamefont {Decelle}}, \bibinfo {author} {\bibfnamefont {Florent}\
  \bibnamefont {Krzakala}}, \bibinfo {author} {\bibfnamefont {Cristopher}\
  \bibnamefont {Moore}}, \ and\ \bibinfo {author} {\bibfnamefont {Lenka}\
  \bibnamefont {Zdeborová}},\ }\bibfield  {title} {\enquote {\bibinfo {title}
  {Phase transition in the detection of modules in sparse networks},}\ }\href
  {http://arxiv.org/abs/1102.1182} {\bibfield  {journal} {\bibinfo  {journal}
  {1102.1182}\ } (\bibinfo {year} {2011}{\natexlab{b}})}\BibitemShut {NoStop}%
\bibitem [{\citenamefont {Cover}\ and\ \citenamefont
  {Thomas}(1991)}]{cover_elements_1991}%
  \BibitemOpen
  \bibfield  {author} {\bibinfo {author} {\bibfnamefont {Thomas~M.}\
  \bibnamefont {Cover}}\ and\ \bibinfo {author} {\bibfnamefont {Joy~A.}\
  \bibnamefont {Thomas}},\ }\href@noop {} {\emph {\bibinfo {title} {Elements of
  {Information} {Theory}}}},\ \bibinfo {edition} {99th}\ ed.\ (\bibinfo
  {publisher} {Wiley-Interscience},\ \bibinfo {year} {1991})\BibitemShut
  {NoStop}%
\bibitem [{\citenamefont {Li}\ and\ \citenamefont
  {Vitányi}(2008)}]{li_introduction_2008}%
  \BibitemOpen
  \bibfield  {author} {\bibinfo {author} {\bibfnamefont {Ming}\ \bibnamefont
  {Li}}\ and\ \bibinfo {author} {\bibfnamefont {Paul M.~B.}\ \bibnamefont
  {Vitányi}},\ }\href@noop {} {{\selectlanguage {english}\emph {\bibinfo
  {title} {An {Introduction} to {Kolmogorov} {Complexity} and {Its}
  {Applications}}}}},\ \bibinfo {edition} {3rd}\ ed.\ (\bibinfo  {publisher}
  {Springer},\ \bibinfo {address} {New York},\ \bibinfo {year}
  {2008})\BibitemShut {NoStop}%
\bibitem [{\citenamefont {Snijders}\ and\ \citenamefont
  {Nowicki}(1997)}]{snijders_estimation_1997}%
  \BibitemOpen
  \bibfield  {author} {\bibinfo {author} {\bibfnamefont {Tom A.~B.}\
  \bibnamefont {Snijders}}\ and\ \bibinfo {author} {\bibfnamefont {Krzysztof}\
  \bibnamefont {Nowicki}},\ }\bibfield  {title} {{\selectlanguage
  {english}\enquote {\bibinfo {title} {Estimation and {Prediction} for
  {Stochastic} {Blockmodels} for {Graphs} with {Latent} {Block} {Structure}},}\
  }}\href {\doibase 10.1007/s003579900004} {\bibfield  {journal} {\bibinfo
  {journal} {Journal of Classification}\ }\textbf {\bibinfo {volume} {14}},\
  \bibinfo {pages} {75--100} (\bibinfo {year} {1997})}\BibitemShut {NoStop}%
\bibitem [{\citenamefont {Nowicki}\ and\ \citenamefont
  {Snijders}(2001)}]{nowicki_estimation_2001}%
  \BibitemOpen
  \bibfield  {author} {\bibinfo {author} {\bibfnamefont {Krzysztof}\
  \bibnamefont {Nowicki}}\ and\ \bibinfo {author} {\bibfnamefont {Tom A.~B}\
  \bibnamefont {Snijders}},\ }\bibfield  {title} {\enquote {\bibinfo {title}
  {Estimation and {Prediction} for {Stochastic} {Blockstructures}},}\ }\href
  {\doibase 10.1198/016214501753208735} {\bibfield  {journal} {\bibinfo
  {journal} {Journal of the American Statistical Association}\ }\textbf
  {\bibinfo {volume} {96}},\ \bibinfo {pages} {1077--1087} (\bibinfo {year}
  {2001})}\BibitemShut {NoStop}%
\bibitem [{\citenamefont {Tallberg}(2004)}]{tallberg_bayesian_2004}%
  \BibitemOpen
  \bibfield  {author} {\bibinfo {author} {\bibfnamefont {Christian}\
  \bibnamefont {Tallberg}},\ }\bibfield  {title} {\enquote {\bibinfo {title} {A
  {Bayesian} {Approach} to {Modeling} {Stochastic} {Blockstructures} with
  {Covariates}},}\ }\href {\doibase 10.1080/00222500590889703} {\bibfield
  {journal} {\bibinfo  {journal} {The Journal of Mathematical Sociology}\
  }\textbf {\bibinfo {volume} {29}},\ \bibinfo {pages} {1--23} (\bibinfo {year}
  {2004})}\BibitemShut {NoStop}%
\bibitem [{\citenamefont {Hastings}(2006)}]{hastings_community_2006}%
  \BibitemOpen
  \bibfield  {author} {\bibinfo {author} {\bibfnamefont {M.~B.}\ \bibnamefont
  {Hastings}},\ }\bibfield  {title} {\enquote {\bibinfo {title} {Community
  detection as an inference problem},}\ }\href {\doibase
  10.1103/PhysRevE.74.035102} {\bibfield  {journal} {\bibinfo  {journal}
  {Physical Review E}\ }\textbf {\bibinfo {volume} {74}},\ \bibinfo {pages}
  {035102} (\bibinfo {year} {2006})}\BibitemShut {NoStop}%
\bibitem [{\citenamefont {Rosvall}\ and\ \citenamefont
  {Bergstrom}(2007)}]{rosvall_information-theoretic_2007}%
  \BibitemOpen
  \bibfield  {author} {\bibinfo {author} {\bibfnamefont {Martin}\ \bibnamefont
  {Rosvall}}\ and\ \bibinfo {author} {\bibfnamefont {Carl~T.}\ \bibnamefont
  {Bergstrom}},\ }\bibfield  {title} {{\selectlanguage {english}\enquote
  {\bibinfo {title} {An information-theoretic framework for resolving community
  structure in complex networks},}\ }}\href {\doibase 10.1073/pnas.0611034104}
  {\bibfield  {journal} {\bibinfo  {journal} {Proceedings of the National
  Academy of Sciences}\ }\textbf {\bibinfo {volume} {104}},\ \bibinfo {pages}
  {7327--7331} (\bibinfo {year} {2007})}\BibitemShut {NoStop}%
\bibitem [{\citenamefont {Airoldi}\ \emph {et~al.}(2008)\citenamefont
  {Airoldi}, \citenamefont {Blei}, \citenamefont {Fienberg},\ and\
  \citenamefont {Xing}}]{airoldi_mixed_2008}%
  \BibitemOpen
  \bibfield  {author} {\bibinfo {author} {\bibfnamefont {Edoardo~M.}\
  \bibnamefont {Airoldi}}, \bibinfo {author} {\bibfnamefont {David~M.}\
  \bibnamefont {Blei}}, \bibinfo {author} {\bibfnamefont {Stephen~E.}\
  \bibnamefont {Fienberg}}, \ and\ \bibinfo {author} {\bibfnamefont {Eric~P.}\
  \bibnamefont {Xing}},\ }\bibfield  {title} {\enquote {\bibinfo {title} {Mixed
  {Membership} {Stochastic} {Blockmodels}},}\ }\href
  {http://dl.acm.org/citation.cfm?id=1390681.1442798} {\bibfield  {journal}
  {\bibinfo  {journal} {J. Mach. Learn. Res.}\ }\textbf {\bibinfo {volume}
  {9}},\ \bibinfo {pages} {1981--2014} (\bibinfo {year} {2008})}\BibitemShut
  {NoStop}%
\bibitem [{\citenamefont {Clauset}\ \emph {et~al.}(2008)\citenamefont
  {Clauset}, \citenamefont {Moore},\ and\ \citenamefont
  {Newman}}]{clauset_hierarchical_2008}%
  \BibitemOpen
  \bibfield  {author} {\bibinfo {author} {\bibfnamefont {Aaron}\ \bibnamefont
  {Clauset}}, \bibinfo {author} {\bibfnamefont {Cristopher}\ \bibnamefont
  {Moore}}, \ and\ \bibinfo {author} {\bibfnamefont {M.~E.~J.}\ \bibnamefont
  {Newman}},\ }\bibfield  {title} {\enquote {\bibinfo {title} {Hierarchical
  structure and the prediction of missing links in networks},}\ }\href
  {\doibase 10.1038/nature06830} {\bibfield  {journal} {\bibinfo  {journal}
  {Nature}\ }\textbf {\bibinfo {volume} {453}},\ \bibinfo {pages} {98--101}
  (\bibinfo {year} {2008})}\BibitemShut {NoStop}%
\bibitem [{\citenamefont {Hofman}\ and\ \citenamefont
  {Wiggins}(2008)}]{hofman_bayesian_2008}%
  \BibitemOpen
  \bibfield  {author} {\bibinfo {author} {\bibfnamefont {Jake~M.}\ \bibnamefont
  {Hofman}}\ and\ \bibinfo {author} {\bibfnamefont {Chris~H.}\ \bibnamefont
  {Wiggins}},\ }\bibfield  {title} {\enquote {\bibinfo {title} {Bayesian
  {Approach} to {Network} {Modularity}},}\ }\href {\doibase
  10.1103/PhysRevLett.100.258701} {\bibfield  {journal} {\bibinfo  {journal}
  {Physical Review Letters}\ }\textbf {\bibinfo {volume} {100}},\ \bibinfo
  {pages} {258701} (\bibinfo {year} {2008})}\BibitemShut {NoStop}%
\bibitem [{\citenamefont {Mørup}\ and\ \citenamefont
  {Hansen}(2009)}]{morup_learning_2009}%
  \BibitemOpen
  \bibfield  {author} {\bibinfo {author} {\bibfnamefont {Morten}\ \bibnamefont
  {Mørup}}\ and\ \bibinfo {author} {\bibfnamefont {Lars~Kai}\ \bibnamefont
  {Hansen}},\ }\bibfield  {title} {\enquote {\bibinfo {title} {Learning latent
  structure in complex networks},}\ }in\ \href@noop {} {\emph {\bibinfo
  {booktitle} {{NIPS} {Workshop} on {Analyzing} {Networks} and {Learning} with
  {Graphs}}}}\ (\bibinfo {year} {2009})\BibitemShut {NoStop}%
\bibitem [{\citenamefont {Boguñá}\ and\ \citenamefont
  {Pastor-Satorras}(2003)}]{boguna_class_2003}%
  \BibitemOpen
  \bibfield  {author} {\bibinfo {author} {\bibfnamefont {Marián}\ \bibnamefont
  {Boguñá}}\ and\ \bibinfo {author} {\bibfnamefont {Romualdo}\ \bibnamefont
  {Pastor-Satorras}},\ }\bibfield  {title} {\enquote {\bibinfo {title} {Class
  of correlated random networks with hidden variables},}\ }\href {\doibase
  10.1103/PhysRevE.68.036112} {\bibfield  {journal} {\bibinfo  {journal}
  {Physical Review E}\ }\textbf {\bibinfo {volume} {68}},\ \bibinfo {pages}
  {036112} (\bibinfo {year} {2003})}\BibitemShut {NoStop}%
\bibitem [{\citenamefont {Bollobás}\ \emph {et~al.}(2007)\citenamefont
  {Bollobás}, \citenamefont {Janson},\ and\ \citenamefont
  {Riordan}}]{bollobas_phase_2007}%
  \BibitemOpen
  \bibfield  {author} {\bibinfo {author} {\bibfnamefont {Béla}\ \bibnamefont
  {Bollobás}}, \bibinfo {author} {\bibfnamefont {Svante}\ \bibnamefont
  {Janson}}, \ and\ \bibinfo {author} {\bibfnamefont {Oliver}\ \bibnamefont
  {Riordan}},\ }\bibfield  {title} {{\selectlanguage {english}\enquote
  {\bibinfo {title} {The phase transition in inhomogeneous random graphs},}\
  }}\href {\doibase 10.1002/rsa.20168} {\bibfield  {journal} {\bibinfo
  {journal} {Random Structures \& Algorithms}\ }\textbf {\bibinfo {volume}
  {31}},\ \bibinfo {pages} {3--122} (\bibinfo {year} {2007})}\BibitemShut
  {NoStop}%
\bibitem [{\citenamefont {Lancichinetti}\ \emph {et~al.}(2008)\citenamefont
  {Lancichinetti}, \citenamefont {Fortunato},\ and\ \citenamefont
  {Radicchi}}]{lancichinetti_benchmark_2008}%
  \BibitemOpen
  \bibfield  {author} {\bibinfo {author} {\bibfnamefont {Andrea}\ \bibnamefont
  {Lancichinetti}}, \bibinfo {author} {\bibfnamefont {Santo}\ \bibnamefont
  {Fortunato}}, \ and\ \bibinfo {author} {\bibfnamefont {Filippo}\ \bibnamefont
  {Radicchi}},\ }\bibfield  {title} {\enquote {\bibinfo {title} {Benchmark
  graphs for testing community detection algorithms},}\ }\href {\doibase
  10.1103/PhysRevE.78.046110} {\bibfield  {journal} {\bibinfo  {journal}
  {Physical Review E}\ }\textbf {\bibinfo {volume} {78}},\ \bibinfo {pages}
  {046110} (\bibinfo {year} {2008})}\BibitemShut {NoStop}%
\bibitem [{\citenamefont {Girvan}\ and\ \citenamefont
  {Newman}(2002)}]{girvan_community_2002}%
  \BibitemOpen
  \bibfield  {author} {\bibinfo {author} {\bibfnamefont {M.}~\bibnamefont
  {Girvan}}\ and\ \bibinfo {author} {\bibfnamefont {M.~E.~J.}\ \bibnamefont
  {Newman}},\ }\bibfield  {title} {\enquote {\bibinfo {title} {Community
  structure in social and biological networks},}\ }\href {\doibase
  10.1073/pnas.122653799} {\bibfield  {journal} {\bibinfo  {journal}
  {Proceedings of the National Academy of Sciences}\ }\textbf {\bibinfo
  {volume} {99}},\ \bibinfo {pages} {7821 --7826} (\bibinfo {year}
  {2002})}\BibitemShut {NoStop}%
\bibitem [{\citenamefont {Lancichinetti}\ and\ \citenamefont
  {Fortunato}(2009)}]{lancichinetti_community_2009}%
  \BibitemOpen
  \bibfield  {author} {\bibinfo {author} {\bibfnamefont {Andrea}\ \bibnamefont
  {Lancichinetti}}\ and\ \bibinfo {author} {\bibfnamefont {Santo}\ \bibnamefont
  {Fortunato}},\ }\bibfield  {title} {\enquote {\bibinfo {title} {Community
  detection algorithms: {A} comparative analysis},}\ }\href {\doibase
  10.1103/PhysRevE.80.056117} {\bibfield  {journal} {\bibinfo  {journal}
  {Physical Review E}\ }\textbf {\bibinfo {volume} {80}},\ \bibinfo {pages}
  {056117} (\bibinfo {year} {2009})}\BibitemShut {NoStop}%
\bibitem [{\citenamefont {Decelle}\ \emph
  {et~al.}(2011{\natexlab{c}})\citenamefont {Decelle}, \citenamefont
  {Krzakala}, \citenamefont {Moore},\ and\ \citenamefont
  {Zdeborová}}]{decelle_inference_2011}%
  \BibitemOpen
  \bibfield  {author} {\bibinfo {author} {\bibfnamefont {Aurelien}\
  \bibnamefont {Decelle}}, \bibinfo {author} {\bibfnamefont {Florent}\
  \bibnamefont {Krzakala}}, \bibinfo {author} {\bibfnamefont {Cristopher}\
  \bibnamefont {Moore}}, \ and\ \bibinfo {author} {\bibfnamefont {Lenka}\
  \bibnamefont {Zdeborová}},\ }\bibfield  {title} {\enquote {\bibinfo {title}
  {Inference and {Phase} {Transitions} in the {Detection} of {Modules} in
  {Sparse} {Networks}},}\ }\href {\doibase 10.1103/PhysRevLett.107.065701}
  {\bibfield  {journal} {\bibinfo  {journal} {Physical Review Letters}\
  }\textbf {\bibinfo {volume} {107}},\ \bibinfo {pages} {065701} (\bibinfo
  {year} {2011}{\natexlab{c}})}\BibitemShut {NoStop}%
\bibitem [{\citenamefont {Newman}(2006)}]{newman_modularity_2006}%
  \BibitemOpen
  \bibfield  {author} {\bibinfo {author} {\bibfnamefont {M.~E.~J.}\
  \bibnamefont {Newman}},\ }\bibfield  {title} {{\selectlanguage
  {english}\enquote {\bibinfo {title} {Modularity and community structure in
  networks},}\ }}\href {\doibase 10.1073/pnas.0601602103} {\bibfield  {journal}
  {\bibinfo  {journal} {Proceedings of the National Academy of Sciences}\
  }\textbf {\bibinfo {volume} {103}},\ \bibinfo {pages} {8577--8582} (\bibinfo
  {year} {2006})}\BibitemShut {NoStop}%
\bibitem [{\citenamefont {Rosvall}\ and\ \citenamefont
  {Bergstrom}(2008)}]{rosvall_maps_2008}%
  \BibitemOpen
  \bibfield  {author} {\bibinfo {author} {\bibfnamefont {Martin}\ \bibnamefont
  {Rosvall}}\ and\ \bibinfo {author} {\bibfnamefont {Carl~T.}\ \bibnamefont
  {Bergstrom}},\ }\bibfield  {title} {{\selectlanguage {english}\enquote
  {\bibinfo {title} {Maps of random walks on complex networks reveal community
  structure},}\ }}\href {\doibase 10.1073/pnas.0706851105} {\bibfield
  {journal} {\bibinfo  {journal} {Proceedings of the National Academy of
  Sciences}\ }\textbf {\bibinfo {volume} {105}},\ \bibinfo {pages} {1118--1123}
  (\bibinfo {year} {2008})}\BibitemShut {NoStop}%
\bibitem [{\citenamefont {Lambiotte}\ \emph {et~al.}(2014)\citenamefont
  {Lambiotte}, \citenamefont {Delvenne},\ and\ \citenamefont
  {Barahona}}]{lambiotte_random_2014}%
  \BibitemOpen
  \bibfield  {author} {\bibinfo {author} {\bibfnamefont {R.}~\bibnamefont
  {Lambiotte}}, \bibinfo {author} {\bibfnamefont {J.~C.}\ \bibnamefont
  {Delvenne}}, \ and\ \bibinfo {author} {\bibfnamefont {M.}~\bibnamefont
  {Barahona}},\ }\bibfield  {title} {\enquote {\bibinfo {title} {Random
  {Walks}, {Markov} {Processes} and the {Multiscale} {Modular} {Organization}
  of {Complex} {Networks}},}\ }\href {\doibase 10.1109/TNSE.2015.2391998}
  {\bibfield  {journal} {\bibinfo  {journal} {IEEE Transactions on Network
  Science and Engineering}\ }\textbf {\bibinfo {volume} {1}},\ \bibinfo {pages}
  {76--90} (\bibinfo {year} {2014})}\BibitemShut {NoStop}%
\bibitem [{\citenamefont {Gelman}\ \emph {et~al.}(2013)\citenamefont {Gelman},
  \citenamefont {Carlin}, \citenamefont {Stern}, \citenamefont {Dunson},
  \citenamefont {Vehtari},\ and\ \citenamefont {Rubin}}]{gelman_bayesian_2013}%
  \BibitemOpen
  \bibfield  {author} {\bibinfo {author} {\bibfnamefont {Andrew}\ \bibnamefont
  {Gelman}}, \bibinfo {author} {\bibfnamefont {John~B.}\ \bibnamefont
  {Carlin}}, \bibinfo {author} {\bibfnamefont {Hal~S.}\ \bibnamefont {Stern}},
  \bibinfo {author} {\bibfnamefont {David~B.}\ \bibnamefont {Dunson}}, \bibinfo
  {author} {\bibfnamefont {Aki}\ \bibnamefont {Vehtari}}, \ and\ \bibinfo
  {author} {\bibfnamefont {Donald~B.}\ \bibnamefont {Rubin}},\ }\href@noop {}
  {{\selectlanguage {english}\emph {\bibinfo {title} {Bayesian {Data}
  {Analysis}}}}},\ \bibinfo {edition} {3rd}\ ed.\ (\bibinfo  {publisher}
  {Chapman and Hall/CRC},\ \bibinfo {address} {Boca Raton},\ \bibinfo {year}
  {2013})\BibitemShut {NoStop}%
\bibitem [{\citenamefont {Bishop}(2011)}]{bishop_pattern_2011}%
  \BibitemOpen
  \bibfield  {author} {\bibinfo {author} {\bibfnamefont {Christopher~M.}\
  \bibnamefont {Bishop}},\ }\href@noop {} {{\selectlanguage {english}\emph
  {\bibinfo {title} {Pattern {Recognition} and {Machine} {Learning}}}}}\
  (\bibinfo  {publisher} {Springer},\ \bibinfo {year} {2011})\BibitemShut
  {NoStop}%
\bibitem [{\citenamefont {Newman}(2018)}]{newman_network_2018-1}%
  \BibitemOpen
  \bibfield  {author} {\bibinfo {author} {\bibfnamefont {M.~E.~J.}\
  \bibnamefont {Newman}},\ }\bibfield  {title} {{\selectlanguage
  {english}\enquote {\bibinfo {title} {Network structure from rich but noisy
  data},}\ }}\href {\doibase 10.1038/s41567-018-0076-1} {\bibfield  {journal}
  {\bibinfo  {journal} {Nature Physics}\ }\textbf {\bibinfo {volume} {14}},\
  \bibinfo {pages} {542--545} (\bibinfo {year} {2018})}\BibitemShut {NoStop}%
\bibitem [{\citenamefont {Martin}\ \emph {et~al.}(2016)\citenamefont {Martin},
  \citenamefont {Ball},\ and\ \citenamefont {Newman}}]{martin_structural_2016}%
  \BibitemOpen
  \bibfield  {author} {\bibinfo {author} {\bibfnamefont {Travis}\ \bibnamefont
  {Martin}}, \bibinfo {author} {\bibfnamefont {Brian}\ \bibnamefont {Ball}}, \
  and\ \bibinfo {author} {\bibfnamefont {M.~E.~J.}\ \bibnamefont {Newman}},\
  }\bibfield  {title} {\enquote {\bibinfo {title} {Structural inference for
  uncertain networks},}\ }\href {\doibase 10.1103/PhysRevE.93.012306}
  {\bibfield  {journal} {\bibinfo  {journal} {Physical Review E}\ }\textbf
  {\bibinfo {volume} {93}},\ \bibinfo {pages} {012306} (\bibinfo {year}
  {2016})}\BibitemShut {NoStop}%
\bibitem [{\citenamefont {Peixoto}(2018)}]{peixoto_reconstructing_2018}%
  \BibitemOpen
  \bibfield  {author} {\bibinfo {author} {\bibfnamefont {Tiago~P.}\
  \bibnamefont {Peixoto}},\ }\bibfield  {title} {\enquote {\bibinfo {title}
  {Reconstructing {Networks} with {Unknown} and {Heterogeneous} {Errors}},}\
  }\href {\doibase 10.1103/PhysRevX.8.041011} {\bibfield  {journal} {\bibinfo
  {journal} {Physical Review X}\ }\textbf {\bibinfo {volume} {8}},\ \bibinfo
  {pages} {041011} (\bibinfo {year} {2018})}\BibitemShut {NoStop}%
\bibitem [{\citenamefont {Guimerà}\ and\ \citenamefont
  {Sales-Pardo}(2009)}]{guimera_missing_2009}%
  \BibitemOpen
  \bibfield  {author} {\bibinfo {author} {\bibfnamefont {Roger}\ \bibnamefont
  {Guimerà}}\ and\ \bibinfo {author} {\bibfnamefont {Marta}\ \bibnamefont
  {Sales-Pardo}},\ }\bibfield  {title} {\enquote {\bibinfo {title} {Missing and
  spurious interactions and the reconstruction of complex networks},}\ }\href
  {\doibase 10.1073/pnas.0908366106} {\bibfield  {journal} {\bibinfo  {journal}
  {Proceedings of the National Academy of Sciences}\ }\textbf {\bibinfo
  {volume} {106}},\ \bibinfo {pages} {22073 --22078} (\bibinfo {year}
  {2009})}\BibitemShut {NoStop}%
\bibitem [{\citenamefont {Hoffmann}\ \emph {et~al.}(2020)\citenamefont
  {Hoffmann}, \citenamefont {Peel}, \citenamefont {Lambiotte},\ and\
  \citenamefont {Jones}}]{hoffmann_community_2020}%
  \BibitemOpen
  \bibfield  {author} {\bibinfo {author} {\bibfnamefont {Till}\ \bibnamefont
  {Hoffmann}}, \bibinfo {author} {\bibfnamefont {Leto}\ \bibnamefont {Peel}},
  \bibinfo {author} {\bibfnamefont {Renaud}\ \bibnamefont {Lambiotte}}, \ and\
  \bibinfo {author} {\bibfnamefont {Nick~S.}\ \bibnamefont {Jones}},\
  }\bibfield  {title} {{\selectlanguage {english}\enquote {\bibinfo {title}
  {Community detection in networks without observing edges},}\ }}\href
  {\doibase 10.1126/sciadv.aav1478} {\bibfield  {journal} {\bibinfo  {journal}
  {Science Advances}\ }\textbf {\bibinfo {volume} {6}},\ \bibinfo {pages}
  {eaav1478} (\bibinfo {year} {2020})},\ \bibinfo {note} {publisher: American
  Association for the Advancement of Science Section: Research
  Article}\BibitemShut {NoStop}%
\bibitem [{\citenamefont {Peixoto}(2019{\natexlab{b}})}]{peixoto_network_2019}%
  \BibitemOpen
  \bibfield  {author} {\bibinfo {author} {\bibfnamefont {Tiago~P.}\
  \bibnamefont {Peixoto}},\ }\bibfield  {title} {\enquote {\bibinfo {title}
  {Network {Reconstruction} and {Community} {Detection} from {Dynamics}},}\
  }\href {\doibase 10.1103/PhysRevLett.123.128301} {\bibfield  {journal}
  {\bibinfo  {journal} {Physical Review Letters}\ }\textbf {\bibinfo {volume}
  {123}},\ \bibinfo {pages} {128301} (\bibinfo {year}
  {2019}{\natexlab{b}})}\BibitemShut {NoStop}%
\bibitem [{\citenamefont {Fosdick}\ \emph {et~al.}(2018)\citenamefont
  {Fosdick}, \citenamefont {Larremore}, \citenamefont {Nishimura},\ and\
  \citenamefont {Ugander}}]{fosdick_configuring_2018}%
  \BibitemOpen
  \bibfield  {author} {\bibinfo {author} {\bibfnamefont {B.}~\bibnamefont
  {Fosdick}}, \bibinfo {author} {\bibfnamefont {D.}~\bibnamefont {Larremore}},
  \bibinfo {author} {\bibfnamefont {J.}~\bibnamefont {Nishimura}}, \ and\
  \bibinfo {author} {\bibfnamefont {J.}~\bibnamefont {Ugander}},\ }\bibfield
  {title} {\enquote {\bibinfo {title} {Configuring {Random} {Graph} {Models}
  with {Fixed} {Degree} {Sequences}},}\ }\href {\doibase 10.1137/16M1087175}
  {\bibfield  {journal} {\bibinfo  {journal} {SIAM Review}\ }\textbf {\bibinfo
  {volume} {60}},\ \bibinfo {pages} {315--355} (\bibinfo {year}
  {2018})}\BibitemShut {NoStop}%
\bibitem [{\citenamefont {Chung}\ and\ \citenamefont
  {Lu}(2002)}]{chung_connected_2002}%
  \BibitemOpen
  \bibfield  {author} {\bibinfo {author} {\bibfnamefont {Fan}\ \bibnamefont
  {Chung}}\ and\ \bibinfo {author} {\bibfnamefont {Linyuan}\ \bibnamefont
  {Lu}},\ }\bibfield  {title} {\enquote {\bibinfo {title} {Connected
  {Components} in {Random} {Graphs} with {Given} {Expected} {Degree}
  {Sequences}},}\ }\href {\doibase 10.1007/PL00012580} {\bibfield  {journal}
  {\bibinfo  {journal} {Annals of Combinatorics}\ }\textbf {\bibinfo {volume}
  {6}},\ \bibinfo {pages} {125--145} (\bibinfo {year} {2002})}\BibitemShut
  {NoStop}%
\bibitem [{\citenamefont {Guimerà}\ \emph {et~al.}(2004)\citenamefont
  {Guimerà}, \citenamefont {Sales-Pardo},\ and\ \citenamefont
  {Amaral}}]{guimera_modularity_2004}%
  \BibitemOpen
  \bibfield  {author} {\bibinfo {author} {\bibfnamefont {Roger}\ \bibnamefont
  {Guimerà}}, \bibinfo {author} {\bibfnamefont {Marta}\ \bibnamefont
  {Sales-Pardo}}, \ and\ \bibinfo {author} {\bibfnamefont {Luís A.~Nunes}\
  \bibnamefont {Amaral}},\ }\bibfield  {title} {\enquote {\bibinfo {title}
  {Modularity from fluctuations in random graphs and complex networks},}\
  }\href {\doibase 10.1103/PhysRevE.70.025101} {\bibfield  {journal} {\bibinfo
  {journal} {Physical Review E}\ }\textbf {\bibinfo {volume} {70}},\ \bibinfo
  {pages} {025101} (\bibinfo {year} {2004})}\BibitemShut {NoStop}%
\bibitem [{\citenamefont {Fortunato}\ and\ \citenamefont
  {Barthélemy}(2007)}]{fortunato_resolution_2007}%
  \BibitemOpen
  \bibfield  {author} {\bibinfo {author} {\bibfnamefont {Santo}\ \bibnamefont
  {Fortunato}}\ and\ \bibinfo {author} {\bibfnamefont {Marc}\ \bibnamefont
  {Barthélemy}},\ }\bibfield  {title} {{\selectlanguage {english}\enquote
  {\bibinfo {title} {Resolution limit in community detection},}\ }}\href
  {\doibase 10.1073/pnas.0605965104} {\bibfield  {journal} {\bibinfo  {journal}
  {Proceedings of the National Academy of Sciences}\ }\textbf {\bibinfo
  {volume} {104}},\ \bibinfo {pages} {36--41} (\bibinfo {year}
  {2007})}\BibitemShut {NoStop}%
\bibitem [{\citenamefont {Good}\ \emph {et~al.}(2010)\citenamefont {Good},
  \citenamefont {de~Montjoye},\ and\ \citenamefont
  {Clauset}}]{good_performance_2010}%
  \BibitemOpen
  \bibfield  {author} {\bibinfo {author} {\bibfnamefont {Benjamin~H.}\
  \bibnamefont {Good}}, \bibinfo {author} {\bibfnamefont {Yves-Alexandre}\
  \bibnamefont {de~Montjoye}}, \ and\ \bibinfo {author} {\bibfnamefont {Aaron}\
  \bibnamefont {Clauset}},\ }\bibfield  {title} {\enquote {\bibinfo {title}
  {Performance of modularity maximization in practical contexts},}\ }\href
  {\doibase 10.1103/PhysRevE.81.046106} {\bibfield  {journal} {\bibinfo
  {journal} {Physical Review E}\ }\textbf {\bibinfo {volume} {81}},\ \bibinfo
  {pages} {046106} (\bibinfo {year} {2010})}\BibitemShut {NoStop}%
\bibitem [{\citenamefont {Newman}(2003)}]{newman_mixing_2003}%
  \BibitemOpen
  \bibfield  {author} {\bibinfo {author} {\bibfnamefont {M.~E.~J.}\
  \bibnamefont {Newman}},\ }\bibfield  {title} {\enquote {\bibinfo {title}
  {Mixing patterns in networks},}\ }\href
  {http://link.aps.org/abstract/PRE/v67/e026126} {\bibfield  {journal}
  {\bibinfo  {journal} {Phys. Rev. E}\ }\textbf {\bibinfo {volume} {67}},\
  \bibinfo {pages} {026126} (\bibinfo {year} {2003})}\BibitemShut {NoStop}%
\bibitem [{\citenamefont {Riolo}\ and\ \citenamefont
  {Newman}(2020)}]{riolo_consistency_2020}%
  \BibitemOpen
  \bibfield  {author} {\bibinfo {author} {\bibfnamefont {Maria~A.}\
  \bibnamefont {Riolo}}\ and\ \bibinfo {author} {\bibfnamefont {M.~E.~J.}\
  \bibnamefont {Newman}},\ }\bibfield  {title} {\enquote {\bibinfo {title}
  {Consistency of community structure in complex networks},}\ }\href {\doibase
  10.1103/PhysRevE.101.052306} {\bibfield  {journal} {\bibinfo  {journal}
  {Physical Review E}\ }\textbf {\bibinfo {volume} {101}},\ \bibinfo {pages}
  {052306} (\bibinfo {year} {2020})}\BibitemShut {NoStop}%
\bibitem [{\citenamefont {Zhang}\ and\ \citenamefont
  {Peixoto}()}]{zhang_preparation}%
  \BibitemOpen
  \bibfield  {author} {\bibinfo {author} {\bibfnamefont {Lizhi}\ \bibnamefont
  {Zhang}}\ and\ \bibinfo {author} {\bibfnamefont {T.~P.}\ \bibnamefont
  {Peixoto}},\ }\bibfield  {title} {\enquote {\bibinfo {title} {Large-scale
  assessment of overfitting, underfitting and model selection for modular
  network structures},}\ }\href@noop {} {\bibinfo  {journal} {in preparation}\
  }\BibitemShut {NoStop}%
\bibitem [{\citenamefont
  {Peixoto}(2020{\natexlab{a}})}]{peixoto_netzschleuder_2020}%
  \BibitemOpen
\bibfield  {journal} {  }\bibfield  {author} {\bibinfo {author} {\bibfnamefont
  {T.~P.}\ \bibnamefont {Peixoto}},\ }\href {https://networks.skewed.de}
  {\enquote {\bibinfo {title} {The {Netzschleuder} network catalogue and
  repository.}}\ } (\bibinfo {year} {2020}{\natexlab{a}}),\ \bibinfo {note}
  {accessible at \url{https://networks.skewed.de}.}\BibitemShut {Stop}%
\bibitem [{\citenamefont {Ghasemian}\ \emph {et~al.}(2019)\citenamefont
  {Ghasemian}, \citenamefont {Hosseinmardi},\ and\ \citenamefont
  {Clauset}}]{ghasemian_evaluating_2019}%
  \BibitemOpen
  \bibfield  {author} {\bibinfo {author} {\bibfnamefont {Amir}\ \bibnamefont
  {Ghasemian}}, \bibinfo {author} {\bibfnamefont {Homa}\ \bibnamefont
  {Hosseinmardi}}, \ and\ \bibinfo {author} {\bibfnamefont {Aaron}\
  \bibnamefont {Clauset}},\ }\bibfield  {title} {\enquote {\bibinfo {title}
  {Evaluating {Overfit} and {Underfit} in {Models} of {Network} {Community}
  {Structure}},}\ }\href {\doibase 10.1109/TKDE.2019.2911585} {\bibfield
  {journal} {\bibinfo  {journal} {IEEE Transactions on Knowledge and Data
  Engineering}\ ,\ \bibinfo {pages} {1--1}} (\bibinfo {year}
  {2019})}\BibitemShut {NoStop}%
\bibitem [{\citenamefont
  {Peixoto}(2014{\natexlab{b}})}]{peixoto_hierarchical_2014}%
  \BibitemOpen
  \bibfield  {author} {\bibinfo {author} {\bibfnamefont {Tiago~P.}\
  \bibnamefont {Peixoto}},\ }\bibfield  {title} {\enquote {\bibinfo {title}
  {Hierarchical {Block} {Structures} and {High}-{Resolution} {Model}
  {Selection} in {Large} {Networks}},}\ }\href {\doibase
  10.1103/PhysRevX.4.011047} {\bibfield  {journal} {\bibinfo  {journal}
  {Physical Review X}\ }\textbf {\bibinfo {volume} {4}},\ \bibinfo {pages}
  {011047} (\bibinfo {year} {2014}{\natexlab{b}})}\BibitemShut {NoStop}%
\bibitem [{\citenamefont {Larremore}\ \emph {et~al.}(2014)\citenamefont
  {Larremore}, \citenamefont {Clauset},\ and\ \citenamefont
  {Jacobs}}]{larremore_efficiently_2014}%
  \BibitemOpen
  \bibfield  {author} {\bibinfo {author} {\bibfnamefont {Daniel~B.}\
  \bibnamefont {Larremore}}, \bibinfo {author} {\bibfnamefont {Aaron}\
  \bibnamefont {Clauset}}, \ and\ \bibinfo {author} {\bibfnamefont
  {Abigail~Z.}\ \bibnamefont {Jacobs}},\ }\bibfield  {title} {\enquote
  {\bibinfo {title} {Efficiently inferring community structure in bipartite
  networks},}\ }\href {\doibase 10.1103/PhysRevE.90.012805} {\bibfield
  {journal} {\bibinfo  {journal} {Physical Review E}\ }\textbf {\bibinfo
  {volume} {90}},\ \bibinfo {pages} {012805} (\bibinfo {year}
  {2014})}\BibitemShut {NoStop}%
\bibitem [{\citenamefont {Zhang}\ \emph {et~al.}(2015)\citenamefont {Zhang},
  \citenamefont {Martin},\ and\ \citenamefont
  {Newman}}]{zhang_identification_2015}%
  \BibitemOpen
  \bibfield  {author} {\bibinfo {author} {\bibfnamefont {Xiao}\ \bibnamefont
  {Zhang}}, \bibinfo {author} {\bibfnamefont {Travis}\ \bibnamefont {Martin}},
  \ and\ \bibinfo {author} {\bibfnamefont {M.~E.~J.}\ \bibnamefont {Newman}},\
  }\bibfield  {title} {\enquote {\bibinfo {title} {Identification of
  core-periphery structure in networks},}\ }\href {\doibase
  10.1103/PhysRevE.91.032803} {\bibfield  {journal} {\bibinfo  {journal}
  {Physical Review E}\ }\textbf {\bibinfo {volume} {91}},\ \bibinfo {pages}
  {032803} (\bibinfo {year} {2015})}\BibitemShut {NoStop}%
\bibitem [{\citenamefont {Zhang}\ and\ \citenamefont
  {Moore}(2014)}]{zhang_scalable_2014}%
  \BibitemOpen
  \bibfield  {author} {\bibinfo {author} {\bibfnamefont {Pan}\ \bibnamefont
  {Zhang}}\ and\ \bibinfo {author} {\bibfnamefont {Cristopher}\ \bibnamefont
  {Moore}},\ }\bibfield  {title} {{\selectlanguage {english}\enquote {\bibinfo
  {title} {Scalable detection of statistically significant communities and
  hierarchies, using message passing for modularity},}\ }}\href {\doibase
  10.1073/pnas.1409770111} {\bibfield  {journal} {\bibinfo  {journal}
  {Proceedings of the National Academy of Sciences}\ }\textbf {\bibinfo
  {volume} {111}},\ \bibinfo {pages} {18144--18149} (\bibinfo {year} {2014})},\
  \bibinfo {note} {publisher: National Academy of Sciences Section: Physical
  Sciences}\BibitemShut {NoStop}%
\bibitem [{\citenamefont {Newman}(2016)}]{newman_equivalence_2016}%
  \BibitemOpen
  \bibfield  {author} {\bibinfo {author} {\bibfnamefont {M.~E.~J.}\
  \bibnamefont {Newman}},\ }\bibfield  {title} {\enquote {\bibinfo {title}
  {Equivalence between modularity optimization and maximum likelihood methods
  for community detection},}\ }\href {\doibase 10.1103/PhysRevE.94.052315}
  {\bibfield  {journal} {\bibinfo  {journal} {Physical Review E}\ }\textbf
  {\bibinfo {volume} {94}} (\bibinfo {year} {2016}),\
  10.1103/PhysRevE.94.052315}\BibitemShut {NoStop}%
\bibitem [{\citenamefont {Reichardt}\ and\ \citenamefont
  {Bornholdt}(2006{\natexlab{a}})}]{reichardt_statistical_2006}%
  \BibitemOpen
  \bibfield  {author} {\bibinfo {author} {\bibfnamefont {Jörg}\ \bibnamefont
  {Reichardt}}\ and\ \bibinfo {author} {\bibfnamefont {Stefan}\ \bibnamefont
  {Bornholdt}},\ }\bibfield  {title} {\enquote {\bibinfo {title} {Statistical
  mechanics of community detection},}\ }\href {\doibase
  10.1103/PhysRevE.74.016110} {\bibfield  {journal} {\bibinfo  {journal}
  {Physical Review E}\ }\textbf {\bibinfo {volume} {74}},\ \bibinfo {pages}
  {016110} (\bibinfo {year} {2006}{\natexlab{a}})}\BibitemShut {NoStop}%
\bibitem [{\citenamefont {Arenas}\ \emph {et~al.}(2008)\citenamefont {Arenas},
  \citenamefont {Fernández},\ and\ \citenamefont
  {Gómez}}]{arenas_analysis_2008}%
  \BibitemOpen
  \bibfield  {author} {\bibinfo {author} {\bibfnamefont {A.}~\bibnamefont
  {Arenas}}, \bibinfo {author} {\bibfnamefont {A.}~\bibnamefont {Fernández}},
  \ and\ \bibinfo {author} {\bibfnamefont {S.}~\bibnamefont {Gómez}},\
  }\bibfield  {title} {{\selectlanguage {english}\enquote {\bibinfo {title}
  {Analysis of the structure of complex networks at different resolution
  levels},}\ }}\href {\doibase 10.1088/1367-2630/10/5/053039} {\bibfield
  {journal} {\bibinfo  {journal} {New Journal of Physics}\ }\textbf {\bibinfo
  {volume} {10}},\ \bibinfo {pages} {053039} (\bibinfo {year}
  {2008})}\BibitemShut {NoStop}%
\bibitem [{\citenamefont {Bickel}\ and\ \citenamefont
  {Chen}(2009)}]{bickel_nonparametric_2009}%
  \BibitemOpen
  \bibfield  {author} {\bibinfo {author} {\bibfnamefont {Peter~J.}\
  \bibnamefont {Bickel}}\ and\ \bibinfo {author} {\bibfnamefont {Aiyou}\
  \bibnamefont {Chen}},\ }\bibfield  {title} {{\selectlanguage
  {english}\enquote {\bibinfo {title} {A nonparametric view of network models
  and {Newman}–{Girvan} and other modularities},}\ }}\href {\doibase
  10.1073/pnas.0907096106} {\bibfield  {journal} {\bibinfo  {journal}
  {Proceedings of the National Academy of Sciences}\ }\textbf {\bibinfo
  {volume} {106}},\ \bibinfo {pages} {21068--21073} (\bibinfo {year}
  {2009})}\BibitemShut {NoStop}%
\bibitem [{\citenamefont {Newman}(2013)}]{newman_spectral_2013}%
  \BibitemOpen
  \bibfield  {author} {\bibinfo {author} {\bibfnamefont {M.~E.~J.}\
  \bibnamefont {Newman}},\ }\bibfield  {title} {\enquote {\bibinfo {title}
  {Spectral methods for community detection and graph partitioning},}\ }\href
  {\doibase 10.1103/PhysRevE.88.042822} {\bibfield  {journal} {\bibinfo
  {journal} {Physical Review E}\ }\textbf {\bibinfo {volume} {88}},\ \bibinfo
  {pages} {042822} (\bibinfo {year} {2013})}\BibitemShut {NoStop}%
\bibitem [{\citenamefont {Massen}\ and\ \citenamefont
  {Doye}(2006)}]{massen_thermodynamics_2006}%
  \BibitemOpen
  \bibfield  {author} {\bibinfo {author} {\bibfnamefont {Claire~P.}\
  \bibnamefont {Massen}}\ and\ \bibinfo {author} {\bibfnamefont {Jonathan
  P.~K.}\ \bibnamefont {Doye}},\ }\bibfield  {title} {\enquote {\bibinfo
  {title} {Thermodynamics of {Community} {Structure}},}\ }\href
  {http://arxiv.org/abs/cond-mat/0610077} {\bibfield  {journal} {\bibinfo
  {journal} {arXiv:cond-mat/0610077}\ } (\bibinfo {year} {2006})}\BibitemShut
  {NoStop}%
\bibitem [{\citenamefont {Lancichinetti}\ and\ \citenamefont
  {Fortunato}(2012)}]{lancichinetti_consensus_2012}%
  \BibitemOpen
  \bibfield  {author} {\bibinfo {author} {\bibfnamefont {Andrea}\ \bibnamefont
  {Lancichinetti}}\ and\ \bibinfo {author} {\bibfnamefont {Santo}\ \bibnamefont
  {Fortunato}},\ }\bibfield  {title} {{\selectlanguage {english}\enquote
  {\bibinfo {title} {Consensus clustering in complex networks},}\ }}\href
  {\doibase 10.1038/srep00336} {\bibfield  {journal} {\bibinfo  {journal}
  {Scientific Reports}\ }\textbf {\bibinfo {volume} {2}},\ \bibinfo {pages}
  {1--7} (\bibinfo {year} {2012})},\ \bibinfo {note} {number: 1 Publisher:
  Nature Publishing Group}\BibitemShut {NoStop}%
\bibitem [{\citenamefont {Reichardt}\ and\ \citenamefont
  {Bornholdt}(2006{\natexlab{b}})}]{reichardt_when_2006}%
  \BibitemOpen
  \bibfield  {author} {\bibinfo {author} {\bibfnamefont {Jörg}\ \bibnamefont
  {Reichardt}}\ and\ \bibinfo {author} {\bibfnamefont {Stefan}\ \bibnamefont
  {Bornholdt}},\ }\bibfield  {title} {\enquote {\bibinfo {title} {When are
  networks truly modular?}}\ }\href {\doibase 10.1016/j.physd.2006.09.009}
  {\bibfield  {journal} {\bibinfo  {journal} {Physica D: Nonlinear Phenomena}\
  }\textbf {\bibinfo {volume} {224}},\ \bibinfo {pages} {20--26} (\bibinfo
  {year} {2006}{\natexlab{b}})}\BibitemShut {NoStop}%
\bibitem [{\citenamefont {Hu}\ \emph {et~al.}(2012)\citenamefont {Hu},
  \citenamefont {Ronhovde},\ and\ \citenamefont {Nussinov}}]{hu_phase_2012}%
  \BibitemOpen
  \bibfield  {author} {\bibinfo {author} {\bibfnamefont {Dandan}\ \bibnamefont
  {Hu}}, \bibinfo {author} {\bibfnamefont {Peter}\ \bibnamefont {Ronhovde}}, \
  and\ \bibinfo {author} {\bibfnamefont {Zohar}\ \bibnamefont {Nussinov}},\
  }\bibfield  {title} {\enquote {\bibinfo {title} {Phase transitions in random
  {Potts} systems and the community detection problem: spin-glass type and
  dynamic perspectives},}\ }\href {\doibase 10.1080/14786435.2011.616547}
  {\bibfield  {journal} {\bibinfo  {journal} {Philosophical Magazine}\ }\textbf
  {\bibinfo {volume} {92}},\ \bibinfo {pages} {406--445} (\bibinfo {year}
  {2012})}\BibitemShut {NoStop}%
\bibitem [{\citenamefont {Kirkley}\ and\ \citenamefont
  {Newman}(2022)}]{kirkley_representative_2022}%
  \BibitemOpen
  \bibfield  {author} {\bibinfo {author} {\bibfnamefont {Alec}\ \bibnamefont
  {Kirkley}}\ and\ \bibinfo {author} {\bibfnamefont {M.~E.~J.}\ \bibnamefont
  {Newman}},\ }\bibfield  {title} {{\selectlanguage {english}\enquote {\bibinfo
  {title} {Representative community divisions of networks},}\ }}\href {\doibase
  10.1038/s42005-022-00816-3} {\bibfield  {journal} {\bibinfo  {journal}
  {Communications Physics}\ }\textbf {\bibinfo {volume} {5}},\ \bibinfo {pages}
  {1--10} (\bibinfo {year} {2022})},\ \bibinfo {note} {number: 1 Publisher:
  Nature Publishing Group}\BibitemShut {NoStop}%
\bibitem [{\citenamefont {Foster}\ \emph {et~al.}(2011)\citenamefont {Foster},
  \citenamefont {Foster}, \citenamefont {Grassberger},\ and\ \citenamefont
  {Paczuski}}]{foster_clustering_2011}%
  \BibitemOpen
  \bibfield  {author} {\bibinfo {author} {\bibfnamefont {David~V.}\
  \bibnamefont {Foster}}, \bibinfo {author} {\bibfnamefont {Jacob~G.}\
  \bibnamefont {Foster}}, \bibinfo {author} {\bibfnamefont {Peter}\
  \bibnamefont {Grassberger}}, \ and\ \bibinfo {author} {\bibfnamefont {Maya}\
  \bibnamefont {Paczuski}},\ }\bibfield  {title} {\enquote {\bibinfo {title}
  {Clustering drives assortativity and community structure in ensembles of
  networks},}\ }\href {\doibase 10.1103/PhysRevE.84.066117} {\bibfield
  {journal} {\bibinfo  {journal} {Physical Review E}\ }\textbf {\bibinfo
  {volume} {84}},\ \bibinfo {pages} {066117} (\bibinfo {year}
  {2011})}\BibitemShut {NoStop}%
\bibitem [{\citenamefont {Lancichinetti}\ and\ \citenamefont
  {Fortunato}(2011)}]{lancichinetti_limits_2011}%
  \BibitemOpen
  \bibfield  {author} {\bibinfo {author} {\bibfnamefont {Andrea}\ \bibnamefont
  {Lancichinetti}}\ and\ \bibinfo {author} {\bibfnamefont {Santo}\ \bibnamefont
  {Fortunato}},\ }\bibfield  {title} {\enquote {\bibinfo {title} {Limits of
  modularity maximization in community detection},}\ }\href {\doibase
  10.1103/PhysRevE.84.066122} {\bibfield  {journal} {\bibinfo  {journal}
  {Physical Review E}\ }\textbf {\bibinfo {volume} {84}},\ \bibinfo {pages}
  {066122} (\bibinfo {year} {2011})}\BibitemShut {NoStop}%
\bibitem [{\citenamefont {Granell}\ \emph {et~al.}(2012)\citenamefont
  {Granell}, \citenamefont {Gómez},\ and\ \citenamefont
  {Arenas}}]{granell_hierarchical_2012}%
  \BibitemOpen
  \bibfield  {author} {\bibinfo {author} {\bibfnamefont {Clara}\ \bibnamefont
  {Granell}}, \bibinfo {author} {\bibfnamefont {Sergio}\ \bibnamefont
  {Gómez}}, \ and\ \bibinfo {author} {\bibfnamefont {Alex}\ \bibnamefont
  {Arenas}},\ }\bibfield  {title} {\enquote {\bibinfo {title} {Hierarchical
  multiresolution method to overcome the resolution limit in complex
  networks},}\ }\href {\doibase 10.1142/S0218127412501714} {\bibfield
  {journal} {\bibinfo  {journal} {International Journal of Bifurcation and
  Chaos}\ }\textbf {\bibinfo {volume} {22}},\ \bibinfo {pages} {1250171}
  (\bibinfo {year} {2012})}\BibitemShut {NoStop}%
\bibitem [{\citenamefont {Kawamoto}\ and\ \citenamefont
  {Rosvall}(2015)}]{kawamoto_estimating_2015}%
  \BibitemOpen
  \bibfield  {author} {\bibinfo {author} {\bibfnamefont {Tatsuro}\ \bibnamefont
  {Kawamoto}}\ and\ \bibinfo {author} {\bibfnamefont {Martin}\ \bibnamefont
  {Rosvall}},\ }\bibfield  {title} {\enquote {\bibinfo {title} {Estimating the
  resolution limit of the map equation in community detection},}\ }\href
  {\doibase 10.1103/PhysRevE.91.012809} {\bibfield  {journal} {\bibinfo
  {journal} {Physical Review E}\ }\textbf {\bibinfo {volume} {91}},\ \bibinfo
  {pages} {012809} (\bibinfo {year} {2015})}\BibitemShut {NoStop}%
\bibitem [{\citenamefont {Peixoto}(2013)}]{peixoto_parsimonious_2013}%
  \BibitemOpen
  \bibfield  {author} {\bibinfo {author} {\bibfnamefont {Tiago~P.}\
  \bibnamefont {Peixoto}},\ }\bibfield  {title} {\enquote {\bibinfo {title}
  {Parsimonious {Module} {Inference} in {Large} {Networks}},}\ }\href {\doibase
  10.1103/PhysRevLett.110.148701} {\bibfield  {journal} {\bibinfo  {journal}
  {Physical Review Letters}\ }\textbf {\bibinfo {volume} {110}},\ \bibinfo
  {pages} {148701} (\bibinfo {year} {2013})}\BibitemShut {NoStop}%
\bibitem [{\citenamefont {Barber}(2007)}]{barber_modularity_2007}%
  \BibitemOpen
  \bibfield  {author} {\bibinfo {author} {\bibfnamefont {Michael~J}\
  \bibnamefont {Barber}},\ }\bibfield  {title} {\enquote {\bibinfo {title}
  {Modularity and community detection in bipartite networks},}\ }\href
  {http://arxiv.org/abs/0707.1616} {\bibfield  {journal} {\bibinfo  {journal}
  {0707.1616}\ } (\bibinfo {year} {2007})}\BibitemShut {NoStop}%
\bibitem [{\citenamefont {MacMahon}\ and\ \citenamefont
  {Garlaschelli}(2015)}]{macmahon_community_2015}%
  \BibitemOpen
  \bibfield  {author} {\bibinfo {author} {\bibfnamefont {Mel}\ \bibnamefont
  {MacMahon}}\ and\ \bibinfo {author} {\bibfnamefont {Diego}\ \bibnamefont
  {Garlaschelli}},\ }\bibfield  {title} {\enquote {\bibinfo {title} {Community
  {Detection} for {Correlation} {Matrices}},}\ }\href {\doibase
  10.1103/PhysRevX.5.021006} {\bibfield  {journal} {\bibinfo  {journal}
  {Physical Review X}\ }\textbf {\bibinfo {volume} {5}},\ \bibinfo {pages}
  {021006} (\bibinfo {year} {2015})}\BibitemShut {NoStop}%
\bibitem [{\citenamefont {Traag}\ and\ \citenamefont
  {Bruggeman}(2009)}]{traag_community_2009}%
  \BibitemOpen
  \bibfield  {author} {\bibinfo {author} {\bibfnamefont {V.~A.}\ \bibnamefont
  {Traag}}\ and\ \bibinfo {author} {\bibfnamefont {Jeroen}\ \bibnamefont
  {Bruggeman}},\ }\bibfield  {title} {\enquote {\bibinfo {title} {Community
  detection in networks with positive and negative links},}\ }\href {\doibase
  10.1103/PhysRevE.80.036115} {\bibfield  {journal} {\bibinfo  {journal}
  {Physical Review E}\ }\textbf {\bibinfo {volume} {80}},\ \bibinfo {pages}
  {036115} (\bibinfo {year} {2009})}\BibitemShut {NoStop}%
\bibitem [{\citenamefont {Expert}\ \emph {et~al.}(2011)\citenamefont {Expert},
  \citenamefont {Evans}, \citenamefont {Blondel},\ and\ \citenamefont
  {Lambiotte}}]{expert_uncovering_2011}%
  \BibitemOpen
  \bibfield  {author} {\bibinfo {author} {\bibfnamefont {Paul}\ \bibnamefont
  {Expert}}, \bibinfo {author} {\bibfnamefont {Tim~S.}\ \bibnamefont {Evans}},
  \bibinfo {author} {\bibfnamefont {Vincent~D.}\ \bibnamefont {Blondel}}, \
  and\ \bibinfo {author} {\bibfnamefont {Renaud}\ \bibnamefont {Lambiotte}},\
  }\bibfield  {title} {{\selectlanguage {english}\enquote {\bibinfo {title}
  {Uncovering space-independent communities in spatial networks},}\ }}\href
  {\doibase 10.1073/pnas.1018962108} {\bibfield  {journal} {\bibinfo  {journal}
  {Proceedings of the National Academy of Sciences}\ }\textbf {\bibinfo
  {volume} {108}},\ \bibinfo {pages} {7663--7668} (\bibinfo {year} {2011})},\
  \bibinfo {note} {publisher: National Academy of Sciences Section: Physical
  Sciences}\BibitemShut {NoStop}%
\bibitem [{\citenamefont {Hric}\ \emph {et~al.}(2016)\citenamefont {Hric},
  \citenamefont {Peixoto},\ and\ \citenamefont
  {Fortunato}}]{hric_network_2016}%
  \BibitemOpen
  \bibfield  {author} {\bibinfo {author} {\bibfnamefont {Darko}\ \bibnamefont
  {Hric}}, \bibinfo {author} {\bibfnamefont {Tiago~P.}\ \bibnamefont
  {Peixoto}}, \ and\ \bibinfo {author} {\bibfnamefont {Santo}\ \bibnamefont
  {Fortunato}},\ }\bibfield  {title} {\enquote {\bibinfo {title} {Network
  {Structure}, {Metadata}, and the {Prediction} of {Missing} {Nodes} and
  {Annotations}},}\ }\href {\doibase 10.1103/PhysRevX.6.031038} {\bibfield
  {journal} {\bibinfo  {journal} {Physical Review X}\ }\textbf {\bibinfo
  {volume} {6}},\ \bibinfo {pages} {031038} (\bibinfo {year}
  {2016})}\BibitemShut {NoStop}%
\bibitem [{\citenamefont {Newman}\ and\ \citenamefont
  {Clauset}(2016)}]{newman_structure_2016}%
  \BibitemOpen
  \bibfield  {author} {\bibinfo {author} {\bibfnamefont {M.~E.~J.}\
  \bibnamefont {Newman}}\ and\ \bibinfo {author} {\bibfnamefont {Aaron}\
  \bibnamefont {Clauset}},\ }\bibfield  {title} {{\selectlanguage
  {english}\enquote {\bibinfo {title} {Structure and inference in annotated
  networks},}\ }}\href {\doibase 10.1038/ncomms11863} {\bibfield  {journal}
  {\bibinfo  {journal} {Nature Communications}\ }\textbf {\bibinfo {volume}
  {7}},\ \bibinfo {pages} {11863} (\bibinfo {year} {2016})}\BibitemShut
  {NoStop}%
\bibitem [{\citenamefont {Peel}\ \emph {et~al.}(2017)\citenamefont {Peel},
  \citenamefont {Larremore},\ and\ \citenamefont {Clauset}}]{peel_ground_2017}%
  \BibitemOpen
  \bibfield  {author} {\bibinfo {author} {\bibfnamefont {Leto}\ \bibnamefont
  {Peel}}, \bibinfo {author} {\bibfnamefont {Daniel~B.}\ \bibnamefont
  {Larremore}}, \ and\ \bibinfo {author} {\bibfnamefont {Aaron}\ \bibnamefont
  {Clauset}},\ }\bibfield  {title} {{\selectlanguage {english}\enquote
  {\bibinfo {title} {The ground truth about metadata and community detection in
  networks},}\ }}\href {\doibase 10.1126/sciadv.1602548} {\bibfield  {journal}
  {\bibinfo  {journal} {Science Advances}\ }\textbf {\bibinfo {volume} {3}},\
  \bibinfo {pages} {e1602548} (\bibinfo {year} {2017})}\BibitemShut {NoStop}%
\bibitem [{\citenamefont {Hu}(2005)}]{hu_efficient_2005}%
  \BibitemOpen
  \bibfield  {author} {\bibinfo {author} {\bibfnamefont {Y.}~\bibnamefont
  {Hu}},\ }\bibfield  {title} {\enquote {\bibinfo {title} {Efficient,
  high-quality force-directed graph drawing},}\ }\href@noop {} {\bibfield
  {journal} {\bibinfo  {journal} {Mathematica Journal}\ }\textbf {\bibinfo
  {volume} {10}},\ \bibinfo {pages} {37--71} (\bibinfo {year}
  {2005})}\BibitemShut {NoStop}%
\bibitem [{\citenamefont {Noack}(2009)}]{noack_modularity_2009}%
  \BibitemOpen
  \bibfield  {author} {\bibinfo {author} {\bibfnamefont {Andreas}\ \bibnamefont
  {Noack}},\ }\bibfield  {title} {\enquote {\bibinfo {title} {Modularity
  clustering is force-directed layout},}\ }\href {\doibase
  10.1103/PhysRevE.79.026102} {\bibfield  {journal} {\bibinfo  {journal}
  {Physical Review E}\ }\textbf {\bibinfo {volume} {79}},\ \bibinfo {pages}
  {026102} (\bibinfo {year} {2009})}\BibitemShut {NoStop}%
\bibitem [{\citenamefont {Wolpert}\ and\ \citenamefont
  {Macready}(1995)}]{wolpert_no_1995}%
  \BibitemOpen
  \bibfield  {author} {\bibinfo {author} {\bibfnamefont {David~H.}\
  \bibnamefont {Wolpert}}\ and\ \bibinfo {author} {\bibfnamefont {William~G.}\
  \bibnamefont {Macready}},\ }\href@noop {} {\emph {\bibinfo {title} {No free
  lunch theorems for search}}},\ \bibinfo {type} {Tech. Rep.}\ (\bibinfo
  {institution} {Technical Report SFI-TR-95-02-010, Santa Fe Institute},\
  \bibinfo {year} {1995})\BibitemShut {NoStop}%
\bibitem [{\citenamefont {Wolpert}(1996)}]{wolpert_lack_1996}%
  \BibitemOpen
  \bibfield  {author} {\bibinfo {author} {\bibfnamefont {David~H.}\
  \bibnamefont {Wolpert}},\ }\bibfield  {title} {\enquote {\bibinfo {title}
  {The {Lack} of {A} {Priori} {Distinctions} {Between} {Learning}
  {Algorithms}},}\ }\href {\doibase 10.1162/neco.1996.8.7.1341} {\bibfield
  {journal} {\bibinfo  {journal} {Neural Computation}\ }\textbf {\bibinfo
  {volume} {8}},\ \bibinfo {pages} {1341--1390} (\bibinfo {year}
  {1996})}\BibitemShut {NoStop}%
\bibitem [{\citenamefont {Wolpert}\ and\ \citenamefont
  {Macready}(1997)}]{wolpert_no_1997}%
  \BibitemOpen
  \bibfield  {author} {\bibinfo {author} {\bibfnamefont {David~H.}\
  \bibnamefont {Wolpert}}\ and\ \bibinfo {author} {\bibfnamefont {William~G.}\
  \bibnamefont {Macready}},\ }\bibfield  {title} {\enquote {\bibinfo {title}
  {No free lunch theorems for optimization},}\ }\href@noop {} {\bibfield
  {journal} {\bibinfo  {journal} {IEEE transactions on evolutionary
  computation}\ }\textbf {\bibinfo {volume} {1}},\ \bibinfo {pages} {67--82}
  (\bibinfo {year} {1997})}\BibitemShut {NoStop}%
\bibitem [{\citenamefont {Schaffer}(1994)}]{schaffer_conservation_1994}%
  \BibitemOpen
  \bibfield  {author} {\bibinfo {author} {\bibfnamefont {Cullen}\ \bibnamefont
  {Schaffer}},\ }\bibfield  {title} {{\selectlanguage {english}\enquote
  {\bibinfo {title} {A {Conservation} {Law} for {Generalization}
  {Performance}},}\ }}in\ \href {\doibase 10.1016/B978-1-55860-335-6.50039-8}
  {{\selectlanguage {english}\emph {\bibinfo {booktitle} {Machine {Learning}
  {Proceedings} 1994}}}},\ \bibinfo {editor} {edited by\ \bibinfo {editor}
  {\bibfnamefont {William~W.}\ \bibnamefont {Cohen}}\ and\ \bibinfo {editor}
  {\bibfnamefont {Haym}\ \bibnamefont {Hirsh}}}\ (\bibinfo  {publisher} {Morgan
  Kaufmann},\ \bibinfo {address} {San Francisco (CA)},\ \bibinfo {year}
  {1994})\ pp.\ \bibinfo {pages} {259--265}\BibitemShut {NoStop}%
\bibitem [{\citenamefont {Streeter}(2003)}]{streeter_two_2003}%
  \BibitemOpen
  \bibfield  {author} {\bibinfo {author} {\bibfnamefont {Matthew~J.}\
  \bibnamefont {Streeter}},\ }\bibfield  {title} {{\selectlanguage
  {english}\enquote {\bibinfo {title} {Two {Broad} {Classes} of {Functions} for
  {Which} a {No} {Free} {Lunch} {Result} {Does} {Not} {Hold}},}\ }}in\ \href
  {\doibase 10.1007/3-540-45110-2_15} {{\selectlanguage {english}\emph
  {\bibinfo {booktitle} {Genetic and {Evolutionary} {Computation} — {GECCO}
  2003}}}},\ \bibinfo {series and number} {Lecture {Notes} in {Computer}
  {Science}},\ \bibinfo {editor} {edited by\ \bibinfo {editor} {\bibfnamefont
  {Erick}\ \bibnamefont {Cantú-Paz}}, \bibinfo {editor} {\bibfnamefont
  {James~A.}\ \bibnamefont {Foster}}, \bibinfo {editor} {\bibfnamefont
  {Kalyanmoy}\ \bibnamefont {Deb}}, \bibinfo {editor} {\bibfnamefont
  {Lawrence~David}\ \bibnamefont {Davis}}, \bibinfo {editor} {\bibfnamefont
  {Rajkumar}\ \bibnamefont {Roy}}, \bibinfo {editor} {\bibfnamefont {Una-May}\
  \bibnamefont {O’Reilly}}, \bibinfo {editor} {\bibfnamefont {Hans-Georg}\
  \bibnamefont {Beyer}}, \bibinfo {editor} {\bibfnamefont {Russell}\
  \bibnamefont {Standish}}, \bibinfo {editor} {\bibfnamefont {Graham}\
  \bibnamefont {Kendall}}, \bibinfo {editor} {\bibfnamefont {Stewart}\
  \bibnamefont {Wilson}}, \bibinfo {editor} {\bibfnamefont {Mark}\ \bibnamefont
  {Harman}}, \bibinfo {editor} {\bibfnamefont {Joachim}\ \bibnamefont
  {Wegener}}, \bibinfo {editor} {\bibfnamefont {Dipankar}\ \bibnamefont
  {Dasgupta}}, \bibinfo {editor} {\bibfnamefont {Mitch~A.}\ \bibnamefont
  {Potter}}, \bibinfo {editor} {\bibfnamefont {Alan~C.}\ \bibnamefont
  {Schultz}}, \bibinfo {editor} {\bibfnamefont {Kathryn~A.}\ \bibnamefont
  {Dowsland}}, \bibinfo {editor} {\bibfnamefont {Natasha}\ \bibnamefont
  {Jonoska}}, \ and\ \bibinfo {editor} {\bibfnamefont {Julian}\ \bibnamefont
  {Miller}}}\ (\bibinfo  {publisher} {Springer},\ \bibinfo {address} {Berlin,
  Heidelberg},\ \bibinfo {year} {2003})\ pp.\ \bibinfo {pages}
  {1418--1430}\BibitemShut {NoStop}%
\bibitem [{\citenamefont {McGregor}(2006)}]{mcgregor_no_2006}%
  \BibitemOpen
  \bibfield  {author} {\bibinfo {author} {\bibfnamefont {Simon}\ \bibnamefont
  {McGregor}},\ }\bibfield  {title} {\enquote {\bibinfo {title} {No free lunch
  and algorithmic randomness},}\ }in\ \href@noop {} {\emph {\bibinfo
  {booktitle} {{GECCO}}}},\ Vol.~\bibinfo {volume} {6}\ (\bibinfo {year}
  {2006})\ pp.\ \bibinfo {pages} {2--4}\BibitemShut {NoStop}%
\bibitem [{\citenamefont {Everitt}(2013)}]{everitt_universal_2013}%
  \BibitemOpen
  \bibfield  {author} {\bibinfo {author} {\bibfnamefont {Tom}\ \bibnamefont
  {Everitt}},\ }\href@noop {} {\enquote {\bibinfo {title} {Universal induction
  and optimisation: {No} free lunch?}}\ } (\bibinfo {year} {2013})\BibitemShut
  {NoStop}%
\bibitem [{\citenamefont {Lattimore}\ and\ \citenamefont
  {Hutter}(2013)}]{lattimore_no_2013}%
  \BibitemOpen
  \bibfield  {author} {\bibinfo {author} {\bibfnamefont {Tor}\ \bibnamefont
  {Lattimore}}\ and\ \bibinfo {author} {\bibfnamefont {Marcus}\ \bibnamefont
  {Hutter}},\ }\bibfield  {title} {{\selectlanguage {english}\enquote {\bibinfo
  {title} {No {Free} {Lunch} versus {Occam}’s {Razor} in {Supervised}
  {Learning}},}\ }}in\ \href {\doibase 10.1007/978-3-642-44958-1_17}
  {{\selectlanguage {english}\emph {\bibinfo {booktitle} {Algorithmic
  {Probability} and {Friends}. {Bayesian} {Prediction} and {Artificial}
  {Intelligence}: {Papers} from the {Ray} {Solomonoff} 85th {Memorial}
  {Conference}, {Melbourne}, {VIC}, {Australia}, {November} 30 – {December}
  2, 2011}}}},\ \bibinfo {series and number} {Lecture {Notes} in {Computer}
  {Science}},\ \bibinfo {editor} {edited by\ \bibinfo {editor} {\bibfnamefont
  {David~L.}\ \bibnamefont {Dowe}}}\ (\bibinfo  {publisher} {Springer},\
  \bibinfo {address} {Berlin, Heidelberg},\ \bibinfo {year} {2013})\ pp.\
  \bibinfo {pages} {223--235}\BibitemShut {NoStop}%
\bibitem [{\citenamefont {Schurz}(2019)}]{schurz_humes_2019}%
  \BibitemOpen
  \bibfield  {author} {\bibinfo {author} {\bibfnamefont {Gerhard}\ \bibnamefont
  {Schurz}},\ }\href@noop {} {{\selectlanguage {english}\emph {\bibinfo {title}
  {Hume's {Problem} {Solved}: {The} {Optimality} of {Meta}-{Induction}}}}},\
  \bibinfo {edition} {illustrated edition}\ ed.\ (\bibinfo  {publisher} {The
  MIT Press},\ \bibinfo {address} {Cambridge, Massachusetts},\ \bibinfo {year}
  {2019})\BibitemShut {NoStop}%
\bibitem [{\citenamefont {Jaynes}(2003)}]{jaynes_probability_2003}%
  \BibitemOpen
  \bibfield  {author} {\bibinfo {author} {\bibfnamefont {E.~T.}\ \bibnamefont
  {Jaynes}},\ }\href@noop {} {{\selectlanguage {english}\emph {\bibinfo {title}
  {Probability {Theory}: {The} {Logic} of {Science}}}}},\ edited by\ \bibinfo
  {editor} {\bibfnamefont {G.~Larry}\ \bibnamefont {Bretthorst}}\ (\bibinfo
  {publisher} {Cambridge University Press},\ \bibinfo {address} {Cambridge, UK
  ; New York, NY},\ \bibinfo {year} {2003})\BibitemShut {NoStop}%
\bibitem [{\citenamefont {Solomonoff}(1964)}]{solomonoff_formal_1964}%
  \BibitemOpen
  \bibfield  {author} {\bibinfo {author} {\bibfnamefont {R.~J.}\ \bibnamefont
  {Solomonoff}},\ }\bibfield  {title} {{\selectlanguage {english}\enquote
  {\bibinfo {title} {A formal theory of inductive inference. {Part} {I}},}\
  }}\href {\doibase 10.1016/S0019-9958(64)90223-2} {\bibfield  {journal}
  {\bibinfo  {journal} {Information and Control}\ }\textbf {\bibinfo {volume}
  {7}},\ \bibinfo {pages} {1--22} (\bibinfo {year} {1964})}\BibitemShut
  {NoStop}%
\bibitem [{\citenamefont {Hutter}(2007)}]{hutter_universal_2007}%
  \BibitemOpen
  \bibfield  {author} {\bibinfo {author} {\bibfnamefont {Marcus}\ \bibnamefont
  {Hutter}},\ }\bibfield  {title} {{\selectlanguage {english}\enquote {\bibinfo
  {title} {On universal prediction and {Bayesian} confirmation},}\ }}\href
  {\doibase 10.1016/j.tcs.2007.05.016} {\bibfield  {journal} {\bibinfo
  {journal} {Theoretical Computer Science}\ }\bibinfo {series} {Theory and
  {Applications} of {Models} of {Computation}},\ \textbf {\bibinfo {volume}
  {384}},\ \bibinfo {pages} {33--48} (\bibinfo {year} {2007})}\BibitemShut
  {NoStop}%
\bibitem [{\citenamefont {Hutter}(2009)}]{hutter_open_2009}%
  \BibitemOpen
  \bibfield  {author} {\bibinfo {author} {\bibfnamefont {Marcus}\ \bibnamefont
  {Hutter}},\ }\bibfield  {title} {{\selectlanguage {english}\enquote {\bibinfo
  {title} {Open {Problems} in {Universal} {Induction} \& {Intelligence}},}\
  }}\href {\doibase 10.3390/a2030879} {\bibfield  {journal} {\bibinfo
  {journal} {Algorithms}\ }\textbf {\bibinfo {volume} {2}},\ \bibinfo {pages}
  {879--906} (\bibinfo {year} {2009})},\ \bibinfo {note} {number: 3 Publisher:
  Molecular Diversity Preservation International}\BibitemShut {NoStop}%
\bibitem [{\citenamefont {Montanez}(2017)}]{montanez_why_2017}%
  \BibitemOpen
  \bibfield  {author} {\bibinfo {author} {\bibfnamefont {George~D.}\
  \bibnamefont {Montanez}},\ }\bibfield  {title} {\enquote {\bibinfo {title}
  {Why machine learning works},}\ }\href {https://www. cs. cmu.
  edu/~gmontane/montanez_dissertation. pdf} {\  (\bibinfo {year}
  {2017})}\BibitemShut {NoStop}%
\bibitem [{\citenamefont {Vallès-Català}\ \emph {et~al.}(2018)\citenamefont
  {Vallès-Català}, \citenamefont {Peixoto}, \citenamefont {Sales-Pardo},\
  and\ \citenamefont {Guimerà}}]{valles-catala_consistencies_2018}%
  \BibitemOpen
  \bibfield  {author} {\bibinfo {author} {\bibfnamefont {Toni}\ \bibnamefont
  {Vallès-Català}}, \bibinfo {author} {\bibfnamefont {Tiago~P.}\ \bibnamefont
  {Peixoto}}, \bibinfo {author} {\bibfnamefont {Marta}\ \bibnamefont
  {Sales-Pardo}}, \ and\ \bibinfo {author} {\bibfnamefont {Roger}\ \bibnamefont
  {Guimerà}},\ }\bibfield  {title} {\enquote {\bibinfo {title} {Consistencies
  and inconsistencies between model selection and link prediction in
  networks},}\ }\href {\doibase 10.1103/PhysRevE.97.062316} {\bibfield
  {journal} {\bibinfo  {journal} {Physical Review E}\ }\textbf {\bibinfo
  {volume} {97}},\ \bibinfo {pages} {062316} (\bibinfo {year}
  {2018})}\BibitemShut {NoStop}%
\bibitem [{\citenamefont {Ghasemian}\ \emph {et~al.}(2020)\citenamefont
  {Ghasemian}, \citenamefont {Hosseinmardi}, \citenamefont {Galstyan},
  \citenamefont {Airoldi},\ and\ \citenamefont
  {Clauset}}]{ghasemian_stacking_2020}%
  \BibitemOpen
  \bibfield  {author} {\bibinfo {author} {\bibfnamefont {Amir}\ \bibnamefont
  {Ghasemian}}, \bibinfo {author} {\bibfnamefont {Homa}\ \bibnamefont
  {Hosseinmardi}}, \bibinfo {author} {\bibfnamefont {Aram}\ \bibnamefont
  {Galstyan}}, \bibinfo {author} {\bibfnamefont {Edoardo~M.}\ \bibnamefont
  {Airoldi}}, \ and\ \bibinfo {author} {\bibfnamefont {Aaron}\ \bibnamefont
  {Clauset}},\ }\bibfield  {title} {{\selectlanguage {english}\enquote
  {\bibinfo {title} {Stacking models for nearly optimal link prediction in
  complex networks},}\ }}\href {\doibase 10.1073/pnas.1914950117} {\bibfield
  {journal} {\bibinfo  {journal} {Proceedings of the National Academy of
  Sciences}\ }\textbf {\bibinfo {volume} {117}},\ \bibinfo {pages}
  {23393--23400} (\bibinfo {year} {2020})}\BibitemShut {NoStop}%
\bibitem [{\citenamefont {Olhede}\ and\ \citenamefont
  {Wolfe}(2014)}]{olhede_network_2014}%
  \BibitemOpen
  \bibfield  {author} {\bibinfo {author} {\bibfnamefont {Sofia~C.}\
  \bibnamefont {Olhede}}\ and\ \bibinfo {author} {\bibfnamefont {Patrick~J.}\
  \bibnamefont {Wolfe}},\ }\bibfield  {title} {{\selectlanguage
  {english}\enquote {\bibinfo {title} {Network histograms and universality of
  blockmodel approximation},}\ }}\href {\doibase 10.1073/pnas.1400374111}
  {\bibfield  {journal} {\bibinfo  {journal} {Proceedings of the National
  Academy of Sciences}\ }\textbf {\bibinfo {volume} {111}},\ \bibinfo {pages}
  {14722--14727} (\bibinfo {year} {2014})}\BibitemShut {NoStop}%
\bibitem [{\citenamefont {Young}\ \emph {et~al.}(2018)\citenamefont {Young},
  \citenamefont {St-Onge}, \citenamefont {Desrosiers},\ and\ \citenamefont
  {Dubé}}]{young_universality_2018}%
  \BibitemOpen
  \bibfield  {author} {\bibinfo {author} {\bibfnamefont {Jean-Gabriel}\
  \bibnamefont {Young}}, \bibinfo {author} {\bibfnamefont {Guillaume}\
  \bibnamefont {St-Onge}}, \bibinfo {author} {\bibfnamefont {Patrick}\
  \bibnamefont {Desrosiers}}, \ and\ \bibinfo {author} {\bibfnamefont
  {Louis~J.}\ \bibnamefont {Dubé}},\ }\bibfield  {title} {\enquote {\bibinfo
  {title} {Universality of the stochastic block model},}\ }\href {\doibase
  10.1103/PhysRevE.98.032309} {\bibfield  {journal} {\bibinfo  {journal}
  {Physical Review E}\ }\textbf {\bibinfo {volume} {98}},\ \bibinfo {pages}
  {032309} (\bibinfo {year} {2018})}\BibitemShut {NoStop}%
\bibitem [{\citenamefont {Hoff}\ \emph {et~al.}(2002)\citenamefont {Hoff},
  \citenamefont {Raftery},\ and\ \citenamefont {Handcock}}]{hoff_latent_2002}%
  \BibitemOpen
  \bibfield  {author} {\bibinfo {author} {\bibfnamefont {Peter~D}\ \bibnamefont
  {Hoff}}, \bibinfo {author} {\bibfnamefont {Adrian~E}\ \bibnamefont
  {Raftery}}, \ and\ \bibinfo {author} {\bibfnamefont {Mark~S}\ \bibnamefont
  {Handcock}},\ }\bibfield  {title} {\enquote {\bibinfo {title} {Latent {Space}
  {Approaches} to {Social} {Network} {Analysis}},}\ }\href {\doibase
  10.1198/016214502388618906} {\bibfield  {journal} {\bibinfo  {journal}
  {Journal of the American Statistical Association}\ }\textbf {\bibinfo
  {volume} {97}},\ \bibinfo {pages} {1090--1098} (\bibinfo {year}
  {2002})}\BibitemShut {NoStop}%
\bibitem [{\citenamefont {Ziv}\ and\ \citenamefont
  {Lempel}(1977)}]{ziv_universal_1977}%
  \BibitemOpen
  \bibfield  {author} {\bibinfo {author} {\bibfnamefont {J.}~\bibnamefont
  {Ziv}}\ and\ \bibinfo {author} {\bibfnamefont {A.}~\bibnamefont {Lempel}},\
  }\bibfield  {title} {\enquote {\bibinfo {title} {A universal algorithm for
  sequential data compression},}\ }\href {\doibase 10.1109/TIT.1977.1055714}
  {\bibfield  {journal} {\bibinfo  {journal} {IEEE Transactions on Information
  Theory}\ }\textbf {\bibinfo {volume} {23}},\ \bibinfo {pages} {337--343}
  (\bibinfo {year} {1977})}\BibitemShut {NoStop}%
\bibitem [{\citenamefont {Gelman}\ \emph {et~al.}(2020)\citenamefont {Gelman},
  \citenamefont {Vehtari}, \citenamefont {Simpson}, \citenamefont {Margossian},
  \citenamefont {Carpenter}, \citenamefont {Yao}, \citenamefont {Kennedy},
  \citenamefont {Gabry}, \citenamefont {Bürkner},\ and\ \citenamefont
  {Modrák}}]{gelman_bayesian_2020}%
  \BibitemOpen
  \bibfield  {author} {\bibinfo {author} {\bibfnamefont {Andrew}\ \bibnamefont
  {Gelman}}, \bibinfo {author} {\bibfnamefont {Aki}\ \bibnamefont {Vehtari}},
  \bibinfo {author} {\bibfnamefont {Daniel}\ \bibnamefont {Simpson}}, \bibinfo
  {author} {\bibfnamefont {Charles~C.}\ \bibnamefont {Margossian}}, \bibinfo
  {author} {\bibfnamefont {Bob}\ \bibnamefont {Carpenter}}, \bibinfo {author}
  {\bibfnamefont {Yuling}\ \bibnamefont {Yao}}, \bibinfo {author}
  {\bibfnamefont {Lauren}\ \bibnamefont {Kennedy}}, \bibinfo {author}
  {\bibfnamefont {Jonah}\ \bibnamefont {Gabry}}, \bibinfo {author}
  {\bibfnamefont {Paul-Christian}\ \bibnamefont {Bürkner}}, \ and\ \bibinfo
  {author} {\bibfnamefont {Martin}\ \bibnamefont {Modrák}},\ }\href {\doibase
  10.48550/arXiv.2011.01808} {\enquote {\bibinfo {title} {Bayesian
  {Workflow}},}\ } (\bibinfo {year} {2020}),\ \bibinfo {note} {number:
  arXiv:2011.01808 arXiv:2011.01808 [stat]}\BibitemShut {NoStop}%
\bibitem [{\citenamefont {Blondel}\ \emph {et~al.}(2008)\citenamefont
  {Blondel}, \citenamefont {Guillaume}, \citenamefont {Lambiotte},\ and\
  \citenamefont {Lefebvre}}]{blondel_fast_2008}%
  \BibitemOpen
  \bibfield  {author} {\bibinfo {author} {\bibfnamefont {Vincent~D.}\
  \bibnamefont {Blondel}}, \bibinfo {author} {\bibfnamefont {Jean-Loup}\
  \bibnamefont {Guillaume}}, \bibinfo {author} {\bibfnamefont {Renaud}\
  \bibnamefont {Lambiotte}}, \ and\ \bibinfo {author} {\bibfnamefont {Etienne}\
  \bibnamefont {Lefebvre}},\ }\bibfield  {title} {{\selectlanguage
  {english}\enquote {\bibinfo {title} {Fast unfolding of communities in large
  networks},}\ }}\href {\doibase 10.1088/1742-5468/2008/10/P10008} {\bibfield
  {journal} {\bibinfo  {journal} {Journal of Statistical Mechanics: Theory and
  Experiment}\ }\textbf {\bibinfo {volume} {2008}},\ \bibinfo {pages} {P10008}
  (\bibinfo {year} {2008})}\BibitemShut {NoStop}%
\bibitem [{\citenamefont {Traag}\ \emph {et~al.}(2019)\citenamefont {Traag},
  \citenamefont {Waltman},\ and\ \citenamefont {van Eck}}]{traag_louvain_2019}%
  \BibitemOpen
  \bibfield  {author} {\bibinfo {author} {\bibfnamefont {V.~A.}\ \bibnamefont
  {Traag}}, \bibinfo {author} {\bibfnamefont {L.}~\bibnamefont {Waltman}}, \
  and\ \bibinfo {author} {\bibfnamefont {N.~J.}\ \bibnamefont {van Eck}},\
  }\bibfield  {title} {{\selectlanguage {english}\enquote {\bibinfo {title}
  {From {Louvain} to {Leiden}: guaranteeing well-connected communities},}\
  }}\href {\doibase 10.1038/s41598-019-41695-z} {\bibfield  {journal} {\bibinfo
   {journal} {Scientific Reports}\ }\textbf {\bibinfo {volume} {9}},\ \bibinfo
  {pages} {5233} (\bibinfo {year} {2019})}\BibitemShut {NoStop}%
\bibitem [{\citenamefont
  {Peixoto}(2014{\natexlab{c}})}]{peixoto_efficient_2014}%
  \BibitemOpen
  \bibfield  {author} {\bibinfo {author} {\bibfnamefont {Tiago~P.}\
  \bibnamefont {Peixoto}},\ }\bibfield  {title} {\enquote {\bibinfo {title}
  {Efficient {Monte} {Carlo} and greedy heuristic for the inference of
  stochastic block models},}\ }\href {\doibase 10.1103/PhysRevE.89.012804}
  {\bibfield  {journal} {\bibinfo  {journal} {Physical Review E}\ }\textbf
  {\bibinfo {volume} {89}},\ \bibinfo {pages} {012804} (\bibinfo {year}
  {2014}{\natexlab{c}})}\BibitemShut {NoStop}%
\bibitem [{\citenamefont
  {Peixoto}(2020{\natexlab{b}})}]{peixoto_merge-split_2020}%
  \BibitemOpen
  \bibfield  {author} {\bibinfo {author} {\bibfnamefont {Tiago~P.}\
  \bibnamefont {Peixoto}},\ }\bibfield  {title} {\enquote {\bibinfo {title}
  {Merge-split {Markov} chain {Monte} {Carlo} for community detection},}\
  }\href {\doibase 10.1103/PhysRevE.102.012305} {\bibfield  {journal} {\bibinfo
   {journal} {Physical Review E}\ }\textbf {\bibinfo {volume} {102}},\ \bibinfo
  {pages} {012305} (\bibinfo {year} {2020}{\natexlab{b}})}\BibitemShut
  {NoStop}%
\bibitem [{\citenamefont {Krzakala}\ \emph {et~al.}(2013)\citenamefont
  {Krzakala}, \citenamefont {Moore}, \citenamefont {Mossel}, \citenamefont
  {Neeman}, \citenamefont {Sly}, \citenamefont {Zdeborová},\ and\
  \citenamefont {Zhang}}]{krzakala_spectral_2013}%
  \BibitemOpen
  \bibfield  {author} {\bibinfo {author} {\bibfnamefont {Florent}\ \bibnamefont
  {Krzakala}}, \bibinfo {author} {\bibfnamefont {Cristopher}\ \bibnamefont
  {Moore}}, \bibinfo {author} {\bibfnamefont {Elchanan}\ \bibnamefont
  {Mossel}}, \bibinfo {author} {\bibfnamefont {Joe}\ \bibnamefont {Neeman}},
  \bibinfo {author} {\bibfnamefont {Allan}\ \bibnamefont {Sly}}, \bibinfo
  {author} {\bibfnamefont {Lenka}\ \bibnamefont {Zdeborová}}, \ and\ \bibinfo
  {author} {\bibfnamefont {Pan}\ \bibnamefont {Zhang}},\ }\bibfield  {title}
  {{\selectlanguage {english}\enquote {\bibinfo {title} {Spectral redemption in
  clustering sparse networks},}\ }}\href {\doibase 10.1073/pnas.1312486110}
  {\bibfield  {journal} {\bibinfo  {journal} {Proceedings of the National
  Academy of Sciences}\ ,\ \bibinfo {pages} {201312486}} (\bibinfo {year}
  {2013})}\BibitemShut {NoStop}%
\bibitem [{\citenamefont {Kawamoto}(2018)}]{kawamoto_algorithmic_2018}%
  \BibitemOpen
  \bibfield  {author} {\bibinfo {author} {\bibfnamefont {Tatsuro}\ \bibnamefont
  {Kawamoto}},\ }\bibfield  {title} {\enquote {\bibinfo {title} {Algorithmic
  detectability threshold of the stochastic block model},}\ }\href {\doibase
  10.1103/PhysRevE.97.032301} {\bibfield  {journal} {\bibinfo  {journal}
  {Physical Review E}\ }\textbf {\bibinfo {volume} {97}},\ \bibinfo {pages}
  {032301} (\bibinfo {year} {2018})}\BibitemShut {NoStop}%
\bibitem [{\citenamefont {Spielman}\ and\ \citenamefont
  {Teng}(2007)}]{spielman_spectral_2007}%
  \BibitemOpen
  \bibfield  {author} {\bibinfo {author} {\bibfnamefont {Daniel~A.}\
  \bibnamefont {Spielman}}\ and\ \bibinfo {author} {\bibfnamefont {Shang-Hua}\
  \bibnamefont {Teng}},\ }\bibfield  {title} {\enquote {\bibinfo {title}
  {Spectral partitioning works: planar graphs and finite element meshes},}\
  }\href {\doibase 10.1016/j.laa.2006.07.020} {\bibfield  {journal} {\bibinfo
  {journal} {Linear Algebra and its Applications}\ }\textbf {\bibinfo {volume}
  {421}},\ \bibinfo {pages} {284--305} (\bibinfo {year} {2007})}\BibitemShut
  {NoStop}%
\bibitem [{\citenamefont {von Luxburg}(2007)}]{von_luxburg_tutorial_2007}%
  \BibitemOpen
  \bibfield  {author} {\bibinfo {author} {\bibfnamefont {Ulrike}\ \bibnamefont
  {von Luxburg}},\ }\bibfield  {title} {\enquote {\bibinfo {title} {A tutorial
  on spectral clustering},}\ }\href {\doibase 10.1007/s11222-007-9033-z}
  {\bibfield  {journal} {\bibinfo  {journal} {Statistics and Computing}\
  }\textbf {\bibinfo {volume} {17}},\ \bibinfo {pages} {395--416} (\bibinfo
  {year} {2007})}\BibitemShut {NoStop}%
\bibitem [{\citenamefont {Rohe}(2011)}]{rohe_spectral_2011}%
  \BibitemOpen
  \bibfield  {author} {\bibinfo {author} {\bibfnamefont {Karl}\ \bibnamefont
  {Rohe}},\ }\bibfield  {title} {{\selectlanguage {english}\enquote {\bibinfo
  {title} {Spectral clustering and the high-dimensional stochastic
  blockmodel},}\ }}\href {\doibase 10.1214/11-AOS887} {\bibfield  {journal}
  {\bibinfo  {journal} {The Annals of Statistics}\ }\textbf {\bibinfo {volume}
  {39}},\ \bibinfo {pages} {1878--1915} (\bibinfo {year} {2011})}\BibitemShut
  {NoStop}%
\bibitem [{\citenamefont {Lehoucq}\ and\ \citenamefont
  {Sorensen}(1996)}]{lehoucq_deflation_1996}%
  \BibitemOpen
  \bibfield  {author} {\bibinfo {author} {\bibfnamefont {R.~B.}\ \bibnamefont
  {Lehoucq}}\ and\ \bibinfo {author} {\bibfnamefont {D.~C.}\ \bibnamefont
  {Sorensen}},\ }\bibfield  {title} {\enquote {\bibinfo {title} {Deflation
  {Techniques} for an {Implicitly} {Restarted} {Arnoldi} {Iteration}},}\ }\href
  {\doibase 10.1137/S0895479895281484} {\bibfield  {journal} {\bibinfo
  {journal} {SIAM Journal on Matrix Analysis and Applications}\ }\textbf
  {\bibinfo {volume} {17}},\ \bibinfo {pages} {789--821} (\bibinfo {year}
  {1996})}\BibitemShut {NoStop}%
\bibitem [{\citenamefont {Fire}\ \emph {et~al.}(2013)\citenamefont {Fire},
  \citenamefont {Puzis},\ and\ \citenamefont {Elovici}}]{fire_link_2013}%
  \BibitemOpen
  \bibfield  {author} {\bibinfo {author} {\bibfnamefont {Michael}\ \bibnamefont
  {Fire}}, \bibinfo {author} {\bibfnamefont {Rami}\ \bibnamefont {Puzis}}, \
  and\ \bibinfo {author} {\bibfnamefont {Yuval}\ \bibnamefont {Elovici}},\
  }\bibfield  {title} {{\selectlanguage {english}\enquote {\bibinfo {title}
  {Link {Prediction} in {Highly} {Fractional} {Data} {Sets}},}\ }}in\ \href
  {\doibase 10.1007/978-1-4614-5311-6_14} {{\selectlanguage {english}\emph
  {\bibinfo {booktitle} {Handbook of {Computational} {Approaches} to
  {Counterterrorism}}}}},\ \bibinfo {editor} {edited by\ \bibinfo {editor}
  {\bibfnamefont {V.S.}\ \bibnamefont {Subrahmanian}}}\ (\bibinfo  {publisher}
  {Springer},\ \bibinfo {address} {New York, NY},\ \bibinfo {year} {2013})\
  pp.\ \bibinfo {pages} {283--300}\BibitemShut {NoStop}%
\bibitem [{\citenamefont {Lehoucq}\ \emph {et~al.}(1998)\citenamefont
  {Lehoucq}, \citenamefont {Sorensen},\ and\ \citenamefont
  {Yang}}]{lehoucq_ARPACK_1998}%
  \BibitemOpen
  \bibfield  {author} {\bibinfo {author} {\bibfnamefont {Richard~B.}\
  \bibnamefont {Lehoucq}}, \bibinfo {author} {\bibfnamefont {Danny~C.}\
  \bibnamefont {Sorensen}}, \ and\ \bibinfo {author} {\bibfnamefont {Chao}\
  \bibnamefont {Yang}},\ }\href@noop {} {\emph {\bibinfo {title} {{ARPACK}
  users' guide: solution of large-scale eigenvalue problems with implicitly
  restarted {Arnoldi} methods}}}\ (\bibinfo  {publisher} {SIAM},\ \bibinfo
  {year} {1998})\BibitemShut {NoStop}%
\bibitem [{\citenamefont {Schwarz}(1978)}]{schwarz_estimating_1978}%
  \BibitemOpen
  \bibfield  {author} {\bibinfo {author} {\bibfnamefont {Gideon}\ \bibnamefont
  {Schwarz}},\ }\bibfield  {title} {{\selectlanguage {english}\enquote
  {\bibinfo {title} {Estimating the {Dimension} of a {Model}},}\ }}\href
  {\doibase 10.1214/aos/1176344136} {\bibfield  {journal} {\bibinfo  {journal}
  {The Annals of Statistics}\ }\textbf {\bibinfo {volume} {6}},\ \bibinfo
  {pages} {461--464} (\bibinfo {year} {1978})}\BibitemShut {NoStop}%
\bibitem [{\citenamefont {Akaike}(1974)}]{akaike_new_1974}%
  \BibitemOpen
  \bibfield  {author} {\bibinfo {author} {\bibfnamefont {H.}~\bibnamefont
  {Akaike}},\ }\bibfield  {title} {\enquote {\bibinfo {title} {A new look at
  the statistical model identification},}\ }\href {\doibase
  10.1109/TAC.1974.1100705} {\bibfield  {journal} {\bibinfo  {journal} {IEEE
  Transactions on Automatic Control}\ }\textbf {\bibinfo {volume} {19}},\
  \bibinfo {pages} {716--723} (\bibinfo {year} {1974})}\BibitemShut {NoStop}%
\bibitem [{\citenamefont {Côme}\ and\ \citenamefont
  {Latouche}(2015)}]{come_model_2015}%
  \BibitemOpen
  \bibfield  {author} {\bibinfo {author} {\bibfnamefont {Etienne}\ \bibnamefont
  {Côme}}\ and\ \bibinfo {author} {\bibfnamefont {Pierre}\ \bibnamefont
  {Latouche}},\ }\bibfield  {title} {{\selectlanguage {english}\enquote
  {\bibinfo {title} {Model selection and clustering in stochastic block models
  based on the exact integrated complete data likelihood},}\ }}\href {\doibase
  10.1177/1471082X15577017} {\bibfield  {journal} {\bibinfo  {journal}
  {Statistical Modelling}\ }\textbf {\bibinfo {volume} {15}},\ \bibinfo {pages}
  {564--589} (\bibinfo {year} {2015})}\BibitemShut {NoStop}%
\bibitem [{\citenamefont {Newman}\ and\ \citenamefont
  {Reinert}(2016)}]{newman_estimating_2016}%
  \BibitemOpen
  \bibfield  {author} {\bibinfo {author} {\bibfnamefont {M.~E.~J.}\
  \bibnamefont {Newman}}\ and\ \bibinfo {author} {\bibfnamefont {Gesine}\
  \bibnamefont {Reinert}},\ }\bibfield  {title} {\enquote {\bibinfo {title}
  {Estimating the {Number} of {Communities} in a {Network}},}\ }\href {\doibase
  10.1103/PhysRevLett.117.078301} {\bibfield  {journal} {\bibinfo  {journal}
  {Physical Review Letters}\ }\textbf {\bibinfo {volume} {117}},\ \bibinfo
  {pages} {078301} (\bibinfo {year} {2016})}\BibitemShut {NoStop}%
\bibitem [{\citenamefont {Burnham}\ and\ \citenamefont
  {Anderson}(2002)}]{burnham_model_2002}%
  \BibitemOpen
  \bibinfo {editor} {\bibfnamefont {Kenneth~P.}\ \bibnamefont {Burnham}}\ and\
  \bibinfo {editor} {\bibfnamefont {David~R.}\ \bibnamefont {Anderson}},\
  eds.,\ \href {\doibase 10.1007/978-0-387-22456-5_1} {{\selectlanguage
  {english}\emph {\bibinfo {title} {Model {Selection} and {Multimodel}
  {Inference}: {A} {Practical} {Information}-{Theoretic} {Approach}}}}}\
  (\bibinfo  {publisher} {Springer},\ \bibinfo {address} {New York, NY},\
  \bibinfo {year} {2002})\BibitemShut {NoStop}%
\end{thebibliography}%
